%% file: SIV_EC25_main.tex
\documentclass[format=acmsmall, review=false]{acmart}

\usepackage{acm-ec-25}
\usepackage{booktabs} 
\usepackage[ruled]{algorithm2e} 

\SetAlFnt{\small}
\SetAlCapFnt{\small}
\SetAlCapNameFnt{\small}
\SetAlCapHSkip{0pt}
\IncMargin{-\parindent}

\usepackage{url}

\input{macros}

\setcopyright{none}
\acmConference{}{}{}
\acmYear{}
\acmDOI{}
\acmISBN{}
\acmVolume{0}
\acmNumber{0}
\acmArticle{0}
\acmMonth{0}

\renewcommand\footnotetextcopyrightpermission[1]{}
\renewcommand{\footnotetextauthorsaddresses}[1]{} 

\setcitestyle{acmnumeric}

\title[Average-Case Analysis of Iterative Voting]{Average-Case Analysis of Iterative Voting}

\author{Joshua Kavner}
\affiliation{
  \institution{Rensselaer Polytechnic Institute}
  \city{Troy}
  \state{New York}
  \country{USA}
}
\email{joshuakavner@gmail.com}

\author{Lirong Xia}
\affiliation{
  \institution{Rutgers University}
  \city{New Brunswick}
  \state{New Jersey}
  \country{USA}
}
\email{xialirong@gmail.com}

\begin{document}


\begin{abstract}
Iterative voting is a natural model of repeated strategic decision-making in social choice theory when agents have the opportunity to update their votes prior to finalizing the group decision. Prior work has analyzed the efficacy of iterative plurality on the welfare of the chosen outcome at equilibrium, relative to the truthful vote profile, via an adaptation of the price of anarchy. However, prior analyses have only studied the worst- and average-case performances when agents' preferences are distributed by the impartial culture. This work extends average-case analysis comprehensively across three alternatives and distinguishes under which of agents' preference distributions iterative plurality improves or degrades asymptotic welfare.
\end{abstract}

\maketitle
\renewcommand{\shortauthors}{}

\vspace{1cm}
\setcounter{tocdepth}{2} 
\tableofcontents





\input{EC_body/my_intro}

\input{EC_body/my_prelims}

\input{EC_body/average_case_discussion}

\input{EC_body/theorem_and_proof}
\input{EC_body/conclusion}

%

%
%
%
%

\bibliographystyle{ACM-Reference-Format}
\bibliography{josh_refs_cleaner}

\newpage

\appendix
\section*{Appendix}

\input{EC_appendix/apx_page_0}
\input{EC_appendix/apx_page_1}

\input{EC_appendix/apx_page_2a}

\input{EC_appendix/apx_page_2b}

\input{EC_appendix/apx_page_2d}

\input{EC_appendix/apx_page_2c}

\input{EC_appendix/apx_page_3a}

\input{EC_appendix/apx_page_3c}

\input{EC_appendix/apx_page_3b}

\input{EC_appendix/apx_page_4}

\input{EC_appendix/apx_page_5a}

\input{EC_appendix/apx_page_5b}

\input{EC_appendix/apx_page_5c}

\input{EC_appendix/apx_page_6}

\input{EC_appendix/apx_page_7}

\input{EC_appendix/apx_page_8}

\end{document}

%% file: macros.tex
\usepackage{balance}
\usepackage{graphicx}
\usepackage{amsmath,amsthm,amsfonts}
\usepackage{xcolor}
\usepackage{url}
\usepackage{natbib}
\usepackage{bbm}

\usepackage{apxproof}
\usepackage{makecell}
\usepackage{multirow}

\newtheorem{thm}{Theorem}
\newtheorem{dfn}{Definition}
\newtheorem{coro}{Corollary}
\newtheorem{lem}{Lemma}
\newtheorem{ex}{Example}
\newtheorem*{remark}{Remark}
\newtheorem{prop}{Proposition}
\newtheorem{claim}{Claim}

\newcommand{\calL}{\mathcal L}
\newcommand{\calA}{\mathcal A}
\newcommand{\calO}{\mathcal O}

\newcommand{\calE}{\mathcal E}

\newcommand{\calH}{\mathcal H}
\newcommand{\calT}{\mathcal T}

\newcommand{\ml}{\mathcal L}

\newcommand{\score}{\text{Score}}

\DeclareMathOperator*{\argmin}{arg\,min}
\DeclareMathOperator*{\argmax}{arg\,max}

\newcommand{\ED}{{\text{EADPoA}}}

\newcommand{\EW}[1]{\text{EW}(#1)}
\newcommand{\SW}[3]{\text{SW}_{#1}(#2, #3)}

\newcommand{\PW}[1]{\text{PW}(#1)}
\newcommand{\AD}{\text{ADPoA}}
\newcommand{\ADS}{\text{D}^+}

\newcommand{\topRank}[1]{\textit{top}(#1)}

\newcommand{\EDS}{\overline{\text{PoA}}}

\newcommand{\bfloor}[1]{\left\lfloor #1 \right\rfloor}
\newcommand{\bceil}[1]{\left\lceil #1 \right\rceil}


%% file: EC_body/my_intro.tex
\section{Introduction}

It is well-known in social choice theory that people may misreport their preferences to improve group decisions in their favor. Consider, for example, Alice, Bob, and Charlie deciding which ice cream flavor to order for a party, and Charlie prefers strawberry to chocolate to vanilla. Given that Alice wants chocolate and Bob wants vanilla, Charlie would be better off voting for chocolate than truthfully (i.e., strawberry), by which vanilla wins as the tie-breaker. This form of strategic behavior is prolific in political science in narrowing the number of political parties (see e.g., Duverger's law \citep{riker1982two}). Still, it is unclear what effect strategic behavior has on the social welfare of chosen outcomes.

Iterative voting (IV) is one model which naturally describes agents' strategic behavior -- in misreporting their truthful preferences -- over time. After agents reveal their truthful preferences initially, they have the opportunity to repeatedly update their votes given information about other agents' votes, before the final decision is reached. \citet{Meir10:Convergence} first proposed IV with plurality and identified many sufficient conditions for IV to converge. This was followed up by a series of work examining various social choice rules, information and behavioral assumptions, and settings to determine when, to what outcomes, and how fast IV converges (see e.g., surveys by \citet{meir2017trends} and \citet{meir2018strategic}).

While most IV research has studied its convergence and equilibrium properties, only a few papers have analyzed its economic performance. 
The empirical literature is mixed -- simulations and lab experiments by \citet{Reijngoud2012VoterRT}, \citet{Grandi13:Restricted}, \citet{bowman14:potential}, and \citet{grandi2022voting} found that IV improves outcome quality, while \citet{Koolyk17:Convergence} and \citet{meir2020strategic} observed the contrary.
\citet{Branzei13:How} analyzed IV theoretically in light of the infamous Gibbard-Satterthwaite theorem \citep{Gibbard73:Manipulation, Satterthwaite75:Strategy}, which guarantees agents' incentive to behave strategically for reasonable voting rules. They defined the \emph{additive dynamic price of anarchy} (ADPOA) as the difference in social welfare between the truthful vote profile and the worst-case equilibrium that is reachable via IV. This notion refines the well-known \emph{price of anarchy} \citep{roughgarden2002bad} for a dynamic setting with myopic agents, with respect to the worst-case preference profile and order of agent improvement steps.
They found the performance is ``very good'' for plurality (with an ADPoA of $1$), ``not bad'' for veto (with a DPoA of $\Omega(m)$ with $m$ alternatives, $m \geq 4$), and ``very bad'' for Borda (with a DPoA of $\Omega(n)$ with $n$ agents). 

Notably, \citet{Branzei13:How}'s theorems assumed that the positional scoring voting rule had the same scoring vector as agents' additive utilities. \citet{Kavner21:strategic} relaxed this assumption to arbitrary utility vectors with respect to iterative plurality. They found the additive DPoA worsened to $\Theta(n)$ in the worst-case. 
While this result bounds the theoretical consequences of IV, it provides little insight into how IV may perform realistically. Upon realizing this poor result, \citet{Kavner21:strategic} took a first step in testing IV's practicality by exploring its average-case performance.
By assuming that agents' preferences are distributed identically and independently, uniformly at random, known as the \emph{impartial culture} (IC), they found the expected additive DPoA to be $-\Omega(1)$. This suggests that IV actually \emph{improves} social welfare over the truthful vote profile on average.

Average-case analysis is traditionally employed in computer science as a way around the intractability of NP-hard problems. This analysis is motivated by the possibility that worst-case results only occur infrequently in practice \citep{bogdanov2006average}. As seen with IV, average-case analysis hopes to provide a less pessimistic measure of an algorithm's performance. Still, the distribution used in the analysis may itself be unrealistic \citep{spielman2009smoothed}. Indeed, IC used by \citet{Kavner21:strategic} is widely understood to be implausible  \citep{Tsetlin2003:The-impartial,regenwetter2006behavioral,van2014empirical}, yet useful perhaps as a benchmark against other analytical results in social choice. This presents an opportunity to advance our understanding of iterative plurality voting beyond IC.

\subsection{Our Contribution}
\label{sec:our_contributions}

We address the limitations of IC by analyzing the expected performance of IV, the \emph{expected additive dynamic price of anarchy} ($\ED$), comprehensively across all strictly positive input preference distributions $\pi$ among $m=3$ alternatives. For any preference profile $P$, we measure the difference in additive social welfare between the equilibrium outcome, due to IV by the worst case scheduler, relative to the truthful winning alternative. Our primary result is a map between independent and identically distributed (i.i.d.) preferences and the asymptotic rate of IV's $\ED$ for large populations of agents $n$.
We describe the extent to which welfare changes, informally, as follows.

\begin{thm}[Informal]
For each strictly positive preference distribution $\pi$ that is i.i.d. across $n$ agents, the expected performance of iterative plurality is either $-\Theta(\sqrt{n})$, $-\Theta(1)$, $\Theta(1)$, or $\Theta(\sqrt{n})$, or it goes to zero at a rate of $\calO \left( \frac{1}{\sqrt{n}} \right)$, $\calO \left( \frac{1}{n} \right)$, or $\calO \left( e^{-\Theta(n)} \right)$.
\end{thm}
\setcounter{thm}{0} 

Negative-valued performance signifies that equilibrium outcomes have \emph{higher} welfare than their corresponding truthful winners. Theorem \ref{thm:main} thus enables us to precisely detail which distributions increase or decrease welfare, due to IV, and to what extent. Surprisingly, for significantly large classes of preference distributions, the $\ED$ is bounded. This suggests that the effect of strategic behavior in social choice among large populations is mild. We may interpret this as follows: given that there is an initial tie between several alternatives, and agents have the power to change the electoral outcome \citep{downs1957economic}, the expected social welfare between the front-runners will be minimal. Still, the different cases of this theorem may explain some variability found across empirical experiments.

Theorem \ref{thm:main} refines both the result of \citet{Kavner21:strategic}, who attained a $-\Omega(1)$ bound on $\ED$ under IC, as well as their techniques. Namely, applying their method directly to our problem would supply a $\calO \left( \frac{1}{\sqrt{n}} \right)$ probability (Corollary \ref{coro:sub_others}, below) against an $\calO(n)$ worst-case value, suggesting only that the $\ED$ is bounded between $-\calO(\sqrt{n})$ and $\calO(\sqrt{n})$. Clearly, this is insufficiently refined. The primary component that goes into the $\ED$ calculation is the case of a two-way  plurality tie (e.g., with alternatives $1$ and $2$). \citet{Kavner21:strategic}'s proof does not distinguish the case when IV transitions the truthful winning alternative of $1$ to an equilibrium winner of $2$ from the other way around, unless the values are exactly the same. When the distribution is not uniformly random, these two cases have a $-\calO(\sqrt{n})$ and $\calO(\sqrt{n})$ finding, respectively. 

In this work, we devise a method for combining these two cases and attain a significantly more nuanced result (detailed in Lemma \ref{lem:sub_12_TLDR}, below). At a high level, our method involves computing the expected value directly using the law of total expectation: we group together similar preference profiles $P$, assessing their likelihood and IV performance value, and determine the asymptotic rate of each group separately. The challenge is in (i) identifying the correct groups of profiles that technically simplify the algebra and enable us to condense like terms (as exemplified in the lemmas of Appendices \ref{apx:two_ties} and \ref{apx:three_ties}), and (ii) using specific techniques for solving each group (as exemplified in the lemmas of Appendices \ref{apx:secondary_equations} -- \ref{apx:helper_lemmas}).
These techniques include \citet{Xia2021:How-Likely}'s smoothed likelihood of ties, local central limit theorems \citep{Petrov+1975}, the Wallis product approximation for the central binomial coefficient \citep{wallis1656,galvin18wallis}, and over a dozen real analysis and binomial lemmas.

More specifically, our analysis makes significant use of the \emph{PMV-in-Polyhedron} theorem from \citet{Xia2021:How-Likely} to characterize the asymptotic likelihood of tied elections. We capture the likelihood that the histogram of a preference profile, which is a Poisson multivariate variable, fits into a polyhedron that specifies either a tied election, by itself (Corollary \ref{coro:sub_others}), and with additional constraints (Lemma \ref{lem:prob_bounds} in Appendix \ref{apx:collision_entropy}). \citet{Xia2021:How-Likely}'s techniques are not directly applicable in our setting because they characterize the likelihood of events occurring, whereas we study the expected value of a function of the histogram of random preference profiles. 
Rather, we devise novel applications of their theorems in our present work.

After discussing related work, Section \ref{sec:prelims} presents iterative voting preliminaries, Section \ref{sec:characterization} presents our main result, and  Section \ref{sec:limitations} concludes. We describe the mapping from agents' preference distributions $\pi$ to asymptotic rates of IV's $\ED$ and provide some detail about how we attained such results. At a high level, the expected performance is partitioned into groups of preference profiles $P$ such that there are two-way plurality ties and three-way plurality ties. 
The key lemmas are presented in Appendix \ref{apx:two_ties}, for two-way ties, and Appendix \ref{apx:three_ties}, for three-way ties. A complete summary of the distribution-to-rate map is presented in Appendix \ref{apx:appendix_contents}. Appendix \ref{apx:smoothed_analysis_full} presents preliminaries from real analysis and smoothed analysis \citep{Xia2020:The-Smoothed, Xia2021:How-Likely} that are used throughout the work. Appendices \ref{apx:secondary_equations} -- \ref{apx:helper_lemmas} provide technical lemmas that support the main lemmas and are grouped together by concept.

\subsection{Related Work}

The present study of IV was initiated by \citet{Meir10:Convergence} who identified that iterative plurality converges when agents sequentially apply best-response updates. This inspired a line of research on sufficient conditions for convergence. For example, \citet{Lev12:Convergence} and \citet{Reyhani12:Best} simultaneously found that iterative veto converges while no other positional scoring rule does. \citet{gourves2016strategic} and \citet{Koolyk17:Convergence} followed up with similar negative results for other voting rules, such as Copeland and STV. In lieu of these negative results, \citet{Grandi13:Restricted}, \citet{Obraztsova15:Convergence}, and \citet{Rabinovich15:Analysis} proved IV's convergence upon imposing stricter assumptions on agents' behavior, such as truth-bias \citep{Thompson13:Empirical, Obraztsova13:Plurality} and voting with abstentions \citep{Desmedt10:Equilibria, elkind2015equilibria}. 

\citet{Reijngoud2012VoterRT} and \citet{Endriss16:Strategic} took a different approach by relaxing assumptions 
about what information agents have access to.
Rather than performing best-response updates, agents make \emph{local dominance improvement} steps that may improve the outcome but cannot degrade the outcome, given their current information \citep{Conitzer11:Dominating}.
\citet{Meir14:Local} and \citet{Meir15:Plurality} characterized convergence of iterative plurality with such local dominance improvements. \citet{kavner2023convergence} extended their model to settings where multiple issues are decided on simultaneously, similar to experiments by \citet{bowman14:potential} and \citet{grandi2022voting}. Relatedly, \citet{sina2015adapting} and \citet{Tsang16:Echo} studied IV with agents embedded in social networks, while \citet{terzopoulou2024iterative} considered partial preferences.

While most IV research focuses on convergence and equilibrium properties, 
\citet{Branzei13:How} quantified the quality of IV via the worst-case DPoA.
\citet{Kavner21:strategic} extended their results for iterative plurality with respect to any additive utility vector and demonstrated an improvement in average social welfare, despite poor worst-case performance. Meanwhile, other synthetic and human subjects experiments have proved inconclusive about the effects IV has on social welfare \citep{Thompson13:Empirical, bowman14:potential, Koolyk17:Convergence, meir2020strategic, grandi2022voting}. 
Other empirical work includes \citet{boudou2022itero}, who developed a user-friendly platform for testing IV in practice, and \citet{baltz2022computer}, who addressed conceptual gaps between computational voting models and real-world elections.
Our present work provides a more comprehensive analysis of IV's economic performance by extending the domain of agents' input preference distributions.
Other forms of sequential and IV include models by \citet{airiau09:iterated}, \citet{Desmedt10:Equilibria}, and \citet{Xia10:Stackelberg}.


Separately, \citet{spielman2004smoothed} introduced \emph{smoothed analysis} as a combination of worst- and average-case analyses to address the issue that instance distributions themselves may not be realistic. Their idea was to measure an algorithm's performance with respect to a worst-case instance subject to a random perturbation. Hence, even if an algorithm has exponential worst-case performance, it may be unlikely to encounter such an instance in practice.  This perspective has since been applied toward a large body of problems (see e.g., surveys by \citet{spielman2009smoothed} and \citet{roughgarden2021beyond}). 
For example, 
\citet{deng17:Smoothed}, \citet{gao19:average}, and 
\citet{deng2022beyond} studied the smoothed performance of the random priority mechanism in matching problems. 
Extensions into social choice were independently proposed by \citet{baumeister2020towards} and \citet{Xia2020:The-Smoothed}. The latter inspired a series of research extending prior results in social choice theory through this lens (e.g., \citet{xia2021semi}, \citet{xia2021smoothed}, \citet{liu2022semi}, \citet{xia2022beyond}, \citet{xia2023semi} and references within). \citet{flanigan2023smoothed} refined Xia's model to provide more standardized proofs about whether common social choice axioms are satisfied with high probability as $n$ increases. Importantly, \citet{Xia2021:How-Likely} studied the smoothed likelihood of ties in elections, which contributes meaningfully toward our primary results. We describe in our conclusions how our contributions may be framed within this perspective.

%% file: EC_body/my_prelims.tex
\section{Preliminaries}
\label{sec:prelims}

\begin{paragraph}{Basic setting.}
Let $\calA = [m] = \{1, \ldots, m\}$ denote the set of $m \geq 3$ \emph{alternatives} and $n \in \mathbb{N}$ denote the number of agents. Unless stated otherwise, we assume that $m=3$ throughout this work. Each agent $j \leq n$ is endowed with a preference \emph{ranking} $R_j \in \calL(\calA)$, the set of strict linear orders over $\calA$. A \emph{preference profile} is denoted $P = (R_1, \ldots, R_n)$. For any pair of alternatives, $c, c' \in \calA$, we use $P[c \succ c']$ to denote the number of agents that prefer $c$ to $c'$ in $P$.

Agents vote by reporting a single alternative $a_j \in \calA$ into the \emph{vote profile} $a = (a_1, \ldots, a_n)$.We use the \emph{plurality} rule defined as
$f(a) = \argmax_{c \in \calA} s_c(a)$, with lexicographical tie-breaking,
where $s_c(a) = |\{j \leq n : a_j = c\}|$ is alternative $c$'s \emph{score}.
A vote $\topRank{R_j}$
is \emph{truthful} if it is agent $j$'s most-favored alternative. We denote the truthful vote profile as $\topRank{P}$.
%
\end{paragraph}

\begin{paragraph}{Rank-based additive utility.}
We take agents with additive utilities characterized by a \emph{rank-based utility vector} $\vec{u} = (u_1, \ldots, u_m) \in \mathbb{R}^m_{\geq 0}$ with $u_1 \geq \ldots u_m$ and $u_1 > u_m$. For example, plurality welfare has $\vec{u} = (1,0, \ldots, 0)$ while Borda welfare has $\vec{u} = (m-1, m-2, \ldots, 0)$. Each agent $j$ gets $\vec{u}(R_j, c) = u_i$ utility for the alternative $c \in \calA$ ranked $i^{th}$ in $R_j$. The additive \emph{social welfare} of $c$ according to preference profile $P$ is $\SW{\vec{u}}{P}{c} = \sum_{j=1}^n \vec{u}(R_j,c)$.
\end{paragraph}

\begin{paragraph}{Iterative plurality voting.}
Given a preference profile $P$, we initialize the vote profile $a(0) = \topRank{P}$ as truthful. We then consider an iterative process of vote profiles $a(t) = (a_1(t), \ldots, a_n(t))$ that describe agents' reported votes over time $t \geq 0$. For each round $t$, a \emph{scheduler} chooses an agent $j$ to make a myopic improvement step over their prior vote 
\citep{Apt12:Classification}. All other votes remain unchanged. 
%
%
%
%
Under \emph{direct best response (BR)} dynamics, $j$ updates their vote to the unique alternative that (i) yields the most-preferred outcome under $f$ with respect to $R_j$, and (ii) will become the winner as a result. 
Specifically, we denote the set of \emph{potential winning} alternatives as those who could become a winner if their plurality score were to increment by one, including the current winner:
\begin{equation*}
\begin{split}
    & \PW{a} = \{f(a)\} \cup \Big\{ c \in \calA~:~
    \begin{cases} s_c(a) = s_{f(a)}(a)-1, & c \text{ is ordered before } f(a) \\ s_{c}(a) = s_{f(a)}(a), & c \text{ is ordered after } f(a) \end{cases} \Big\} 
\end{split}
\end{equation*}
where the ordering is lexicographical for tie-breaking. We call these alternatives \emph{approximately-tied}. BR dynamics from the truthful profile stipulate that agents
change their vote from a non-winner to their favorite alternative in $\PW{a}$ \citep{Branzei13:How}.
%
%
\citet{Reyhani12:Best} proved that $\forall t\geq 0$, $\PW{a(t+1)} \subseteq \PW{a(t)}$, so every BR sequence converges to a \emph{Nash equilibrium} (NE) in $\calO(nm)$ rounds.
We denote the set of \emph{equilibrium winning} alternatives as those corresponding to any NE reachable from $a$ via some BR sequence:
\begin{equation*}
\begin{split}
    \EW{a} = & \{f(\tilde{a})~:~\exists \text{ a BR sequence from } a 
    \text{ leading to the NE profile } \tilde{a}\}.
\end{split}
\end{equation*}
%
\end{paragraph}

\subsection{Dynamic Price of Anarchy}

The performance of IV is commonly measured by a worst-case comparison in social welfare between the truthful vote profile and the equilibrium that are reachable via the dynamics. This captures the impact that IV has against the outcome that would take place without agents' strategic manipulation of their votes. Moreover, it does not assume that the order agents make their improvement steps is controlled; the measure is over the worst-case scheduler. In the following definitions, we consider this performance measure according to the worst- and average-case preference profiles.

\begin{dfn}[Additive Dynamic Price of Anarchy (ADPoA) \citep{Branzei13:How}] 
Given $n \in \mathbb{N}$, a utility vector $\vec{u}$, and a preference profile $P$, the \emph{adversarial loss} starting from the truthful vote profile $\topRank{P}$ is
\[
\ADS_{\vec{u}}(P) = \SW{\vec{u}}{P}{f(\topRank{P})} - \min\nolimits_{c \in \EW{\topRank{P}}} \SW{\vec{u}}{P}{c}.
\]
The \emph{additive dynamic price of anarchy (ADPoA)} is
\[
\AD_{\vec{u}} = \max\nolimits_{P \in \calL(\calA)^n} \ADS_{\vec{u}}(P).
\]    
\label{def:adpoa}
\end{dfn}
\citet{Branzei13:How} proved that the ADPoA of plurality is $1$ when $\vec{u} = (1, 0, \ldots, 0)$ is the plurality welfare. \citet{Kavner21:strategic} proved the ADPoA is $\Theta(n)$ when $\vec{u}$ is otherwise. 
Upon realizing this negative result, they then studied the average-case adversarial loss. Rather than assuming a single-input profile $P$ in the adversarial loss, they considered distributions over agents' preferences that were
identical and independently distributed (i.i.d.).

\begin{dfn}[Expected Additive DPoA (EADPoA) \citep{Kavner21:strategic}]
Given $n \in \mathbb{N}$, a utility vector $\vec{u}$, and a distribution over agents' preferences $\vec{\pi} \in \Delta(\calL(\calA))^n$, the \emph{expected additive dynamic price of anarchy (EADPoA)} is
\[
    \ED_{\vec{u}}(\vec{\pi}) = \mathbb{E}_{P \sim \vec{\pi}} \left[ \ADS_{\vec{u}}(P) \right].
\]
\label{def:eadpoa}
\end{dfn}
In particular, \citet{Kavner21:strategic} focused on the \emph{impartial culture} (IC) where preference rankings are 
i.i.d. uniformly over $\mathcal{L}(\mathcal{A})$. They found the $\ED$ to be $-\Omega(1)$, suggesting that IV improves social welfare on average, even if it degrades welfare in the worst-case.
%
%

In what follows, we denote rankings $R_{xyz} = (x \succ y \succ z)$ and corresponding probabilities by $\pi_{xyz}$.
We denote agents' joint i.i.d. preferences by $P \sim \pi^n = (\pi, \pi, \ldots, \pi)$. A distribution $\pi \in \Delta(\calL(\calA))$ is called \emph{strictly positive} if $\pi_j > 0,~\forall j \in [m!]$.

%% file: EC_body/average_case_discussion.tex
\section{Characterization of Average-Case Iterative Voting}
\label{sec:characterization}

Our main result extends the $\ED$ beyond \citet{Kavner21:strategic}'s study of IC toward general classes of single-agent preference distributions. With IC, each agent has an equal probability of voting for each alternative, truthfully, and equal likelihood of preferring $c \succ c'$ or $c' \succ c$ for any $c,c' \in \calA$. It was realized that these two concepts led $P \sim IC$ to be concentrated around profiles $P$ that yielded a negative adversarial loss $\ADS(P)$, leading to an $\ED = -\Omega(1)$ conclusion.

In this work, we find $\ED$ to be significantly sensitive to the preference distribution $\pi$. For example, consider any $\pi$ such that $\pi_{123} = \pi_{231} > 2\pi_{321} = 2\pi_{312} > 0$ and $\pi_{132} = \pi_{213} = 0$. 
This distribution is designed to have equal probability for agents preferring alternatives $1$ and $2$ most and for preferring either $1 \succ 2$ or $2 \succ 1$. This maximizes the likelihood of a $\{1,2\}$-tie and ensures that the likelihood of any other-way tie (i.e., $\PW{\topRank{P}} = W \subseteq 2^\calA \backslash \{1,2\}$, $|W| \geq 2$) is exponentially small (Corollary \ref{coro:sub_others}). With a $\{1,2\}$-tie, IV will then be characterized by the \emph{third-party} agents, those with rankings $R_{321}$ and $R_{312}$, alternatively switching their votes for alternatives $1$ and $2$ until convergence 
\citep[Lemma 1]{Kavner21:strategic}. 
This entail that
each agent with ranking $R_{123}$ adds $u_1 - u_2$ to $\ADS(P)$ while each agent with ranking $R_{231}$ subtracts $u_1 - u_3$ from $\ADS(P)$. Hence, we must keep track of how many agents have each of these rankings in our analysis (i.e., $\frac{n}{2}-q$ in Lemma \ref{lem:sub_12_TLDR}). With IC (i.e., $\pi_{123} = \ldots = \pi_{321}$), the average contribution that agents with ranking $R_{123}$ make to $\ADS(P)$ cancel out with those with ranking $R_{132}$; likewise, the contributions that agents with rankings $R_{231}$ and $R_{213}$ cancel out \citep[Equation 3]{Kavner21:strategic}. This distinction significantly complicates our proof over \citet{Kavner21:strategic}'s and yields different asymptotic values for different values of $\pi_{123}$ (Lemma \ref{lem:sub_12_TLDR}).

As demonstrated by this example, there are two conceptual tiers that describe the dependency of $\ED$ on $\pi$. First, we must consider which alternatives have the highest likelihood of agents truthfully voting for them. We denote this set by $W^*(\pi) \subseteq \calA$ below. In this example, alternatives $W^*(\pi) = \{1,2\}$ had the highest probability. When $P \sim \pi^n$, this suggests that the truthful vote profiles $\topRank{P}$ will be asymptotically concentrated around those with two-way ties $\PW{\topRank{P}} = \{1,2\}$. Since the equilibrium winning set $\EW{\topRank{P}}$ is a subset of $\PW{\topRank{P}}$, we can mathematically describe how IV dynamics will unfold using \citep[Lemma 1]{Kavner21:strategic}. A second conceptual tier depicts the subtle dependence of $\ED$ on $\pi$ within the class of $\{\pi \in \Delta(\calL(\calA))~:~W^*(\pi) 
= \{1,2\}\}$, as exemplified above, but this will be characteristically different if $W^*(\pi)$ is $\{1,3\}$, $\{2,3\}$, or $\{1,2,3\}$.

Our main result is a map between $\pi$ and the asymptotic rate of $\ED_{\vec u}(\pi^n)$ for a given utility vector $\vec{u}$ and certain subsequences of $(n)_{n \in \mathbb{N}}$. Due to the complex nature of this map, we only explicitly state the span of possible rates on $\ED$ as a function of $\vec{u}$ in the statement of Theorem \ref{thm:main}. This complication is because the aggregate map is an additive composite of four separate maps between $\pi$ and certain asymptotic rates, which depict the character of $W^*(\pi) = \{1,2\}$, $\{1,3\}$, $\{2,3\}$, and $\{1,2,3\}$, respectively. Our method for deriving these four maps is described comprehensively in the theorem proof, while the complete map is summarized in Appendix \ref{apx:concise_summary}.
The primary techniques used to derive the four component maps involve \citet{Xia2021:How-Likely}'s smoothed likelihood of ties, local central limit theorems \citep{Petrov+1975}, the Wallis product approximation for the central binomial coefficient \citep{wallis1656,galvin18wallis}, and over a dozen real analysis and binomial lemmas.

%
%



\begin{remark}
Let $f, g, h$ be real-valued functions of $n$. We denote by $f(n) = \calO(g(n))$ if $\exists N > 0$ and $C \geq 0$ such that $\forall n > N$, $0 \leq f(n) \leq C g(n)$. We use this descriptor for all $n > N$ or the specific subsequences of $(n)_{n > N}$ that are even or odd. We denote by $f(n) = \pm \calO(g(n))$ if  $0 \leq |f(n)| \leq C g(n)$. We denote by $f(n) = \pm \Theta(g(n))$ if $f(n) \in \{\Theta(g(n)), - \Theta(g(n))\}$. Finally, we denote by $f(n) = \pm \calO(e^{-\Theta(n)})$ if $f(n) = \pm \calO(g(n))$ and $g(n) = e^{-h(n)}$ for some $h(n) = \Theta(n)$.
This is described further in Appendix \ref{sec:asymptotic_analysis}.
\end{remark}

%% file: EC_body/theorem_and_proof.tex
\begin{thm}
First, if $u_1 = u_2 = u_3$, then $\ED_{\vec{u}}(\pi^n) = 0$. 
Second, if $u_1 > u_2 = u_3$, then $\exists N > 0$ such that $\forall n > N$,
\[
\ED_{\vec{u}}(\pi^n) \in \left\{\pm \calO \left( \frac{1}{\sqrt{n}} \right), \pm \calO \left( \frac{1}{n} \right), \pm \calO\left(e^{-\Theta(n)}\right) \right\}
\]
depending on $\pi$.
Third, if $u_1 \geq u_2 > u_3$, $\exists N > 0$ such that $\forall n > N$,
\[
\ED_{\vec{u}}(\pi^n) \in \left\{-\Theta(\sqrt{n}), -\Theta(1), \pm \calO \left( \frac{1}{\sqrt{n}} \right), \pm \calO\left(e^{-\Theta(n)}\right), \Theta(1), \Theta(\sqrt{n}) \right\}
\]
depending on $\pi$ and the subsequence of $(n)_{n > N}$ that is even or odd.
\label{thm:main}
\end{thm}

\begin{proof}
Clearly if $u_1 = u_2 = u_3$, then $\ED_{\vec{u}}(\pi^n) = 0$ since every alternative would have the same social welfare, regardless of preference profile $P$. Henceforth, assume this is not the case.

We prove the theorem by partitioning $\calL(\calA)^n$ based on the possible potential winner sets $\PW{\topRank{P}}$ and applying the law of total expectation to sum $\ED$ across these disjoint partitions. 
Specifically, for every $W \subseteq \calA$ we define
\begin{align*}
    \EDS(W) & = \Pr\nolimits_{P \sim \pi^n}(\PW{\topRank{P}} = W)
    \times \mathbb{E}_{P \sim \pi^n}[\ADS(P)~\vert~\PW{\topRank{P}}=W]
\end{align*}
where $\topRank{P}$ is the truthful vote profile of preference profile $P$. This entails
\begin{align}
    \ED_{\vec{u}}(\pi^n) & = \sum_{c \in \calA} \EDS(\{c\})
    +  \EDS(\{1,2\}) + \EDS(\{1,3\}) + \EDS(\{2,3\}) + \EDS(\{1,2,3\}). \label{eq:main_partition} 
\end{align}
It is clear that $\sum_{c \in \calA} \EDS(\{c\}) = 0$ since any preference profile $P$ with $|\PW{\topRank{P}}| =1$ is an equilibrium, so the adversarial loss $\ADS(P) = 0$. Rather, our determination of $\EDS(W)$ for any other $W \subseteq \calA, |W| \geq 2$, depends on the preference distribution $\pi$. In particular, it depends on the likelihood of a tied plurality election among $|W| \leq m$ alternatives, as exemplified in the following corollary.
Let $\lambda_i(\pi) = \sum_{j:\topRank{R_j} = i} \pi_j$ be the likelihood of an agent truthfully voting for alternative $i$ and $W^*(\pi) = \argmax_{i \in [m]} \lambda_i(\pi)$ be a set.

\begin{coro}
Fix $m \geq 3$ and strictly positive distribution $\pi \in \Delta(\calL(\calA))$. Then
\begin{align*}
    & \Pr\nolimits_{P \sim \pi^n}(\PW{\topRank{P}} = W) 
    = \begin{cases}
    \Theta \left( n^{-\frac{|W|-1}{2}} \right), & W \subseteq W^*(\pi) \\
    \calO(e^{-\Theta(n)}),& W \nsubseteq W^*(\pi).
\end{cases}
\end{align*}
\label{coro:sub_others}
\end{coro}

Corollary \ref{coro:sub_others} generalizes \citet[Corollary 1]{Xia2021:How-Likely}, the likelihood of $k$-way plurality ties under IC, to distributions beyond IC. It follows directly from the proof of \citet[Theorem 3]{Xia2021:How-Likely}, especially Claim 4(ii) in their appendix, and is discussed further in Appendix \ref{apx:smoothed_prelims}. For example, consider $\pi' = (0.35, 0.25, 0.1, 0.1, 0.05, 0.15)$, corresponding with the rankings $(R_{123}, R_{231}, R_{321}, R_{312}, R_{132}, R_{213})$, when there are $m=3$ alternatives. Then $(\lambda_1(\pi'), \lambda_2(\pi'), \lambda_3(\pi')) = (0.4, 0.4, 0.2)$ which entails that $W^*(\pi') = \{1,2\}$. It follows that $\Pr(\PW{\topRank{P}} = \{1,2\}) = \Theta \left( \frac{1}{\sqrt{n}} \right)$ while $\Pr(\PW{\topRank{P}} = W) = \calO(e^{-\Theta(n)})$ for any other $W \subseteq \calA \backslash \{1,2\}, |W| \geq 2$. 
This observation yields the following proposition.


\begin{prop}
For any $m \geq 3$ and strictly positive distribution $\pi \in \Delta(\calL(\calA))$
such that $\{\lambda_1(\pi), \ldots, \lambda_m(\pi)\}$ has a unique maximum,
\begin{equation*}            
    \ED_{\vec{u}}(\pi^n) = \pm \calO(e^{-\Theta(n)}).
\end{equation*}
\label{prop:other_distributions}
\end{prop}

\begin{proof}
\begin{align}
    \left| \ED_{\vec{u}}(\pi^n) \right| & = \sum_{c \in \calA} \EDS(\{c\}) + \sum_{ \substack{W \subseteq \calA, |W| \geq 2 }} \left| \EDS(W) \right| \notag \\
    & = \sum_{ \substack{W \subseteq \calA, |W| \geq 2 }} \Pr ( \PW{\topRank{P}} = W)  \left| \mathbb{E}[\ADS(P)~|~\PW{\topRank{P}} = W] \right| \notag \\
    & \leq \calO(n) \sum_{ \substack{W \subseteq \calA, |W| \geq 2 }} \Pr ( \PW{\topRank{P}} = W) \label{eq:lem1v2_worstcase} \\
    & = \calO(e^{-\calO(n)}). \label{eq:lem1v2_mainstep}
\end{align} 
Equation (\ref{eq:lem1v2_worstcase}) follows from $\max_P |\ADS(P)| = \calO(n)$ since each agent contributes only a constant amount to $\ADS(P)$ \citep[Theorem 1]{Kavner21:strategic}. Equation (\ref{eq:lem1v2_mainstep}) follows from Corollary \ref{coro:sub_others} and the fact that $|2^\calA| = 2^m$ is constant for fixed $m$. 
\end{proof}




Proposition \ref{prop:other_distributions} proves that each of the four non-zero terms of Equation (\ref{eq:main_partition}) 
is exponentially small when $\pi$ has a unique maximum among $\{\lambda_1(\pi), \lambda_2(\pi), \lambda_3(\pi)\}$. There are four other high-level classes of $\pi$, depending on whether $W^*(\pi)$ is $\{1,2\}$, $\{1,3\}$, $\{2,3\}$, or $\{1,2,3\}$. These cases translate directly to each of our solutions for $\EDS(W), W \subseteq \calA, |W| \geq 2$.
For the duration of this proof, we discuss each of the four non-zero terms of Equation (\ref{eq:main_partition}) in turn. The $\ED_{\vec{u}}(\pi^n)$ is the sum of each $\EDS(W), W \subseteq \calA, |W| \geq 2$ for a given distribution $\pi$.

%
%
%
For instance, consider $\EDS(\{1,2\})$. It is clear that $\pi_{123} + \pi_{132} = \pi_{231} + \pi_{213} \geq \pi_{321} + \pi_{312}$ implies that $W^*(\pi) = \{1,2\}$ (when the inequality is strict) or $W^*(\pi) = \{1,2,3\}$ (when the equality holds). 
Then we prove that 
\[
    \EDS(\{1,2\}) \in \left\{-\Theta(\sqrt{n}), -\Theta(1), \pm \calO \left( \frac{1}{\sqrt{n}} \right), \pm \calO \left( \frac{1}{n} \right), \pm \calO\left(e^{-\Theta(n)}\right), \Theta(1), \Theta(\sqrt{n}) \right\}
\]
depending on (i) additional criteria on the distribution $\pi$, (ii) whether $u_1 \geq u_2 > u_3$ or $u_1 > u_2 = u_3$ in $\vec{u}$, and (iii) whether $n$ is even or odd.
%
%
%
Specifically, Table \ref{tab:eadpoa_case_12_tldr} demonstrates a mapping from $\pi$ to the asymptotic rate of $\EDS(\{1,2\})$ when $u_1 \geq u_2 > u_3$, for both even and odd $n$.
We read 
this table
column-wise to identify the appropriate conditions on $\pi$, and then row-wise to determine the asymptotic rate. 
For example, if both $\pi_{321} = \pi_{312}$ and $4 \pi_{123} + \pi_{231} + 3 \pi_{132} > 2$, then $\EDS(\{1,2\})$ is $\Theta(1)$ if $n$ is even and $-\Theta(1)$ if $n$ is odd.
On the other hand, if both $\pi_{321} > \pi_{312}$ and $\pi_{123} + 2 \pi_{312} < \pi_{231} + 2 \pi_{321}$, then $\EDS(\{1,2\}) = \Theta(\sqrt{n})$.
When $u_1 > u_2 = u_3$, we prove that $\EDS(\{1,2\}) = \pm \calO \left( \frac{1}{n} \right)$ if $\pi_{321} \leq \pi_{312}$ and $\pm \calO \left( e^{-\Theta(n)} \right)$ otherwise.
Otherwise, if the inequality $\pi_{123} + \pi_{132} = \pi_{231} + \pi_{213} \geq \pi_{321} + \pi_{312}$ fails to hold, then $\{1,2\} \nsubseteq W^*(\pi)$ and $\EDS(\{1,2\}) = \pm \calO(e^{-\Theta(n)})$ by Corollary \ref{coro:sub_others}.


Our claim on $\EDS(\{1,2\})$ is proved by Lemmas \ref{lem:sub_12} and \ref{lem:sub_12_odd} in Appendix \ref{apx:two_ties} about the subsequences for which $n$ is even or odd, respectively. For conciseness in the main body of this paper, we provide a simplistic proof in Lemma \ref{lem:sub_12_TLDR}. This lemma is a corollary of Lemma \ref{lem:sub_12} and holds after making some additional assumptions on $\pi$ and $\vec{u}$.


\begin{table*}[t]
    \centering
    \small
    \begin{tabular}{|c|c|}
        \hline
         \makecell{$\EDS(\{1,2\})$ when \\$\pi_{123} + \pi_{132} = \pi_{231} + \pi_{213} \geq \pi_{321} + \pi_{312}$} & Asymptotic Rate \\
          \hline \hline
        \makecell{$\begin{cases}\pi_{321} = \pi_{312} \\ 4\pi_{123} + \pi_{231} + 3 \pi_{132} > 2 \end{cases}$} & $\begin{cases}\Theta(1),& n \text{ is even} \\-\Theta(1), & n \text{ is odd}\end{cases}$\\
        \hline
        \makecell{$\begin{cases}\pi_{321} = \pi_{312} \\ 4\pi_{123} + \pi_{231} + 3 \pi_{132} < 2 \end{cases}$} & $-\Theta(1)$\\
        \hline
        \makecell{$\begin{cases}\pi_{321} = \pi_{312} \\ 4\pi_{123} + \pi_{231} + 3 \pi_{132} = 2 \end{cases}$} & $\begin{cases}\pm \calO \left( \frac{1}{\sqrt{n}} \right),& n \text{ is even} \\ - \Theta(1),& n \text{ is odd}\end{cases}$ \\
        \hline
        \makecell{$\begin{cases}\pi_{321} \neq \pi_{312} \\ \pi_{123} + 2\pi_{312} = \pi_{231} + 2 \pi_{321} \end{cases}$} & $\pm \calO \left( \frac{1}{\sqrt{n}}\right)$  \\
        \hline
        \makecell{$\begin{cases}\pi_{321} > \pi_{312} \\ \pi_{123} + 2\pi_{312} < \pi_{231} + 2 \pi_{321} \end{cases}$ or $\begin{cases}\pi_{321} < \pi_{312} \\ \pi_{123} + 2\pi_{312} > \pi_{231} + 2 \pi_{321} \end{cases}$} & $\Theta(\sqrt{n})$ \\
        \hline
        \makecell{$\begin{cases}\pi_{321} > \pi_{312} \\ \pi_{123} + 2\pi_{312} > \pi_{231} + 2 \pi_{321} \end{cases}$ or $\begin{cases}\pi_{321} < \pi_{312} \\ \pi_{123} + 2\pi_{312} < \pi_{231} + 2 \pi_{321} \end{cases}$} & $-\Theta(\sqrt{n})$ \\
        \hline
    \end{tabular}
    \caption{Asymptotic rate of $\EDS(\{1,2\})$ given conditions on $\pi$ when $u_1 \geq u_2 > u_3$.}
    \label{tab:eadpoa_case_12_tldr}
\end{table*}

\input{EC_body/main_lemma_tldr}


Notably, the techniques used to prove Lemma \ref{lem:sub_12_TLDR} are almost identical to those for every other major lemma used to prove Theorem \ref{thm:main}. Lemmas \ref{lem:sub_12} (when $n$ is even) and \ref{lem:sub_12_odd} (when $n$ is odd) are strictly more complicated due to relaxing the assumptions that $\pi_{123} = \pi_{231} > 2 \pi_{321} = 2 \pi_{312} > 0$ and $\pi_{132} = \pi_{213} = 0$ in Lemma \ref{lem:sub_12_TLDR}. They include two more index variables $e$ and $f$ to distinguish how many agents have rankings $R_{123}$ or $R_{132}$, and rankings $R_{231}$ or $R_{213}$, respectively.
The cases of $\EDS(\{1,3\})$ and $\EDS(\{2,3\})$ are 
proved in Lemmas \ref{lem:sub_13} and \ref{lem:sub_23}, respectively, in Appendix \ref{apx:two_ties_other_cases}. The proofs of these lemmas essentially detail a permutation of the preference distribution $\pi$ that may then be applied to Lemmas \ref{lem:sub_12} and \ref{lem:sub_12_odd}.


Finally, consider $\EDS(\{1,2,3\})$. 
It follows from Corollary \ref{coro:sub_others} that $\EDS(\{1,2,3\}) = \pm \calO(e^{-\Theta(n)})$ unless $\pi_{123} + \pi_{132} = \pi_{231} + \pi_{213} = \pi_{321} + \pi_{312} = \frac{1}{3}$. When this equality does hold, there are three cases for $\EDS(\{1,2,3\})$: (i) when $n$ is divisible by $3$, (ii) when $n-2$ is divisible by $3$, and (iii) when $n-1$ is divisible by $3$.
These cases correspond to the three possibilities of $\PW{\topRank{P}}$ when $f(\topRank{P}) = 1$, $3$, or $2$, respectively.
%
The first of these cases is proved by Lemma \ref{lem:three_tie_priority1}. We provide a proof sketch here for completeness, while the full proof is presented in Appendix \ref{apx:three_ties}.

\input{EC_body/three_ties_lemma_tldr}

The other two cases of $\EDS(\{1,2,3\})$ are proved very similarly in Lemmas \ref{lem:three_tie_priority3} and \ref{lem:three_tie_priority2}, respectively, in Appendix \ref{apx:three_ties}. In Appendix \ref{apx:concise_summary} we provide a concise summary of the main results that prove Theorem \ref{thm:main}. This summary includes $\EDS(\{1,2\})$ from Lemmas \ref{lem:sub_12} (when $n$ is even) and \ref{lem:sub_12_odd} (when $n$ is odd), $\EDS(\{1,3\})$ from Lemma \ref{lem:sub_13}, $\EDS(\{2,3\})$ from Lemma \ref{lem:sub_23}, and $\EDS(\{1,2,3\})$ from Lemmas \ref{lem:three_tie_priority1}, \ref{lem:three_tie_priority3} and \ref{lem:three_tie_priority2}.
This concludes the proof of Theorem \ref{thm:main}.
\end{proof}


%
%

%% file: EC_body/main_lemma_tldr.tex



\begin{lem}
Suppose that $\pi_{123} = \pi_{231} > 2\pi_{321} = 2\pi_{312} > 0$ and $\pi_{132} = \pi_{213} = 0$. Furthermore, let $u_1 \geq u_2 > u_3$ in $\vec{u}$. Then $\exists N > 0$ such that $\forall n > N$ that are even, 
\[
\EDS(\{1,2\}) = \begin{cases}
    \Theta(1),& 0.4 < \pi_{123} < \frac{1}{2} \\
    - \Theta(1),& \frac{1}{3} < \pi_{123} < 0.4 \\
    \pm \calO \left( \frac{1}{\sqrt{n}} \right), & \pi_{123} = 0.4.
\end{cases}
\]
\label{lem:sub_12_TLDR}
\end{lem}

\begin{proof}

We prove the lemma by summing up the adversarial loss $\ADS(P)$ of every preference profile $P \in \calL(\calA)^n$ such that the potential winning set $\PW{P} = \{1,2\}$, weighted by their likelihood of occurrence. 
Recall that iterative plurality starting from the truthful vote profile $\topRank{P}$ consists of agents changing their votes from alternatives that were not already winning to those that then become the winner \citep{Branzei13:How}. This occurs until no agent has an incentive to change their vote. \citet{Lev12:Convergence} demonstrated that the equilibrium winning set $\EW{\topRank{P}}$ is a subset of the initial potential winning set $\PW{P}$. Subsequently, \citet[Lemma 1]{Kavner21:strategic} proved that $\EW{\topRank{P}}$ is the unique alternative with more agents preferring it (subject to lexicographical tie-breaking), when $|\PW{P}| = 2$. Under the lemma's conditions, the equilibrium winner is therefore determined by whether $P[1 \succ 2] \geq P[2 \succ 1]$ or not.
There are thus four cases we must consider: alternatives $1$ and $2$ may individually be either the truthful or equilibrium winners, or both. 

Clearly, for any $P$ where the equilibrium winning alternative is the same as the truthful one, $\ADS(P) = 0$, following its definition. This leaves two cases: (Case 1) where alternative $1$ is the truthful winner and $2$ is the equilibrium winner, and (Case 2) where alternative $2$ is the truthful winner and $1$ is the equilibrium winner. We define $\calE_1$ and $\calE_2$ to represent these cases, as follows:
\begin{itemize}
    \item $\calE_1 = \{P \in \calL(\calA)^n~:~s_1(\topRank{P}) = s_2(\topRank{P}) > s_3(\topRank{P}) \text{ and } P[2 \succ 1] > P[1 \succ 2]\}$,
    \item $\calE_2 = \{P \in \calL(\calA)^n~:~s_1(\topRank{P})+1 = s_2(\topRank{P}) > s_3(\topRank{P}) \text{ and } P[1 \succ 2] \geq P[2 \succ 1]\}$.
\end{itemize}
%
This suggests the following partition:
\begin{align}
    \EDS(\{1,2\}) & = \Pr\nolimits_{P \sim \pi^n}(P \in \mathcal{E}_1) \times \mathbb{E}_{P \sim \pi^n}[\ADS(P)~|~P \in \mathcal{E}_1] \notag \\
    & + \Pr\nolimits_{P \sim \pi^n}(P \in \mathcal{E}_2) \times \mathbb{E}_{P \sim \pi^n}[\ADS(P)~|~P \in \mathcal{E}_2]. \label{eq:tldr_0}
\end{align}

It follows from Corollary \ref{coro:sub_others} that $\Pr\nolimits_{P \sim \pi^n}(P \in \mathcal{E}_1)$ and $\Pr\nolimits_{P \sim \pi^n}(P \in \mathcal{E}_2)$ are both $\Theta \left( \frac{1}{\sqrt{n}} \right)$. From \citet[Theorem 1]{Kavner21:strategic} we have $|\mathbb{E}_{P \sim \pi^n}[\ADS(P)~|~P \in \mathcal{E}_1]| = \calO(n)$, while $\mathbb{E}_{P \sim \pi^n}[\ADS(P)~|~P \in \mathcal{E}_2]$ has the same asymptotic rate but a negated sign. This follows since $\calE_1$ describes iterative voting sequences from $\topRank{P}$, where alternative $1$ is winning, to equilibria where alternative $2$ is winning; $\calE_2$ is the inverse. Equation (\ref{eq:tldr_0}) using these broad substitutions would yield $\calO(\sqrt{n}) - \calO(\sqrt{n})$ which is bounded between $- \calO(\sqrt{n})$ and $\calO(\sqrt{n})$. This yields too general of bounds for Lemma \ref{lem:sub_12_TLDR}, so we must analyze these conditional expected values more precisely. 

In Step 1, we characterize the $\calE_1$ case by detailing the number of agents with each ranking in any preference profile within the set. That is, any $P \in \calE_1$ has $\left( \frac{n}{2}-q, \frac{n}{2}-q, \beta, 2q-\beta\right)$ agents with rankings $(R_{123}, R_{231}, R_{321}, R_{312})$ respectively, for some $q, \beta \in \mathbb{N}$. We use the fact that $\calE_1$ is the disjoint union every $P$, characterized by $q$ and $\beta$ that span certain ranges, in order to devise a closed-form solution for $\Pr\nolimits_{P \sim \pi^n}(\mathcal{E}_1) \times \mathbb{E}_{P \sim \pi^n}[\ADS(P)~|~\mathcal{E}_1]$. In Step 2, we follow the same procedure for the $\calE_2$ case. In Step 3, we re-combine these two cases back into Equation (\ref{eq:tldr_0}). We rearrange certain terms and demonstrate how the aggregate summations yield Lemma \ref{lem:sub_12_TLDR}'s conclusion.


\begin{paragraph}{Step 1: Characterize the $\calE_1$ case.}
We begin by characterizing the set of profiles $P \in \calE_1$ in terms of how many agents have each ranking in the profile. This case covers the events where alternative $1$ is the truthful winner with the most truthful votes, which is equal to alternative $2$ and greater than those for alternative $3$. 
%
%
Let $\beta, q \in \mathbb{N}$. Given that $n \in \mathbb{N}$ is even, we take throughout this step:
\begin{itemize}
    \item $\frac{n}{2}-q$ agents with ranking $R_{123}$,
    \item $\frac{n}{2}-q$ agents with ranking $R_{231}$,
    \item $2q$ agents with either $R_{321}$ or $R_{312}$: with $\beta$ for $R_{321}$ and $2q-\beta$ for $R_{312}$.
\end{itemize}
%

The minimum of $q$ is $1$. Otherwise, if $q=0$, then there are no \emph{third-party} agents (i.e., agents with rankings $R_{321}$ or $R_{312}$), so there is not iterative plurality dynamics. The adversarial loss $\ADS(P)$ for any such $P$, indexed by $q=0$, is then clearly zero. The maximum of $q$ is $q^* = \max\{q \in \mathbb{Z}~:~\left(\frac{n}{2}-q\right) > 2q\}$, so that $q^* \in \{\frac{n}{6}-1, \bfloor{\frac{n}{6}}+1, \bfloor{\frac{n}{6}}+3\}$ depending if $n~mod~6 \in \{0, 2, 4\}$ respectively.
%
In order to uphold the condition that $P[2 \succ 1] > P[1 \succ 2]$, so that alternative $2$ is the equilibrium winner, we must have $\beta > q$.

\begin{table}[t]
    \centering
    \begin{tabular}{|c|c|c|c|}
        \hline
         Ranking & Probability & Frequency & Loss per Agent \\
          \hline \hline
         $R_{123} = (1 \succ 2 \succ 3)$ & $\pi_{123}$ & $\frac{n}{2}-q$ & $u_1 - u_2$ \\
         $R_{231} = (2 \succ 3 \succ 1)$ & $\pi_{231}$ & $\frac{n}{2}-q$ & $-u_1 + u_3$ \\
         $R_{321} = (3 \succ 2 \succ 1)$ & $\pi_{321}$ & $\beta$ & $-u_2 + u_3$ \\
         $R_{312} = (3 \succ 1 \succ 2)$ & $\pi_{312}$ & $2q-\beta$ & $u_2 - u_3$ \\
         $R_{132} = (1 \succ 3 \succ 2)$ & $0$ & $0$ & $u_1 - u_3$ \\
         $R_{213} = (2 \succ 1 \succ 3)$ & $0$ & $0$ & $-u_1 + u_2$ \\
        \hline
    \end{tabular}
    \caption{Character of profiles $P$ for $\PW{P} = \{1,2\}$ and even $n$ such that the truthful and equilibrium winners are $1$ and $2$, respectively. Here, we assume $\pi_{123} = \pi_{231} > 2 \pi_{321} = 2 \pi_{321} > 0$ and $\pi_{132} = \pi_{213} = 0$ in Lemma \ref{lem:sub_12_TLDR}.}
    \label{tab:character_profiles_E1_TLDR}
\end{table}

When $\calE_1$ holds, each agent with ranking $R_j$ in $P$ contributes some amount of utility to the adversarial loss function $\ADS(P)$. For instance, each agent with ranking $R_{123}$ contributes $\vec{u}(R_{123}, 1) - \vec{u}(R_{123}, 2) = u_1 - u_2$. Recall our use of rank-based utility $\vec{u} = (u_1, u_2, u_3)$. These amounts are also summarized by Table \ref{tab:character_profiles_E1_TLDR}.
Put together, we get the equation
\begin{align}
    \Pr\nolimits_{P \sim \pi^n}(\mathcal{E}_1) \times \mathbb{E}_{P \sim \pi^n}[\ADS(P)~|~\mathcal{E}_1]
    & = \sum_{q = 1}^{q^*} \sum_{\beta = q+1}^{2q} \mathcal{P}^1_{\vec{\pi},n}(q,\beta) \cdot \mathcal{V}^1_{\vec{u},n}(q,\beta) \label{eq:tldr_1}
\end{align}
where we define
\[
    \mathcal{P}^1_{\vec{\pi},n}(q,\beta) = \binom{n}{\frac{n}{2}-q, \frac{n}{2}-q, \beta, 2q-\beta} \pi_{123}^{\frac{n}{2}-q} \pi_{231}^{\frac{n}{2}-q} \pi_{321}^{\beta} \pi_{312}^{2q-\beta}
\]
and 
\begin{align*}
    \mathcal{V}^1_{\vec{u},n}(e,f,\beta,q) & = \begin{pmatrix}
        \frac{n}{2}-q, & \frac{n}{2}-q, & \beta, & 2q-\beta \end{pmatrix}
    \cdot \begin{pmatrix}
        u_1 - u_2, & -u_1+u_3, & -u_2+u_3, & u_2-u_3 \end{pmatrix}.
\end{align*}
Without loss of generality, we will assume for the duration of the proof that $q^* = \bfloor{\frac{n}{6}} -1$, taking the case that $n$ is divisible by $6$. It is easy to show that for a constant number of terms in Equation (\ref{eq:main1}) such that $q = \Theta(n)$, the objective is exponentially small and hence does not affect the result of this lemma.
We begin by factoring the probability term, which equals:
\begin{align*}
    \mathcal{P}^1_{\vec{\pi},n}(e,f,\beta,q)
    &=  \binom{n}{\frac{n}{2}-q, \frac{n}{2}-q, 2q} \pi_{123}^{\frac{n}{2}-q} \pi_{231}^{\frac{n}{2}-q} \times \binom{2q}{\beta} \pi_{321}^\beta \pi_{312}^{2q-\beta} \notag \\
    & = \binom{n}{\frac{n}{2}-q, \frac{n}{2}-q, q,q} \pi_{123}^{n-2q} \pi_{321}^{2q} \frac{2^{2q}}{\binom{2q}{q}} \times \binom{2q}{\beta} \frac{1}{2^{2q}} 
\end{align*}
using the assumption that $\pi_{123} = \pi_{231}$ and $\pi_{321} = \pi_{312}$.
Next, the value factor is
\begin{align*}
    \mathcal{V}^1_{\vec{u},n}(e,f,\beta,q) 
    & = (u_2 - u_3) \left( -\frac{n}{2} + 3q - 2\beta \right). 
\end{align*}
Put together, Equation (\ref{eq:tldr_1}) is
\begin{align}
    & (u_2 - u_3) \sum_{q=1}^{\frac{n}{6}-1} \binom{n}{\frac{n}{2}-q, \frac{n}{2}-q, q,q} \pi_{123}^{n-2q} \pi_{321}^{2q} \frac{2^{2q}}{\binom{2q}{q}} \sum_{\beta = q+1}^{2q} \binom{2q}{\beta} \frac{1}{2^{2q}}  \left( -\frac{n}{2} + 3q - 2\beta \right). \label{eq:tldr_2}
\end{align}
Since $u_2 > u_3$ by assumption, we henceforth forego writing $(u_2 - u_3)$ for ease of notation.
In order to simplify Equation (\ref{eq:tldr_2}), we employ the following lemma, which we prove in Appendix \ref{apx:concentration_inequalities}.
\begin{lem}
The following equations hold.
\begin{enumerate}
    \item 
    \begin{align*}
        \sum_{\beta = q+1}^{2q} \binom{2q}{\beta} \frac{1}{2^{2q}} 
        = 
        \frac{1}{2} - \frac{1}{2^{2q}} \binom{2q-1}{q-1},
    \end{align*}
    \item 
    \begin{align*}
        \sum_{\beta = q+1}^{2q} \binom{2q}{\beta} \frac{1}{2^{2q}} \beta 
        = 
        \frac{q}{2}.
    \end{align*}
\end{enumerate}
\end{lem}
Therefore Equation (\ref{eq:tldr_2}) may be written as
\begin{align}
    & \sum_{q=1}^{\frac{n}{6}-1} \binom{n}{\frac{n}{2}-q, \frac{n}{2}-q, q,q} \pi_{123}^{n-2q} \pi_{321}^{2q} \frac{2^{2q}}{\binom{2q}{q}} \left( \left( -\frac{n}{2} + 3q \right) \left( \frac{1}{2} - \frac{1}{2^{2q}} \binom{2q-1}{q-1} \right) - q \right) \notag \\
    & =\sum_{q=1}^{\frac{n}{6}-1} \binom{n}{\frac{n}{2}-q, \frac{n}{2}-q, q,q} \pi_{123}^{n-2q} \pi_{321}^{2q}  \left( \frac{\left( -\frac{n}{2} + q \right) 2^{2q-1}}{\binom{2q}{q}}  - \frac{\left( -\frac{n}{2}+3q \right)}{2}  \right). \label{eq:tldr_2b}
\end{align}
using the fact that $2\binom{2q-1}{q-1} = \binom{2q}{q}$. 

As described above, we observe that Equation (\ref{eq:tldr_2b}) is $-\calO(\sqrt{n})$. Since $q = \calO(n)$, it follows from Stirling's approximation (Proposition \ref{prop:stirling}, below) that $\binom{2q}{q} = \calO \left(\frac{2^n}{\sqrt{n}} \right)$. This entails $\frac{\left( -\frac{n}{2} + q \right) 2^{2q-1}}{\binom{2q}{q}}  - \frac{\left( -\frac{n}{2}+3q \right)}{2} = -\calO(n^{1.5})$. Meanwhile, it is shown in Lemma \ref{lem:prob_bounds} in Appendix \ref{apx:helper_lemmas} that $\sum_{q=1}^{\frac{n}{6}-1} \binom{n}{\frac{n}{2}-q, \frac{n}{2}-q, q,q} \pi_{123}^{n-2q} \pi_{321}^{2q} = \calO \left( \frac{1}{n} \right)$. We require this $-\calO(\sqrt{n})$ finding, in combination with Step 2, in order to yield more precise bounds.
\end{paragraph}

\begin{paragraph}{Step 2: Characterize the $\calE_2$ case.}
We next repeat the above process for the $\calE_2$ case. This case covers the events where alternative $2$ is the truthful winner with the most truthful votes, which is one more than alternative $1$ and greater than those for alternative $3$. Given that $n \in \mathbb{N}$ is even, we take throughout this step:
\begin{itemize}
    \item $\frac{n}{2}-1-q$ agents with ranking $R_{123}$ or $R_{132}$, 
    \item $\frac{n}{2}-q$ agents with ranking $R_{231}$,
    \item $2q+1$ agents with either $R_{321}$ or $R_{312}$: with $\beta$ for $R_{321}$ and $2q+1-\beta$ for $R_{312}$. 
\end{itemize}
The minimum of $q$ is $0$, while its maximum is $q^* = \max\{q \in \mathbb{Z}~:~\left(\frac{n}{2}-q\right) > 2q+1\}$, so that $q^* \in \{\bfloor{\frac{n}{6}}-1, \bfloor{\frac{n}{6}}\}$ depending on whether $n~mod~6 \in \{0\}$ or $\{2,4\}$, respectively.
Like in Step 1, we will assume $q^* = \frac{n}{6}-1$ without loss of generality, taking the case that $n$ is divisible by $6$.
In order to uphold the condition that $P[1 \succ 2] \geq P[2 \succ 1]$, so that alternative $1$ is the equilibrium winner, we must have $\beta \leq q$.
While $\calE_2$ holds, it should be clear that the values per agent are the negative of those presented in Table \ref{tab:character_profiles_E1_TLDR}. Put together, we get the equation
\begin{align}
    \Pr\nolimits_{P \sim \pi^n}(P \in \mathcal{E}_2) \times \mathbb{E}_{P \sim \pi^n}[\ADS(P)~|~P \in \mathcal{E}_2]
    & = -\sum_{q = 0}^{\frac{n}{6}-1} \sum_{\beta = 0}^{q} \mathcal{P}^2_{\vec{\pi},n}(q,\beta) \cdot \mathcal{V}^2_{\vec{u},n}(q,\beta) \label{eq:tldr_3}
\end{align}
where
\[
    \mathcal{P}^2_{\vec{\pi},n}(q,\beta) = \binom{n}{\frac{n}{2}-1-q, \frac{n}{2}-q, \beta, 2q+1-\beta} \pi_{123}^{\frac{n}{2}-1-q} \pi_{231}^{\frac{n}{2}-q} \pi_{321}^{\beta} \pi_{312}^{2q+1-\beta} 
\]
and
\begin{align*}
    \mathcal{V}^2_{\vec{u},n}(q,\beta) & = 
        \begin{pmatrix}
            \frac{n}{2}-1-q, & \frac{n}{2}-q,& \beta, & 2q+1-\beta,
        \end{pmatrix} 
    \cdot \begin{pmatrix} u_1 - u_2, & -u_1+u_3, & -u_2+u_3, & u_2-u_3 
        \end{pmatrix}.
\end{align*}
We begin by factoring the probability term, which equals
\begin{align*}
    \mathcal{P}^2_{\vec{\pi},n}(q,\beta)
    &= \binom{n}{\frac{n}{2}-1-q, \frac{n}{2}-q, 2q+1} \pi_{123}^{\frac{n}{2}-1-q} \pi_{231}^{\frac{n}{2}-q} \times \binom{2q+1}{\beta} \pi_{321}^\beta \pi_{312}^{2q+1-\beta} \notag \\
    & = \binom{n}{\frac{n}{2}-q, \frac{n}{2}-q, q,q} \pi_{123}^{n-2q} \pi_{321}^{2q} \cdot \frac{\pi_{321} (\frac{n}{2}-q) 2^{2q+1}}{\pi_{123} (2q+1) \binom{2q}{q}} \times \binom{2q+1}{\beta} \frac{1}{2^{2q+1}} 
\end{align*}
making use of the facts that $\pi_{123} = \pi_{231}$ and $\pi_{321} = \pi_{312}$.
The value term may be written as
\begin{align*}
    \mathcal{V}^2_{\vec{u},n}(e,f,\beta,q) 
    & = (u_2 - u_3) \left( -\frac{n}{2} + 3 - 2\beta \right) + \left(-u_1 + 2u_2 - u_3 \right).
\end{align*}
Substituting the latter constant terms into Equation (\ref{eq:tldr_3}) yields $(-u_1 + 2u_2 - u_3) \sum_{q = 0}^{\frac{n}{6}-1} \sum_{\beta = 0}^{q} \mathcal{P}^2_{\vec{\pi},n}(q,\beta).$ This is $\pm \calO \left( \frac{1}{\sqrt{n}} \right)$ by Corollary \ref{coro:sub_others}, since it is the probability of a two-way tie for plurality voting under i.i.d. preferences.



Since $u_2 > u_3$ by assumption, we henceforth forego writing $(u_2 - u_3)$ for ease of notation. Putting the remainder of the value terms with the probability above, Equation (\ref{eq:tldr_3}) is
\begin{align}
    & -\sum_{q=0}^{\frac{n}{6}-1} \binom{n}{\frac{n}{2}-q, \frac{n}{2}-q, q,q} \pi_{123}^{n-2q} \pi_{321}^{2q} \frac{\pi_{321} \left( \frac{n}{2}-q \right) 2^{2q+1}}{\pi_{123} (2q+1) \binom{2q}{q}} \sum_{\beta = 0}^{q} \binom{2q+1}{\beta} \frac{1}{2^{2q+1}}  \left( -\frac{n}{2} + 3q - 2\beta \right). \label{eq:tldr_4}
\end{align}
Briefly consider Equation (\ref{eq:tldr_4}) when $q=0$. Unlike the prior case in Step 1, now there is a single third-party agent with ranking $R_{312}$. This is represented with $\beta=0$ as
\begin{align*}
    & - \binom{n}{\frac{n}{2}-1, \frac{n}{2}, 1}  \pi_{123}^{\frac{n}{2}-1} \pi_{231}^{\frac{n}{2}} \binom{1}{0} \pi_{312} \left( -\frac{n}{2} \right) \\
    & = -\frac{\pi_{123}^{n-1} \pi_{321} n^2}{4} \binom{n}{\frac{n}{2}} \\
    & = - \calO(n^{1.5}) (2 \pi_{123})^n \\
    & = - \calO \left( e^{-\Theta(n)} \right)
\end{align*}
by Stirling's approximation.
\begin{prop}[Stirling's approximation]
    Stirling's approximation says that $n! \sim \sqrt{2 \pi n} \left( \frac{n}{e} \right)^n$. Therefore, we have $\binom{2n}{n} \sim \frac{2^{2n}}{\sqrt{n \pi}}$.
    \label{prop:stirling}
\end{prop}
This proposition is discussed further in Appendix \ref{apx:stirling_and_wallis}. Next, we make use of the following binomial identities.
\setcounter{lem}{1}
\begin{lem}
The following equations hold.
\begin{enumerate}
    \item[(3)]
    \begin{align*}
        \sum_{\beta = 0}^{q} \binom{2q+1}{\beta} \frac{1}{2^{2q+1}} 
        = \frac{1}{2},
    \end{align*}
    \item[(4)]
    \begin{align*}
        \sum_{\beta = 0}^{q} \binom{2q+1}{\beta} \frac{1}{2^{2q+1}} \beta 
        = 
        \left( \frac{2q+1}{4} \right) - \frac{2q+1}{2^{2q+1}} \binom{2q-1}{q-1}.
    \end{align*}
\end{enumerate}
\end{lem}
Therefore Equation (\ref{eq:tldr_4}) may be written as
\begin{align}
    & -\sum_{q=1}^{\frac{n}{6}-1} \binom{n}{\frac{n}{2}-q, \frac{n}{2}-q, q,q} \pi_{123}^{n-2q} \pi_{321}^{2q} \frac{\pi_{321} \left( \frac{n}{2}-q \right) 2^{2q+1}}{\pi_{123} (2q+1) \binom{2q}{q}} \notag \\
    & \quad \quad \times \left( \frac{1}{2} \left( -\frac{n}{2} + 3q \right) - 2 \left( \frac{2q+1}{4} - \frac{2q+1}{2^{2q+1}} \binom{2q-1}{q-1} \right) \right) \notag \\
    & =  - \sum_{q=1}^{\frac{n}{6}-1} \binom{n}{\frac{n}{2}-q, \frac{n}{2}-q, q,q} \pi_{123}^{n-2q} \pi_{321}^{2q} \left( \frac{\pi_{321} \left( -\frac{n}{2}+q-1 \right) \left( -\frac{n}{2}+q \right) 2^{2q}}{\pi_{123} (2q+1) \binom{2q}{q}} + \frac{\pi_{321} \left( \frac{n}{2}-q \right)}{\pi_{123}} \right). \label{eq:tldr_5}
\end{align}
As in Step 1 above, the absolute value of Equation (\ref{eq:tldr_5}) is $|\calO \left( \frac{1}{n} \right) \cdot \calO \left( n^{1.5} \right)| = \calO \left( \sqrt{n} \right)$.
\end{paragraph}

\begin{paragraph}{Step 3: Putting the pieces back together.}
Recall that our original problem began as Equation (\ref{eq:tldr_0}) which we initially split into Equations (\ref{eq:tldr_1}) and (\ref{eq:tldr_3}). Through a sequence of steps we transformed these equations into Equations (\ref{eq:tldr_2b}) and (\ref{eq:tldr_5}) and an additional $\pm \calO \left( \frac{1}{\sqrt{n}} \right)$ term. Recombining these simplified equations yields
\begin{align}
     & \sum_{q = 1}^{\frac{n}{6}-1} \binom{n}{\frac{n}{2}-q, \frac{n}{2}-q, q,q} \pi_{123}^{n-2q} \pi_{321}^{2q} \left( - \frac{1}{2} \left( -\frac{n}{2}+3q \right) + \frac{\pi_{321}}{\pi_{123}} \left( -\frac{n}{2}+q \right)  \right) \notag \\
     & + \sum_{q = 1}^{\frac{n}{6}-1} \binom{n}{\frac{n}{2}-q, \frac{n}{2}-q, q,q} \pi_{123}^{n-2q} \pi_{321}^{2q}  \frac{\left( -\frac{n}{2} + q \right) 2^{2q}}{\binom{2q}{q}} \left( \frac{1}{2}   - \frac{\pi_{321} \left( -\frac{n}{2}+q-1 \right)}{\pi_{123} (2q+1)}  \right).   \label{eq:tldr_6}
\end{align}
We introduce Lemma \ref{lem:soln_part_a} in Appendix \ref{apx:secondary_equations} to prove that the first summation of Equation (\ref{eq:tldr_6}) is $\Theta(1)$ if $\pi_{123} < 0.4$, $-\Theta(1)$ if $\pi_{123} > 0.4$, and $\pm \calO \left( \frac{1}{n} \right)$ otherwise (i.e., if $\pi_{123}=0.4$). Notice that
\begin{align*}
    & \left( -\frac{n}{2} +q \right) \left( \frac{1}{2} + \frac{\pi_{321} ( - \frac{n}{2} + q -1)}{\pi_{123} (2q+1)} \right)
    = \frac{\left( -\frac{n}{2} + q \right)}{2 \pi_{123} (2q+1)}  \left( q - \pi_{321} n + \pi_{123} - 2 \pi_{321} \right) 
\end{align*}
using the fact that $\pi_{123} + \pi_{321} = \frac{1}{2}$.
We prove Lemma \ref{lem:soln_part_b} in Appendix \ref{apx:secondary_equations} that the second summation of Equation (\ref{eq:tldr_6}) is therefore $\pm \calO \left( \frac{1}{\sqrt{n}} \right)$. This concludes the proof of Lemma \ref{lem:sub_12_TLDR}.
\end{paragraph}
\end{proof}

%% file: EC_body/three_ties_lemma_tldr.tex
\begin{table}[t]
    \centering
    \begin{tabular}{|c||c|c|c|}
        \hline
         & $\pi_{231} = \pi_{213}$ & $\pi_{231} > \pi_{213}$ & $\pi_{231} < \pi_{213}$ \\
          \hline \hline
            $\pi_{321} = \pi_{312}$ & 
            $\pi_{123} - \pi_{132}$ &  
            $2 \pi_{123} + \pi_{213} - 3 \pi_{231}$ & 
            $\pi_{213} - \pi_{123}$ \\[0.25em]
          \hline
            $\pi_{321} > \pi_{312}$ & 
            $2\pi_{123} + 2\pi_{132} + \pi_{312} - 5\pi_{321}$ &  
            $\pi_{123} + \pi_{213} - \pi_{231} - \pi_{321}$ & 
            $\pi_{312} + \pi_{213} - \pi_{123} - \pi_{321}$ \\[0.25em]
          \hline 
            $\pi_{321} < \pi_{312}$ & 
            $\pi_{123} - \pi_{321}$ &  
            $\pi_{123} + \pi_{213} - \pi_{231} - \pi_{321}$ & 
            \makecell{N/A} \\
         \hline
    \end{tabular}
    \caption{Values of $f^1(\pi)$ given conditions on $\pi$ for Lemma \ref{lem:three_tie_priority1_TLDR}.}
    \label{tab:complexity_three_sides_A_TLDR}
\end{table}

\setcounter{lem}{7}
\begin{lem}
Suppose that $\pi_{123} + \pi_{132} = \pi_{231} + \pi_{213} = \pi_{321} + \pi_{312} = \frac{1}{3}$ and $\pi_i > 0, ~\forall i \in [6].$ Furthermore, let $u_1 \geq u_2 > u_3$ in $\vec{u}$. 
Then $\exists N > 0$ such that $\forall n > N$ that are divisible by $3$, 
\[
\EDS(\{1,2, 3\}) = \begin{cases}
    \pm \calO \left( e^{-\Theta(n)} \right), & \pi_{231} < \pi_{213}, \pi_{321} < \pi_{312} \\
    f^1(\pi) \Theta(1) + g^1_n(\pi), & \text{otherwise}
\end{cases}
\]
where $f^1(\pi)$ is determined by Table \ref{tab:complexity_three_sides_A_TLDR} and
\begin{align*}
    g^1_n(\pi) = \begin{cases}
    \Theta(1), & \pi_{231} + \pi_{132} > \pi_{123} + \pi_{321} \\
    \pm \calO \left( \frac{1}{\sqrt{n}} \right), & \pi_{231} + \pi_{132} \leq \pi_{123} + \pi_{321}.
    \end{cases}
\end{align*}
\label{lem:three_tie_priority1_TLDR}
\end{lem}

Assuming that $\pi_{231} \geq \pi_{213}$ and $\pi_{321} \geq \pi_{312}$, we read Lemma \ref{lem:three_tie_priority1_TLDR} as the sum of two terms: $f^1(\pi) \Theta(1)$ and $g^1_n(\pi)$. First, $f^1(\pi)$ may be less than zero, zero, or greater than zero. This determines whether $f^1(\pi) \Theta(1) \in \{-\Theta(1), 0, \Theta(1)\}$, respectively. Depending on what $g^1_n(\pi)$ is, this yields a solution in $\{-\Theta(1) + \Theta(1), \Theta(1), \pm \calO \left( \frac{1}{\sqrt{n}} \right)\}$, where $\Theta(1) - \Theta(1) \in (-\calO(1), \calO(1))$.

\begin{proof}[Proof Sketch]
We prove this lemma similar to Lemma \ref{lem:sub_12_TLDR} by summing up the adversarial loss $\ADS(P)$ of every preference profile $P \in \calL(\calA)^n$ such that the potential winning set $\PW{P} = \{1,2,3\}$, weighted by their likelihood of occurrence. Since $n$ is assumed to be divisible by $3$, this covers the case where there are exactly $\frac{n}{3}$ agents that vote for each of the alternatives $1$, $2$, and $3$. 
We first must discuss what the equilibrium winning set $\EW{\topRank{P}}$ is for any profile $P$.

Recall that iterative plurality starting from the truthful vote profile $\topRank{P}$ consists of agents changing their votes from alternatives that were not already winning to those that then become the winner \citep{Branzei13:How}. Therefore any improvement step from alternative $c \in \calA$ to another $c' \in \calA$ means that neither $c$ nor $c'$ could have been the winner, prior to this step. Hence, after this step, no agent will change their vote to $c$, since doing so would not make it the winner. Since there are $m=3$ alternatives, it follows that the first improvement step determines which two alternatives are in the run-off to be the equilibrium winner. By \citet[Lemma 1]{Kavner21:strategic}, the winner is then whichever more agents prefer out of the entire agent pool.

For example, if all agents in $P$ have preference rankings $R_{123}$, $R_{132}$, $R_{213}$, or $R_{312}$, then no agent has an incentive to change their vote and alternative $1$ is both the equilibrium and truthful winners. Now suppose that there is at least one agent $j$ with ranking $R_{231} = (2 \succ 3 \succ 1)$. If agent $j$ switches their vote first, then the plurality scores of the alternatives would be $( \frac{n}{3}, \frac{n}{3}-1, \frac{n}{3}+1)$. From this vote profile, alternative $2$ cannot become the winner, so no agent will henceforth switch their vote to $2$. Iterative plurality thereafter consists of agents that were voting for alternative $2$ iteratively switching their votes to either alternatives $1$ or $3$. The winner is whichever alternative more agents prefer (subject to tie-breaking) \citep[Lemma 1]{Kavner21:strategic}. We conclude that $2 \in \EW{\topRank{P}}$ if $R_{321} \in P$ and $P[2 \succ 1] > P[1 \succ 2]$, whereas $3 \in \EW{\topRank{P}}$ if $R_{231} \in P$ and $P[3 \succ 1] > P[1 \succ 3]$. This yields three cases for whether either or both of these are the case. We define $\calE_2$, $\calE_3$, and $\calE_{2,3}$ as follows:
\begin{itemize}
    \item $\calE_2 = \{P \in \calL(\calA)^n~:~R_{321} \in P \text{ and } P[2 \succ 1] > P[1 \succ 2] \text{, and either } R_{231} \notin P \text{ or } P[1 \succ 3] \geq P[3 \succ 1]\}$,
    \item $\calE_3 = \{P \in \calL(\calA)^n~:~R_{231} \in P \text{ and } P[3 \succ 1] > P[1 \succ 3] \text{, and either } R_{321} \notin P \text{ or } P[1 \succ 2] \geq P[2 \succ 1]\}$,
    \item $\calE_{2,3} = \{P \in \calL(\calA)^n~:~R_{231}, R_{321} \in P \text{ and } P[2 \succ 1] > P[1 \succ 2] \text{ and } P[3 \succ 1] > P[1 \succ 3]\}$.
\end{itemize}
Let $a, b, c \in [0, \frac{n}{3}]$. Given that $n \in \mathbb{N}$ is divisible by $3$, we take throughout this proof:
\begin{itemize}
    \item $\frac{n}{3}$ agents with rankings either $R_{123}$ or $R_{132}$: with $a$ for $R_{123}$ and $\frac{n}{3}-a$ for $R_{132}$,
    \item $\frac{n}{3}$ agents with rankings either $R_{231}$ or $R_{213}$: with $b$ for $R_{231}$ and $\frac{n}{3}-b$ for $R_{213}$,
    \item $\frac{n}{3}$ agents with rankings either $R_{321}$ or $R_{312}$: with $c$ for $R_{321}$ and $\frac{n}{3}-c$ for $R_{312}$.
\end{itemize}
%
We begin by characterizing the $\calE_2$ case. Namely, $P \in \calE_2$ if the following ranges are satisfied. First, $a \in [0, \frac{n}{3}]$ has its full range. Second, $b \leq \frac{n}{6}$, so that there are at least as many agents preferring $R_{213}$ than $R_{231}$, which entails $3 \notin \EW{\topRank{P}}$. Third, $c > \frac{n}{6}$, so that there are more agents preferring $R_{321}$ than $R_{312}$, which entails $2 \in \EW{\topRank{P}}$. The value per agent and probability of each ranking is summarized by Table \ref{tab:character_profiles_E1} (see Appendix \ref{apx:three_ties}).
Put together, we get the equation
\begin{align}
    & \Pr\nolimits_{P \sim \pi^n}(P \in \mathcal{E}_2) \times \mathbb{E}_{P \sim \pi^n}[\ADS(P)~|~P \in \mathcal{E}_2]
    = \sum_{a=0}^{\frac{n}{3}} \sum_{b=0}^{\bfloor{\frac{n}{6}}} \sum_{c=\bfloor{\frac{n}{6}}+1}^{\frac{n}{3}} \mathcal{P}_{\vec{\pi},n}(a,b,c) \cdot \mathcal{V}^2_{\vec{u},n}(a,b,c) \label{eq:three_tldr_1}
\end{align}
where we define
\begin{align*}
    \mathcal{P}_{\vec{\pi},n}(a,b,c) & = \binom{n}{a, b, c, \frac{n}{3}-c, \frac{n}{3}-a, \frac{n}{3}-b} \pi_{123}^a \pi_{231}^b \pi_{321}^{c} \pi_{312}^{\frac{n}{3}-c} \pi_{132}^{\frac{n}{3}-a} \pi_{213}^{\frac{n}{3}-b} \\
    & = \binom{n}{\frac{n}{3}, \frac{n}{3}, \frac{n}{3}} \frac{1}{3^n}   \times  \binom{\frac{n}{3}}{a} \left( \frac{\pi_{123}}{\pi_{123}+\pi_{132}} \right)^a \left( \frac{\pi_{132}}{\pi_{123}+\pi_{132}} \right) ^{\frac{n}{3}-a} \\
    & \quad  \quad \times \binom{\frac{n}{3}}{b} \left( \frac{\pi_{231}}{\pi_{231}+\pi_{213}} \right)^b \left( \frac{\pi_{213}}{\pi_{231}+\pi_{213}} \right)^{\frac{n}{3}-b} \times \binom{\frac{n}{3}}{c} \left( \frac{\pi_{321}}{\pi_{321}+\pi_{312}} \right)^c \left( \frac{\pi_{312}}{\pi_{321}+\pi_{312}} \right)^{\frac{n}{3}-c}
\end{align*}
and 
\begin{align*}
    \mathcal{V}^2_{\vec{u},n}(a, b, c) & = \begin{pmatrix}
        a, & b, & c, & \frac{n}{3}-c, & \frac{n}{3}-a, & \frac{n}{3}-b \end{pmatrix} \\
    & \quad \cdot \begin{pmatrix}
        u_1 - u_2, & -u_1+u_3, & -u_2+u_3, & u_2-u_3, & u_1-u_3, & -u_1+u_2 \end{pmatrix} \\
    & = (u_2 - u_3) \left( \frac{2n}{3} -a -b-2c \right).
\end{align*}
In the full proof, we use two main techniques to simplify Equation (\ref{eq:three_tldr_1}). First, we employ Lemma \ref{lem:stirling_trinom} in Appendix \ref{apx:helper_lemmas}, which demonstrates that $\binom{3n}{n, n, n} \frac{1}{3^{3n}} = \Theta \left( \frac{1}{n} \right)$.
%
%
Second, we can see Equation (\ref{eq:three_tldr_1}) may be broken down into several separable summations each of the template $\sum_{t \in T} \binom{m}{t} p^t (1-p)^{m-t}$ or $\sum_{t \in T} \binom{m}{t} p^t (1-p)^{m-t} t$, corresponding to some contiguous domain $T \subseteq [m]$ for a binomial random variable $Bin(m,p)$. 
By Lemma \ref{lem:bin_theorems_approx} in Appendix \ref{apx:concentration_inequalities}, it follows that each summation is either $\Theta(1) \pm \calO \left( \frac{1}{\sqrt{m}} \right)$ or $\Theta(m) \pm \calO \left( \sqrt{m} \right)$, if $mp \in T$, and $\calO \left( e^{-\Theta(m)} \right)$ otherwise. For instance, we have
\begin{align*}
    \sum_{c=\bfloor{\frac{n}{6}}+1}^{\frac{n}{3}}  \binom{\frac{n}{3}}{c} \left( \frac{\pi_{321}}{\pi_{321}+\pi_{312}} \right)^c \left( \frac{\pi_{312}}{\pi_{321}+\pi_{312}} \right)^{\frac{n}{3}-c} = \begin{cases} 
        \frac{1}{2} \pm \calO \left( \frac{1}{\sqrt{n}} \right), & \pi_{321} = \pi_{312} \\
        1 - \calO \left( e^{-\Theta(n)} \right), & \pi_{321} > \pi_{312} \\
        \calO \left( e^{-\Theta(n)} \right), & \pi_{321} < \pi_{312}
    \end{cases}
\end{align*}
and
\begin{align*}
    \sum_{c=\bfloor{\frac{n}{6}}+1}^{\frac{n}{3}}  \binom{\frac{n}{3}}{c} \left( \frac{\pi_{321}}{\pi_{321}+\pi_{312}} \right)^c \left( \frac{\pi_{312}}{\pi_{321}+\pi_{312}} \right)^{\frac{n}{3}-c} c = \begin{cases} 
        \frac{n}{12} \pm \calO \left( \frac{1}{\sqrt{n}} \right), & \pi_{321} = \pi_{312} \\
        \frac{\tilde{\pi}_3 n}{3} \left(  1- \calO \left( e^{-\Theta(n)} \right) \right), & \pi_{321} > \pi_{312} \\
        \calO \left( e^{-\Theta(n)} \right), & \pi_{321} < \pi_{312}.
    \end{cases}
\end{align*}
These two techniques enable us to identify closed form solutions for Equation (\ref{eq:three_tldr_1}) in the form of $f^1(\pi) \Theta(1) \pm \calO \left( \frac{1}{\sqrt{n}} \right)$, where $f^1(\pi)$ is a function of the distribution $\pi$.

The $\calE_3$ case is quite similar to what we described above and yields an equation like Equation (\ref{eq:three_tldr_1}), except with different ranges for $b$ and $c$ and the value function $\mathcal{V}^3_{\vec{u},n}(a, b, c) = (u_2 - u_3) \left( \frac{n}{3} +a -2b -c \right)$. The $\calE_{2,3}$, rather, is more complicated and requires techniques above and beyond those described above and used in Lemma \ref{lem:sub_12_TLDR}.
Recall that the definition of adversarial loss for a preference profile $P$, against truthful vote profile $\topRank{P}$, is $\ADS_{\vec{u}}(P) = \SW{\vec{u}}{P}{f(\topRank{P})} - \min\nolimits_{c \in \EW{\topRank{P}}} \SW{\vec{u}}{P}{c}$. 
Since $|\EW{\topRank{P}}| = 2$ for this case, we must apply nuance in determining $\ADS(P)$, depending on number of agents with each ranking in $P$ (i.e., the values of $a$, $b$, and $c$). That is, the loss is the maximum of $\mathcal{V}^2_{\vec u, n}(a,b,c)$ and $\mathcal{V}^3_{\vec u, n}(a,b,c)$:
\begin{align*}
    & (u_2 - u_3) \max \left\{ \frac{2n}{3}-a-b-2c, \frac{n}{3}+a-2b-c \right\} \\
    & = (u_2 - u_3) \left( \frac{n}{3} +a -2b -c + \max\left\{\frac{n}{3}-2a+b-c, 0\right\}\right).
\end{align*}
Solving the problem $\Pr\nolimits_{P \sim \pi^n}(P \in \mathcal{E}_{2,3}) \times \mathbb{E}_{P \sim \pi^n}[\ADS(P)~|~P \in \mathcal{E}_{2,3}]$ for $\frac{n}{3}+a -2b +c$ is almost identical to the $\calE_3$ case described above. Solving the problem for $\max\left\{\frac{n}{3}-2a+b-c, 0\right\}$ requires an additional technique, which we describe in the full version of the proof in Appendix \ref{apx:three_ties}. There, we prove that as long as $\pi_{231} \geq \pi_{213}$ and $\pi_{321} \geq \pi_{312}$, then this term is 
\[
\begin{cases}
    \Theta(1), & \pi_{231} + \pi_{132} > \pi_{123} + \pi_{321} \\
    \calO \left( \frac{1}{\sqrt{n}} \right), & \pi_{231} + \pi_{132} = \pi_{123} + \pi_{321} \\
    \calO \left( e^{-\Theta(n)} \right), & \pi_{231} + \pi_{132} < \pi_{123} + \pi_{321}.
    \end{cases}
\]
This concludes the proof sketch of Lemma \ref{lem:three_tie_priority1_TLDR}.
\end{proof}

%% file: EC_body/conclusion.tex
\section{Discussion and Future Directions}
\label{sec:limitations}

Iterative voting is a naturalistic model for strategic behavior over time. It relaxes strict rationality and information assumptions commonly utilized in game theoretic models and incorporates more information about agents' higher-order preferences than standard social choice rules. Analyzing this model provides descriptive value by better understanding electoral behavior in applications, such as Doodle and political opinion polls, where agents can update their votes prior to finalizing the group decision. It also provides prescriptive value from an artificial intelligence perspective, which studies agent behaviors that maximize long-term reward, and an economics perspective, which studies behaviors that yield socially good outcomes.

Our work has demonstrated the average-case analysis of iterative plurality's economic performance across agents' preference distributions. We have contributed several novel binomial and multinomial lemmas that may be useful for future study of IV, and we have extended \citet{Xia2021:How-Likely}'s theorems to expectations of random functions, rather than the likelihood of events.
Our bounds on $\ED$ are significantly tighter than \citet{Kavner21:strategic}, who only provided an asymptotic upper-bound $-\Omega(1)$ with respect to IC. Plugging in $\pi_{123} = \ldots = \pi_{321}$ into our Theorem \ref{thm:main} confirms their finding, with $\ED = -\Theta(1)$ when $m=3$.

Our work may be interpreted within the smoothed analysis framework put forth by \citet{Xia2020:The-Smoothed} and \citet{Xia2021:How-Likely}. Namely, Xia expressed the smoothed likelihood of an event as the supremum (and infimum) expectation of an indicator function, representing the worst- (and best-) average-case analysis where input distributions are sampled from a set $\Pi \subseteq \Delta(\calL(\calA))$. A comparable ``smoothed additive dynamic price of anarchy'' notion would define $\Pi$ and study $\sup_{\pi \in \Pi}$ (and $\inf_{\pi \in \Pi}$) $\ED_{\vec{u}}(\pi^n)$. Our work provides insights into these values for $m=3$ alternatives.


Our work is limited technically in that we define a separate mapping from $\pi$ for each of the four non-zero values of Equation (\ref{eq:main_partition}): $\EDS(\{1,2\})$, $\EDS(\{1,3\})$, $\EDS(\{2,3\})$, and $\EDS(\{1,2,3\})$. This leaves some room for refinement when we attain incomparable results, such as $\Theta(1) - \Theta(1)$, in the solution. Our finding that $\ED$ depends on whether $n$ is even or odd will need further study.
%
We expect that extending our results to $m > 3$ will be the most involved. In order to apply our methods of partitioning $\ED$ by the potential winning sets, the set $\calE_1$ in Lemma \ref{lem:sub_12} (and similar sets in comparable lemmas) would need to be adapted to suggest 
$
s_1(\topRank{P}) = s_2(\topRank{P}) > s_{\ell}(\topRank{P}), ~ \forall \ell \geq 3
$,
which would significantly complicate our already-extensive analysis. 
Our present work contributes techniques that may assist this future direction.
Furthermore, 
our analysis may contribute to solving 
the more general problem of computing expected 
social welfare for any piecewise linear loss function of the histogram of preference profiles.


Another avenue of future work is testing the empirical significance of our theoretical results, as with the experiments by \citet{Zou15:Doodle}, \citet{tal2015study}, and \citet{meir2020strategic}. Understanding the extent to which strategic behavior actually affects electoral outcome quality would help mechanism designers elicit more authentic preferences.
This could be tested, for example, by treating peoples' preferences to align with the assumptions of Lemma \ref{lem:sub_12_TLDR} and varying $\pi_{123}$ across the dichotomy threshold of $0.4$. It is still uncertain how well the iterative plurality protocol models real-world strategic behavior.
While we assume myopic
best responses in this work, peoples’ actual behavior through an IV procedure may yield
different quality results, even while fixing their preferences.
It may further be fruitful to test whether 
people would trust the use of proxy agents, acting on their behalf, who follow best-response dynamics
\citep{rossi2016moral, xia2017improving}.

%% file: EC_appendix/apx_page_0.tex
\section{Appendix Contents and Disclaimer}
\label{apx:appendix_contents}


The appendices of this paper are organized so that the material of each appendix is used to prove the lemmas introduced in prior appendices, while they depend on the lemmas of later appendices. In particular, the primary results used to prove Theorem \ref{thm:main} include the two-tie case lemmas of Appendix \ref{apx:two_ties} -- Lemmas \ref{lem:sub_12}, \ref{lem:sub_12_odd}, \ref{lem:sub_12_non_34_equal}, \ref{lem:sub_13}, and \ref{lem:sub_23} -- and the three-tie case lemmas of Appendix \ref{apx:three_ties} -- Lemmas \ref{lem:three_tie_priority1}, \ref{lem:three_tie_priority3}, and \ref{lem:three_tie_priority2}. 
In order to prove these claims, we make use of several supplementary multinomial results, which are depicted in Appendix \ref{apx:secondary_equations}, including  Lemmas \ref{lem:soln_part_a} and \ref{lem:soln_part_b}.
These lemmas, in turn, depend on several technical lemmas that appear in the subsequent appendix, Appendix \ref{apx:collision_entropy}, including Lemmas \ref{lem:dec23_expected_demoivre} -- \ref{lem:real_expected_demoivre}. The technical lemmas that are used to prove these are included in the remaining appendices: Appendix \ref{apx:stirling_and_wallis} discusses Stirling's approximation and an approximation for the central binomial coefficient, Appendix \ref{apx:concentration_inequalities} discusses concentration inequalities, and Appendix \ref{apx:helper_lemmas} covers the remainder.

First, we offer a brief disclaimer about the notation used throughout the appendices. The results of the primary lemmas are summarized in Appendix \ref{apx:concise_summary}, below. Appendix \ref{apx:smoothed_analysis_full} then provides a primer about the asymptotic rates of sequences of functions and smoothed analysis, with respect to the likelihood that large elections are tied \citep{Xia2020:The-Smoothed, Xia2021:How-Likely}. 


Throughout these appendices, we make use of the following notational correspondence.
\begin{remark}
Preference distribution 
$\pi = (\pi_1, \ldots, \pi_6)$ corresponds to the rankings
\begin{align*}
        & R_1 = (1\succ 2\succ 3); & R_5 = (1\succ 3\succ 2) \\
        & R_2 = (2\succ 3\succ 1); & R_6 = (2\succ 1\succ 3) \\
        & R_3 = (3\succ 2\succ 1); & R_4 = (3\succ 1\succ 2). 
    \end{align*}
\end{remark}

Note we make use of different sets of variable nomenclature throughout these appendices in order to help readability. This is an unfortunate consequence due to the amount of content in this manuscript. Many of the $26$ lemmas are quoted within the proof of another and proved in a separate appendix. Changing the nomenclature therefore does not affect our results. For each proof, we used the nomenclature that was the most consistent with other lemmas that expressed similar ideas. 

First, in Appendices \ref{apx:two_ties} and \ref{apx:three_ties}, the variables $q, \beta, e, f, a, b$, and $c$ count the number of agents with each ranking $R \in \calL(\calA)$. Agents' preference distribution is denoted by $\pi = (\pi_1, \pi_2, \pi_3, \pi_4, \pi_5, \pi_6)$. In Definition \ref{dfn:tilde_definitions} in Lemma \ref{lem:sub_12}, we introduce a new notation of $\tilde{\pi}_1 = \frac{\pi_1}{\pi_1 + \pi_5}$, $\tilde{\pi}_3 = \frac{\pi_3}{\pi_3 + \pi_4}$, $\Pi_1 = \pi_1 + \pi_5$ and $\Pi_3 = \pi_3 + \pi_4$. This notation is used throughout the two-tie and three-tie proofs.

Second, the lemmas included in Appendix \ref{apx:secondary_equations} deal with the asymptotic rate of functions resembling expected values based on a symmetric multinomial distribution $\sum_{q=1}^{\frac{n}{6}-1} \binom{n}{\frac{n}{2}-q, \frac{n}{2}-q, q, q} \pi_1^{n-2q} \pi_3^{2q} f_n(q)$ for some function $f_n(q)$. The four indices correspond to agents with rankings $R_1$ or $R_5$, $R_2$ or $R_6$, $R_3$, and $R_4$ respectively, with corresponding probabilities $(\pi_1, \pi_1, \pi_3, \pi_3)$. We continue to use $q$ as our index variable, but use the lower-case notation $\pi_1$ and $\pi_3$ instead of the upper-case $\Pi_1$ and $\Pi_3$ notation of the prior appendices.

Third, in Appendix \ref{apx:collision_entropy}, it serves us to make several variable substitutions. We describe these substitutions formally in Appendix \ref{sec:expected_prelims}. Namely, we take $S_n \sim Bin(n,p)$ for some $p \in (0, \frac{2}{3})$ and $x_{n,k} = \frac{k-np}{\sqrt{np(1-p)}}$. The lemmas of this appendix are used by those of Appendix \ref{apx:secondary_equations} after making the change of variables
$\begin{pmatrix}
    \frac{n}{2},&
    q,&
    2\pi_3,&
    2\pi_1
\end{pmatrix} \mapsto \begin{pmatrix}
    n,&
    k,&
    p,&
    q
\end{pmatrix}$, where $2 \pi_3 + 2 \pi_1 = 1$. This follows because $\Pi_1 + \Pi_3 = \frac{1}{2}$ by assumption of Lemma \ref{lem:sub_12}, using the nomenclature of Definition \ref{dfn:tilde_definitions}.

Finally, Appendices \ref{apx:stirling_and_wallis}, \ref{apx:concentration_inequalities}, and \ref{apx:helper_lemmas} provide lemmas that are agnostic to circumstance, so they often use more standard notation. In these appendices, $n$ is the variable that is scaled, $p \in (0,1)$ is a probability, $q = 1-p$, and $k \in [0,n]$ is an index variable.


\subsection{Concise Summary of Main Results}
\label{apx:concise_summary}

In this appendix, we depict a concise summary of $\ED$ by compiling the results from Appendix \ref{apx:two_ties} -- $\EDS(\{1,2\})$ from Lemma \ref{lem:sub_12}  ($n$ is even) and Lemma \ref{lem:sub_12_odd} ($n$ is odd),  $\EDS(\{1,3\})$ from Lemma \ref{lem:sub_13}, and $\EDS(\{2,3\})$ from Lemma \ref{lem:sub_23} -- and Appendix \ref{apx:three_ties} -- $\EDS(\{1,2,3\})$ from Lemmas \ref{lem:three_tie_priority1}, \ref{lem:three_tie_priority3}, and \ref{lem:three_tie_priority2} ($n$, $n-2$, and $n-1$ is divisible by $3$, respectively).
As described by Equation (\ref{eq:main_partition}), we have
\[
\ED_{\vec{u}}(\pi^n) = \EDS(\{1,2\}) + \EDS(\{1,3\}) + \EDS(\{2,3\}) + \EDS(\{1,2,3\}). 
\]
For any $\pi \in \Delta( \calL(\calA) )$ and $W \subseteq \calA, |W| \geq 2$, we provide a map from $\EDS(W)$ to
\[
\left\{ \pm \Theta(\sqrt{n}), \pm \Theta(1), \pm \calO(1), \pm \calO \left( \frac{1}{\sqrt{n}} \right), \pm \calO \left( \frac{1}{n} \right), \pm \calO \left( e^{-\Theta(n)} \right) \right\}.
\]
\begin{paragraph}{Step 1: One-way ties}
First, consider any $\pi$ such that $|W^*(\pi)| = 1$. Recall from the proof of Theorem \ref{thm:main} in the main text that $W^*(\pi) = \argmax_{i \in [m]} \lambda_i(\pi)$ and $\lambda_i(\pi) = \sum_{j:\topRank{R_j} = i} \pi_j$ is the likelihood of an agent truthfully voting for alternative $i$. From Corollary \ref{coro:sub_others} and Proposition \ref{prop:other_distributions}, we know $\EDS(W) = \pm \calO \left( e^{-\Theta(n)} \right)$ for each $W \subseteq \calA, |W| \geq 2$. Therefore $\ED_{\vec{u}}(\pi^n) = \pm \calO \left( e^{-\Theta(n)} \right)$ for any such distribution $\pi$.
\end{paragraph}

\begin{paragraph}{Step 2: Two-way ties}
Next, consider any $\pi$ such that $W^*(\pi) = \{1,2\}$. Using the notation from the above remark, this entails that $\pi_1 + \pi_5 = \pi_2 + \pi_6 > \pi_3 + \pi_4$. From the above discussion, this immediately implies $\EDS(W) = \pm \calO \left( e^{-\Theta(n)} \right)$ for each $W \in 2^{\calA} \backslash \{1,2\}, |W| \geq 2$. The asymptotic rate of the term $\EDS(\{1,2\})$ is presented in Table \ref{tab:eadpoa_full}, when $u_1 \geq u_2 > u_3$, and Table \ref{tab:eadpoa_full_u1main}, when $u_1 > u_2 \geq u_3$, for both even and odd subsequence of $(n)_{n \in \mathbb{N}}$. This is due to Lemma \ref{lem:sub_12} for even $n$ and Lemma \ref{lem:sub_12_odd} for odd $n$. Note that Lemma \ref{lem:sub_12_non_34_equal} serves as a supplementary lemma that proves the cases when $\pi_3 \neq \pi_4$.

We read Table \ref{tab:eadpoa_full} (when $u_1 \geq u_2 > u_3$) column-wise to identify the appropriate conditions on $\pi$, and then row-wise to determine the asymptotic rate. For the case of $\EDS(\{1,2\})$, we only read the left-most and right-most columns; this conveys the information as Table \ref{tab:eadpoa_case_12_tldr} in the main text and Table \ref{tab:eadpoa_case_12} in Appendix \ref{apx:two_ties}.
For example, if both $\pi_3 = \pi_4$ and $4 \pi_1 + \pi_2 + 3 \pi_5 > 2$, then $\EDS(\{1,2\})$ is $\Theta(1)$ if $n$ is even and $-\Theta(1)$ if $n$ is odd.
On the other hand, if both $\pi_3 > \pi_4$ and $\pi_1 + 2 \pi_4 < \pi_2 + 2 \pi_3$, then $\EDS(\{1,2\}) = \Theta(\sqrt{n})$.
Note that this table is exhaustive.
Table \ref{tab:eadpoa_full_u1main} (when $u_1 > u_2 = u_3$) is read similarly, but is much simpler. Specifically, $\EDS(\{1,2\}) = \pm \calO \left( \frac{1}{n} \right)$ if $\pi_3 \leq \pi_4$ and $\pm \calO \left( e^{-\Theta(n)} \right)$ otherwise.

\begin{table*}[t]
    \centering
    \small
    \begin{tabular}{|c|c|c|c|}
        \hline
         \makecell{$\EDS(\{1,2\})$ when \\$\pi_1 + \pi_5 = \pi_2 + \pi_6$ \\ $\geq \pi_3 + \pi_4$} & \makecell{$\EDS(\{1,3\})$ when \\ $\pi_1 + \pi_5 = \pi_3 + \pi_4$ \\ $\geq \pi_2 + \pi_6$} & \makecell{$\EDS(\{2,3\})$ when \\ $\pi_2 + \pi_6 = \pi_3 + \pi_4$ \\ $\geq \pi_1 + \pi_5$} & Asymptotic Rate \\
          \hline \hline
        \makecell{$\begin{cases}\pi_3 = \pi_4 \\ 4\pi_1 + \pi_2 + 3 \pi_5 > 2 \end{cases}$} & \makecell{$\begin{cases}\pi_2 = \pi_6 \\ 4\pi_5 + \pi_3 + 3 \pi_1 > 2 \end{cases}$} & \makecell{$\begin{cases}\pi_5 = \pi_1 \\ 4\pi_2 + \pi_4 + 3 \pi_6 > 2 \end{cases}$} & $\begin{cases}\Theta(1),& n \text{ is even} \\-\Theta(1), & n \text{ is odd}\end{cases}$\\
        \hline
        \makecell{$\begin{cases}\pi_3 = \pi_4 \\ 4\pi_1 + \pi_2 + 3 \pi_5 < 2 \end{cases}$} & \makecell{$\begin{cases}\pi_2 = \pi_6 \\ 4\pi_5 + \pi_3 + 3 \pi_1 < 2 \end{cases}$} & \makecell{$\begin{cases}\pi_5 = \pi_1 \\ 4\pi_2 + \pi_4 + 3 \pi_6 < 2 \end{cases}$} & $-\Theta(1)$\\
        \hline
        \makecell{$\begin{cases}\pi_3 = \pi_4 \\ 4\pi_1 + \pi_2 + 3 \pi_5 = 2 \end{cases}$} & \makecell{$\begin{cases}\pi_2 = \pi_6 \\ 4\pi_5 + \pi_3 + 3 \pi_1 = 2 \end{cases}$} & \makecell{$\begin{cases}\pi_5 = \pi_1 \\ 4\pi_2 + \pi_4 + 3 \pi_6 = 2 \end{cases}$} & $\begin{cases}\pm \calO \left( \frac{1}{\sqrt{n}} \right),& n \text{ is even} \\ - \Theta(1),& n \text{ is odd}\end{cases}$ \\
        \hline
        \makecell{$\begin{cases}\pi_3 \neq \pi_4 \\ \pi_1 + 2\pi_4 = \pi_2 + 2 \pi_3 \end{cases}$} & \makecell{$\begin{cases}\pi_2 \neq \pi_6 \\ \pi_5 + 2\pi_6 = \pi_3 + 2 \pi_2 \end{cases}$} & \makecell{$\begin{cases}\pi_5 \neq \pi_1 \\ \pi_2 + 2\pi_1 = \pi_4 + 2 \pi_5 \end{cases}$} & $\pm \calO \left( \frac{1}{\sqrt{n}}\right)$  \\
        \hline
        \makecell{$\begin{cases}\pi_3 > \pi_4 \\ \pi_1 + 2\pi_4 < \pi_2 + 2 \pi_3 \end{cases}$ \\ or \\ $\begin{cases}\pi_3 < \pi_4 \\ \pi_1 + 2\pi_4 > \pi_2 + 2 \pi_3 \end{cases}$} & \makecell{$\begin{cases}\pi_2 > \pi_6 \\ \pi_5 + 2\pi_6 < \pi_3 + 2 \pi_2 \end{cases}$ \\ or \\ $\begin{cases}\pi_2 < \pi_6 \\ \pi_5 + 2\pi_6 > \pi_3 + 2 \pi_2 \end{cases}$} & \makecell{$\begin{cases}\pi_5 > \pi_1 \\ \pi_2 + 2\pi_1 < \pi_4 + 2 \pi_5 \end{cases}$ \\ or \\ $\begin{cases}\pi_5 < \pi_1 \\ \pi_2 + 2\pi_1 > \pi_4 + 2 \pi_5 \end{cases}$} & $\Theta(\sqrt{n})$ \\
        \hline
        \makecell{$\begin{cases}\pi_3 > \pi_4 \\ \pi_1 + 2\pi_4 > \pi_2 + 2 \pi_3 \end{cases}$ \\ or \\ $\begin{cases}\pi_3 < \pi_4 \\ \pi_1 + 2\pi_4 < \pi_2 + 2 \pi_3 \end{cases}$} & \makecell{$\begin{cases}\pi_2 > \pi_6 \\ \pi_5 + 2\pi_6 > \pi_3 + 2 \pi_2 \end{cases}$ \\ or \\ $\begin{cases}\pi_2 < \pi_6 \\ \pi_5 + 2\pi_6 < \pi_3 + 2 \pi_2 \end{cases}$} & \makecell{$\begin{cases}\pi_5 > \pi_1 \\ \pi_2 + 2\pi_1 > \pi_4 + 2 \pi_5 \end{cases}$ \\ or \\ $\begin{cases}\pi_5 < \pi_1 \\ \pi_2 + 2\pi_1 < \pi_4 + 2 \pi_5 \end{cases}$} & $-\Theta(\sqrt{n})$ \\
        \hline
    \end{tabular}
    \caption{Asymptotic rate of $\EDS(\{1,2\})$, $\EDS(\{1,3\})$, and $\EDS(\{2,3\})$ given conditions on $\pi$ when $u_1 \geq u_2 > u_3$. The stated rate applies to each column, separately, when the corresponding conditions hold.}
    \label{tab:eadpoa_full}
\end{table*}

\begin{table*}[t]
    \centering
    \small
    \begin{tabular}{|c|c|c|c|}
        \hline
         \makecell{$\EDS(\{1,2\})$ when \\$\pi_1 + \pi_5 = \pi_2 + \pi_6$ \\ $\geq \pi_3 + \pi_4$} & \makecell{$\EDS(\{1,3\})$ when \\ $\pi_1 + \pi_5 = \pi_3 + \pi_4$ \\ $\geq \pi_2 + \pi_6$} & \makecell{$\EDS(\{2,3\})$ when \\ $\pi_2 + \pi_6 = \pi_3 + \pi_4$ \\ $\geq \pi_1 + \pi_5$} & Asymptotic Rate \\
          \hline \hline
        $\pi_3 \leq \pi_4$ & ${\pi}_2 \leq {\pi}_6$ & ${\pi}_5 \leq {\pi}_1$ & $\pm \calO \left( \frac{1}{\sqrt{n}} \right)$\\
        \hline
        $\pi_3 > \pi_4$ & ${\pi}_2 > {\pi}_6$ & ${\pi}_5 > {\pi}_1$ & $\pm \calO \left( e^{-\Theta(n)} \right)$\\
        \hline
    \end{tabular}
    \caption{Asymptotic rate of $\EDS(\{1,2\})$, $\EDS(\{1,3\})$, and $\EDS(\{2,3\})$ given conditions on $\pi$ when $u_1 > u_2=u_3$. The stated rate applies to each column, separately, when the corresponding conditions hold.} 
    \label{tab:eadpoa_full_u1main}
\end{table*}

The case of $\pi$ such that $W^*(\pi) = \{1,3\}$ is very similar to that of $W^*(\pi) = \{1,2\}$. In Lemma \ref{lem:sub_13} in Appendix \ref{apx:two_ties_other_cases}, we demonstrate a permutation $\pi \mapsto \pi'$ of the preference distribution that applies Lemmas \ref{lem:sub_12} and \ref{lem:sub_12_odd} to $W^*(\pi) = \{1,3\}$ in the same manner as $W^*(\pi') = \{1,2\}$, as described above. Specifically, consider the probability distribution $\pi$ such that $\pi_1 + \pi_5 = \pi_3 + \pi_4 > \pi_2 + \pi_6$. Let $\pi' = (\pi_1', \pi_2', \pi_3', \pi_4', \pi_5', \pi_6')$ be defined according to the permutation:
\[
\begin{pmatrix}
    \pi_5 \\ \pi_3 \\ \pi_2 \\ \pi_6 \\ \pi_1 \\ \pi_4
\end{pmatrix}
\mapsto
\begin{pmatrix}
    \pi_1' \\ \pi_2' \\ \pi_3' \\ \pi_4' \\ \pi_5' \\ \pi_6'
\end{pmatrix}.
\]
Then, in Lemma \ref{lem:sub_13}, we prove that Lemmas \ref{lem:sub_12} and \ref{lem:sub_12_odd} hold for $\pi'$.

The result of Lemma \ref{lem:sub_13} is that, when $\pi_1 + \pi_5 = \pi_3 + \pi_4 > \pi_2 + \pi_6$, the same rates described in Tables \ref{tab:eadpoa_full} and \ref{tab:eadpoa_full_u1main} that applied to $W^*(\pi) = \{1,2\}$ before now apply to $W^*(\pi) = \{1,3\}$, subject to this permutation over the preference distribution. Tables \ref{tab:eadpoa_full} and \ref{tab:eadpoa_full_u1main} are now read by their second and fourth columns and applied to $\EDS(\{1,3\})$. For example, if both $\pi_2 = \pi_6$ and $4 \pi_5 + \pi_3 + 3 \pi_1 < 2$, then $\EDS(\{1,3\})$ is $\Theta(1)$ if $n$ is even and $-\Theta(1)$ if $n$ is odd. We have $\EDS(W) = \pm \calO \left( e^{-\Theta(n)} \right)$ for each $W \in 2^{\calA} \backslash \{1,3\}, |W| \geq 2$ by Corollary \ref{coro:sub_others}.

The case of $\pi$ such that $W^*(\pi) = \{2,3\}$ is very similar to that of $W^*(\pi) = \{1,2\}$. In Lemma \ref{lem:sub_23} in Appendix \ref{apx:two_ties_other_cases}, we demonstrate a permutation $\pi \mapsto \pi'$ of the preference distribution that applies Lemmas \ref{lem:sub_12} and \ref{lem:sub_12_odd} to $W^*(\pi) = \{2,3\}$ in the same manner as $W^*(\pi') = \{1,2\}$, as described above. Specifically, consider the probability distribution $\pi$ such that $\pi_1 + \pi_5 = \pi_3 + \pi_4 > \pi_2 + \pi_6$. Let $\pi' = (\pi_1', \pi_2', \pi_3', \pi_4', \pi_5', \pi_6')$ be defined according to the permutation:
\[
\begin{pmatrix}
    \pi_2 \\ \pi_4 \\ \pi_5 \\ \pi_1 \\ \pi_6 \\ \pi_3
\end{pmatrix}
\mapsto
\begin{pmatrix}
    \pi_1' \\ \pi_2' \\ \pi_3' \\ \pi_4' \\ \pi_5' \\ \pi_6'
\end{pmatrix}.
\]
Then, in Lemma \ref{lem:sub_23}, we prove that Lemmas \ref{lem:sub_12} and \ref{lem:sub_12_odd} hold for $\pi'$. The corresponding result for $\EDS(\{2,3\})$ is read by the third and fourth columns of Tables \ref{tab:eadpoa_full} and \ref{tab:eadpoa_full_u1main}.

\end{paragraph}

\begin{paragraph}{Step 3: Three-way ties}
Finally, consider any $\pi$ such that $W^*(\pi) = \{1,2,3\}$. This entails that $\pi_1 + \pi_5 = \pi_2 + \pi_6 = \pi_3 + \pi_4 = \frac{1}{3}$. Under this class of preference distributions, \emph{any} of the above results that hold for $\EDS(\{1,2\})$, $\EDS(\{1,3\})$, or $\EDS(\{2,3\})$ from Tables \ref{tab:eadpoa_full} and \ref{tab:eadpoa_full_u1main} may be valid.

\begin{table}[t]
    \centering
    \footnotesize{}
    \begin{tabular}{|c|c||c|c|c|}
        \hline
         \multicolumn{2}{|c||}{Conditions} & $\pi_3 = \pi_4$ & $\pi_3 > \pi_4$ & $\pi_3 < \pi_4$ \\
          \hline \hline
          
          \multirow{3}{*}{$\pi_1 = \pi_5$}
        & $\pi_2 = \pi_6$ & 
        \makecell{$\begin{cases}f^1 = \pi_1 - \pi_5 \\ f^2 = \pi_6 - \pi_2 \\ f^3 = 3 \pi_4 - 5 \pi_3\end{cases}$} & 
        \makecell{$\begin{cases}f^1 = 2 \pi_1 + 2 \pi_5 + \pi_4 - 5 \pi_3 \\ f^2 = \pi_1 + 2 \pi_3 - 2 \pi_2 - \pi_5 \\ f^3 = 3 \pi_4 - 5 \pi_3\end{cases}$} & 
        \makecell{$\begin{cases}f^1 = \pi_1 - \pi_3 \\ f^2 = 2 \pi_2 + \pi_3 - 3 \pi_4 \\ f^3 = 3 \pi_4 - 5 \pi_3\end{cases}$} \\
        \cline{2-5}
        & $\pi_2 > \pi_6$ & 
        \makecell{$\begin{cases}f^1 = 2 \pi_1 + \pi_6 - 3 \pi_2 \\ f^2 = \pi_6 - \pi_2 \\ f^3 = \pi_2 - 2 \pi_1 - 3 \pi_6\end{cases}$} & 
        \makecell{$\begin{cases}f^1 = \pi_1 + \pi_6 - \pi_2 - \pi_3 \\ f^2 = \pi_1 + 2 \pi_3 - 2 \pi_2 - \pi_5 \\ f^3 = \pi_2 - 2 \pi_1 - 3 \pi_6\end{cases}$} & 
        \makecell{$\begin{cases}f^1 = \pi_1 + \pi_6 - \pi_2 - \pi_3 \\ f^2 = \pi_2 + \pi_3 - 3 \pi_4 \\ f^3 = \pi_2 - 2 \pi_1 - 3 \pi_6\end{cases}$} \\
        \cline{2-5}
        & $\pi_2 < \pi_6$ & 
        \makecell{$\begin{cases}f^1 = \pi_6 - \pi_1 \\ f^2 = \pi_6 - \pi_2 \\ f^3 = \pi_2 - \pi_3\end{cases}$} & 
        \makecell{$\begin{cases}f^1 = \pi_4 + \pi_6 - \pi_1 - \pi_3 \\ f^2 = \pi_1 + 2 \pi_3 - 2 \pi_2 - \pi_5 \\ f^3 = \pi_2 - \pi_3\end{cases}$} & 
        \makecell{$\begin{cases}f^1 = \text{N/A} \\ f^2 = \pi_2 + \pi_3 - 3 \pi_4 \\ f^3 = \pi_2 - \pi_3\end{cases}$} \\
         \hline  \hline

    \multirow{3}{*}{$\pi_1 > \pi_5$}
        & $\pi_2 = \pi_6$ & 
        \makecell{$\begin{cases}f^1 = \pi_1 - \pi_5 \\ f^2 = \pi_2 - \pi_4 \\ f^3 = \pi_4 - \pi_3 \end{cases}$} & 
        \makecell{$\begin{cases}f^1 = 2 \pi_1 + 2 \pi_5 + \pi_4 - 5 \pi_3 \\ f^2 = \text{N/A} \\ f^3 = \pi_4 - \pi_3 \end{cases}$} & 
        \makecell{$\begin{cases}f^1 = \pi_1 - \pi_3 \\ f^2 = \pi_1 + \pi_2 - 2 \pi_4 \\ f^3 = \pi_4 - \pi_3 \end{cases}$} \\
        \cline{2-5}
        & $\pi_2 > \pi_6$ & 
        \makecell{$\begin{cases}f^1 = 2 \pi_1 + \pi_6 - 3 \pi_2 \\ f^2 = \pi_2 - \pi_4 \\ f^3 = \pi_2 + \pi_6 - 2 \pi_1 \end{cases}$} & 
        \makecell{$\begin{cases}f^1 = \pi_1 + \pi_6 - \pi_2 - \pi_3 \\ f^2 = \text{N/A} \\ f^3 = \pi_2 + \pi_6 - 2 \pi_1 \end{cases}$} & 
        \makecell{$\begin{cases}f^1 = \pi_1 + \pi_6 - \pi_2 - \pi_3 \\ f^2 = \pi_1 + \pi_2 - 2 \pi_4 \\ f^3 = \pi_2 + \pi_6 - 2 \pi_1 \end{cases}$} \\
        \cline{2-5}
        & $\pi_2 < \pi_6$ & 
        \makecell{$\begin{cases}f^1 = \pi_6 - \pi_1 \\ f^2 = \pi_2 - \pi_4 \\ f^3 = \pi_2 + \pi_5 - 2 \pi_1 \end{cases}$} & 
        \makecell{$\begin{cases}f^1 = \pi_4 + \pi_6 - \pi_1 - \pi_3 \\ f^2 = \text{N/A} \\ f^3 = \pi_2 + \pi_5 - 2 \pi_1 \end{cases}$} & 
        \makecell{$\begin{cases}f^1 = \text{N/A} \\ f^2 = \pi_1 + \pi_2 - 2 \pi_4 \\ f^3 = \pi_2 + \pi_5 - 2 \pi_1 \end{cases}$} \\
         \hline \hline


    \multirow{3}{*}{$\pi_1 < \pi_5$}
        & $\pi_2 = \pi_6$ & 
        \makecell{$\begin{cases}f^1 = \pi_1 - \pi_5 \\ f^2 = 3 \pi_1 - 2 \pi_2 - \pi_5 \\ f^3 = \pi_1 - \pi_3 \end{cases}$} & 
        \makecell{$\begin{cases}f^1 = 2 \pi_1 + 2 \pi_5 + \pi_4 - 5 \pi_3 \\ f^2 = 2 \pi_1 - \pi_2 - \pi_4 \\ f^3 = \pi_1 - \pi_3 \end{cases}$} & 
        \makecell{$\begin{cases}f^1 = \pi_1 - \pi_3 \\ f^2 = 2 \pi_1 - \pi_2 - \pi_4 \\ f^3 = \pi_1 - \pi_3 \end{cases}$} \\
        \cline{2-5}
        & $\pi_2 > \pi_6$ & 
        \makecell{$\begin{cases}f^1 = 2 \pi_1 + \pi_6 - 3 \pi_2 \\ f^2 = 3 \pi_1 - 2 \pi_2 - \pi_5 \\ f^3 = \pi_2 - \pi_6 - 2 \pi_1 \end{cases}$} & 
        \makecell{$\begin{cases}f^1 = \pi_1 + \pi_6 - \pi_2 - \pi_3 \\ f^2 = 2 \pi_1 - \pi_2 - \pi_4 \\ f^3 = \pi_2 - \pi_6 - 2 \pi_1 \end{cases}$} & 
        \makecell{$\begin{cases}f^1 = \pi_1 + \pi_6 - \pi_2 - \pi_3 \\ f^2 = 2 \pi_1 - \pi_2 - \pi_4 \\ f^3 = \pi_2 - \pi_6 - 2 \pi_1 \end{cases}$} \\
        \cline{2-5}
        & $\pi_2 < \pi_6$ & 
        \makecell{$\begin{cases}f^1 = \pi_6 - \pi_1 \\ f^2 = 3 \pi_1 - 2 \pi_2 - \pi_5 \\ f^3 = \text{N/A} \end{cases}$} & 
        \makecell{$\begin{cases}f^1 = \pi_4 + \pi_6 - \pi_1 - \pi_3 \\ f^2 = 2 \pi_1 - \pi_2 - \pi_4 \\ f^3 = \text{N/A} \end{cases}$} & 
        \makecell{$\begin{cases}f^1 = \text{N/A} \\ f^2 = 2 \pi_1 - \pi_2 - \pi_4 \\ f^3 = \text{N/A} \end{cases}$} \\
         \hline
    \end{tabular}
    \caption{Values of $f^i(\pi_1, \ldots \pi_6)$, $i \in [3]$, given the relative values of $\pi_1$ to $\pi_5$, $\pi_2$ to $\pi_6$, and $\pi_3$ to $\pi_4$.}
    \label{tab:complexity_three_sides_main_body}
\end{table}

For example, consider the preference distribution IC where $\pi_1 = \ldots, \pi_6$. This satisfies the requirement that $\pi_1 + \pi_5 = \pi_2 + \pi_6 = \pi_3 + \pi_4 = \frac{1}{3}$. Then we know each of the following statements hold:
\begin{itemize}
    \item $\begin{cases}
        \pi_3 = \pi_4, \\
        4 \pi_1 + \pi_2 + 3 \pi_5 < 2,
    \end{cases}$
    \item $\begin{cases}
        \pi_2 = \pi_6, \\
        4 \pi_5 + \pi_3 + 3 \pi_1 < 2,
    \end{cases}$
    \item $\begin{cases}
        \pi_5 = \pi_1, \\
        4 \pi_2 + \pi_4 + 3 \pi_6 < 2
    \end{cases}$
\end{itemize}
since $\frac{1}{6}(4 + 1 + 3) = \frac{8}{6} < 2$. Therefore, when $u_1 \geq u_2 > u_3$, we have $\EDS(\{1,2\}) = - \Theta(1)$, $\EDS(\{1,3\}) = - \Theta(1)$, and $\EDS(\{2,3\}) = - \Theta(1)$.

In addition to these three terms, we also have $\EDS(\{1,2,3\})$ to consider. As discussed in the proof of Theorem \ref{thm:main}, there are three cases for $\EDS(\{1,2,3\})$: 
\begin{itemize}
    \item (i) when $n$ is divisible by $3$; this corresponds to the case of $\PW{\topRank{P}}$ when $f(\topRank{P}) = 1$ and is proved in Lemma \ref{lem:three_tie_priority1} in Appendix \ref{apx:three_ties_P1}; we denote this case $i=1$;
    \item (ii) when $n-2$ is divisible by $3$; this corresponds to the case of $\PW{\topRank{P}}$ when $f(\topRank{P}) = 2$ and is proved in Lemma \ref{lem:three_tie_priority3} in Appendix \ref{apx:three_ties_P3}; we denote this case $i=2$;
    \item (iii) when $n-1$ is divisible by $3$; this corresponds to the case of $\PW{\topRank{P}}$ when $f(\topRank{P}) = 3$ and is proved in Lemma \ref{lem:three_tie_priority2} in Appendix \ref{apx:three_ties_P2}; we denote this case $i=3$.
\end{itemize}
%
If $u_1 > u_2 = u_3$, we prove that 
\[
\EDS(\{1,2,3\}) = \begin{cases}
    
    \pm \calO \left( \frac{1}{n} \right),& i=2, \pi_3 \leq \pi_4 \\
    \pm \calO \left( \frac{1}{n} \right),& i=3, \{\pi_1 \geq \pi_5 \text{ and } \pi_2 \leq \pi_6\} \text{ or } \{\pi_1 \leq \pi_5 \text{ and } \pi_2 \geq \pi_6\} \\
    \pm \calO \left( e^{-\Theta(n)} \right) ,& \text{otherwise}.
\end{cases}
\]
If $u_1 \geq u_2 > u_3$, we prove that 
\[
 \begin{cases}
    \pm \calO \left( e^{-\Theta(n)} \right),& i=1, \pi_2 < \pi_6, \text{ and } \pi_3 < \pi_4 \\
    \pm \calO \left( e^{-\Theta(n)} \right),& i=2, \pi_5 < \pi_1, \text{ and } \pi_4 < \pi_3 \\
    \pm \calO \left( e^{-\Theta(n)} \right),& i=3, \pi_1 < \pi_5, \text{ and } \pi_2 < \pi_6 \\
    f^i(\pi) \Theta(1) + g^i_n(\pi),& \text{otherwise}
\end{cases}
\]
where $f^i(\pi_1, \ldots, \pi_6)$ is presented in Table \ref{tab:complexity_three_sides_main_body} and 
\begin{itemize}
    \item $g^1_n(\pi_1, \ldots, \pi_6) = \begin{cases}
    \Theta(1), & \pi_1 + \pi_3 < \pi_2 + \pi_5 \\
    \pm \calO \left( \frac{1}{\sqrt{n}} \right), & \pi_1 + \pi_3 \geq \pi_2 + \pi_5,
    \end{cases}$
    \item $g^2_n(\pi_1, \ldots, \pi_6) = \begin{cases}
    \Theta(1), & \pi_2 + \pi_3 < \pi_1 + \pi_6 \\
    \pm \calO \left( \frac{1}{\sqrt{n}} \right), & \pi_2 + \pi_3 \geq \pi_1 + \pi_6,
    \end{cases}$
    \item $g^3_n(\pi_1, \ldots, \pi_6) = \pm \calO \left( \frac{1}{\sqrt{n}} \right)$.
\end{itemize}
%
Table \ref{tab:complexity_three_sides_main_body} is read by comparing the relative values of $\pi_1$ to $\pi_5$, $\pi_2$ to $\pi_6$, and $\pi_3$ to $\pi_4$. For example, if $\pi_1 = \pi_5$, $\pi_2 > \pi_6$, and $\pi_3 < \pi_4$, then 
\begin{itemize}
    \item $f^1(\pi) = \pi_1 + \pi_6 - \pi_2 - \pi_3$, 
    \item $f^2(\pi) = \pi_2 + \pi_3 - 3 \pi_4$,
    \item $f^3(\pi) = \pi_2 - 2 \pi_1 - 3 \pi_6$.
\end{itemize}
If $n$ is divisible by $3$ (so that $i=1$) 
and $\pi_1 + \pi_6 - \pi_2 - \pi_3 = 0$, then $\EDS(\{1,2,3\}) = g^1_n(\pi)$. Otherwise, $\EDS(\{1,2,3\}) \in \{\Theta(1), -\Theta(1)\}$ depending on whether $\pi_1 + \pi_6 - \pi_2 - \pi_3 > 0$ or not.

This concludes the summary of Theorem \ref{thm:main}.
\end{paragraph}



%% file: EC_appendix/apx_page_1.tex
\newpage
\section{Smoothed Analysis Lemmas}
\label{apx:smoothed_analysis_full}



Lemma \ref{lem:prob_bounds} in Appendix \ref{apx:helper_lemmas} is based on the \emph{smoothed likelihood of ties}, as introduced by \citet{Xia2021:How-Likely}. There, we represent the likelihood that a two-way tied plurality election occurs out of $m=3$ alternatives, subject to an additional constraint that there are the same number of agents with rankings for the third alternative. This additional constraint reduces the likelihood from $\Theta \left( \frac{1}{\sqrt{n}} \right)$, without the additional constraint, to $\Theta \left( \frac{1}{n} \right)$, as long as the summation region contains the probability distribution $\pi$; the likelihood is exponentially small otherwise. 
In the lemma, we sum an index $q$ over the probabilities of attaining $\frac{n}{2}-q$ agents with rankings $R_1$ or $R_5$ (with probability $\pi_1$), $\frac{n}{2}-q$ agents with rankings $R_2$ or $R_6$ (with probability $\pi_1$), and $q$ agents for either ranking $R_3$ and $R_4$ (with probability $\pi_3$ each).
That is, we sum over $\binom{n}{\frac{n}{2}-q, \frac{n}{2}-q, q, q} \pi_1^{n-2q} \pi_3^{2q}$ to constitute a tied election between alternatives $1$ and $2$ with the additional constraint equalizing the number of rankings $R_3$ and $R_4$.
This represents a summation over four indices, rather than the six rankings when there are three alternatives. 
The stated condition, for whether the summation is polynomially small or exponentially small, holds because $q=\pi_3 n$ is the center of this multinomial mass function.

Since this lemma and Corollary \ref{coro:sub_others}, in the paper's main body,
depend heavily on results from \citet{Xia2021:How-Likely}, we restate their main results (without proof) and their necessary preliminaries for completeness. We first recall some notation about the asymptotic behavior of sequences from calculus.




\subsection{Asymptotic Analysis}
\label{sec:asymptotic_analysis}


In this paper we explore the long-term behavior of sequences in the limit of large numbers of agents $n \in \mathbb{N}$. 
We aim to quantify how quickly sequences converge to certain values or diverge to $\pm \infty$, or if sequences are bounded, so that we may compare them. For example, the sequence $(\log n)_{n \in \mathbb{N}}$ diverges slower than $(n^2)_{n \in \mathbb{N}}$, which diverges slower than $(e^n)_{n \in \mathbb{N}}$. The nomenclature of Big-$\calO$ notation enables us to make these comparisons.

\begin{dfn}
Let $f$ and $g$ be real-valued functions. We say that $f(n) = \calO(g(n))$ if $\exists N > 0$ and $C \geq 0$ such that $\forall n > N$, $0 \leq f(n) \leq C g(n)$.
\end{dfn}

For example, $f(n) = n^2 + 2n = \calO(n^2)$ since $f(n) \leq 2 n^2$, $\forall n > 2$. 
One useful application of big-$\calO$ notation is to describe Maclaurin series. For example, $e^x = \sum_{n=0}^{\infty} \frac{x^n}{n!}$.
Hence,
\begin{equation}
    e^{-\frac{1}{n}} = 1 - \frac{1}{n} + \calO \left( \frac{1}{n^2} \right).
\end{equation}
Big-$\calO$ is often used to evaluate the asymptotic runtime of algorithms. In our case, we use it to describe the asymptotic economic efficiency of IV. Hence, $f(n)$ may be non-positive. We use the following notation to describe combined positive and negative bounds on $f(n)$.


\begin{dfn}
Let $f$ and $g$ be real-valued functions.
We say that $f(n) = \pm \calO(g(n))$ if $\exists N > 0$ and $C \geq 0$ such that $\forall n > N$, $|f(n)| \leq C g(n)$.
\end{dfn}

Equivalently, we have that $|f(n)| = \calO(g(n))$. For example, $f(n) = n \cdot cos(n) = \pm\calO(n)$ since $-n \leq f(n) \leq  n$, $\forall n > 0$.
The next two definitions describe asymptotic lower-and tight-bounds on functions.

\begin{dfn}
Let $f$ and $g$ be real-valued functions.
We say that $f(n) = \Omega(g(n))$ if $\exists N > 0$ and $C \geq 0$ such that $\forall n > N$, $f(n) \geq C g(n) \geq 0$.
\end{dfn}

For example, $f(n) = n^2 + 2n = \Omega(n)$ since $f(n) \geq 2 n$, $\forall n > 0$. Notice also that saying $f(n) = -\Omega(g(n))$ is equivalent to $-f(n) =\Omega(g(n))$.

\begin{dfn}
Let $f$ and $g$ be real-valued functions.
We say that $f(n) = \Theta(g(n))$ if $f(n) = \calO(g(n))$ and $f(n) = \Omega(g(n))$.
\end{dfn}

This entails that $f(n) = \pm \Theta(g(n))$ if $f(n) \in \{\Theta(g(n)), -\Theta(g(n))\}$. Notice that Big-$\calO$ and Big-$\Omega$ notation do not describe smallest-upper-bounds or largest-lower-bounds like the \emph{supremum} and \emph{infimum} attributes. Hence, we have that $f(n) = n^2 + 2n = \Theta(n^2)$ since $f = \calO(n^2)$ and $f = \Omega(n^2)$.

Finally, we write $\calO \left( e^{-\Theta(n)} \right)$ for $\calO(e^{-g(n)})$, where $g(n) = \Theta(n)$. A function $f(n)$ is then $\pm \calO \left( e^{-\Theta(n)} \right)$ if $|f(n)| \leq \calO(e^{-g(n)})$. We write $f(n) \sim g(n)$ if $\lim_{n \rightarrow \infty} \frac{f(n)}{g(n)} = 1$.

\subsubsection{Little-$o$ Notation}

Little-$o$ notation  compares the asymptotic rate of functions such that one pales in comparison to another.
\begin{dfn}
Let $f$ and $g$ be real-valued functions.
We say that $f(n) = o(g(n))$ if $\forall \epsilon > 0$, $\exists N > 0$ such that $\forall n > N$, $|f(n)| \leq \epsilon |g(n)|$.
When $g(n)$ does not vanish, we may write
$\lim_{n \rightarrow \infty} \frac{f(n)}{g(n)} = 0.$
\end{dfn}

For example, $f(n) = \sqrt{np(1-p)}$ for $p \in (0,1)$ is $o(n)$ since $\lim_{n \rightarrow \infty} \frac{\sqrt{np(1-p)}}{n} = 0$.

\subsubsection{Asymptotic Multiplication}

Let $f_1(n) = \calO(g_1(n))$, $f_2(n) = \calO(g_2(n))$, $f_3(n) = \Theta(g_3(n))$ and $f_4(n) = \Theta(g_4(n))$. Then by these definitions we have 
\begin{itemize}
    \item $f_1(n) \cdot f_2(n) = \calO(g_1(n) \cdot g_2(n))$,
    \item $f_3(n) \cdot f_4(n) = \Theta(g_3(n) \cdot g_4(n))$,
    \item $f_1(n) \cdot f_3(n) = \calO(g_1(n) \cdot g_3(n))$.
\end{itemize}
To be more precise, we give the examples of  $f_1(n) = n^2 + 2n$, $f_2(n) = \log(n)$, and $f_3(n) = \calO(1)$. It is clear that $f_1(n) \cdot f_2(n) = \Theta(n^2 \log(n))$. We can say that $f_1(n) \cdot f_3(n) = \calO(n^2)$ but not that it is $\Theta(n^2)$. This is because we do not have enough information about the lower-bound $\Omega(f_3(n))$. It holds that $f_1(n) \cdot f_3(n) = \Theta(n^2)$ if $f_3(n) = \Theta(1)$, whereas $f_1(n) \cdot f_3(n) = \Theta(n)$ if $f_3(n) = \Theta\left( \frac{1}{n} \right)$, and thirdly $f_1(n) \cdot f_3(n) = \Theta(1)$ if $f_3(n) = \Theta\left( \frac{1}{n^2} \right)$.

\subsection{Smoothed Likelihood of Ties}
\label{apx:smoothed_prelims}


A tied election is a characterization on the histogram of a preference profile satisfying certain criterion. With positional scoring rules $f_{\vec s}$, for instance, a \emph{$W$-way tie} (i.e., a \emph{$k$-way tie} between the alternatives $W \subseteq \calA$, $|W|=k$) is the event that these alternatives have the same score and that this score is strictly greater than those of other alternatives (subject to minor variation due to the tie-breaking rule). 
This may be characterized as a system of linear constraints on the multiplicity of rankings in $P$, as described by \citet{Xia2021:How-Likely}, as follows.

\begin{dfn}[\bf Score difference vector]\label{dfn:varcons} For any scoring vector $\vec s$ and pair $u,v \in \calA$, let $\score_{u,v}^{\vec s}$ denote the $m!$-dimensional vector indexed by rankings in $\ml(\calA)$ such that $\forall R\in\ml(\calA)$, the $R$-component of $\score_{u,v}^{\vec s}$ is $s_{R[u]} - s_{R[v]}$, where $R[c]$ is the index of $c$ in $R$. 
\end{dfn}

Let $Hist(P) = (x_{R}:R\in\ml(\calA))$ denote the vector of $m!$ variables, each of which represents the multiplicity of a linear order in a profile $P$. Therefore, $\score_{u,v}^{\vec s}\cdot Hist(P)$ represents the score difference between $u$ and $v$ in $P$. 
For any $W\subseteq \calA$, we define the polyhedron $\calH^{\vec s,W}$ as follows.

\begin{dfn}\label{dfn:convscore}
Let $\mathbf{E}^{\vec s, W}$ denote the matrix whose row vectors are $\{\score_{u,v}^{\vec s}: u\in W, v\in W, u\neq v\}$. 
Let $\mathbf{S}^{\vec s, W}$ denote the matrix whose row vectors are $\{\score_{u,v}^{\vec s}: u\not\in W, v\in W\}$. Let 
$
    \mathbf{A}^{\vec s, W} = \begin{bmatrix} \mathbf{E}^{\vec s, W} \\ 
    \mathbf{S}^{\vec s, T}
    \end{bmatrix}
$,  
$
    \vec b = \begin{bmatrix}
        \vec 0 \\
        -\vec 1
    \end{bmatrix}
$, and let 
$
    \calH^{\vec s, W} = \{\vec{x} \in \mathbb{R}^{m!}~:~ \mathbf{A}^{\vec s, W} \vec{x} \leq \vec{b}\}
$ denote the corresponding polyhedron.
\end{dfn}

It follows that the alternatives $W$ are tied in $f_{\vec s}(P)$ (notwithstanding any tie-breaking) if and only if $Hist(P) \in \calH^{\vec s,W}$.
The following example characterizes a plurality tie
between alternatives $1$ and $2$ with this polyhedral representation. We denote the plurality score vector by $\vec{s}_{plu} = (1,0,\ldots,0)$.


\begin{ex}[Polyhedral representation of a $\{1,2\}$-way plurality tie]
Let $m=3$ and consider the truthful vote profile $\topRank{P}$ for some $P$.
Then a $W$-way tied plurality election for $W = \{1,2\}$ occurs if and only if $Hist(P)$ is in a polyhedron $\calH^{\vec{s}_{plu}, W}$ represented by the following inequalities:
\begin{equation}
\begin{split}
    x_{123} + x_{132} - x_{213} - x_{231} & \leq 0; \\
    -x_{123} - x_{132} + x_{213} + x_{231} & \leq 0; \\
    - x_{123} - x_{132} + x_{312} + x_{321} & \leq -1; \\
    - x_{213} - x_{231} + x_{312} + x_{321} & \leq -1.
\end{split}
\end{equation}
The variables are $\vec{x} = (x_{123}, x_{132}, x_{213}, x_{231}, x_{312}, x_{321})$ where $x_{xyz}$ corresponds to the number of rankings in $P$ with ranking $(x \succ y \succ z)$.
The first two inequalities state that alternatives $1$ and $2$ have the same plurality score, while the last two inequality states that alternative $3$ has a strictly smaller plurality score than alternatives $1$ and $2$. This suggests that $\calH^{\vec{s}_{plu}, W} = \{ \vec{x} \in \mathbb{R}^{6} ~:~ \mathbf{A}^{\vec{s}_{plu}, W}  \vec{x} \leq \vec{b} \}$ where
\begin{equation}
\mathbf{A}^{\vec{s}_{plu}, W} = \begin{bmatrix}
    1 & 1 & -1 & -1 & 0 & 0 \\
    -1 & -1 & 1 & 1 & 0 & 0 \\
    -1 & -1 & 0 & 0 & 1 & 1 \\
    0 & 0 & -1 & -1 & 1 & 1 \\
\end{bmatrix}, \quad \vec{b} = \begin{bmatrix}
    0 \\
    0 \\
    -1 \\
    -1
\end{bmatrix}.
\end{equation}
\end{ex}
Following this example, for the plurality score vector $\vec{s}_{plu}$, general $m \geq 3$, and $W \subseteq \calA$, the polyhedron $\calH^{\vec s_{plu},W}$ is represented by the following inequalities:
\begin{align}
    & \forall \{i_1,i_2\}\subseteq W\text{ s.t. }i_1\neq i_2, 
    \sum\nolimits_{R:\topRank{R} = i_1} x_R - \sum\nolimits_{R:\topRank{R} = i_2} x_R \le 0; \notag \\
    & \forall i_1\in W, i_2\in \calA \backslash W, 
    \sum\nolimits_{R:\topRank{R} = i_2} x_R - \sum\nolimits_{R:\topRank{R} = i_1} x_R \le -1.
\end{align}
These inequalities cover the case of $\PW{\topRank{P}} = W$ such that all alternatives in $W$ have the same score. In fact, there are $|W|$ possible cases depending on which alternative $f(\topRank{P}) \in \PW{\topRank{P}}$ is the winner. The other cases may be characterized by modifying $\vec{b}$ accordingly. For example, if $m > 3$, consider $\PW{\topRank{P}} = W$ with $W = \{1,2,3\}$ such that $s_1(\topRank{P}) +1 = s_2(\topRank{P}) = s_3(\topRank{P})$. Then $\calH^{\vec s_{plu},W}$ would be represented by the inequalities:
\begin{equation}
\begin{split}
     & \sum\nolimits_{R:\topRank{R} = 1} x_R - \sum\nolimits_{R:\topRank{R} = 2} x_R \le -1;\\
     & \sum\nolimits_{R:\topRank{R} = 2} x_R - \sum\nolimits_{R:\topRank{R} = 1} x_R \le 1;\\
     & \sum\nolimits_{R:\topRank{R} = 2} x_R - \sum\nolimits_{R:\topRank{R} = 3} x_R \le 0;\\
     & \sum\nolimits_{R:\topRank{R} = 3} x_R - \sum\nolimits_{R:\topRank{R} = 2} x_R \le 0;\\
    \forall  i\in [4,m],&\sum\nolimits_{R:\topRank{R} = i} x_R - \sum\nolimits_{R:\topRank{R} = 2} x_R \le -1.
\end{split}
\end{equation}
This accounts for any variation due to the tie-breaking rule. 







This polyhedral representation of agents' preferences is described in the \emph{smoothed analysis} work of \citet{Xia2020:The-Smoothed} and \citet{Xia2021:How-Likely}. Xia studied how likely large elections are tied according to several voting rules when preferences are independently (but not necessarily identically) distributed: $P \sim \vec{\pi}$ where $\forall j \leq n$, $R_j \sim \vec{\pi}(j) \in \Delta(\calL(\calA))$, the probability simplex over $\calL(\calA)$. This problem has been studied extensively in the public choice literature (see e.g., \citet{beck1975note}, \citet{Gillett1977:Collective}, \citet{margolis1977probability}, \citet{gillett1980comparative}, \citet{chamberlain1981note}, and \citet{marchant2001probability}).
\citet{Xia2021:How-Likely} solved this problem beyond the prior work by recognizing that the histogram of a randomly generated preference profile is a \emph{Poisson multivariate variable (PMV)}. 
A tied election of the alternatives $W$, then, is that PMV occurring within the polyhedron $\calH^{\vec{s}, W}$.
To determine the likelihood of this event, \citet{Xia2021:How-Likely} defined the \emph{PMV-in-polyhedron problem} as $\Pr_{P \sim {\vec{\pi}}}( Hist(P) \in \calH)$ for any polyhedron $\calH$, taken in supremum or infimum over distributions $\vec{\pi} \in \Pi^n$, and proved a dichotomy theorem for conditions on this likelihood. The following definitions are used to formally describe his main result.

\begin{dfn}[Poisson multivariate variables (PMVs)]
    Given $\mu, n \in \mathbb{N}$ and distribution $\vec{\pi}$ over $[\mu]$, let $\vec{X}_{\vec{\pi}}$ denote the $(n,\mu)\text{-PMV}$ that corresponds to $\vec{\pi}$. That is, let $Y_1, \ldots, Y_n$ denote $n$ independent random variables over $[\mu]$ such that for any $j \leq n$, $Y_j$ is distributed as $\vec{\pi}(j)$. For any $i \in [\mu]$, the $i$-th component of $\vec{X}_{\vec{\pi}}$ is the number of $Y_j$'s that take value $i$.
    \label{dfn:pmv}
\end{dfn}

Given $\mu, L, n \in \mathbb{N}$, an $L \times \mu$ matrix $\mathbf{A}$, and an $L$-dimensional vector $\vec{b}$, we define $\calH, \calH_{\leq 0}, \calH_n$ and $\calH_n^{\mathbb{Z}}$ as follows:
\begin{align}
    & \calH = \left\{ \vec{x} \in \mathbb{R}^\mu ~:~ \mathbf{A} \vec{x} \leq \vec{b} \right\};
    & \calH_{\leq 0} = \left\{ \vec{x} \in \mathbb{R}^\mu ~:~ \mathbf{A} \vec{x} \leq \vec{0} \right\}; \notag \\
    & \calH_n = \left\{ \vec{x} \in \calH \cap \mathbb{R}^\mu_{\geq 0} ~:~ \vec{x} \cdot \vec{1} = n \right\};
    & \calH^{\mathbb{Z}}_n = \calH_n \cap \mathbb{Z}^\mu_{\geq 0}.
\end{align}
That is, $\calH$ is the polyhedron represented by $\mathbf{A}$ and $\vec{b}$; $\calH_{\leq 0}$ is the \emph{characteristic cone} of $\calH$, $\calH_n$ consists of non-negative vectors in $\calH$ whose $L_1$ norm is $n$, and $\calH^{\mathbb{Z}}_n$ consists of non-negative integer vectors in $\calH_n$. By definition, $\calH^{\mathbb{Z}}_n \subseteq \calH_n \subseteq \calH$. Let $\text{dim}(\calH_{\leq 0})$ denote the dimension of $\calH_{\leq 0}$, i.e., the dimension of the minimal linear subspace of $\mathbb{R}^\mu$ that contains $\calH_{\leq 0}$.
%
For a set $\Pi$ of distributions over $[\mu]$, $CH(\Pi)$ denotes the convex hull of $\Pi$. $\Pi$ is called \emph{strictly positive} (by $\epsilon > 0$) if $\forall \vec{\pi} \in \Pi, \forall j \in [\mu], \vec{\pi}(j) > \epsilon$.


\setcounter{thm}{1}
\begin{thm}[\citet{Xia2021:How-Likely}, Theorem 1]
Given any $\mu \in \mathbb{N}$, any closed and strictly positive $\Pi$ over $[\mu]$, and any polyhedron $\calH$ characterized by a matrix $\mathbf{A}$, for any $n \in \mathbb{N}$,
\begin{align*}
    & \sup\nolimits_{\vec{\pi} \in \Pi^n} \Pr \left(\vec{X}_{\vec{\pi}} \in \calH \right) =  \begin{cases}
        0, & \text{if } \calH^{\mathbb{Z}}_n = \emptyset \\
        \calO \left( e^{-\Theta(n)} \right), & \text{if } \calH^{\mathbb{Z}}_n \neq \emptyset \text{ and } \calH_{\leq 0} \cap CH(\Pi) = \emptyset\\
        \Theta \left( n^{\frac{dim(\calH_{\leq 0}) - \mu}{2}}\right), &  
        \text{otw. }(\text{i.e.}, \calH^{\mathbb{Z}}_n \neq \emptyset,~ \calH_{\leq 0} \cap CH(\Pi) \neq \emptyset);
    \end{cases}
\end{align*}
\begin{align*}
    & \inf\nolimits_{\vec{\pi} \in \Pi^n} \Pr \left(\vec{X}_{\vec{\pi}} \in \calH \right) =  \begin{cases}
        0, & \text{if } \calH^{\mathbb{Z}}_n = \emptyset \\
        \calO \left( e^{-\Theta(n)} \right), & \text{if } \calH^{\mathbb{Z}}_n \neq \emptyset \text{ and } CH(\Pi) \nsubseteq \calH_{\leq 0} \\
        \Theta \left( n^{\frac{dim(\calH_{\leq 0}) - \mu}{2}}\right), & \text{otw. } (\text{i.e.}, \calH^{\mathbb{Z}}_n \neq \emptyset,~ CH(\Pi) \subseteq \calH_{\leq 0}).
    \end{cases}
\end{align*}
\label{thm:smoothed-likelihood}
\end{thm}
%
\citet{Xia2021:How-Likely} used this theorem to depict the likelihood of $k$-way ties according to several voting rules. In particular, the likelihood of $k$-way plurality ties with i.i.d. preferences corresponds to $\Pr_{P \sim \pi^n}(Hist(P) \in \calH^k)$, where $\pi^n = (\pi, \pi, \ldots, \pi)$ and $\calH^k = \bigcup\nolimits_{W \subseteq 2^{\calA}:|W|=k} \calH^{\vec{s}_{plu}, W}$. In this case, $\Pi = \{\pi\}$ consists of a single distribution $\pi \in \Delta(\calL(\calA))$ and the two probabilities of Theorem \ref{thm:smoothed-likelihood} coincide;
$Hist(P)$ then follows the multinomial distribution \citep{daskalakis2016size}.
The following corollary holds for either $\calH^k$ or $\calH^{\vec{s}_{plu}, W}$ that corresponds to any case of $\PW{\topRank{P}} = W$ with $|W|=k$.

\begin{coro}[\citet{Xia2021:How-Likely}, Corollary 1]
Fix $m \geq 3$ and let $n \in \mathbb{N}$ agents' preferences be i.i.d. according to IC. Then the likelihood of a $k$-way plurality tied election is $\Theta \left( n^{-{\frac{k-1}{2}}} \right)$. 
\label{coro:smoothed_ties}
\end{coro}

The probability of a $2$- or $3$-way tie with respect to IC is therefore $\Theta \left( \frac{1}{\sqrt{n}} \right)$ or $\Theta \left( \frac{1}{n} \right)$, respectively.

%% file: EC_appendix/apx_page_2a.tex
\newpage

\section{Two-way tie proofs}
\label{apx:two_ties}

This appendix introduces the primary lemmas for two-way ties that are used to prove Theorem \ref{thm:main}. First, in Appendix \ref{apx:two_ties_primary}, we prove Lemma \ref{lem:sub_12}, which demonstrates the asymptotic rate of $\EDS(\{1,2\})$ when $n$ is even. This is the more general version of Lemma \ref{lem:sub_12_TLDR} in the main text. Second, in Appendix \ref{apx:two_ties_n_odd}, we prove Lemma \ref{lem:sub_12_odd}, which covers the case when $n$ is odd. Finally, in Appendix \ref{apx:two_ties_other_cases}, we prove the other cases for two-way ties. This includes $\EDS(\{1,3\})$ in Lemma \ref{lem:sub_13}, $\EDS(\{2,3\})$ in Lemma \ref{lem:sub_23}, and a technical supplementary lemma for the $\EDS(\{1,2\})$ case when $\pi_3 \neq \pi_4$ in Lemma \ref{lem:sub_12_non_34_equal}.

First, recall the correspondence introduced in Appendix \ref{apx:appendix_contents}.
\begin{remark}
Preference distribution 
$\pi = (\pi_1, \ldots, \pi_6)$ corresponds to the rankings
\begin{align*}
        & R_1 = (1\succ 2\succ 3); & R_5 = (1\succ 3\succ 2) \\
        & R_2 = (2\succ 3\succ 1); & R_6 = (2\succ 1\succ 3) \\
        & R_3 = (3\succ 2\succ 1); & R_4 = (3\succ 1\succ 2). 
    \end{align*}
\end{remark}

\subsection{Primary lemma for two-way ties}
\label{apx:two_ties_primary}

\begin{table*}[t]
    \centering
    \small
    \begin{tabular}{|c|c|}
        \hline
         \makecell{$\EDS(\{1,2\})$ when \\$\pi_1 + \pi_5 = \pi_2 + \pi_6 \geq \pi_3 + \pi_4$} & Asymptotic Rate \\
          \hline \hline
        \makecell{$\begin{cases}\pi_3 = \pi_4 \\ 4\pi_1 + \pi_2 + 3 \pi_5 > 2 \end{cases}$} & $\begin{cases}\Theta(1),& n \text{ is even} \\-\Theta(1), & n \text{ is odd}\end{cases}$\\
        \hline
        \makecell{$\begin{cases}\pi_3 = \pi_4 \\ 4\pi_1 + \pi_2 + 3 \pi_5 < 2 \end{cases}$} & $-\Theta(1)$\\
        \hline
        \makecell{$\begin{cases}\pi_3 = \pi_4 \\ 4\pi_1 + \pi_2 + 3 \pi_5 = 2 \end{cases}$} & $\begin{cases}\pm \calO \left( \frac{1}{\sqrt{n}} \right),& n \text{ is even} \\ - \Theta(1),& n \text{ is odd}\end{cases}$ \\
        \hline
        \makecell{$\begin{cases}\pi_3 \neq \pi_4 \\ \pi_1 + 2\pi_4 = \pi_2 + 2 \pi_3 \end{cases}$} & $\pm \calO \left( \frac{1}{\sqrt{n}}\right)$  \\
        \hline
        \makecell{$\begin{cases}\pi_3 > \pi_4 \\ \pi_1 + 2\pi_4 < \pi_2 + 2 \pi_3 \end{cases}$ or $\begin{cases}\pi_3 < \pi_4 \\ \pi_1 + 2\pi_4 > \pi_2 + 2 \pi_3 \end{cases}$} & $\Theta(\sqrt{n})$ \\
        \hline
        \makecell{$\begin{cases}\pi_3 > \pi_4 \\ \pi_1 + 2\pi_4 > \pi_2 + 2 \pi_3 \end{cases}$ or $\begin{cases}\pi_3 < \pi_4 \\ \pi_1 + 2\pi_4 < \pi_2 + 2 \pi_3 \end{cases}$} & $-\Theta(\sqrt{n})$ \\
        \hline
    \end{tabular}
    \caption{Asymptotic rate of $\EDS(\{1,2\})$ given conditions on $\pi$ when $u_1 \geq u_2 > u_3$.}
    \label{tab:eadpoa_case_12}
\end{table*}

\setcounter{lem}{2}
\begin{lem}
Suppose that $\pi_1 + \pi_5 = \pi_2 + \pi_6 \geq \pi_3 + \pi_4$ and $\pi_i > 0,~\forall i \in [6]$. Furthermore, let $u_1 \geq u_2 > u_3$ in $\vec{u}$. 
Then $\exists N > 0$ such that $\forall n > N$ that are even, $\EDS(\{1,2\})$ is determined by the mapping from $\pi$ to asymptotic rates as described by Table \ref{tab:eadpoa_case_12}.
Both conditions on $\pi$ need to hold; note that this table is exhaustive.
%
%

If $u_1 > u_2 = u_3$, then $\EDS(\{1,2\}) = \pm \calO\left( \frac{1}{\sqrt{n}} \right)$ if $\pi_3 \leq \pi_4$ and $\pm \calO \left( e^{-\Theta(n)} \right)$ otherwise.
\label{lem:sub_12}
\end{lem}

\begin{proof}

We prove the lemma by summing up the adversarial loss $\ADS(P)$ of every preference profile $P \in \calL(\calA)^n$ such that the potential winning set $\PW{P} = \{1,2\}$, weighted by their likelihood of occurrence. 
Recall that iterative plurality starting from the truthful vote profile $\topRank{P}$ consists of agents changing their votes from alternatives that were not already winning to those that then become the winner \citep{Branzei13:How}. This occurs until no agent has an incentive to change their vote. \citet{Lev12:Convergence} demonstrated that the equilibrium winning set $\EW{\topRank{P}}$ is a subset of the initial potential winning set $\PW{P}$. Subsequently, \citet[Lemma 1]{Kavner21:strategic} proved that $\EW{\topRank{P}}$ is the unique alternative with more agents preferring it (subject to lexicographical tie-breaking), when $|\PW{P}| = 2$. Under the lemma's conditions, the equilibrium winner is therefore determined by whether $P[1 \succ 2] \geq P[2 \succ 1]$ or not.
There are thus four cases we must consider: alternatives $1$ and $2$ may individually be either the truthful or equilibrium winners, or both. 

Clearly, for any $P$ where the equilibrium winning alternative is the same as the truthful one, $\ADS(P) = 0$, following its definition. This leaves two cases: (Case 1) where alternative $1$ is the truthful winner and $2$ is the equilibrium winner, and (Case 2) where alternative $2$ is the truthful winner and $1$ is the equilibrium winner. We define $\calE_1$ and $\calE_2$ to represent these cases, as follows:
\begin{itemize}
    \item $\calE_1 = \{P \in \calL(\calA)^n~:~s_1(\topRank{P}) = s_2(\topRank{P}) > s_3(\topRank{P}) \text{ and } P[2 \succ 1] > P[1 \succ 2]\}$,
    \item $\calE_2 = \{P \in \calL(\calA)^n~:~s_1(\topRank{P})+1 = s_2(\topRank{P}) > s_3(\topRank{P}) \text{ and } P[1 \succ 2] \geq P[2 \succ 1]\}$.
\end{itemize}
%
This suggests the following partition:
\begin{align}
    \EDS(\{1,2\}) & = \Pr\nolimits_{P \sim \pi^n}(P \in \mathcal{E}_1) \times \mathbb{E}_{P \sim \pi^n}[\ADS(P)~|~P \in \mathcal{E}_1] \notag \\
    & + \Pr\nolimits_{P \sim \pi^n}(P \in \mathcal{E}_2) \times \mathbb{E}_{P \sim \pi^n}[\ADS(P)~|~P \in \mathcal{E}_2]. \label{eq:eit_line0}
\end{align}

It follows from Corollary \ref{coro:sub_others} that both $\Pr\nolimits_{P \sim \pi^n}(P \in \mathcal{E}_1)$ and $\Pr\nolimits_{P \sim \pi^n}(P \in \mathcal{E}_2)$ are $\Theta \left( \frac{1}{\sqrt{n}} \right)$. From \citet[Theorem 1]{Kavner21:strategic} we have $|\mathbb{E}_{P \sim \pi^n}[\ADS(P)~|~P \in \mathcal{E}_1]| = \calO(n)$, while $\mathbb{E}_{P \sim \pi^n}[\ADS(P)~|~P \in \mathcal{E}_2]$ has the same asymptotic rate but a negated sign. This follows since $\calE_1$ describes iterative voting sequences from $\topRank{P}$, where alternative $1$ is winning, to equilibria where alternative $2$ is winning; $\calE_2$ is the inverse. Equation (\ref{eq:tldr_0}) using these broad substitutions would yield $\calO(\sqrt{n}) - \calO(\sqrt{n})$ which is bounded between $- \calO(\sqrt{n})$ and $\calO(\sqrt{n})$. This yields too general of bounds for Lemma \ref{lem:sub_12}, so we must analyze these conditional expected values more precisely. 

In Step 1, we characterize the $\calE_1$ case by detailing the number of agents with each ranking in any preference profile within the set. That is, any $P \in \calE_1$ has $\left(e, f, \beta, 2q-\beta, \frac{n}{2}-q-e, \frac{n}{2}-q-f, \right)$ agents with rankings $(R_1, R_2, R_3, R_4, R_5, R_6)$ respectively, for some $e, f, \beta, q \in \mathbb{N}$. We use the fact that $\calE_1$ is the disjoint union every $P$, characterized by $e$, $f$, $\beta$ and $q$ that span certain ranges, in order to devise a closed-form solution for $\Pr\nolimits_{P \sim \pi^n}(\mathcal{E}_1) \times \mathbb{E}_{P \sim \pi^n}[\ADS(P)~|~\mathcal{E}_1]$. In Step 2, we follow the same procedure for the $\calE_2$ case. In Step 3, we re-combine these two cases back into Equation (\ref{eq:tldr_0}). We rearrange certain terms and demonstrate how the aggregate summations yield Lemma \ref{lem:sub_12}'s conclusion.

\begin{paragraph}{Step 1: Characterize the $\calE_1$ case.}
We begin by characterizing the set of profiles $P \in \calE_1$ in terms of how many agents have each ranking in the preference profile. This case covers the events where alternative $1$ is the truthful winner with the most truthful votes, which is equal to alternative $2$ and greater than those for alternative $3$. 
%
%
Let $e, f, \beta, q \in \mathbb{N}$. Given that $n \in \mathbb{N}$ is even, we take throughout this step:
\begin{itemize}
    \item $\frac{n}{2}-q$ agents with either $R_1$ or $R_5$: with $e$ for $R_1$ and $\frac{n}{2}-q-e$ for $R_5$,
    \item $\frac{n}{2}-q$ agents with either $R_2$ or $R_6$: with $f$ for $R_2$ and $\frac{n}{2}-q-f$ for $R_6$,
    \item $2q$ agents with either $R_3$ or $R_4$: with $\beta$ for $R_3$ and $2q-\beta$ for $R_4$.
\end{itemize}
%
The frequency of each ranking is tabulated succinctly in Table \ref{tab:character_profiles_E1}.
The minimum of $q$ is $1$. Otherwise, if $q=0$, then there are no \emph{third-party} agents (i.e., agents with rankings $R_3$ or $R_4$), so there is not iterative plurality dynamics. The adversarial loss $\ADS(P)$ for any such $P$, indexed by $q=0$, is then clearly zero. The maximum of $q$ is $q^* = \max\{q \in \mathbb{Z}~:~\left(\frac{n}{2}-q\right) > 2q\}$, so that 
\[
    q^* = \begin{cases}
        \frac{n}{6}-1, & n~mod~6 = 0 \\
        \bfloor{\frac{n}{6}}+1, & n~mod~6 = 2 \\
        \bfloor{\frac{n}{6}}+3, & n~mod~6 = 4.
    \end{cases}
\]
%
We next take $e \in [0, \frac{n}{2}-q]$, $f  \in [0, \frac{n}{2}-q]$, and $\beta \in [0, 2q]$.
In order to uphold the condition that $P[2 \succ 1] > P[1 \succ 2]$, so that alternative $2$ is the equilibrium winner, we must have $\beta > q$.

\begin{table}[t]
    \centering
    \begin{tabular}{|c|c|c|c|}
        \hline
         Ranking & Probability & Frequency & Loss per Agent \\
          \hline \hline
         $R_1 = (1 \succ 2 \succ 3)$ & $\pi_1$ & $e$ & $u_1 - u_2$ \\
         $R_2 = (2 \succ 3 \succ 1)$ & $\pi_2$ & $f$ & $-u_1 + u_3$ \\
         $R_3 = (3 \succ 2 \succ 1)$ & $\pi_3$ & $\beta$ & $-u_2 + u_3$ \\
         $R_4 = (3 \succ 1 \succ 2)$ & $\pi_4$ & $2q-\beta$ & $u_2 - u_3$ \\
         $R_5 = (1 \succ 3 \succ 2)$ & $\pi_5$ & $\frac{n}{2}-q-e$ & $u_1 - u_3$ \\
         $R_6 = (2 \succ 1 \succ 3)$ & $\pi_6$ & $\frac{n}{2}-q-f$ & $-u_1 + u_2$ \\
        \hline
    \end{tabular}
    \caption{Character of profiles $P$ for $\PW{P} = \{1,2\}$ and even $n$ such that the truthful and equilibrium winners are $1$ and $2$, respectively.}
    \label{tab:character_profiles_E1}
\end{table}

When $\calE_1$ holds, each agent with ranking $R_j$ in $P$ contributes some amount of utility to the adversarial loss function $\ADS(P)$. For instance, each agent with ranking $R_1$ contributes $\vec{u}(R_1, 1) - \vec{u}(R_1, 2) = u_1 - u_2$. Recall our use of rank-based utility $\vec{u} = (u_1, u_2, u_3)$. These amounts are also summarized by Table \ref{tab:character_profiles_E1}.
Put together, we get the equation
\begin{align}
    & \Pr\nolimits_{P \sim \pi^n}(P \in \mathcal{E}_1) \times \mathbb{E}_{P \sim \pi^n}[\ADS(P)~|~P \in \mathcal{E}_1] \notag \\
    & = \sum_{q = 0}^{q^*} \sum_{e=0}^{\frac{n}{2}-q} \sum_{f=0}^{\frac{n}{2}-q} \sum_{\beta = q+1}^{2q} \mathcal{P}^1_{\vec{\pi},n}(e,f,\beta,q) \cdot \mathcal{V}^1_{\vec{u},n}(e,f,\beta,q) \label{eq:main1}
\end{align}
where we define
\[
    \mathcal{P}^1_{\vec{\pi},n}(e,f,\beta,q) = \binom{n}{e, f, \beta, 2q-\beta, \frac{n}{2}-q-e, \frac{n}{2}-q-f} \pi_1^e \pi_2^f \pi_3^{\beta} \pi_4^{2q-\beta} \pi_5^{\frac{n}{2}-q-e} \pi_6^{\frac{n}{2}-q-f}
\]
and 
\begin{align*}
    \mathcal{V}^1_{\vec{u},n}(e,f,\beta,q) & = \begin{pmatrix}
        e, & f, & \beta, & 2q-\beta, & \frac{n}{2}-q-e, & \frac{n}{2}-q-f \end{pmatrix} \\
    & \quad \cdot \begin{pmatrix}
        u_1 - u_2, & -u_1+u_3, & -u_2+u_3, & u_2-u_3, & u_1-u_3, & -u_1+u_2 \end{pmatrix}.
\end{align*}
%
%
%
%
Without loss of generality, we will assume for the duration of the proof that $q^* = \bfloor{\frac{n}{6}} -1$, taking the case that $n$ is divisible by $6$. It is easy to show that for a constant number of terms in Equation (\ref{eq:main1}) such that $q = \Theta(n)$, the objective is exponentially small and hence does not affect the result of this lemma.
We begin by factoring the probability term, which equals:
\begin{align}
    & \mathcal{P}^1_{\vec{\pi},n}(e,f,\beta,q) \notag \\
    &=  \binom{n}{\frac{n}{2}-q, \frac{n}{2}-q, 2q} \times  \binom{\frac{n}{2}-q}{e} \pi_1^e \pi_5^{\frac{n}{2}-q-e} \times \binom{\frac{n}{2}-q}{f} \pi_2^e \pi_6^{\frac{n}{2}-q-f} \times \binom{2q}{\beta} \pi_3^\beta \pi_4^{2q-\beta} \notag \\
    & = \binom{n}{\frac{n}{2}-q, \frac{n}{2}-q, 2q} (\pi_1+\pi_5)^{\frac{n}{2}-q} (\pi_2+\pi_6)^{\frac{n}{2}-q} (\pi_3+\pi_4)^{2q} \notag \\
    & \quad \quad \times  \binom{\frac{n}{2}-q}{e} \tilde{\pi}_1^e \tilde{\pi}_5^{\frac{n}{2}-q-e} \times \binom{\frac{n}{2}-q}{f} \tilde{\pi}_2^e \tilde{\pi}_6^{\frac{n}{2}-q-f} \times \binom{2q}{\beta} \tilde{\pi}_3^\beta \tilde{\pi}_4^{2q-\beta} \notag \\
    & = \binom{n}{\frac{n}{2}-q, \frac{n}{2}-q, q,q} \Pi_1^{n-2q} \Pi_3^{2q} \frac{2^{2q}}{\binom{2q}{q}} \notag \\
    & \quad \quad \times  \binom{\frac{n}{2}-q}{e} \tilde{\pi}_1^e \tilde{\pi}_5^{\frac{n}{2}-q-e} \times \binom{\frac{n}{2}-q}{f} \tilde{\pi}_2^e \tilde{\pi}_6^{\frac{n}{2}-q-f} \times \binom{2q}{\beta} \tilde{\pi}_3^\beta \tilde{\pi}_4^{2q-\beta}. \label{eq:prob_main}
\end{align}
This equation uses the following variable definitions.

\begin{dfn}
Given probability distribution $\pi = (\pi_1, \ldots, \pi_6)$ over the six corresponding preference rankings $R_1, \ldots, R_6$, we define
\begin{itemize}
    \item $\tilde{\pi}_1 = \frac{\pi_1}{\pi_1+\pi_5}$ and $\tilde{\pi}_5 = \frac{\pi_5}{\pi_1+\pi_5}$,
    \item $\tilde{\pi}_2 = \frac{\pi_2}{\pi_2+\pi_6}$ and $\tilde{\pi}_6 = \frac{\pi_6}{\pi_2+\pi_6}$,
    \item $\tilde{\pi}_3 = \frac{\pi_3}{\pi_3+\pi_4}$ and $\tilde{\pi}_4 = \frac{\pi_4}{\pi_3+\pi_4}$.
\end{itemize}
Furthermore, let $\Pi_1 = \pi_1 + \pi_5 = \pi_2 + \pi_6$ and $\Pi_3 = \frac{\pi_3 + \pi_4}{2}$.
\label{dfn:tilde_definitions}
\end{dfn}

Notice that $\Pi_1 + \Pi_3 = \frac{1}{2}$.
Next, the value factor of Equation (\ref{eq:main1}) may be written as
\begin{align}
    \mathcal{V}^1_{\vec{u},n}(e,f,\beta,q) 
    & = (u_2 - u_3) \left( \frac{n}{2} + q - 2\beta -e -f \right). \label{eq:value_main}
\end{align}
%
%
We next combine Equations (\ref{eq:prob_main}) and (\ref{eq:value_main}), back into Equation (\ref{eq:main1}), and employ a separation of variables technique to yield 
\begin{align}
    (u_2 - u_3) \sum_{q = 1}^{\frac{n}{6}-1} \binom{n}{\frac{n}{2}-q, \frac{n}{2}-q, q,q} \Pi_1^{n-2q} \Pi_3^{2q} \frac{2^{2q}}{\binom{2q}{q}} C^1_q \label{eq:main2}
\end{align}
where we define
\begin{align*}
    C^1_q & = \sum_{\beta = q+1}^{2q} \binom{2q}{\beta} \tilde{\pi}_3^{\beta} \tilde{\pi}_4^{2q-\beta} B^1_{q,\beta};
\end{align*}
\begin{align*}
    B^1_{q,\beta} & = \sum_{f=0}^{\frac{n}{2}-q} \binom{\frac{n}{2}-q}{f} \tilde{\pi}_2^f \tilde{\pi}_6^{\frac{n}{2}-q-f} A^1_{q,\beta,f};
\end{align*}
\begin{align*}
    A^1_{q,\beta,f} & = \sum_{e=0}^{\frac{n}{2}-q} \binom{\frac{n}{2}-q}{e} \tilde{\pi}_1^e \tilde{\pi}_5^{\frac{n}{2}-q-e} \left( \frac{n}{2} + q - 2\beta -e -f \right).
\end{align*}
We simplify these terms as follows.
First, we have
\begin{align}
    A^1_{q,\beta,f} & = \left( \frac{n}{2} + q - 2\beta -f \right) - \left( \frac{n}{2} -q \right) \tilde{\pi}_1 \notag \\
    & = \left( \frac{n}{2}-q \right) \left( 1 - \tilde{\pi}_1 \right) +2q -2\beta -f \notag \\
    & = \left( \frac{n}{2}-q \right) \tilde{\pi}_5 +2q -2\beta -f \notag
\end{align}
by definition of binomial probability and expectation. Second, we have
\begin{align}
    B^1_{q,\beta} & = \left( \left( \frac{n}{2}-q \right) \tilde{\pi}_5 +2q -2\beta \right) \sum_{f=0}^{\frac{n}{2}-q} \binom{\frac{n}{2}-q}{f} \tilde{\pi}_2^f \tilde{\pi}_6^{\frac{n}{2}-q-f} \notag \\
    & \quad \quad - \sum_{f=0}^{\frac{n}{2}-q} \binom{\frac{n}{2}-q}{f} \tilde{\pi}_2^f \tilde{\pi}_6^{\frac{n}{2}-q-f} f \notag \\
    & = \left( \left( \frac{n}{2}-q \right) \tilde{\pi}_5 +2q -2\beta \right) - \left( \frac{n}{2}-q \right) \tilde{\pi}_2 \notag \\
    & = \left( \frac{n}{2}-q \right) \left( \tilde{\pi}_5 - \tilde{\pi}_2 \right) +2q -2\beta \notag
\end{align}
also by definition of binomial probability and expectation. Third, we have
\begin{align}
    C^1_q & =  \left( \left( \frac{n}{2}-q \right) \left( \tilde{\pi}_5 - \tilde{\pi}_2 \right) +2q \right) \sum_{\beta = q+1}^{2q} \binom{2q}{\beta} \tilde{\pi}_3^{\beta} \tilde{\pi}_4^{2q-\beta}
    -2  \sum_{\beta = q+1}^{2q} \binom{2q}{\beta} \tilde{\pi}_3^{\beta} \tilde{\pi}_4^{2q-\beta} \beta. \label{eq:c_base}
\end{align}
Notice that the summations in Equation (\ref{eq:c_base}) correspond to $\Pr\left(S_n > \frac{n}{2} \right)$ and $\mathbb{E}[S_n \cdot \mathbbm{1}\{S_n > \frac{n}{2}\}]$ for a random variable $S_n \sim Bin(n,p)$. It is easy to see by Hoeffding's inequality that these are exponentially small (or one minus an exponentially small value) if $p < \frac{1}{2}$ or ($p > \frac{1}{2}$). Moreover, if $p=\frac{1}{2}$, a well-known finding from the Berry-Esseen theorem suggests that $\left| \Pr\left(S_n > \frac{n}{2} \right) - \frac{1}{2} \right| = \calO \left( \frac{1}{\sqrt{n}} \right)$ (see e.g., \citet{durrett2019probability}). These broad estimates are captured by Lemma \ref{lem:bin_theorems_approx} in Appendix \ref{apx:concentration_inequalities}. However, we need more precise closed-form estimates in order to prove Lemma \ref{lem:sub_12}.
We therefore employ the following lemma, proved in Appendix \ref{apx:concentration_inequalities}.
\setcounter{lem}{1}
\begin{lem}
Let $p \in (0,1)$.
The following equations hold.
\begin{enumerate}
    \item 
    \begin{align*}
        & \sum_{\beta = q+1}^{2q} \binom{2q}{\beta} p^{\beta} (1-p)^{2q-\beta} 
        = \begin{cases}
        \frac{1}{2} - \frac{1}{2^{2q}} \binom{2q-1}{q-1}, & p = \frac{1}{2} \\
        \exp(-\Theta(q)), & p < \frac{1}{2} \\ 
        1 - \exp(-\Theta(q)), & p > \frac{1}{2},
        \end{cases}
    \end{align*}
    \item 
    \begin{align*}
        & \sum_{\beta = q+1}^{2q} \binom{2q}{\beta} p^{\beta} (1-p)^{2q-\beta} \beta 
        = \begin{cases}
        \frac{q}{2}, & p = \frac{1}{2} \\
        \exp(-\Theta(q)), & p < \frac{1}{2} \\ 
        2qp (1 - \exp(-\Theta(q))), & p > \frac{1}{2}.
        \end{cases}
    \end{align*}
\end{enumerate}
\end{lem}
Since $\tilde{\pi}_3 = \tilde{\pi}_4$ is assumed, we have
\begin{align}
    C^1_q 
    & = \left( \left( \frac{n}{2}-q \right) \left( \tilde{\pi}_5 - \tilde{\pi}_2 \right) +2q \right) \left( \frac{1}{2} - \frac{1}{2^{2q}} \binom{2q-1}{q-1} \right) - q \notag \\
    & = \frac{1}{2} \left( \frac{n}{2}-q \right) \left( \tilde{\pi}_5 - \tilde{\pi}_2 \right) - \frac{ \left( \frac{n}{2}-q \right) \left( \tilde{\pi}_5 - \tilde{\pi}_2 \right) +2q }{2^{2q}} \binom{2q-1}{q-1}. \label{eq:main_c_base1_1}
\end{align}
We continue with $C^1_q$ in Step 3, below, and defer the cases where $\tilde{\pi}_3 \neq \tilde{\pi}_4$ to Lemma \ref{lem:sub_12_non_34_equal}.

As described above, we observe that Equation (\ref{eq:main_c_base1_1}) is $\pm\calO(\sqrt{n})$. Since $q = \calO(n)$, it follows from Stirling's approximation (Proposition \ref{prop:stirling}, below) that $\binom{2q}{q} = \calO \left(\frac{2^n}{\sqrt{n}} \right)$. This entails $\frac{2^{2q}}{\binom{2q}{q}} |C^1_q| = \calO(\sqrt{n}) \calO(n) = \calO(n^{1.5})$. Meanwhile, it is shown in Lemma \ref{lem:prob_bounds} (discussed in Appendix \ref{apx:helper_lemmas} and introduced formally, later on) that $\sum_{q=1}^{\frac{n}{6}-1} \binom{n}{\frac{n}{2}-q, \frac{n}{2}-q, q,q} \pi_1^{n-2q} \pi_3^{2q} = \calO \left( \frac{1}{n} \right)$. We require this $\pm \calO(\sqrt{n})$ finding, in combination with Step 2, in order to yield more precise bounds.
%
%
\end{paragraph}




\begin{paragraph}{Step 2: Characterize the $\calE_2$ case.}
We next repeat the above process for the $\calE_2$ case. This case covers the events where alternative $2$ is the truthful winner with the most truthful votes, which is one more than alternative $1$ and greater than those for alternative $3$. Given that $n \in \mathbb{N}$ is even, we take throughout this step:
\begin{itemize}
    \item $\frac{n}{2}-1-q$ agents with either $R_1$ or $R_5$: with $e$ for $R_1$ and $\frac{n}{2}-1-q-e$ for $R_5$, 
    \item $\frac{n}{2}-q$ agents with rankings either $R_2$ or $R_5$: with $f$ for $R_2$ and $\frac{n}{2}-q-f$ for $R_6$,
    \item $2q+1$ agents with $R_3$ or $R_4$: with $\beta$ for $R_3$ and $2q+1-\beta$ for $R_4$. 
\end{itemize}
The minimum of $q$ is $0$, while its maximum is $q^* = \max\{q \in \mathbb{Z}~:~\left(\frac{n}{2}-q\right) > 2q+1\}$, so that 
\[
    q^* = \begin{cases}
        \bfloor{\frac{n}{6}}-1, & n~mod~6 = 0 \\
        \bfloor{\frac{n}{6}}, & n~mod~6 \in \{2, 4\}.
    \end{cases}
\]
Like in Step 1, we will assume $q^* = \frac{n}{6}-1$ without loss of generality, taking the case that $n$ is divisible by $6$.
In order to uphold the condition that $P[1 \succ 2] \geq P[2 \succ 1]$, so that alternative $1$ is the equilibrium winner, we must have $\beta \leq q$.
While $\calE_2$ holds, it should be clear that the values per agent are the negative of those presented in Table \ref{tab:character_profiles_E1}. Put together, we get the equation
\begin{align}
    & \Pr\nolimits_{P \sim \pi^n}(P \in \mathcal{E}_2) \times \mathbb{E}_{P \sim \pi^n}[\ADS(P)~|~P \in \mathcal{E}_2] \notag \\
    & = -\sum_{q = 0}^{\frac{n}{6}-1} \sum_{e=0}^{\frac{n}{2}-1-q} \sum_{f=0}^{\frac{n}{2}-q} \sum_{\beta = 0}^{q} \mathcal{P}^2_{\vec{\pi},n}(e,f,\beta,q) \cdot \mathcal{V}^2_{\vec{u},n}(e,f,\beta,q) \label{eq:main4}
\end{align}
where
\[
    \mathcal{P}^2_{\vec{\pi},n}(e,f,\beta,q) = \binom{n}{e, f, \beta, 2q+1-\beta, \frac{n}{2}-1-q-e, \frac{n}{2}-q-f} \pi_1^e \pi_2^f \pi_3^{\beta} \pi_4^{2q+1-\beta} \pi_5^{\frac{n}{2}-1-q-e} \pi_6^{\frac{n}{2}-q-f} 
\]
and
\begin{align*}
    \mathcal{V}^2_{\vec{u},n}(e,f,\beta,q) & = 
        \begin{pmatrix}
            e, & f, & \beta, & 2q+1-\beta, & \frac{n}{2}-1-q-e, & \frac{n}{2}-q-f 
        \end{pmatrix} \\
    & \quad \cdot \begin{pmatrix} u_1 - u_2, & -u_1+u_3, & -u_2+u_3, & u_2-u_3, & u_1-u_3, & -u_1+u_2 
        \end{pmatrix}.
\end{align*}
We begin by factoring the probability term, which equals
\begin{align}
    & \mathcal{P}^2_{\vec{\pi},n}(e,f,\beta,q) \notag \\
    &= \binom{n}{\frac{n}{2}-1-q, \frac{n}{2}-q, 2q+1} \times  \binom{\frac{n}{2}-1-q}{e} \pi_1^e \pi_5^{\frac{n}{2}-1-q-e} \times \binom{\frac{n}{2}-q}{f} \pi_2^e \pi_6^{\frac{n}{2}-q-f} \notag \\
    & \quad \quad \times \binom{2q+1}{\beta} \pi_3^\beta \pi_4^{2q+1-\beta} \notag \\
    & = \binom{n}{\frac{n}{2}-1-q, \frac{n}{2}-q, 2q+1} (\pi_1+\pi_5)^{\frac{n}{2}-1-q} (\pi_2+\pi_6)^{\frac{n}{2}-q} (\pi_3+\pi_4)^{2q+1} \notag \\
    & \quad \quad \times  \binom{\frac{n}{2}-1-q}{e} \tilde{\pi}_1^e \tilde{\pi}_5^{\frac{n}{2}-1-q-e} \times \binom{\frac{n}{2}-q}{f} \tilde{\pi}_2^e \tilde{\pi}_6^{\frac{n}{2}-q-f} \times \binom{2q+1}{\beta} \tilde{\pi}_3^\beta \tilde{\pi}_4^{2q+1-\beta} \notag \\
    & = \binom{n}{\frac{n}{2}-q, \frac{n}{2}-q, q,q} \Pi_1^{n-2q} \Pi_3^{2q} \cdot \frac{\Pi_3 (\frac{n}{2}-q) 2^{2q+1}}{\Pi_1 (2q+1) \binom{2q}{q}} \notag \\
    & \quad \quad \times  \binom{\frac{n}{2}-1-q}{e} \tilde{\pi}_1^e \tilde{\pi}_5^{\frac{n}{2}-1-q-e} \times \binom{\frac{n}{2}-q}{f} \tilde{\pi}_2^e \tilde{\pi}_6^{\frac{n}{2}-q-f} \times \binom{2q+1}{\beta} \tilde{\pi}_3^\beta \tilde{\pi}_4^{2q+1-\beta} \label{eq:prob_main2}
\end{align}
where $\tilde{\pi}_1, \ldots, \tilde{\pi}_6$, $\Pi_1$, and $\Pi_3$ are defined as above.
The value terms of Equation (\ref{eq:main4}) may be written as
\begin{align}
    \mathcal{V}^2_{\vec{u},n}(e,f,\beta,q) 
    & = (u_2 - u_3) \left( \frac{n}{2} + q - 2\beta -e -f \right) -u_1 + u_2. \label{eq:value_main2}
\end{align}
%
We next combine and simplify Equations (\ref{eq:prob_main2}) and (\ref{eq:value_main2}) using a separation of variables technique,
as follows, so that Equation (\ref{eq:main4}) is equal to
\begin{align}
    -\sum_{q = 0}^{\frac{n}{6}-1} \binom{n}{\frac{n}{2}-q, \frac{n}{2}-q, q,q} \Pi_1^{n-2q} \Pi_3^{2q}  \frac{\Pi_3 (\frac{n}{2}-q) 2^{2q+1}}{\Pi_1 (2q+1) \binom{2q}{q}} C^2_q \label{eq:main5}
\end{align}
where we define
\begin{align*}
    C^2_q & = \sum_{\beta = 0}^{q} \binom{2q+1}{\beta} \tilde{\pi}_3^{\beta} \tilde{\pi}_4^{2q+1-\beta} B^2_{q,\beta};
\end{align*}
\begin{align*}
    B^2_{q,\beta} & = \sum_{f=0}^{\frac{n}{2}-q} \binom{\frac{n}{2}-q}{f} \tilde{\pi}_2^f \tilde{\pi}_6^{\frac{n}{2}-q-f} A^2_{q,\beta,f};
\end{align*}
\begin{align*}
    A^2_{q,\beta,f} & = \sum_{e=0}^{\frac{n}{2}-1-q} \binom{\frac{n}{2}-1-q}{e} \tilde{\pi}_1^e \tilde{\pi}_5^{\frac{n}{2}-1-q-e} \notag \\
    & \quad \quad \times \left( (u_2 - u_3) \left( \frac{n}{2} + q - 2\beta -e -f \right) -u_1 + u_2 \right).
\end{align*}
We simplify these variables as follows. First, we have
\begin{align}
    A^2_{q,\beta,f} & =  (u_2 - u_3) \left( \frac{n}{2} + q - 2\beta -f - \left( \frac{n}{2} -1-q \right) \tilde{\pi}_1 \right) -u_1+u_2 \notag \\
    & = (u_2 - u_3) \left( \left( \frac{n}{2}-q \right) \left( 1 - \tilde{\pi}_1 \right) +2q -2\beta -f \right) + (u_2 - u_3) \tilde{\pi}_1 -u_1 + u_2 \label{eq:A2_finish}
\end{align}
by definition of binomial probability and expectation.
%
%
Consider the constant terms of Equation (\ref{eq:A2_finish}). To this point, from Equation (\ref{eq:main5}), we have
\begin{align}
    & - \left( (u_2 - u_3) \tilde{\pi}_1 -u_1 + u_2 \right) \sum_{q = 0}^{\frac{n}{6}-1} \binom{n}{\frac{n}{2}-1-q, \frac{n}{2}-q, 2q+1} \Pi_1^{n-1-q} (2\Pi_3)^{2q+1} \notag \\
    & \quad \quad \times \sum_{\beta = 0}^{q} \binom{2q+1}{\beta} \tilde{\pi}_3^{\beta} \tilde{\pi}_4^{2q+1-\beta} \sum_{f=0}^{\frac{n}{2}-q} \binom{\frac{n}{2}-q}{f} \tilde{\pi}_2^f \tilde{\pi}_6^{\frac{n}{2}-q-f}. \label{eq:const_terms_pre} 
\end{align}
The $f$-summation of Equation (\ref{eq:const_terms_pre}) is clearly $\Theta(1)$ by definition of binomial probability. By Lemma \ref{lem:new_comb} (below), the $q$-summation of Equation (\ref{eq:const_terms_pre}) is either $\Theta(1)$, if $\tilde{\pi}_3 \leq \tilde{\pi}_4$, and $\calO \left( e^{-\Theta(n)} \right)$ otherwise. Finally, the $q$-summation is the probability of a two-way tie for plurality voting, with three alternatives under i.i.d preferences. By Corollary \ref{coro:sub_others}, Equation (\ref{eq:const_terms_pre}) is therefore
\begin{align}
    \begin{cases}
        \pm \calO\left( \frac{1}{\sqrt{n}} \right), & \pi_3 \leq \pi_4 \\
        \pm \calO \left( e^{-\Theta(n)} \right), & \pi_3 > \pi_4.
    \end{cases} \label{eq:const_terms}
\end{align}
%
%
Now consider the non-constant terms of Equation (\ref{eq:A2_finish}). 
For ease of notation, we will continue without writing $(u_2 - u_3)$ as a factor in front of every remaining term; this will be implicitly pulled outside the summation of Equation (\ref{eq:main5}).
Continuing with $B^2_{q,\beta}$, we have
\begin{align}
    B^2_{q,\beta} & = \left( \left( \frac{n}{2}-q \right) \tilde{\pi}_5 +2q -2\beta \right) \sum_{f=0}^{\frac{n}{2}-q} \binom{\frac{n}{2}-q}{f} \tilde{\pi}_2^f \tilde{\pi}_6^{\frac{n}{2}-q-f} 
    - \sum_{f=0}^{\frac{n}{2}-q} \binom{\frac{n}{2}-q}{f} \tilde{\pi}_2^f \tilde{\pi}_6^{\frac{n}{2}-q-f} f \notag \\
    & = \left( \left( \frac{n}{2}-q \right) \tilde{\pi}_5 +2q -2\beta  \right) - \left( \frac{n}{2}-q \right) \tilde{\pi}_2 \notag \\
    & =\left( \frac{n}{2}-q \right) \left( \tilde{\pi}_5 - \tilde{\pi}_2 \right) +2q -2\beta \notag
\end{align}
also by definition of binomial probability and expectation.

Before proceeding to simplify $C^2_q$, we must consider the case when $q=0$. Unlike case $\calE_1$, now when $q=0$ there is a single agent with ranking $R_4$. This is illustrated in Equation (\ref{eq:main5}) with $q=\beta=0$ as
\begin{align}
    & -\binom{n}{\frac{n}{2}-1, \frac{n}{2}, 1} \Pi_1^{n-1} (\pi_3 + \pi_4) \times \binom{1}{0} \tilde{\pi}_4 B^2_{0,0} \notag \\
    & = - \frac{n}{2} \binom{n}{\frac{n}{2}} \Pi_1^{n-1} \pi_4 \times \frac{n}{2} (\tilde{\pi}_5 - \tilde{\pi}_2) \notag \\
    & = \pm \calO(n^{1.5}) (2 \Pi_1)^n \notag \\
    & = \pm \calO \left( e^{-\Theta(n)} \right) \label{eq:main_Qzero}
\end{align}
by Stirling's approximation, where $\Pi_1 = (\pi_1 + \pi_5) = (\pi_2 + \pi_6) < \frac{1}{2}$. This holds as long as $\tilde{\pi}_2 \neq \tilde{\pi}_5$ and $\pi_4 \neq 0$; otherwise, Equation (\ref{eq:main_Qzero}) is zero.
\setcounter{prop}{1}
\begin{prop}[Stirling's approximation]
    Stirling's approximation says that $n! \sim \sqrt{2 \pi n} \left( \frac{n}{e} \right)^n$. Therefore, we have $\binom{2n}{n} \sim \frac{2^{2n}}{\sqrt{n \pi}}$.
\end{prop}
This proposition is discussed further in Appendix \ref{apx:stirling_and_wallis}.
%
%
Continuing with $C^2_q$ when $q>0$, we have
\begin{align}
    C^2_q & =  \left( \left( \frac{n}{2}-q \right) \left( \tilde{\pi}_5 - \tilde{\pi}_2 \right) +2q \right) \sum_{\beta = 0}^{q} \binom{2q+1}{\beta} \tilde{\pi}_3^{\beta} \tilde{\pi}_4^{2q+1-\beta} 
    -2  \sum_{\beta = 0}^{q} \binom{2q+1}{\beta} \tilde{\pi}_3^{\beta} \tilde{\pi}_4^{2q+1-\beta} \beta. \label{eq:c_base2}
\end{align}
We next employ the following lemma, proved in Appendix \ref{apx:concentration_inequalities}.
\setcounter{lem}{1}
\begin{lem}
Let $p \in (0,1)$.
The following equations hold.
\begin{enumerate}
    \item[(3)]
    \begin{align*}
        & \sum_{\beta = 0}^{q} \binom{2q+1}{\beta} p^{\beta} (1-p)^{2q+1-\beta} 
        = \begin{cases}
        \frac{1}{2}, & p = \frac{1}{2} \\
        1-\exp(-\Theta(q)), & p < \frac{1}{2} \\ 
        \exp(-\Theta(q)), & p > \frac{1}{2},
    \end{cases}
    \end{align*}
    \item[(4)]
    \begin{align*}
        & \sum_{\beta = 0}^{q} \binom{2q+1}{\beta} p^{\beta} (1-p)^{2q+1-\beta} \beta 
        = \begin{cases}
        \left( \frac{2q+1}{4} \right) - \frac{2q+1}{2^{2q+1}} \binom{2q-1}{q-1}, & p = \frac{1}{2} \\
        (2q+1)p \left( 1 - \exp(-\Theta(q)) \right), & p < \frac{1}{2} \\ 
        \exp(-\Theta(p)), & p > \frac{1}{2}.
        \end{cases}
    \end{align*}
\end{enumerate}
\end{lem}
Since $\tilde{\pi}_3 = \tilde{\pi}_4$ is assumed, we have
\begin{align}
    C^2_q 
    & = \left( \left( \frac{n}{2}-q \right) \left( \tilde{\pi}_5 - \tilde{\pi}_2 \right) +2q \right) \left( \frac{1}{2} \right) - \left( \frac{2q+1}{2} \right) + \frac{2q+1}{2^{2q}} \binom{2q-1}{q-1}  \notag \\
    & = \frac{1}{2} \left( \left( \frac{n}{2}-q \right) \left( \tilde{\pi}_5 - \tilde{\pi}_2 \right) -1 \right) + \frac{2q+1}{2^{2q}} \binom{2q-1}{q-1}. \label{eq:main_c_base2_1}
\end{align}
We continue with $C^2_q$ in Step 3, below, and defer the cases where $\tilde{\pi}_3 \neq \tilde{\pi}_4$ to Lemma \ref{lem:sub_12_non_34_equal}.
\end{paragraph}

\begin{paragraph}{Step 3: Putting the pieces back together.}
Recall that our original problem began as Equation (\ref{eq:eit_line0}) which we initially split into Equations (\ref{eq:main1}) and (\ref{eq:main4}). Through a sequence of steps we transformed these equations into Equations (\ref{eq:main2}) and (\ref{eq:main5}) and an additional $+$ or $-\Theta \left( \frac{1}{\sqrt{n}} \right)$ term; 
recall Equation (\ref{eq:const_terms}). Recombining these simplified equations yields
\begin{align}
    (u_2 - u_3) \sum_{q = 1}^{\frac{n}{6}-1} \binom{n}{\frac{n}{2}-q, \frac{n}{2}-q, q,q} \Pi_1^{n-2q} \Pi_3^{2q}  \frac{2^{2q}}{\binom{2q}{q}} \left( C^1_q - \frac{2 \Pi_3 (\frac{n}{2}-q)}{\Pi_1 (2q+1)} C^2_q \right). \label{eq:main7}
\end{align}
%
%
%
Plugging in Equations (\ref{eq:main_c_base1_1}) and (\ref{eq:main_c_base2_1}) into Equation (\ref{eq:main7}) yields
\begin{align}
    & \frac{2^{2q}}{\binom{2q}{q}} \left( C^1_q - \frac{2 \Pi_3 (\frac{n}{2}-q)}{\Pi_1 (2q+1)} C^2_q \right) \notag \\
    & = \frac{2^{2q}}{\binom{2q}{q}} \left ( 
        \frac{1}{2} \left( \frac{n}{2}-q \right) \left( \tilde{\pi}_5 - \tilde{\pi}_2 \right) - \frac{\left( \frac{n}{2}-q \right) \left( \tilde{\pi}_5 - \tilde{\pi}_2 \right) +2q }{2^{2q}} \binom{2q-1}{q-1} 
    \right) \notag \\
    & \quad \quad - \frac{2^{2q+1}}{\binom{2q}{q}} \frac{\Pi_3 (\frac{n}{2}-q)}{\Pi_1 (2q+1)} \left ( 
        \frac{1}{2} \left( \left( \frac{n}{2}-q \right) \left( \tilde{\pi}_5 - \tilde{\pi}_2 \right) -1 \right) + \frac{2q+1}{2^{2q}} \binom{2q-1}{q-1}
    \right) \notag \\
    & = \left( -\frac{1}{2} \left( \left( -\frac{n}{2}+q \right) \left( \tilde{\pi}_2 - \tilde{\pi}_5 \right) +2q \right) + \frac{\Pi_3}{\Pi_1} \left( -\frac{n}{2}+q \right) \right) \notag \\
    & \quad \quad + \frac{\left( -\frac{n}{2} +q \right) 2^{2q}}{\binom{2q}{q}} \left( \frac{\tilde{\pi}_2 - \tilde{\pi}_5}{2} + \frac{\Pi_3 \left( \left( -\frac{n}{2}+q \right) (\tilde{\pi}_2 - \tilde{\pi}_5) -1 \right)}{\Pi_1 (2q+1)} \right) \label{eq:main8_sub} \\
    & = \left( -\frac{1}{2} \left( \left( -\frac{n}{2}+q \right) \left( \tilde{\pi}_2 - \tilde{\pi}_5 \right) +2q \right) + \frac{\Pi_3}{\Pi_1} \left( -\frac{n}{2}+q \right) \right) \notag \\
    & \quad \quad + (\tilde{\pi}_2 - \tilde{\pi}_5) \frac{\left( -\frac{n}{2} +q \right) 2^{2q}}{\binom{2q}{q}} \left( \frac{1}{2} + \frac{\Pi_3 \left( -\frac{n}{2}+q  -1 \right)}{\Pi_1 (2q+1)} \right) \notag \\
    & \quad \quad + \frac{\Pi_3 (\tilde{\pi}_2 - \tilde{\pi}_5 -1)}{\Pi_1} \frac{\left( -\frac{n}{2} +q \right) 2^{2q}}{(2q+1) \binom{2q}{q}} \label{eq:main8}
\end{align}
To get Equation (\ref{eq:main8_sub}), we used the fact that $2 \binom{2q-1}{q-1} = \binom{2q}{q}$ and rearranged certain signs. To get Equation (\ref{eq:main8}), we factored $(\tilde{\pi}_2 - \tilde{\pi}_5)$ from the second term of the prior equation.
Regarding the first summation of Equation (\ref{eq:main8}), we introduce Lemma \ref{lem:soln_part_a} in Appendix \ref{apx:secondary_equations} to prove that 
\begin{align*}
    & \sum_{q = 1}^{\frac{n}{6}-1} \binom{n}{\frac{n}{2}-q, \frac{n}{2}-q, q,q} \Pi_1^{n-2q} \Pi_3^{2q}
    \left( -\frac{1}{2} \left( \left( -\frac{n}{2}+q \right) \left( \tilde{\pi}_2 - \tilde{\pi}_5 \right) +2q \right) + \frac{\Pi_3}{\Pi_1} \left( -\frac{n}{2}+q \right) \right) \\
    & = \begin{cases}
        \Theta(1), & \Pi_1 > \frac{2}{4+\tilde{\pi}_2 - \tilde{\pi}_5} \\
        -\Theta(1), & \Pi_1 < \frac{2}{4+\tilde{\pi}_2 - \tilde{\pi}_5} \\
        \pm \calO \left( \frac{1}{n} \right), & \Pi_1 = \frac{2}{4+\tilde{\pi}_2 - \tilde{\pi}_5}. 
    \end{cases}
\end{align*}
For the second summation of Equation (\ref{eq:main8}), observe that
\begin{align*}
    & \left( -\frac{n}{2} +q \right) \left( \frac{1}{2} + \frac{\Pi_3 ( - \frac{n}{2} + q -1)}{\Pi_1 (2q+1)} \right) \\
    & = \frac{1}{2 \Pi_1 (2q+1)} \left( -\frac{n}{2} + q \right) \left( q - \Pi_3 n + \Pi_1 - 2 \Pi_3 \right) 
\end{align*}
using the fact that $\Pi_1 + \Pi_3 = \frac{1}{2}$. 
We introduce Lemma \ref{lem:soln_part_b} in Appendix \ref{apx:secondary_equations} to prove that
\begin{align*}
    & \frac{(\tilde{\pi}_2 - \tilde{\pi}_5)}{2 \Pi_1} 
    \sum_{q = 1}^{\frac{n}{6}-1} \binom{n}{\frac{n}{2}-q, \frac{n}{2}-q, q,q} \Pi_1^{n-2q} \Pi_3^{2q}  \frac{\left( -\frac{n}{2}+q \right) f_n(q) 2^{2q}}{(2q+1) \binom{2q}{q}}
    = \pm \calO \left( \frac{1}{\sqrt{n}} \right).
\end{align*}
where $f_n(q) = q - \Pi_3 n + \Pi_1 - 2 \Pi_3$. Note that, in Appendix \ref{apx:secondary_equations}, Lemma \ref{lem:soln_part_b} is stated generally to encompass several use-cases for conciseness. For this specific instance, we plug in $\tau_1 = 1$, $\tau_2 = \Pi_1 - 2 \Pi_3$, and $f_n(q) = \frac{-\frac{n}{2}+q}{2q+1}$ into that lemma.

%
%

The third summation of Equation (\ref{eq:main8}) is proportional to
\begin{align*}
    & 
    \sum_{q = 1}^{\frac{n}{6}-1} \binom{n}{\frac{n}{2}-1-q, \frac{n}{2}-q, 2q+1} \Pi_1^{n-1-2q} (2\Pi_3)^{2q+1}
    = \pm \calO \left( \frac{1}{\sqrt{n}} \right)
\end{align*}
by Corollary \ref{coro:sub_others}, since it is proportional to the probability of a two-way plurality tie under i.i.d. preferences.
This concludes the proof of Lemma \ref{lem:sub_12}.
\end{paragraph}
\end{proof}

%% file: EC_appendix/apx_page_2b.tex
\subsection{Two-tie case when $n$ is odd}
\label{apx:two_ties_n_odd}

\setcounter{lem}{3}
\begin{lem}


Suppose that $\pi_1 + \pi_5 = \pi_2 + \pi_6 \geq \pi_3 + \pi_4$ and $\pi_i > 0,~\forall i \in [6]$. Furthermore, let $u_1 \geq u_2 > u_3$ in $\vec{u}$. Then $\exists N > 0$ such that $\forall n > N$ that are odd, $\EDS(\{1,2\})$ is determined by the mapping from $\pi$ to asymptotic rates as described by Table \ref{tab:eadpoa_case_12}.
Both conditions on $\pi$ need to hold; note that this table is exhaustive.

If $u_1 > u_2 = u_3$, then $\EDS(\{1,2\}) = \pm \calO\left( \frac{1}{\sqrt{n}} \right)$ if $\pi_3 \leq \pi_4$ and $\pm \calO \left( e^{-\Theta(n)} \right)$ otherwise.

\label{lem:sub_12_odd}
\end{lem}

\begin{proof}

This lemma's proof follows almost identically to that of Lemma \ref{lem:sub_12}, where $n$ is even, except for how $\calE_1$ and $\calE_2$ are defined, on account of $n$ being odd. Our subsequent analysis therefore yields a different conclusion than that lemma.

For any preference profile $P$ where the equilibrium winning alternative is the same as the truthful one, $\ADS(P) = 0$, following its definition. This leaves two cases: (Case 1) where alternative $1$ is the truthful winner and $2$ is the equilibrium winner, and (Case 2) where alternative $2$ is the truthful winner and $1$ is the equilibrium winner. We define $\calE_1$ and $\calE_2$ to represent these cases, as follows:
\begin{itemize}
    \item $\calE_1 = \{P \in \calL(\calA)^n~:~s_1(\topRank{P}) = s_2(\topRank{P}) > s_3(\topRank{P}) \text{ and } P[2 \succ 1] > P[1 \succ 2]\}$,
    \item $\calE_2 = \{P \in \calL(\calA)^n~:~s_1(\topRank{P})+1 = s_2(\topRank{P}) > s_3(\topRank{P}) \text{ and } P[1 \succ 2] \geq P[2 \succ 1]\}$.
\end{itemize}
This suggests the following partition:
\begin{align}
    \EDS(\{1,2\}) & = \Pr\nolimits_{P \sim \pi^n}(P \in \mathcal{E}_1) \times \mathbb{E}_{P \sim \pi^n}[\ADS(P)~|~P \in \mathcal{E}_1] \notag \\
    & + \Pr\nolimits_{P \sim \pi^n}(P \in \mathcal{E}_2) \times \mathbb{E}_{P \sim \pi^n}[\ADS(P)~|~P \in \mathcal{E}_2]. \label{eq:odd_eit_line0}
\end{align}

\begin{paragraph}{Step 1: Characterize the $\calE_1$ case.}
We begin by characterizing the set of profiles $P \in \calE_1$ in terms of how many agents have each ranking in the preference profile. This case covers the events where alternative $1$ is the truthful winner with the most truthful votes, which is equal to alternative $2$ and greater than those for alternative $3$. 
Let $e, f, \beta, q \in \mathbb{N}$. Given that $n \in \mathbb{N}$ is odd, we take throughout this step:
\begin{itemize}
    \item $\frac{n-1}{2}-q$ agents with either $R_1$ or $R_5$: with $e$ for $R_1$ and $\frac{n}{2}-q-e$ for $R_5$,
    \item $\frac{n-1}{2}-q$ agents with either $R_2$ or $R_6$: with $f$ for $R_2$ and $\frac{n}{2}-q-f$ for $R_6$,
    \item $2q+1$ agents with either $R_3$ or $R_4$: with $\beta$ for $R_3$ and $2q-\beta$ for $R_4$.
\end{itemize}
The frequency of each ranking is tabulated succinctly in Table \ref{tab:character_profiles_E4}.
The minimum of $q$ is $0$, 
while it's maximum is $q^* = \max\{q \in \mathbb{Z}~:~\left(\frac{n-1}{2}-q\right) > 2q+1\}$, so that 
\[
    q^* = \begin{cases}
        \bfloor{\frac{n}{6}}-1, & n~mod~6 \in \{1,3\} \\
        \bfloor{\frac{n}{6}}, & n~mod~6 = 5.
    \end{cases}
\]
Without loss of generality, we will assume for the duration of the proof that $q^* = \bfloor{\frac{n}{6}} -1$. It is easy to show that for a constant number of terms in Equation (\ref{eq:odd_main1}) (below) such that $q = \Theta(n)$, the objective is exponentially small and hence does not affect the result of this lemma.
We then take $e \in [0, \frac{n-1}{2}-q]$, $f  \in [0, \frac{n-1}{2}-q]$, and $\beta \in [0, 2q+1]$.
In order to uphold the condition that $P[2 \succ 1] > P[1 \succ 2]$, so that alternative $2$ is the equilibrium winner, we must have $\beta > q$.

\begin{table}[t]
    \centering
    \begin{tabular}{|c|c|c|c|}
        \hline
         Ranking & Probability & Frequency & Loss per Agent \\
          \hline \hline
         $R_1 = (1 \succ 2 \succ 3)$ & $\pi_1$ & $e$ & $u_1 - u_2$ \\
         $R_2 = (2 \succ 3 \succ 1)$ & $\pi_2$ & $f$ & $-u_1 + u_3$ \\
         $R_3 = (3 \succ 2 \succ 1)$ & $\pi_3$ & $\beta$ & $-u_2 + u_3$ \\
         $R_4 = (3 \succ 1 \succ 2)$ & $\pi_4$ & $2q+1-\beta$ & $u_2 - u_3$ \\
         $R_5 = (1 \succ 3 \succ 2)$ & $\pi_5$ & $\frac{n-1}{2}-q-e$ & $u_1 - u_3$ \\
         $R_6 = (2 \succ 1 \succ 3)$ & $\pi_6$ & $\frac{n-1}{2}-q-f$ & $-u_1 + u_2$ \\
        \hline
    \end{tabular}
    \caption{Character of profiles $P$ for $\PW{P} = \{1,2\}$ and odd $n$ such that the truthful and equilibrium winners are $1$ and $2$, respectively.}
    \label{tab:character_profiles_E4}
\end{table}

When $\calE_1$ holds, each agent with ranking $R_j$ in $P$ contributes some amount of utility to the adversarial loss function $\ADS(P)$. These amounts are also summarized by Table \ref{tab:character_profiles_E4}.
Put together, we get the equation
\begin{align}
    & \Pr\nolimits_{P \sim \pi^n}(P \in \mathcal{E}_1) \times \mathbb{E}_{P \sim \pi^n}[\ADS(P)~|~P \in \mathcal{E}_1] \notag \\
    & = \sum_{q = 0}^{\bfloor{\frac{n}{6}}-1} \sum_{e=0}^{\frac{n-1}{2}-q} \sum_{f=0}^{\frac{n-1}{2}-q} \sum_{\beta = q+1}^{2q+1} \mathcal{P}^1_{\vec{\pi},n}(e,f,\beta,q) \cdot \mathcal{V}^1_{\vec{u},n}(e,f,\beta,q) \label{eq:odd_main1}
\end{align}
where we define
\[
    \mathcal{P}^1_{\vec{\pi},n}(e,f,\beta,q) = \binom{n}{e, f, \beta, 2q+1-\beta, \frac{n-1}{2}-q-e, \frac{n-1}{2}-q-f} \pi_1^e \pi_2^f \pi_3^{\beta} \pi_4^{2q+1-\beta} \pi_5^{\frac{n-1}{2}-q-e} \pi_6^{\frac{n-1}{2}-q-f}
\]
and 
\begin{align*}
    \mathcal{V}^1_{\vec{u},n}(e,f,\beta,q) & = \begin{pmatrix}
        e, & f, & \beta, & 2q+1-\beta, & \frac{n-1}{2}-q-e, & \frac{n-1}{2}-q-f \end{pmatrix} \\
    & \quad \cdot \begin{pmatrix}
        u_1 - u_2, & -u_1+u_3, & -u_2+u_3, & u_2-u_3, & u_1-u_3, & -u_1+u_2 \end{pmatrix}.
\end{align*}
%
We begin by factoring the probability term, which equals:
\begin{align}
    & \mathcal{P}^1_{\vec{\pi},n}(e,f,\beta,q) \notag \\
    &=  \binom{n}{\frac{n-1}{2}-q, \frac{n-1}{2}-q, 2q+1} \times  \binom{\frac{n-1}{2}-q}{e} \pi_1^e \pi_5^{\frac{n-1}{2}-q-e} \notag \\
    & \quad \quad \times \binom{\frac{n-1}{2}-q}{f} \pi_2^e \pi_6^{\frac{n-1}{2}-q-f} \times \binom{2q+1}{\beta} \pi_3^\beta \pi_4^{2q+1-\beta} \notag \\
    & = \binom{n}{\frac{n-1}{2}-q, \frac{n-1}{2}-q, 2q+1} (\pi_1+\pi_5)^{\frac{n-1}{2}-q} (\pi_2+\pi_6)^{\frac{n-1}{2}-q} (\pi_3+\pi_4)^{2q+1} \notag \\
    & \quad \quad \times  \binom{\frac{n-1}{2}-q}{e} \tilde{\pi}_1^e \tilde{\pi}_5^{\frac{n-1}{2}-q-e} \times \binom{\frac{n-1}{2}-q}{f} \tilde{\pi}_2^e \tilde{\pi}_6^{\frac{n-1}{2}-q-f} \times \binom{2q+1}{\beta} \tilde{\pi}_3^\beta \tilde{\pi}_4^{2q+1-\beta} \notag \\
    & = \binom{n-1}{\frac{n-1}{2}-q, \frac{n-1}{2}-q, q,q} \Pi_1^{n-1-2q} \Pi_3^{2q+1} \frac{n 2^{2q+1}}{(2q+1) \binom{2q}{q}} \notag \\
    & \quad \quad \times  \binom{\frac{n-1}{2}-q}{e} \tilde{\pi}_1^e \tilde{\pi}_5^{\frac{n-1}{2}-q-e} \times \binom{\frac{n-1}{2}-q}{f} \tilde{\pi}_2^e \tilde{\pi}_6^{\frac{n-1}{2}-q-f} \times \binom{2q+1}{\beta} \tilde{\pi}_3^\beta \tilde{\pi}_4^{2q+1-\beta}. \label{eq:odd_prob_main}
\end{align}
This equation uses the same definitions of $\tilde{\pi}_1, \ldots, \tilde{\pi}_6, \Pi_1$ and $\Pi_3$ from Definition \ref{dfn:tilde_definitions}, as in Lemma \ref{lem:sub_12}.
Next, the value factor of Equation (\ref{eq:odd_main1}) may be written as
\begin{align}
    \mathcal{V}^1_{\vec{u},n}(e,f,\beta,q) 
    & = (u_2 - u_3) \left( \frac{n+1}{2} + q - 2\beta -e -f \right). \label{eq:odd_value_main}
\end{align}
We next combine Equations (\ref{eq:odd_prob_main}) and (\ref{eq:odd_value_main}), back into Equation (\ref{eq:odd_main1}), and employ a separation of variables technique to yield
\begin{align}
    (u_2 - u_3) \sum_{q = 0}^{\bfloor{\frac{n}{6}}-1} \binom{n-1}{\frac{n-1}{2}-q, \frac{n-1}{2}-q, q,q} \Pi_1^{n-1-2q} \Pi_3^{2q+1} \frac{n 2^{2q+1}}{(2q+1) \binom{2q}{q}} C^1_q \label{eq:odd_main2}
\end{align}
where we define
\begin{align*}
    C^1_q & = \sum_{\beta = q+1}^{2q+1} \binom{2q+1}{\beta} \tilde{\pi}_3^{\beta} \tilde{\pi}_4^{2q+1-\beta} B^1_{q,\beta};
\end{align*}
\begin{align*}
    B^1_{q,\beta} & = \sum_{f=0}^{\frac{n-1}{2}-q} \binom{\frac{n-1}{2}-q}{f} \tilde{\pi}_2^f \tilde{\pi}_6^{\frac{n-1}{2}-q-f} A^1_{q,\beta,f};
\end{align*}
\begin{align*}
    A^1_{q,\beta,f} & = \sum_{e=0}^{\frac{n-1}{2}-q} \binom{\frac{n-1}{2}-q}{e} \tilde{\pi}_1^e \tilde{\pi}_5^{\frac{n-1}{2}-q-e} \left( \frac{n+1}{2} + q - 2\beta -e -f \right).
\end{align*}
We simplify these terms as follows.
First, we have
\begin{align}
    A^1_{q,\beta,f} & = \left( \frac{n+1}{2} + q - 2\beta -f \right) - \left( \frac{n-1}{2} -q \right) \tilde{\pi}_1 \notag \\
    & = \left( \frac{n-1}{2}-q \right) \left( 1 - \tilde{\pi}_1 \right) +2q -2\beta -f +1 \notag \\
    & = \left( \frac{n-1}{2}-q \right) \tilde{\pi}_5 +2q -2\beta -f +1 \notag
\end{align}
by definition of binomial probability and expectation. 
Second, we have
\begin{align}
    B^1_{q,\beta} & = \left( \left( \frac{n-1}{2}-q \right) \tilde{\pi}_5 +2q -2\beta +1\right) \sum_{f=0}^{\frac{n-1}{2}-q} \binom{\frac{n-1}{2}-q}{f} \tilde{\pi}_2^f \tilde{\pi}_6^{\frac{n-1}{2}-q-f} \notag \\
    & \quad \quad - \sum_{f=0}^{\frac{n-1}{2}-q} \binom{\frac{n-1}{2}-q}{f} \tilde{\pi}_2^f \tilde{\pi}_6^{\frac{n-1}{2}-q-f} f \notag \\
    & = \left( \left( \frac{n-1}{2}-q \right) \tilde{\pi}_5 +2q -2\beta +1\right) - \left( \frac{n-1}{2}-q \right) \tilde{\pi}_2 \notag \\
    & = \left( \frac{n-1}{2}-q \right) \left( \tilde{\pi}_5 - \tilde{\pi}_2 \right) +2q -2\beta +1. \notag
\end{align}
also by definition of binomial probability and expectation.

%
%
Before proceeding to simplify $C^1_q$, we must consider the case when $q=0$. This is similar to case $\calE_2$ of Lemma \ref{lem:sub_12}, where there is a single agent with ranking $R_4$. This is illustrated in Equation (\ref{eq:odd_main2}) with $q=\beta=0$ as
\begin{align}
    & \binom{n}{\frac{n-1}{2}, \frac{n-1}{2}, 1} \Pi_1^{n-1} (\pi_3 + \pi_4) \times \binom{1}{0} \tilde{\pi}_4 B^1_{0,0} \notag \\
    & = n \binom{n-1}{\frac{n-1}{2}} \Pi_1^{n-1} \pi_4 \times \left( \frac{n-1}{2} (\tilde{\pi}_5 - \tilde{\pi}_2) +1 \right) \notag \\
    & = \pm \calO(n^{1.5}) (2 \Pi_1)^n \notag \\
    & = \pm \calO \left( e^{-\Theta(n)} \right) \label{eq:odd_main_Qzero}
\end{align}
by Stirling's approximation, where $\Pi_1 < \frac{1}{2}$ by definition. This holds as long as $\tilde{\pi}_2 \neq \tilde{\pi}_5$ and $\pi_4 \neq 0$; otherwise, Equation (\ref{eq:odd_main_Qzero}) is zero.
%
%

Continuing with $C^1_q$ when $q>0$, we have
\begin{align}
    C^1_q & =  \left( \left( \frac{n-1}{2}-q \right) \left( \tilde{\pi}_5 - \tilde{\pi}_2 \right) +2q +1\right) \sum_{\beta = q+1}^{2q+1} \binom{2q+1}{\beta} \tilde{\pi}_3^{\beta} \tilde{\pi}_4^{2q+1-\beta} \notag \\
    & \quad \quad -2  \sum_{\beta = q+1}^{2q+1} \binom{2q+1}{\beta} \tilde{\pi}_3^{\beta} \tilde{\pi}_4^{2q+1-\beta} \beta. \label{eq:odd_c_base}
\end{align}
We next employ the following lemma, proved in Appendix \ref{apx:concentration_inequalities}.
\setcounter{lem}{1}
\begin{lem}
Let $p \in (0,1)$.
The following equations hold.
\begin{enumerate}
    \item[(5)] 
    \begin{align*}
        & \sum_{\beta = q+1}^{2q+1} \binom{2q+1}{\beta} p^{\beta} (1-p)^{2q+1-\beta} 
        = \begin{cases}
        \frac{1}{2}, & p = \frac{1}{2} \\
        \exp(-\Theta(q)), & p < \frac{1}{2} \\ 
        1 - \exp(-\Theta(q)), & p > \frac{1}{2},
        \end{cases}
    \end{align*}
    \item[(6)] 
    \begin{align*}
        & \sum_{\beta = q+1}^{2q+1} \binom{2q+1}{\beta} p^{\beta} (1-p)^{2q+1-\beta} \beta 
        = \begin{cases}
        \frac{2q+1}{4} + \frac{2q+1}{2^{2q+1}} \binom{2q-1}{q-1}, & p = \frac{1}{2} \\
        \exp(-\Theta(q)), & p < \frac{1}{2} \\ 
        (2q+1)p (1 - \exp(-\Theta(q))), & p > \frac{1}{2}.
        \end{cases}
    \end{align*}
\end{enumerate}
\end{lem}
Since $\tilde{\pi}_3 = \tilde{\pi}_4$ is assumed, we have
\begin{align}
    C^1_q 
    & = \frac{1}{2} \left( \left( \frac{n-1}{2}-q \right) \left( \tilde{\pi}_5 - \tilde{\pi}_2 \right) +2q +1\right) -  \left( \frac{2q+1}{2} + \frac{2q+1}{2^{2q}} \binom{2q-1}{q-1} \right)  \notag \\
    & = \frac{1}{2} \left( \frac{n-1}{2}-q \right) \left( \tilde{\pi}_5 - \tilde{\pi}_2 \right) - \frac{ 2q+1 }{2^{2q}} \binom{2q-1}{q-1}. \label{eq:odd_main_c_base1_1}
\end{align}
We continue with $C^1_q$ in Step 3, below, and defer the cases where $\tilde{\pi}_3 \neq \tilde{\pi}_4$ to Lemma \ref{lem:sub_12_non_34_equal}.
\end{paragraph}

\begin{paragraph}{Step 2: Characterize the $\calE_2$ case.}
%
We next repeat the above process for the $\calE_2$ case. This case covers the events where alternative $2$ is the truthful winner with the most truthful votes, which is one more than alternative $1$ and greater than those for alternative $3$. Given that $n \in \mathbb{N}$ is odd, we take throughout this step:
\begin{itemize}
    \item $\frac{n-1}{2}-q$ agents with either $R_1$ or $R_5$: with $e$ for $R_1$ and $\frac{n-1}{2}-q-e$ for $R_5$, 
    \item $\frac{n+1}{2}-q$ agents with rankings either $R_2$ or $R_5$: with $f$ for $R_2$ and $\frac{n+1}{2}-q-f$ for $R_6$,
    \item $2q$ agents with $R_3$ or $R_4$: with $\beta$ for $R_3$ and $2q-\beta$ for $R_4$. 
\end{itemize}
The minimum of $q$ is $1$. Otherwise, if $q=0$, then there are no \emph{third-party} agents (i.e., agents with rankings $R_3$ or $R_4$), so there is not iterative plurality dynamics. The adversarial loss $\ADS(P)$ for any such $P$, indexed by $q=0$, is then clearly zero.
The maximum of is $q^* = \max\{q \in \mathbb{Z}~:~\left(\frac{n+1}{2}-q\right) > 2q\}$, so that $q^* = \bfloor{\frac{n}{6}}$ for any $n~mod~6 \in \{1,3,5\}$.
To keep in line with the notation of the first case, in Step 1, and with Lemma \ref{lem:sub_12}, we will assume $q^* = \bfloor{\frac{n}{6}}-1$ without loss of generality. It is easy to show that the case of Equation (\ref{eq:odd_main4}) (below) for $q = \bfloor{\frac{n}{6}}$ is exponentially small.
In order to uphold the condition that $P[1 \succ 2] \geq P[2 \succ 1]$, so that alternative $1$ is the equilibrium winner, we must have $\beta < q$.
While $\calE_2$ holds, it should be clear that the values per agent are the negative of those presented in Table \ref{tab:character_profiles_E4}. Put together, we get the equation
\begin{align}
    & \Pr\nolimits_{P \sim \pi^n}(P \in \mathcal{E}_2) \times \mathbb{E}_{P \sim \pi^n}[\ADS(P)~|~P \in \mathcal{E}_2] \notag \\
    & = -\sum_{q = 0}^{\bfloor{\frac{n}{6}}-1} \sum_{e=0}^{\frac{n-1}{2}-q} \sum_{f=0}^{\frac{n+1}{2}-q} \sum_{\beta = 0}^{q-1} \mathcal{P}^2_{\vec{\pi},n}(e,f,\beta,q) \cdot \mathcal{V}^2_{\vec{u},n}(e,f,\beta,q) \label{eq:odd_main4}
\end{align}
where
\[
    \mathcal{P}^2_{\vec{\pi},n}(e,f,\beta,q) = \binom{n}{e, f, \beta, 2q-\beta, \frac{n-1}{2}-q-e, \frac{n+1}{2}-q-f} \pi_1^e \pi_2^f \pi_3^{\beta} \pi_4^{2q-\beta} \pi_5^{\frac{n-1}{2}-q-e} \pi_6^{\frac{n+1}{2}-q-f} 
\]
and
\begin{align*}
    \mathcal{V}^2_{\vec{u},n}(e,f,\beta,q) & = 
        \begin{pmatrix}
            e, & f, & \beta, & 2q-\beta, & \frac{n-1}{2}-q-e, & \frac{n+1}{2}-q-f 
        \end{pmatrix} \\
    & \quad \cdot \begin{pmatrix} u_1 - u_2, & -u_1+u_3, & -u_2+u_3, & u_2-u_3, & u_1-u_3, & -u_1+u_2 
        \end{pmatrix}.
\end{align*}
We begin by factoring the probability term, which equals 
\begin{align}
    & \mathcal{P}^2_{\vec{\pi},n}(e,f,\beta,q) \notag \\
    &= \binom{n}{\frac{n-1}{2}-q, \frac{n+1}{2}-q, 2q} \times  \binom{\frac{n-1}{2}-q}{e} \pi_1^e \pi_5^{\frac{n-1}{2}-q-e} \times \binom{\frac{n+1}{2}-q}{f} \pi_2^e \pi_6^{\frac{n+1}{2}-q-f} \notag \\
    & \quad \quad \times \binom{2q}{\beta} \pi_3^\beta \pi_4^{2q-\beta} \notag \\
    & = \binom{n}{\frac{n-1}{2}-q, \frac{n+1}{2}-q, 2q} (\pi_1+\pi_5)^{\frac{n-1}{2}-q} (\pi_2+\pi_6)^{\frac{n+1}{2}-q} (\pi_3+\pi_4)^{2q} \notag \\
    & \quad \quad \times  \binom{\frac{n-1}{2}-q}{e} \tilde{\pi}_1^e \tilde{\pi}_5^{\frac{n-1}{2}-q-e} \times \binom{\frac{n+1}{2}-q}{f} \tilde{\pi}_2^e \tilde{\pi}_6^{\frac{n+1}{2}-q-f} \times \binom{2q}{\beta} \tilde{\pi}_3^\beta \tilde{\pi}_4^{2q-\beta} \notag \\
    & = \binom{n-1}{\frac{n-1}{2}-q, \frac{n-1}{2}-q, q,q} \Pi_1^{n-2q} \Pi_3^{2q} \frac{n 2^{2q}}{(\frac{n+1}{2}-q) \binom{2q}{q}} \notag \\
    & \quad \quad \times  \binom{\frac{n-1}{2}-q}{e} \tilde{\pi}_1^e \tilde{\pi}_5^{\frac{n-1}{2}-q-e} \times \binom{\frac{n+1}{2}-q}{f} \tilde{\pi}_2^f \tilde{\pi}_6^{\frac{n+1}{2}-q-f} \times \binom{2q}{\beta} \tilde{\pi}_3^\beta \tilde{\pi}_4^{2q-\beta} \label{eq:odd_prob_main2}
\end{align}
where $\tilde{\pi}_1, \ldots, \tilde{\pi}_6$, $\Pi_1$, and $\Pi_3$ are defined as above.
The value terms of Equation (\ref{eq:odd_main4}) may be written as
\begin{align}
    \mathcal{V}^2_{\vec{u},n}(e,f,\beta,q) 
    & = (u_2 - u_3) \left( \frac{n+1}{2} + q - 2\beta -e -f \right) -u_1 + u_3. \label{eq:odd_value_main2}
\end{align}
We next combine and simplify Equations (\ref{eq:odd_prob_main2}) and (\ref{eq:odd_value_main2}) using a separation of variables technique,
as follows, so that Equation (\ref{eq:odd_main4}) is equal to
\begin{align}
    -\sum_{q = 0}^{\bfloor{\frac{n}{6}}-1} \binom{n-1}{\frac{n-1}{2}-q, \frac{n-1}{2}-q, q,q} \Pi_1^{n-2q} \Pi_3^{2q}  \frac{n 2^{2q}}{ \left( \frac{n+1}{2}-q \right) \binom{2q}{q}} C^2_q \label{eq:odd_main5}
\end{align}
where we define
\begin{align*}
    C^2_q & = \sum_{\beta = 0}^{q-1} \binom{2q}{\beta} \tilde{\pi}_3^{\beta} \tilde{\pi}_4^{2q-\beta} B^2_{q,\beta};
\end{align*}
\begin{align*}
    B^2_{q,\beta} & = \sum_{f=0}^{\frac{n+1}{2}-q} \binom{\frac{n+1}{2}-q}{f} \tilde{\pi}_2^f \tilde{\pi}_6^{\frac{n+1}{2}-q-f} A^2_{q,\beta,f};
\end{align*}
\begin{align*}
    A^2_{q,\beta,f} & = \sum_{e=0}^{\frac{n-1}{2}-q} \binom{\frac{n-1}{2}-q}{e} \tilde{\pi}_1^e \tilde{\pi}_5^{\frac{n-1}{2}-q-e} \notag \\
    & \quad \quad \times \left( (u_2 - u_3) \left( \frac{n+1}{2} + q - 2\beta -e -f \right) -u_1 + u_3 \right).
\end{align*}
We simplify these variables as follows. First, we have 
\begin{align}
    A^2_{q,\beta,f} & =  (u_2 - u_3) \left( \frac{n+1}{2} + q - 2\beta -f - \left( \frac{n-1}{2}-q \right) \tilde{\pi}_1 \right) -u_1+u_3 \notag \\
    & = (u_2 - u_3) \left( \left( \frac{n+1}{2}-q \right) \left( 1 - \tilde{\pi}_1 \right) +2q -2\beta -f \right) + (u_2 - u_3) \tilde{\pi}_1 -u_1 + u_3 \label{eq:odd_A2_finish}
\end{align}
by definition of binomial probability and expectation.
Consider the constant terms of Equation (\ref{eq:odd_A2_finish}). To this point, from Equation (\ref{eq:odd_main5}), we have
\begin{align}
    & \left(  (-u_2 + u_3) \tilde{\pi}_1 +u_1 - u_3 \right) \sum_{q = 0}^{\bfloor{\frac{n}{6}}-1} \binom{n}{\frac{n-1}{2}-q, \frac{n+1}{2}-q, 2q} \Pi_1^{n-2q} (2\Pi_3)^{2q} \notag \\
    & \quad \quad \times \sum_{\beta = 0}^{q-1} \binom{2q}{\beta} \tilde{\pi}_3^{\beta} \tilde{\pi}_4^{2q-\beta} \sum_{f=0}^{\frac{n+1}{2}-q} \binom{\frac{n+1}{2}-q}{f} \tilde{\pi}_2^f \tilde{\pi}_6^{\frac{n+1}{2}-q-f}. \label{eq:odd_const_terms_pre} 
\end{align}
The $f$-summation of Equation (\ref{eq:odd_const_terms_pre}) is clearly $\Theta(1)$ by definition of binomial probability. By Lemma \ref{lem:new_comb}, the $q$-summation of Equation (\ref{eq:odd_const_terms_pre}) is either $\Theta(1)$, if $\tilde{\pi}_3 \leq \tilde{\pi}_4$, and $\calO \left( e^{-\Theta(n)} \right)$ otherwise. Finally, the $q$-summation is the probability of a two-way tie for plurality voting, with three alternatives under i.i.d preferences. By Corollary \ref{coro:sub_others}, Equation (\ref{eq:odd_const_terms_pre}) is therefore
\begin{align}
    \begin{cases}
        \pm \calO\left( \frac{1}{\sqrt{n}} \right), & \pi_3 \leq \pi_4 \\
        \pm \calO \left( e^{-\Theta(n)} \right), & \pi_3 > \pi_4.
    \end{cases} \label{eq:odd_const_terms}
\end{align}
Now consider the non-constant terms of Equation (\ref{eq:odd_A2_finish}). 
For ease of notation, we will continue without writing $(u_2 - u_3)$ as a factor in front of every remaining term; this will be implicitly pulled outside the summation of Equation (\ref{eq:odd_main5}).
Continuing with $B^2_{q,\beta}$, we have
\begin{align}
    B^2_{q,\beta} & = \left( \left( \frac{n+1}{2}-q \right) \tilde{\pi}_5 +2q -2\beta \right) \sum_{f=0}^{\frac{n+1}{2}-q} \binom{\frac{n+1}{2}-q}{f} \tilde{\pi}_2^f \tilde{\pi}_6^{\frac{n+1}{2}-q-f} \notag \\
    & \quad \quad - \sum_{f=0}^{\frac{n+1}{2}-q} \binom{\frac{n}{2}-q}{f} \tilde{\pi}_2^f \tilde{\pi}_6^{\frac{n+1}{2}-q-f} f \notag \\
    & = \left( \left( \frac{n-1}{2}-q \right) \tilde{\pi}_5 +2q -2\beta  \right) - \left( \frac{n+1}{2}-q \right) \tilde{\pi}_2 \notag \\
    & =\left( \frac{n+1}{2}-q \right) \left( \tilde{\pi}_5 - \tilde{\pi}_2 \right) +2q -2\beta.
\end{align}
by definition of binomial probability and expectation.
Third, we have
\begin{align}
    C^2_q & =  \left( \left( \frac{n+1}{2}-q \right) \left( \tilde{\pi}_5 - \tilde{\pi}_2 \right) +2q \right) \sum_{\beta = 0}^{q-1} \binom{2q}{\beta} \tilde{\pi}_3^{\beta} \tilde{\pi}_4^{2q-\beta}
    -2  \sum_{\beta = 0}^{q-1} \binom{2q}{\beta} \tilde{\pi}_3^{\beta} \tilde{\pi}_4^{2q-\beta} \beta. \label{eq:odd_c_base2}
\end{align}
We next employ the following lemma, proved in Appendix \ref{apx:concentration_inequalities}.
\setcounter{lem}{1}
\begin{lem}
Let $p \in (0,1)$.
The following equations hold.
\begin{enumerate}
    \item[(7)]
    \begin{align*}
        & \sum_{\beta = 0}^{q-1} \binom{2q}{\beta} p^{\beta} (1-p)^{2q-\beta} 
        = \begin{cases}
        \frac{1}{2} - \frac{1}{2^{2q+1}} \binom{2q}{q}, & p = \frac{1}{2} \\
        1-\exp(-\Theta(q)), & p < \frac{1}{2} \\ 
        \exp(-\Theta(q)), & p > \frac{1}{2}.
    \end{cases}
    \end{align*}
    \item[(8)]
    \begin{align*}
        & \sum_{\beta = 0}^{q-1} \binom{2q}{\beta} p^{\beta} (1-p)^{2q-\beta} \beta 
        = \begin{cases}
        \frac{q}{2} - \frac{q}{2^{2q}} \binom{2q}{q}, & p = \frac{1}{2} \\
        2qp \left( 1 - \exp(-\Theta(q)) \right), & p < \frac{1}{2} \\ 
        \exp(-\Theta(p)), & p > \frac{1}{2}.
        \end{cases}
    \end{align*}
\end{enumerate}
\end{lem}
Since $\tilde{\pi}_3 = \tilde{\pi}_4$ is assumed, we have
\begin{align}
    C^2_q 
    & = \left( \left( \frac{n+1}{2}-q \right) \left( \tilde{\pi}_5 - \tilde{\pi}_2 \right) +2q \right) \left( \frac{1}{2} - \frac{1}{2^{2q+1}} \binom{2q}{q} \right) - q + \frac{2q}{2^{2q}} \binom{2q}{q}  \notag \\
    & = \frac{1}{2} \left( \frac{n+1}{2} -q \right) (\tilde{\pi}_5 - \tilde{\pi}_2) - \frac{\left( \left( \frac{n+1}{2}-q \right) \left( \tilde{\pi}_5 - \tilde{\pi}_2 \right) -2q \right)}{2^{2q+1}} \binom{2q}{q} \label{eq:odd_main_c_base2_1}
\end{align}
We continue with $C^2_q$ in Step 3, below, and defer the cases where $\tilde{\pi}_3 \neq \tilde{\pi}_4$ to Lemma \ref{lem:sub_12_non_34_equal}.
\end{paragraph}

\begin{paragraph}{Step 3: Putting the pieces back together.}
Recall that our original problem began as Equation (\ref{eq:odd_eit_line0}) which we initially split into Equations (\ref{eq:odd_main1}) and (\ref{eq:odd_main4}). Through a sequence of steps we transformed these equations into Equations (\ref{eq:odd_main2}) and (\ref{eq:odd_main5}) and an additional $+$ or $-\Theta \left( \frac{1}{\sqrt{n}} \right)$ term; recall Equation (\ref{eq:odd_const_terms}). Recombining these simplified equations yields
\begin{align}
    (u_2 - u_3) \sum_{q = 1}^{\bfloor{\frac{n}{6}}-1} \binom{n-1}{\frac{n-1}{2}-q, \frac{n-1}{2}-q, q,q} \Pi_1^{n-1-2q} \Pi_3^{2q} \frac{n 2^{2q}}{\binom{2q}{q}} \left(   \frac{2 \Pi_3 C^1_q}{(2q+1)} - \frac{\Pi_1 C^2_q}{(\frac{n+1}{2}-q)} \right). \label{eq:odd_main7}
\end{align}
By plugging in Equations (\ref{eq:odd_main_c_base1_1}) and (\ref{eq:odd_main_c_base2_1}) into Equation (\ref{eq:odd_main7}), we get 
\begin{align}
    & \frac{2^{2q}}{\binom{2q}{q}} \left( \frac{2 \Pi_3 C^1_q}{(2q+1)} - \frac{\Pi_1 C^2_q}{(\frac{n+1}{2}-q)} \right) \notag \\
    & = \frac{\Pi_3 2^{2q+1}}{(2q+1)\binom{2q}{q}} \left( 
        \frac{1}{2} \left( \frac{n-1}{2}-q \right) \left( \tilde{\pi}_5 - \tilde{\pi}_2 \right) - \frac{ 2q+1 }{2^{2q}} \binom{2q-1}{q-1}
    \right) \notag \\
    & \quad \quad - \frac{\Pi_1 2^{2q}}{(\frac{n+1}{2}-q) \binom{2q}{q}} \left(
         \frac{1}{2} \left( \frac{n+1}{2} -q \right) (\tilde{\pi}_5 - \tilde{\pi}_2) - \frac{\left( \left( \frac{n+1}{2}-q \right) \left( \tilde{\pi}_5 - \tilde{\pi}_2 \right) -2q \right)}{2^{2q+1}} \binom{2q}{q}
    \right) \notag \\
    & = \frac{\Pi_3 (\tilde{\pi}_5 - \tilde{\pi}_2) (\frac{n-1}{2}-q) 2^{2q}}{(2q+1)\binom{2q}{q}} - \Pi_3  - \frac{\Pi_1 (\tilde{\pi}_5 - \tilde{\pi}_2) 2^{2q-1}}{\binom{2q}{q}} + \frac{\Pi_1}{2}\left( \tilde{\pi}_5 - \tilde{\pi}_2 \right)  - \frac{2q \Pi_1}{2(\frac{n+1}{2}-q)} \notag \\ 
    & = \frac{(\tilde{\pi}_5 - \tilde{\pi}_2) \Pi_1 2^{2q}}{\binom{2q}{q}} \left( \frac{\Pi_3 (\frac{n-1}{2} -q)}{\Pi_1 (2q+1)} - \frac{1}{2} \right) - \Pi_3 + \frac{\Pi_1}{2} (\tilde{\pi}_5 - \tilde{\pi}_2) - \frac{q \Pi_1}{(\frac{n+1}{2} -q)}. \label{eq:odd_main8}
\end{align}
When $u_2 > u_3$, this yields the sum of equations
\begin{align}
    n (\tilde{\pi}_5 - \tilde{\pi}_2) \sum_{q = 1}^{\bfloor{\frac{n}{6}}-1} \binom{n-1}{\frac{n-1}{2}-q, \frac{n-1}{2}-q, q,q} \Pi_1^{n-2q} \Pi_3^{2q} \frac{2^{2q}}{\binom{2q}{q}} \left( \frac{\Pi_3 (\frac{n-1}{2} -q)}{\Pi_1 (2q+1)} - \frac{1}{2} \right)  \label{eq:odd_main8a}
\end{align}
and 
\begin{align}
    n \sum_{q = 1}^{\frac{n}{6}-1} \binom{n-1}{\frac{n-1}{2}-q, \frac{n-1}{2}-q, q,q} \Pi_1^{n-1-2q} \Pi_3^{2q} \left( - \Pi_3 + \frac{\Pi_1}{2} (\tilde{\pi}_5 - \tilde{\pi}_2) - \frac{q \Pi_1}{(\frac{n+1}{2} -q)} \right). \label{eq:odd_main8b}
\end{align}
For Equation (\ref{eq:odd_main8a}), observe that
\begin{align*}
    & \frac{\Pi_3 (\frac{n-1}{2} -q)}{\Pi_1 (2q+1)} - \frac{1}{2}
    = \frac{1}{2 \Pi_1 (2q+1)} \left( \left( \Pi_3 (n-1) -q \right) - \Pi_1 \right).
\end{align*}
Then
\begin{align*}
    & \sum_{q = 1}^{\bfloor{\frac{n}{6}}-1} \binom{n-1}{\frac{n-1}{2}-q, \frac{n-1}{2}-q, q,q} \Pi_1^{n-2q} \Pi_3^{2q} \frac{n f_n(q) 2^{2q}}{(2q+1) \binom{2q}{q}} = \pm \calO \left( \frac{1}{\sqrt{n}} \right)
\end{align*}
by Lemma \ref{lem:soln_part_b}, where $f_n(q) = q - \Pi_3 (n-1) + \Pi_1$. Note that Lemma \ref{lem:soln_part_b}, in Appendix \ref{apx:secondary_equations}, is stated where $n$ is even. Thus, to use that lemma, we transform $n-1 \mapsto n$ and plug in $\tau_1 = 1$, $\tau_2 = \Pi_1$, $f_n(q) = \frac{n+1}{2q+1}$.

Equation (\ref{eq:odd_main8b}) is $-\Theta(1)$ by Lemma \ref{lem:prob_bounds}, which follows after realizing that
\[
\left( - \Pi_3 + \frac{\Pi_1}{2} (\tilde{\pi}_5 - \tilde{\pi}_2) - \frac{q \Pi_1}{(\frac{n+1}{2} -q)} \right) = -\Theta(1)
\]
for each $q$ in its domain.
This concludes the proof of Lemma \ref{lem:sub_12_odd}.
\end{paragraph}
\end{proof}

%% file: EC_appendix/apx_page_2d.tex


\subsection{Proof of other cases for two-way ties}
\label{apx:two_ties_other_cases}

\begin{table*}[t]
    \centering
    \small
    \begin{tabular}{|c|c|}
        \hline
         \makecell{$\EDS(\{1,3\})$ when \\ $\pi_1 + \pi_5 = \pi_3 + \pi_4 \geq \pi_2 + \pi_6$} & Asymptotic Rate \\
          \hline \hline
        \makecell{$\begin{cases}\pi_2 = \pi_6 \\ 4\pi_5 + \pi_3 + 3 \pi_1 > 2 \end{cases}$} & $\begin{cases}\Theta(1),& n \text{ is even} \\-\Theta(1), & n \text{ is odd}\end{cases}$\\
        \hline
        \makecell{$\begin{cases}\pi_2 = \pi_6 \\ 4\pi_5 + \pi_3 + 3 \pi_1 < 2 \end{cases}$} & $-\Theta(1)$\\
        \hline
        \makecell{$\begin{cases}\pi_2 = \pi_6 \\ 4\pi_5 + \pi_3 + 3 \pi_1 = 2 \end{cases}$} & $\begin{cases}\pm \calO \left( \frac{1}{\sqrt{n}} \right),& n \text{ is even} \\ - \Theta(1),& n \text{ is odd}\end{cases}$ \\
        \hline
        \makecell{$\begin{cases}\pi_2 \neq \pi_6 \\ \pi_5 + 2\pi_6 = \pi_3 + 2 \pi_2 \end{cases}$} & $\pm \calO \left( \frac{1}{\sqrt{n}}\right)$  \\
        \hline
        \makecell{$\begin{cases}\pi_2 > \pi_6 \\ \pi_5 + 2\pi_6 < \pi_3 + 2 \pi_2 \end{cases}$ or $\begin{cases}\pi_2 < \pi_6 \\ \pi_5 + 2\pi_6 > \pi_3 + 2 \pi_2 \end{cases}$} & $\Theta(\sqrt{n})$ \\
        \hline
        \makecell{$\begin{cases}\pi_2 > \pi_6 \\ \pi_5 + 2\pi_6 > \pi_3 + 2 \pi_2 \end{cases}$ or $\begin{cases}\pi_2 < \pi_6 \\ \pi_5 + 2\pi_6 < \pi_3 + 2 \pi_2 \end{cases}$} & $-\Theta(\sqrt{n})$ \\
        \hline
    \end{tabular}
    \caption{Asymptotic rate of $\EDS(\{1,3\})$ given conditions on $\pi$ when $u_1 \geq u_2 > u_3$.}
    \label{tab:eadpoa_case_13}
\end{table*}

\setcounter{lem}{4}
\begin{lem}

Suppose that $\pi_1 + \pi_5 = \pi_3 + \pi_4 \geq \pi_2 + \pi_6$ and $\pi_i > 0,~\forall i \in [6]$. Furthermore, let $u_1 \geq u_2 > u_3$ in $\vec{u}$. Then $\exists N > 0$ such that $\forall n > N$, $\EDS(\{1,3\})$ is determined by the mapping from $\pi$ to asymptotic rates as described by Table \ref{tab:eadpoa_case_13}.
Both conditions on $\pi$ need to hold; note that this table is exhaustive.

If $u_1 > u_2 = u_3$, then $\EDS(\{1,3\}) = \pm \calO\left( \frac{1}{\sqrt{n}} \right)$ if $\pi_2 \leq \pi_6$ and $\pm \calO \left( e^{-\Theta(n)} \right)$ otherwise.


\label{lem:sub_13}
\end{lem}

\begin{proof}

\begin{table}[t]
    \centering
    \begin{tabular}{|c|c|c|c|}
        \hline
         Ranking & Probability & Frequency & Utility per Agent \\
          \hline \hline
         $R_5 = (1 \succ 3 \succ 2)$ & $\pi_5$ & $e$ & $u_1 - u_2$ \\
         $R_3 = (3 \succ 2 \succ 1)$ & $\pi_3$ & $f$ & $-u_1 + u_3$ \\
         $R_2 = (2 \succ 3 \succ 1)$ & $\pi_2$ & $\beta$ & $-u_2 + u_3$ \\
         $R_6 = (2 \succ 1 \succ 3)$ & $\pi_6$ & $2q-\beta$ & $u_2 - u_3$ \\
         $R_1 = (1 \succ 2 \succ 3)$ & $\pi_1$ & $\frac{n}{2}-q-e$ & $u_1 - u_3$ \\
         $R_4 = (3 \succ 1 \succ 2)$ & $\pi_4$ & $\frac{n}{2}-q-f$ & $-u_1 + u_2$ \\
        \hline
    \end{tabular}
    \caption{Character of profiles $P$ for $\PW{P} = \{1,3\}$ such that the truthful and equilibrium winners are $1$ and $3$, respectively.}
    \label{tab:character_profiles_E3}
\end{table}

We can immediately tell that this lemma can be proved in an identical manner to that of Lemma \ref{lem:sub_12} (when $n$ is even) and Lemma \ref{lem:sub_12_odd} (when $n$ is odd), except for a reassignment of values, probabilities, and labels of frequencies to each ranking. Here, we will demonstrate that this is the case. There are two cases for profiles $P$ such that the equilibrium winning alternative is different than the truthful one, when $\PW{P} = \{1,3\}$: (Case 1) where alternative $1$ is the truthful winner and $3$ is the equilibrium winner, and (Case 2) where alternative $3$ is the truthful winner and $1$ is the equilibrium winner. We define $\calE_1$ and $\calE_3$ to represent these cases, as follows:
\begin{itemize}
    \item $\calE_1 = \{P \in \calL(\calA)^n~:~s_1(\topRank{P}) = s_3(\topRank{P}) > s_2(\topRank{P}) \text{ and } P[3 \succ 1] > P[1 \succ 3]\}$,
    \item $\calE_3 = \{P \in \calL(\calA)^n~:~s_1(\topRank{P})+1 = s_3(\topRank{P}) > s_2(\topRank{P})+1 \text{ and } P[1 \succ 3] \geq P[3 \succ 1]\}$.
\end{itemize}
%
This suggests the following partition:
\begin{align}
    \EDS(\{1,3\}) & = \Pr\nolimits_{P \sim \pi^n}(P \in \mathcal{E}_1) \times \mathbb{E}_{P \sim \pi^n}[\ADS(P)~|~P \in \mathcal{E}_1] \notag \\
    & + \Pr\nolimits_{P \sim \pi^n}(P \in \mathcal{E}_3) \times \mathbb{E}_{P \sim \pi^n}[\ADS(P)~|~P \in \mathcal{E}_3]. \label{eq:solve13_line1}
\end{align}
Let $e, f, \beta, q \in \mathbb{N}$. For Case $\calE_1$, we take when $n \in \mathbb{N}$ is even:
\begin{itemize}
    \item $\frac{n}{2}-q$ agents with either $R_1$ or $R_5$: with $e$ for $R_1$ and $\frac{n}{2}-q-e$ for $R_5$,
    \item $\frac{n}{2}-q$ agents with either $R_3$ or $R_4$: with $f$ for $R_3$ and $\frac{n}{2}-q-f$ for $R_4$,
    \item $2q$ agents with either $R_2$ or $R_6$: with $\beta$ for $R_2$ and $2q-\beta$ for $R_6$.
\end{itemize}
The exact frequencies are tabulated in Table \ref{tab:character_profiles_E3}. In order to uphold the condition that $P[3 \succ 1] > P[1 \succ 3]$, so that alternative $3$ is the equilibrium winner, we must have $\beta > q$. 

Compare this table with Table \ref{tab:character_profiles_E1} from Lemma \ref{lem:sub_12} (when $n$ is even). We see that the only change that differentiates the computation of Equation (\ref{eq:solve13_line1}) from Equation (\ref{eq:main1}) in Lemma \ref{lem:sub_12} is a permutation of the rankings associated with each pair of frequencies and utilities per agent. This permutation is made so that the index variables $e$, $f$, and $\beta$ correspond to the same utility-per-agent as designated in the original Table \ref{tab:character_profiles_E1}. That is, $\pi_2$ and $\pi_6$ swap with $\pi_3$ and $\pi_4$, while $\pi_1$ and $\pi_5$ switch variables.

An identical argument may be made about Case $\calE_3$ or when $n$ is odd. Although there is minor variation in how $\calE_1$ and $\calE_3$ are defined here, as opposed to in Lemma \ref{lem:sub_12}, the maximum of $q$ is still $\approx \frac{n}{6}$. This variation, with some constant numbers of $q$ that may be different, does not affect the result. It therefore follows that the findings of Lemmas \ref{lem:sub_12} and \ref{lem:sub_12_odd} translate to this lemma upon making the following correspondence. 

Consider the probability distribution $\pi$ such that $\pi_1 + \pi_5 = \pi_3 + \pi_4 \geq \pi_2 + \pi_6$. Let $\pi' = (\pi_1', \pi_2', \pi_3', \pi_4', \pi_5', \pi_6')$ be defined according to the permutation:
\[
\begin{pmatrix}
    \pi_5 \\ \pi_3 \\ \pi_2 \\ \pi_6 \\ \pi_1 \\ \pi_4
\end{pmatrix}
\mapsto
\begin{pmatrix}
    \pi_1' \\ \pi_2' \\ \pi_3' \\ \pi_4' \\ \pi_5' \\ \pi_6'
\end{pmatrix}.
\]
Then Lemmas \ref{lem:sub_12} and \ref{lem:sub_12_odd} hold for $\pi'$. This concludes the proof of Lemma \ref{lem:sub_13}.
\end{proof}


\begin{table*}[t]
    \centering
    \small
    \begin{tabular}{|c|c|}
        \hline
         \makecell{$\EDS(\{2,3\})$ when \\ $\pi_2 + \pi_6 = \pi_3 + \pi_4 \geq \pi_1 + \pi_5$} & Asymptotic Rate \\
          \hline \hline
        \makecell{$\begin{cases}\pi_5 = \pi_1 \\ 4\pi_2 + \pi_4 + 3 \pi_6 > 2 \end{cases}$} & $\begin{cases}\Theta(1),& n \text{ is even} \\-\Theta(1), & n \text{ is odd}\end{cases}$\\
        \hline
        \makecell{$\begin{cases}\pi_5 = \pi_1 \\ 4\pi_2 + \pi_4 + 3 \pi_6 < 2 \end{cases}$} & $-\Theta(1)$\\
        \hline
        \makecell{$\begin{cases}\pi_5 = \pi_1 \\ 4\pi_2 + \pi_4 + 3 \pi_6 = 2 \end{cases}$} & $\begin{cases}\pm \calO \left( \frac{1}{\sqrt{n}} \right),& n \text{ is even} \\ - \Theta(1),& n \text{ is odd}\end{cases}$ \\
        \hline
        \makecell{$\begin{cases}\pi_5 \neq \pi_1 \\ \pi_2 + 2\pi_1 = \pi_4 + 2 \pi_5 \end{cases}$} & $\pm \calO \left( \frac{1}{\sqrt{n}}\right)$  \\
        \hline
        \makecell{$\begin{cases}\pi_5 > \pi_1 \\ \pi_2 + 2\pi_1 < \pi_4 + 2 \pi_5 \end{cases}$ or $\begin{cases}\pi_5 < \pi_1 \\ \pi_2 + 2\pi_1 > \pi_4 + 2 \pi_5 \end{cases}$} & $\Theta(\sqrt{n})$ \\
        \hline
        \makecell{$\begin{cases}\pi_5 > \pi_1 \\ \pi_2 + 2\pi_1 > \pi_4 + 2 \pi_5 \end{cases}$ or $\begin{cases}\pi_5 < \pi_1 \\ \pi_2 + 2\pi_1 < \pi_4 + 2 \pi_5 \end{cases}$} & $-\Theta(\sqrt{n})$ \\
        \hline
    \end{tabular}
    \caption{Asymptotic rate of $\EDS(\{2,3\})$ given conditions on $\pi$ when $u_1 \geq u_2 > u_3$.}
    \label{tab:eadpoa_case_23}
\end{table*}

\setcounter{lem}{5}
\begin{lem}

Suppose that $\pi_2 + \pi_6 = \pi_3 + \pi_4 \geq \pi_1 + \pi_5$ and $\pi_i > 0,~\forall i \in [6]$. Furthermore, let $u_2 > u_3$ in $\vec{u}$. Then $\exists N > 0$ such that $\forall n > N$, $\EDS(\{2,3\})$ is determined by the mapping from $\pi$ to asymptotic rates as described by Table \ref{tab:eadpoa_case_23}.
Both conditions on $\pi$ need to hold; note that this table is exhaustive.

If $u_1 > u_2 = u_3$, then $\EDS(\{2,3\}) = \pm \calO\left( \frac{1}{\sqrt{n}} \right)$ if $\pi_5 \leq \pi_1$ and $\pm \calO \left( e^{-\Theta(n)} \right)$ otherwise.




\label{lem:sub_23}
\end{lem}

\begin{proof}


This lemma follows from Lemma \ref{lem:sub_12} (when $n$ is even) and Lemma \ref{lem:sub_12_odd} (when $n$ is odd), just like in Lemma \ref{lem:sub_13}. 
There are two cases for profiles $P$ such that the equilibrium winning alternative is different than the truthful one, when $\PW{P} = \{2,3\}$: (Case 1) where alternative $2$ is the truthful winner and $3$ is the equilibrium winner, and (Case 2) where alternative $3$ is the truthful winner and $2$ is the equilibrium winner. We define $\calE_2$ and $\calE_3$ to represent these cases, as follows:
\begin{itemize}
    \item $\calE_2 = \{P \in \calL(\calA)^n~:~s_2(\topRank{P}) = s_3(\topRank{P}) > s_1(\topRank{P})+1 \text{ and } P[3 \succ 2] > P[2 \succ 3]\}$,
    \item $\calE_3 = \{P \in \calL(\calA)^n~:~s_2(\topRank{P})+1 = s_3(\topRank{P}) > s_1(\topRank{P})+1 \text{ and } P[2 \succ 3] \geq P[3 \succ 2]\}$.
\end{itemize}
The correspondence of rankings and probabilities to frequencies and utilities per agent, in this case, is tabulated in Table \ref{tab:character_profiles_E5}.
Like in Lemma \ref{lem:sub_13}, this permutation is made so that the index variables $e$, $f$, and $\beta$ correspond to the same utility-per-agent as designated in the original Table \ref{tab:character_profiles_E1}. That is, $\pi_2$ and $\pi_6$ take the place of $\pi_1$ and $\pi_5$, which take the place of $\pi_4$ and $\pi_3$, which take the place of $\pi_2$ and $\pi_6$. Furthermore, although there is minor variation in how $\calE_2$ and $\calE_3$ are defined here, as opposed to in Lemma \ref{lem:sub_12}, this does not significantly affect the main result.

Consider the case of $\calE_2$, where alternative $2$ and $3$ are the truthful and equilibrium winners, respectively. Then we see that alternatives $2$ and $3$ have the same number of agents voting for them (i.e., $\frac{n}{2}-q$) and alternative $3$ has more agents preferring it when $\beta > q$. This confirms the context that motivated the proofs of Lemmas \ref{lem:sub_12} and \ref{lem:sub_12_odd}.

Consider the probability distribution $\pi$ such that $\pi_1 + \pi_5 = \pi_3 + \pi_4 \geq \pi_2 + \pi_6$. Let $\pi' = (\pi_1', \pi_2', \pi_3', \pi_4', \pi_5', \pi_6')$ be defined according to the permutation:
\[
\begin{pmatrix}
    \pi_2 \\ \pi_4 \\ \pi_5 \\ \pi_1 \\ \pi_6 \\ \pi_3
\end{pmatrix}
\mapsto
\begin{pmatrix}
    \pi_1' \\ \pi_2' \\ \pi_3' \\ \pi_4' \\ \pi_5' \\ \pi_6'
\end{pmatrix}.
\]
Then Lemmas \ref{lem:sub_12} and \ref{lem:sub_12_odd} hold for $\pi'$. This concludes the proof of Lemma \ref{lem:sub_23}.

\begin{table}[t]
    \centering
    \begin{tabular}{|c|c|c|c|}
        \hline
         Ranking & Probability & Frequency & Utility per Agent \\
          \hline \hline
         $R_2 = (2 \succ 3 \succ 1)$ & $\pi_2$ & $e$ & $u_1 - u_2$ \\
         $R_4 = (3 \succ 1 \succ 2)$ & $\pi_4$ & $f$ & $-u_1 + u_3$ \\
         $R_5 = (1 \succ 3 \succ 2)$ & $\pi_5$ & $\beta$ & $-u_2 + u_3$ \\
         $R_1 = (1 \succ 2 \succ 3)$ & $\pi_1$ & $2q-\beta$ & $u_2 - u_3$ \\
         $R_6 = (2 \succ 1 \succ 3)$ & $\pi_6$ & $\frac{n}{2}-q-e$ & $u_1 - u_3$ \\
         $R_3 = (3 \succ 2 \succ 1)$ & $\pi_3$ & $\frac{n}{2}-q-f$ & $-u_1 + u_2$ \\
        \hline
    \end{tabular}
    \caption{Character of profiles $P$ for $\PW{P} = \{2,3\}$ such that the truthful and equilibrium winners are $2$ and $3$, respectively.}
    \label{tab:character_profiles_E5}
\end{table}

\end{proof}

%% file: EC_appendix/apx_page_2c.tex



\setcounter{lem}{6}
\begin{lem}

Suppose that $\pi_1 + \pi_5 = \pi_2 + \pi_6 \geq \pi_3 + \pi_4$ and $\pi_i > 0,~\forall i \in [6]$. Furthermore, let $u_2 > u_3$ in $\vec{u}$ and $\pi_3 \neq \pi_4$. Then $\exists N > 0$ such that $\forall n > N$, $\EDS(\{1,2\})$ is determined by the mapping from $\pi$ to asymptotic rates as described by Table \ref{tab:eadpoa_case_12}.

\label{lem:sub_12_non_34_equal}
\end{lem}

\begin{proof}
This proof differentiates covers both cases for whether $n$ is even or odd.

\begin{paragraph}{Step 1 ($n$ is even):}
This proof continues that of Lemma \ref{lem:sub_12} when $n$ is even, but $\pi_3 \neq \pi_4$. We make use of the same definitions of $\tilde{\pi}_1, \ldots, \tilde{\pi}_6$, $\Pi_1$, and $\Pi_3$ as in Definition \ref{dfn:tilde_definitions} from that lemma.
Our objective is to simplify Equation (\ref{eq:main7}):
\begin{align*}
    (u_2 - u_3) \sum_{q = 1}^{\frac{n}{6}-1} \binom{n}{\frac{n}{2}-q, \frac{n}{2}-q, q,q} \Pi_1^{n-2q} \Pi_3^{2q}  \frac{2^{2q}}{\binom{2q}{q}} \left( C^1_q - \frac{2 \Pi_3 (\frac{n}{2}-q)}{\Pi_1 (2q+1)} C^2_q \right).
\end{align*}
In order to proceed when $\pi_3 \neq \pi_4$, we must first split this 
into separate parts:
\begin{align}
    & \sum_{q =1}^{\bfloor{\frac{\Pi_3 n}{4}}-1} \binom{n}{\frac{n}{2}-q, \frac{n}{2}-q, q,q} \Pi_1^{n-2q} \Pi_3^{2q}  \frac{2^{2q}}{\binom{2q}{q}} \left( C^1_q - \frac{2 \Pi_3 (\frac{n}{2}-q)}{\Pi_1 (2q+1)} C^2_q \right) \notag \\
    & + \sum_{q=\bfloor{\frac{\Pi_3 n}{4}}}^{\frac{n}{6}-1} \binom{n}{\frac{n}{2}-q, \frac{n}{2}-q, q,q} \Pi_1^{n-2q} \Pi_3^{2q}  \frac{2^{2q}}{\binom{2q}{q}} \left( C^1_q - \frac{2 \Pi_3 (\frac{n}{2}-q)}{\Pi_1 (2q+1)} C^2_q \right). \label{eq:split_main8}
\end{align}
where we recall from Equations (\ref{eq:c_base}) and (\ref{eq:c_base2}) that
\begin{align*}
    C^1_q & =  \left( \left( \frac{n}{2}-q \right) \left( \tilde{\pi}_5 - \tilde{\pi}_2 \right) +2q \right) \sum_{\beta = q+1}^{2q} \binom{2q}{\beta} \tilde{\pi}_3^{\beta} \tilde{\pi}_4^{2q-\beta} 
    -2  \sum_{\beta = q+1}^{2q} \binom{2q}{\beta} \tilde{\pi}_3^{\beta} \tilde{\pi}_4^{2q-\beta} \beta.
\end{align*}
and
\begin{align*}
    C^2_q & =  \left( \left( \frac{n}{2}-q \right) \left( \tilde{\pi}_5 - \tilde{\pi}_2 \right) +2q \right) \sum_{\beta = 0}^{q} \binom{2q+1}{\beta} \tilde{\pi}_3^{\beta} \tilde{\pi}_4^{2q+1-\beta} 
    -2  \sum_{\beta = 0}^{q} \binom{2q+1}{\beta} \tilde{\pi}_3^{\beta} \tilde{\pi}_4^{2q+1-\beta} \beta.
\end{align*}
By Lemma \ref{lem:new_comb},
%
%
%
%
it follows that if $\tilde{\pi}_3 < \tilde{\pi}_4$, then 
\begin{align}
    C^1_q = \calO(n) \exp(-\Theta(q)). 
    \notag
\end{align}
and
\begin{align}
    C^2_q 
    & = \left( \left( \frac{n}{2}-q \right) \left( \tilde{\pi}_5 - \tilde{\pi}_2 \right) +2q \right) \left( 1 - \exp(-\Theta(q)) \right) - (4q+2) \tilde{\pi}_3 \left( 1 - \exp(-\Theta(q)) \right) \notag \\
    & = \left( \frac{n}{2}-q \right) \left( \tilde{\pi}_5 - \tilde{\pi}_2 \right) +2q( \tilde{\pi}_4 - \tilde{\pi}_3) - 2 \tilde{\pi}_3  - \calO(n) \exp(-\Theta(q)). \notag 
\end{align}
On the other hand, if $\tilde{\pi}_3 > \tilde{\pi}_4$, then we have
%
%
%
\begin{align}
    C^1_q 
    & = \left( \left( \frac{n}{2}-q \right) \left( \tilde{\pi}_5 - \tilde{\pi}_2 \right) +2q \right) \left( 1 - \exp(-\Theta(q)) \right) - (4q \tilde{\pi}_3) \left( 1 - \exp(-\Theta(q)) \right) \notag \\
    & =  \left( \frac{n}{2}-q \right) \left( \tilde{\pi}_5 - \tilde{\pi}_2 \right) +2q(\tilde{\pi}_4 -\tilde{\pi}_3)  - \calO(n) \exp(-\Theta(q)). \notag
\end{align}
and
%
\begin{align}
    C^2_q = \calO(n) \exp(-\Theta(q)). \notag
\end{align}
Regarding the first summand of Equation (\ref{eq:split_main8}), it is therefore easy to see 
\[
    \left| \frac{2^{2q}}{\binom{2q}{q}} \left( C^1_q - \frac{2 \Pi_3 (\frac{n}{2}-q)}{\Pi_1 (2q+1)} C^2_q \right) \right| \leq \calO(n^{2.5}).
\]
since $\frac{2^{2q}}{\binom{2q}{q}} = \calO(\sqrt{q})$ by Stirling's approximation (Proposition \ref{prop:stirling}), $q = \calO(n)$ along its domain, $|C^1_q| \leq \calO(n)$ and $|C^2_q| \leq \calO(n)$.
We conclude that the first summand is $\calO \left( e^{-\Theta(n)} \right)$ by 
the following lemma. 

\setcounter{lem}{11}
\begin{lem}
    Fix $a,b \in (0,\frac{1}{6})$, $a<b$. Let $\Pi_3 \in (0, \frac{1}{3}]$ and $\Pi_1 = \frac{1}{2} - \Pi_3$.
    Then
\begin{align*}
    & \sum_{q=\bfloor{a n}}^{\frac{n}{6} -1} \binom{n}{\frac{n}{2}-q, \frac{n}{2}-q, q, q} \Pi_1^{n-2q} \Pi_3^{2q}
    = \begin{cases}
    \Theta\left( \frac{1}{n} \right), & \Pi_3 \geq a \\
    \calO \left( e^{-\Theta(n)} \right), & \text{otherwise}
    \end{cases}
\end{align*}
and
\begin{align*}
    & \sum_{q=1}^{\bceil{b n}} \binom{n}{\frac{n}{2}-q, \frac{n}{2}-q, q, q} \Pi_1^{n-2q} \Pi_3^{2q}
    = \begin{cases}
    \Theta\left( \frac{1}{n} \right), & \Pi_3 \leq b \\
    \calO \left( e^{-\Theta(n)} \right), & \text{otherwise}.
    \end{cases}
\end{align*}
\end{lem}
Lemma \ref{lem:prob_bounds} is a direct application of \citet[Theorem 1]{Xia2021:How-Likely} and is proved in Appendix \ref{apx:helper_lemmas}.
We henceforth may focus on the second summand of Equation (\ref{eq:split_main8}). 

\begin{paragraph}{Case where $\pi_3 < \pi_4$:}
When $\pi_3 < \pi_4$, we have
\begin{align*}
    & C^1_q - \frac{2 \Pi_3 (\frac{n}{2}-q)}{\Pi_1 (2q+1)} C^2_q \\
    & = \calO(n) \exp(-\Theta(q)) \\
    & \quad \quad - \frac{2 \Pi_3 (\frac{n}{2}-q)}{\Pi_1 (2q+1)} \left ( 
        \left( \frac{n}{2}-q \right) \left( \tilde{\pi}_5 - \tilde{\pi}_2 \right) +2q( \tilde{\pi}_4 - \tilde{\pi}_3)  - 2 \tilde{\pi}_3  - \calO(n) \exp(-\Theta(q))
    \right) \\
    & = \pm \calO \left( e^{-\Theta(n)} \right)  - \frac{2 \Pi_3 (\frac{n}{2}-q)}{\Pi_1 (2q+1)} ((\Pi_3 n-q)(\tilde{\pi}_5 - \tilde{\pi}_2 - 2 \tilde{\pi}_4 + 2 \tilde{\pi}_3) - 2 \tilde{\pi}_3) \notag \\
    & \quad \quad  - \frac{2 \Pi_3 n (\frac{n}{2}-q)}{\Pi_1 (2q+1)}  (\Pi_1 (\tilde{\pi}_5 - \tilde{\pi}_2 + 2 \Pi_3 (\tilde{\pi}_4 - \tilde{\pi}_3))
\end{align*}
where the exponential term 
is found by Stirling's approximation (Proposition \ref{prop:stirling}), and noting that $q = \Theta(n)$ for every term in the summation. 
Therefore, Equation (\ref{eq:split_main8}) is
\begin{align*}
    & \pm \calO \left( e^{-\Theta(n)} \right) \notag \\
    & - \frac{2 \Pi_3}{\Pi_1} \sum_{q=\bfloor{\frac{\Pi_3 n}{4}}}^{\frac{n}{6}-1} \binom{n}{\frac{n}{2}-q, \frac{n}{2}-q, q,q} \Pi_1^{n-2q} \Pi_3^{2q} \frac{\left( -\frac{n}{2}+q \right) f_n(q) 2^{2q}}{(2q+1) \binom{2q}{q}} \notag \\
    & - n \left( \pi_5 - \pi_2 + 2 \pi_4 - 2\pi_3 \right) 
    \sum_{q=\bfloor{\frac{\Pi_3 n}{4}}}^{\frac{n}{6}-1} \binom{n}{\frac{n}{2}-1-q, \frac{n}{2}-q, 2q+1} \Pi_1^{n-1-2q} (2\Pi_3)^{2q+1} \\ 
    & = \pm \calO \left( \frac{1}{\sqrt{n}} \right) - \left( \pi_5 - \pi_2 + 2 \pi_4 - 2\pi_3 \right) \calO(\sqrt{n})
\end{align*}
by Lemma \ref{lem:soln_part_b} and Corollary \ref{coro:sub_others}, where $f_n(q) = 
        \left( q - \Pi_3 n \right) \left( \tilde{\pi}_5 - \tilde{\pi}_2 - 2 \tilde{\pi}_4 + 2 \tilde{\pi}_3 \right)   + 2 \tilde{\pi}_3$.
The $\calO(\sqrt{n})$ sign is positive if $\pi_5 + 2 \pi_4 < \pi_2 + 2 \pi_3$.
\end{paragraph}

\begin{paragraph}{Case where $\pi_3 > \pi_4$:}
When $\pi_3 > \pi_4$, we have
\begin{align*}
    & C^1_q - \frac{2 \Pi_3 (\frac{n}{2}-q)}{\Pi_1 (2q+1)} C^2_q \\
    & = \left( \frac{n}{2}-q \right) \left( \tilde{\pi}_5 - \tilde{\pi}_2 \right) +2q(\tilde{\pi}_4 -\tilde{\pi}_3)  - \calO(n) \exp(-\Theta(q)) \\
    & = \pm \calO \left( e^{-\Theta(n)} \right) +  (\Pi_3 n-q)(\tilde{\pi}_5 - \tilde{\pi}_2 - 2 \tilde{\pi}_4 + 2 \tilde{\pi}_3) \\
    & \quad \quad + n (\Pi_1 (\tilde{\pi}_5 - \tilde{\pi}_2 + 2 \Pi_3 (\tilde{\pi}_4 - \tilde{\pi}_3))
\end{align*}
where the exponential term 
is found by Stirling's approximation (Proposition \ref{prop:stirling}), and noting that $q = \Theta(n)$ for every term in the summation. Therefore Equation (\ref{eq:split_main8}) is
\begin{align*}
    & \pm \calO \left( e^{-\Theta(n)} \right) \notag \\
    & + \left( \tilde{\pi}_5 - \tilde{\pi}_2 - 2 \tilde{\pi}_4 + 2 \tilde{\pi}_3 \right) 
    \sum_{q=\bfloor{\frac{\Pi_3 n}{4}}}^{\frac{n}{6}-1} \binom{n}{\frac{n}{2}-q, \frac{n}{2}-q, q,q} \Pi_1^{n-2q} \Pi_3^{2q} \frac{f_n(q) 2^{2q}}{(2q+1) \binom{2q}{q}} \notag \\
    & + n \left( \pi_5 - \pi_2 + 2 \pi_4 - 2 \pi_3 \right) 
    \sum_{q=\bfloor{\frac{\Pi_3 n}{4}}}^{\frac{n}{6}-1} \binom{n}{\frac{n}{2}-q, \frac{n}{2}-q, 2q} \Pi_1^{n-2q} (2\Pi_3)^{2q}  \\ 
    & = \pm \calO \left( \frac{1}{\sqrt{n}} \right) + \left( \pi_5 - \pi_2 + 2 \pi_4 - 2 \pi_3 \right) \calO(\sqrt{n})
\end{align*}
by Lemma \ref{lem:soln_part_b} and Corollary \ref{coro:sub_others},
where 
\begin{align*}
    f_n(q) & = \left( \Pi_3 n-q \right) (2q+1) \\
    & = 2\left(-\frac{n}{2}+q\right)(q-\Pi_3 n) - (n+1)(q-\Pi_3 n).
\end{align*}
The $\calO(\sqrt{n})$ sign is positive if $\pi_5 + 2 \pi_4 > \pi_2 + 2 \pi_3$.
\end{paragraph}
This concludes the proof of Lemma \ref{lem:sub_12_non_34_equal} where $n$ is even.
\end{paragraph}





\begin{paragraph}{Step 2 ($n$ is odd):}
This step of the proof continues that of Lemma \ref{lem:sub_12_odd} when $n$ is odd, but $\pi_3 \neq \pi_4$.
Our objective is to simplify Equation (\ref{eq:odd_main7}):
\begin{align*}
    (u_2 - u_3) \sum_{q = 1}^{\bfloor{\frac{n}{6}}-1} \binom{n-1}{\frac{n-1}{2}-q, \frac{n-1}{2}-q, q,q} \Pi_1^{n-1-2q} \Pi_3^{2q} \frac{n 2^{2q}}{\binom{2q}{q}} \left(   \frac{2 \Pi_3 C^1_q}{(2q+1)} - \frac{\Pi_1 C^2_q}{(\frac{n+1}{2}-q)} \right).
\end{align*}
In order to proceed when $\pi_3 \neq \pi_4$, we must first split this 
into separate parts:
\begin{align}
    & \sum_{q =1}^{\bfloor{\frac{\Pi_3 n}{4}}-1} 
    \binom{n-1}{\frac{n-1}{2}-q, \frac{n-1}{2}-q, q,q} \Pi_1^{n-1-2q} \Pi_3^{2q} \frac{n 2^{2q}}{\binom{2q}{q}} \left(   \frac{2 \Pi_3 C^1_q}{(2q+1)} - \frac{\Pi_1 C^2_q}{(\frac{n+1}{2}-q)} \right) \notag \\
    & + \sum_{q=\bfloor{\frac{\Pi_3 n}{4}}}^{\bfloor{\frac{n}{6}}-1} 
    \binom{n-1}{\frac{n-1}{2}-q, \frac{n-1}{2}-q, q,q} \Pi_1^{n-1-2q} \Pi_3^{2q} \frac{n 2^{2q}}{\binom{2q}{q}} \left(   \frac{2 \Pi_3 C^1_q}{(2q+1)} - \frac{\Pi_1 C^2_q}{(\frac{n+1}{2}-q)} \right). \label{eq:odd_split_main8}
\end{align}
where we recall from Equations (\ref{eq:odd_c_base}) and (\ref{eq:odd_c_base2}) that
\begin{align*}
    C^1_q & =  \left( \left( \frac{n-1}{2}-q \right) \left( \tilde{\pi}_5 - \tilde{\pi}_2 \right) +2q +1\right) \sum_{\beta = q+1}^{2q+1} \binom{2q+1}{\beta} \tilde{\pi}_3^{\beta} \tilde{\pi}_4^{2q+1-\beta} \notag \\
    & \quad \quad -2  \sum_{\beta = q+1}^{2q+1} \binom{2q+1}{\beta} \tilde{\pi}_3^{\beta} \tilde{\pi}_4^{2q+1-\beta} \beta.
\end{align*}
and
\begin{align*}
    C^2_q & =  \left( \left( \frac{n+1}{2}-q \right) \left( \tilde{\pi}_5 - \tilde{\pi}_2 \right) +2q \right) \sum_{\beta = 0}^{q-1} \binom{2q}{\beta} \tilde{\pi}_3^{\beta} \tilde{\pi}_4^{2q-\beta} 
    -2  \sum_{\beta = 0}^{q-1} \binom{2q}{\beta} \tilde{\pi}_3^{\beta} \tilde{\pi}_4^{2q-\beta} \beta.
\end{align*}
By Lemma \ref{lem:new_comb},
%
%
%
%
it follows that if $\tilde{\pi}_3 < \tilde{\pi}_4$, then 
\begin{align}
    C^1_q = \calO(n) \exp(-\Theta(q)). 
    \notag
\end{align}
and
\begin{align}
    C^2_q & = \left( \left( \frac{n+1}{2}-q \right) \left( \tilde{\pi}_5 - \tilde{\pi}_2 \right) +2q \right) \left( 1 - \exp(-\Theta(q)) \right) - (4q) \tilde{\pi}_3 \left( 1 - \exp(-\Theta(q)) \right) \notag \\
    & = \left( \frac{n+1}{2}-q \right) \left( \tilde{\pi}_5 - \tilde{\pi}_2 \right) +2q( \tilde{\pi}_4 - \tilde{\pi}_3)- \calO(n) \exp(-\Theta(q)). \notag 
\end{align}
On the other hand, if $\tilde{\pi}_3 > \tilde{\pi}_4$, then we have 
\begin{align}
    C^1_q & = \left( \left( \frac{n-1}{2}-q \right) \left( \tilde{\pi}_5 - \tilde{\pi}_2 \right) +2q +1 \right) \left( 1 - \exp(-\Theta(q)) \right) - (4q+2) \tilde{\pi}_3 \left( 1 - \exp(-\Theta(q)) \right) \notag \\
    & =  \left( \frac{n-1}{2}-q \right) \left( \tilde{\pi}_5 - \tilde{\pi}_2 \right) +(2q+1)(\tilde{\pi}_4 -\tilde{\pi}_3)  - \calO(n) \exp(-\Theta(q)). \notag
\end{align}
and
\begin{align}
    C^2_q = \calO(n) \exp(-\Theta(q)). \notag
\end{align}
Regarding the first summand of Equation (\ref{eq:odd_split_main8}), it is therefore easy to see
\[
    \left| \frac{2^{2q}}{\binom{2q}{q}} \left( \frac{2 \Pi_3 C^1_q}{(2q+1)} - \frac{\Pi_1 C^2_q}{(\frac{n+1}{2}-q)} \right) \right| \leq \calO(n^{2.5}).
\]
since $\frac{2^{2q}}{\binom{2q}{q}} = \calO(\sqrt{q})$ by Stirling's approximation (Proposition \ref{prop:stirling}), $q = \calO(n)$ along its domain, $|C^1_q| \leq \calO(n)$ and $|C^2_q| \leq \calO(n)$.
The first summand is therefore $\calO \left( e^{-\Theta(n)} \right)$ by Lemma \ref{lem:prob_bounds}.
We henceforth may focus on the second summand of Equation (\ref{eq:odd_split_main8}). 

\begin{paragraph}{Case where $\pi_3 < \pi_4$:}
When $\pi_3 < \pi_4$, we have
\begin{align*}
    & \frac{2 \Pi_3 C^1_q}{(2q+1)} - \frac{\Pi_1 C^2_q}{(\frac{n+1}{2}-q)} \\
    & = \calO(n) \exp(-\Theta(q))  \\
    & \quad \quad - \frac{\Pi_1}{(\frac{n+1}{2}-q)} \left( 
        \left( \frac{n+1}{2}-q \right) \left( \tilde{\pi}_5 - \tilde{\pi}_2 \right) +2q( \tilde{\pi}_4 - \tilde{\pi}_3)   - \calO(n) \exp(-\Theta(q))
    \right) \\
    & = \pm \calO \left( e^{-\Theta(n)} \right) - \frac{\Pi_1}{(\frac{n+1}{2}-q)} \left( (\Pi_3 (n-1)-q)(\tilde{\pi}_5 - \tilde{\pi}_2 - 2 \tilde{\pi}_4 + 2 \tilde{\pi}_3) + (\tilde{\pi}_5 - \tilde{\pi}_2) \right) \\
    & \quad \quad - \frac{\Pi_1 (n-1)}{(\frac{n+1}{2}-q)} (\Pi_1 (\tilde{\pi}_5 - \tilde{\pi}_2 + 2 \Pi_3 (\tilde{\pi}_4 - \tilde{\pi}_3))
\end{align*}
where the exponential term 
is found by Stirling's approximation (Proposition \ref{prop:stirling}), and noting that $q = \Theta(n)$ for every term in the summation. Therefore Equation (\ref{eq:odd_split_main8}) is
\begin{align}
    & \pm \calO \left( e^{-\Theta(n)} \right) \notag \\
    & + \sum_{q=\bfloor{\frac{\Pi_3 n}{4}}}^{\bfloor{\frac{n}{6}}-1} \binom{n-1}{\frac{n-1}{2}-q, \frac{n-1}{2}-q, q, q} \Pi_1^{n-2q} \Pi_3^{2q} \frac{n f_n(q) 2^{2q}}{(\frac{n+1}{2}-q) \binom{2q}{q}} \notag \\
    & - (n-1) \left( \pi_5 - \pi_2 + 2 \pi_4 - 2\pi_3 \right)
    \sum_{q=\bfloor{\frac{\Pi_3 n}{4}}}^{\bfloor{\frac{n}{6}}-1} \binom{n}{\frac{n-1}{2}-q, \frac{n+1}{2}-q, 2q} \Pi_1^{n-2q} (2\Pi_3)^{2q} \\ 
    & = \pm \calO \left( \frac{1}{\sqrt{n}} \right) - \left( \pi_5 - \pi_2 + 2 \pi_4 - 2\pi_3 \right) \calO \left( \sqrt{n} \right)
\end{align}
by Lemma \ref{lem:soln_part_b} and Corollary \ref{coro:sub_others},
where $f_n(q) =  (q-\Pi_3 (n-1))(\tilde{\pi}_5 - \tilde{\pi}_2 - 2 \tilde{\pi}_4 + 2 \tilde{\pi}_3) + \tilde{\pi}_2 - \tilde{\pi}_5$.
The $\calO(\sqrt{n})$ sign is positive if $\pi_5 + 2 \pi_4 < \pi_2 + 2 \pi_3$.

\end{paragraph}

\begin{paragraph}{Case where $\pi_3 > \pi_4$:}
When $\pi_3 > \pi_4$, we have
\begin{align*}
    & \frac{2 \Pi_3 C^1_q}{(2q+1)} - \frac{\Pi_1 C^2_q}{(\frac{n+1}{2}-q)} \\
    & = \frac{2 \Pi_3}{(2q+1)} \left( 
        \left( \frac{n-1}{2}-q \right) \left( \tilde{\pi}_5 - \tilde{\pi}_2 \right) +(2q+1)( \tilde{\pi}_4 - \tilde{\pi}_3)   - \calO(n) \exp(-\Theta(q))
    \right) \\
    & \quad \quad - \calO(n) \exp(-\Theta(q)) \\
    & = \pm \calO \left( e^{-\Theta(n)} \right) + \frac{2 \Pi_3}{(2q+1)} \left( (\Pi_3 (n-1)-q)(\tilde{\pi}_5 - \tilde{\pi}_2 - 2 \tilde{\pi}_4 + 2 \tilde{\pi}_3) + \frac{1}{2}(\tilde{\pi}_2 + 2 \tilde{\pi}_3 -2 \tilde{\pi}_4 - \tilde{\pi}_5) \right) \\
    & \quad \quad + \frac{2 \Pi_3 (n-1)}{(2q+1)} (\Pi_1 (\tilde{\pi}_5 - \tilde{\pi}_2 + 2 \Pi_3 (\tilde{\pi}_4 - \tilde{\pi}_3))
\end{align*}
where the exponential term is found by Stirling's approximation (Proposition \ref{prop:stirling}), and noting that $q = \Theta(n)$ for every term in the summation. Therefore, Equation (\ref{eq:odd_split_main8}) is 
\begin{align*}
    & \pm \calO \left( e^{-\Theta(n)} \right)  \\
    & - 2 \Pi_3 \sum_{q=\bfloor{\frac{\Pi_3 n}{4}}}^{\bfloor{\frac{n}{6}}-1} \binom{n-1}{\frac{n-1}{2}-q, \frac{n-1}{2}-q, q, q} \Pi_1^{n-1-2q} \Pi_3^{2q} \frac{n f_n(q) 2^{2q}}{(2q+1) \binom{2q}{q}} \\
    & + (n-1) \left( \pi_5 - \pi_2 + 2 \pi_4 - 2\pi_3 \right) 
    \sum_{q=\bfloor{\frac{\Pi_3 n}{4}}}^{\bfloor{\frac{n}{6}}-1} \binom{n}{\frac{n-1}{2}-q, \frac{n-1}{2}-q, 2q+1} \Pi_1^{n-1-2q} (2\Pi_3)^{2q+1} \\ 
    & = \pm \calO \left( \frac{1}{\sqrt{n}} \right) + \left( \pi_5 - \pi_2 + 2 \pi_4 - 2\pi_3 \right) \calO(\sqrt{n})
\end{align*}
by Lemma \ref{lem:soln_part_b} and Corollary \ref{coro:sub_others}, where $f_n(q) = (q-\Pi_3 (n-1))(\tilde{\pi}_5 - \tilde{\pi}_2 - 2 \tilde{\pi}_4 + 2 \tilde{\pi}_3) - \frac{1}{2}(\tilde{\pi}_2 + 2 \tilde{\pi}_3-2 \tilde{\pi}_4 - \tilde{\pi}_5)$. 
The $\calO(\sqrt{n})$ sign is positive if $\pi_5 + 2 \pi_4 > \pi_2 + 2 \pi_3$.
\end{paragraph}
This concludes the proof of Lemma \ref{lem:sub_12_non_34_equal} where $n$ is odd.
\end{paragraph}
\end{proof}

%% file: EC_appendix/apx_page_3a.tex
\newpage

\section{Three-way tie proofs}
\label{apx:three_ties}

This appendix proves the three-way ties cases about $\EDS(\{1,2, 3\})$ that are used to prove Theorem \ref{thm:main}. Lemma \ref{lem:three_tie_priority1} covers the case where $n$ is divisible by $3$ in Appendix \ref{apx:three_ties_P1}, Lemma \ref{lem:three_tie_priority3} covers the case where $n-2$ is divisible by $3$ in Appendix \ref{apx:three_ties_P3}, and Lemma \ref{lem:three_tie_priority2} covers the case where $n-1$ is divisible by $3$ in Appendix \ref{apx:three_ties_P2}.

First, recall the correspondence introduced in Appendix \ref{apx:appendix_contents}.
\begin{remark}
Preference distribution 
$\pi = (\pi_1, \ldots, \pi_6)$ corresponds to the rankings
\begin{align*}
        & R_1 = (1\succ 2\succ 3); & R_5 = (1\succ 3\succ 2) \\
        & R_2 = (2\succ 3\succ 1); & R_6 = (2\succ 1\succ 3) \\
        & R_3 = (3\succ 2\succ 1); & R_4 = (3\succ 1\succ 2). 
    \end{align*}
\end{remark}

\subsection{Case when alternative $1$ wins}
\label{apx:three_ties_P1}

\begin{table}[t]
    \centering
    \begin{tabular}{|c||c|c|c|}
        \hline
         & $\pi_2 = \pi_6$ & $\pi_2 > \pi_6$ & $\pi_2 < \pi_6$ \\
          \hline \hline
            $\pi_3 = \pi_4$ & 
            $\pi_1 - \pi_5$ &  
            $2 \pi_1 + \pi_6 - 3 \pi_2$ & 
            $\pi_6 - \pi_1$ \\[0.25em]
          \hline
            $\pi_3 > \pi_4$ & 
            $2\pi_1 + 2\pi_5 + \pi_4 - 5\pi_3$ &  
            $\pi_1 + \pi_6 - \pi_2 - \pi_3$ & 
            $\pi_4 + \pi_6 - \pi_1 - \pi_3$ \\[0.25em]
          \hline 
            $\pi_3 < \pi_4$ & 
            $\pi_1 - \pi_3$ &  
            $\pi_1 + \pi_6 - \pi_2 - \pi_3$ & 
            \makecell{N/A} \\
         \hline
    \end{tabular}
    \caption{Values of $f^1(\pi_1, \ldots, \pi_6)$ given conditions on $\pi$ for Lemma \ref{lem:three_tie_priority1}.}
    \label{tab:complexity_three_sides_A_final}
\end{table}

\setcounter{lem}{7}
\begin{lem}
Suppose that $\pi_1 + \pi_5 = \pi_2 + \pi_6 = \pi_3 + \pi_4 = \frac{1}{3}$ and $\pi_i > 0, ~\forall i \in [6].$ Furthermore, let $u_1 \geq u_2 > u_3$ in $\vec{u}$. 
Then $\exists N > 0$ such that $\forall n > N$ that are divisible by $3$, 
\[
\EDS(\{1,2, 3\}) = \begin{cases}
    \pm \calO \left( e^{-\Theta(n)} \right), & \pi_2 < \pi_6, \pi_3 < \pi_4 \\
    f^1(\pi_1, \ldots, \pi_6) \Theta(1) + g^1_n(\pi_1, \ldots, \pi_6), & \text{ otherwise}
\end{cases}
\]
where $f^1(\pi_1, \ldots, \pi_6)$ is determined by Table \ref{tab:complexity_three_sides_A_final} and
\begin{align*}
    g^1_n(\pi_1, \ldots, \pi_6) = \begin{cases}
    \Theta(1), & \pi_2 + \pi_5 > \pi_1 + \pi_3 \\
    \pm \calO \left( \frac{1}{\sqrt{n}} \right), & \pi_2 + \pi_5 \leq \pi_1 + \pi_3.
    \end{cases}
\end{align*}
If $u_1 > u_2 = u_3$, then $\EDS(\{1,2, 3\}) = \pm \calO \left( e^{-\Theta(n)} \right)$.
\label{lem:three_tie_priority1}
\end{lem}

\begin{proof}
We prove this lemma similar to Lemma \ref{lem:sub_12} by summing up the adversarial loss $\ADS(P)$ of every preference profile $P \in \calL(\calA)^n$ such that the potential winning set $\PW{P} = \{1,2,3\}$, weighted by their likelihood of occurrence. Since $n$ is assumed to be divisible by $3$, this covers the case where there are exactly $\frac{n}{3}$ agents that vote for each of the alternatives $1$, $2$, and $3$. Were there to be different number of votes for the alternatives, rather, then either $n$ would not be divisible by $3$ or $\PW{P} \neq \{1,2,3\}$. We first must discuss what the equilibrium winning set $\EW{\topRank{P}}$ is for any profile $P$, with respect to its truthful vote profile $\topRank{P}$. Recall that this is the set of equilibrium winning alternatives following \emph{any} sequence of agents changing their votes to the best response of all other agents' votes.

Recall that iterative plurality starting from the truthful vote profile $\topRank{P}$ consists of agents changing their votes from alternatives that were not already winning to those that then become the winner \citep{Branzei13:How}. Therefore any improvement step from alternative $c \in \calA$ to another $c' \in \calA$ means that neither $c$ nor $c'$ could have been the winner, prior to this step. Hence, after this step, no agent will change their vote to $c$, since doing so would not make it the winner. Since there are $m=3$ alternatives, it follows that the first improvement step determines which two alternatives are in the run-off to be the equilibrium winner. By \citet[Lemma 1]{Kavner21:strategic}, the winner is then whichever more agents prefer out of the entire agent pool.

For example, if all agents in $P$ have preference rankings $R_1 = (1 \succ 2 \succ 3)$, $R_5 = (1 \succ 3 \succ 2)$, $R_6 = (2 \succ 1 \succ 3)$, or $R_4 = (3 \succ 1 \succ 2)$, then no agent has an incentive to change their vote and alternative $1$ is both the equilibrium and truthful winners. Now suppose that there is at least one agent $j$ with ranking $R_2 = (2 \succ 3 \succ 1)$. If agent $j$ switches their vote first, then the plurality scores of the alternatives would be $( \frac{n}{3}, \frac{n}{3}-1, \frac{n}{3}+1)$. From this vote profile, alternative $2$ cannot become the winner, so no agent will henceforth switch their vote to $2$. Iterative plurality thereafter consists of agents that were voting for alternative $2$ iteratively switching their votes to either alternatives $1$ or $3$. The winner is whichever alternative more agents prefer (subject to tie-breaking) \citep[Lemma 1]{Kavner21:strategic}. We conclude that $2 \in \EW{\topRank{P}}$ if $R_3 \in P$ and $P[2 \succ 1] > P[1 \succ 2]$, whereas $3 \in \EW{\topRank{P}}$ if $R_2 \in P$ and $P[3 \succ 1] > P[1 \succ 3]$. This yields three cases for whether either or both of these are the case. We define $\calE_2$, $\calE_3$, and $\calE_{2,3}$ as follows:
\begin{itemize}
    \item $\calE_2 = \{P \in \calL(\calA)^n~:~R_3 \in P \text{ and } P[2 \succ 1] > P[1 \succ 2] \text{, and either } R_2 \notin P \text{ or } P[1 \succ 3] \geq P[3 \succ 1]\}$,
    \item $\calE_3 = \{P \in \calL(\calA)^n~:~R_2 \in P \text{ and } P[3 \succ 1] > P[1 \succ 3] \text{, and either } R_3 \notin P \text{ or } P[1 \succ 2] \geq P[2 \succ 1]\}$,
    \item $\calE_{2,3} = \{P \in \calL(\calA)^n~:~R_2, R_3 \in P \text{ and } P[2 \succ 1] > P[1 \succ 2] \text{ and } P[3 \succ 1] > P[1 \succ 3]\}$.
\end{itemize}
Implicitly, we note $s_1(\topRank{P}) = s_2(\topRank{P}) = s_3(\topRank{P})$ for each of these cases.
The subscript denotes which alternatives (among $2$ and $3$, excluding $1$) appears in the equilibrium winning set $\EW{\topRank{P}}$, for ease of readability.
Let $a, b, c \in [0, \frac{n}{3}]$. Given that $n \in \mathbb{N}$ is divisible by $3$, we take throughout this proof:
\begin{itemize}
    \item $\frac{n}{3}$ agents with rankings either $R_1$ or $R_5$: with $a$ for $R_1$ and $\frac{n}{3}-a$ for $R_5$,
    \item $\frac{n}{3}$ agents with rankings either $R_2$ or $R_6$: with $b$ for $R_2$ and $\frac{n}{3}-b$ for $R_6$,
    \item $\frac{n}{3}$ agents with rankings either $R_3$ or $R_4$: with $c$ for $R_3$ and $\frac{n}{3}-c$ for $R_4$.
\end{itemize}

\begin{paragraph}{Step 1: Characterize the $\calE_2$ case.}
We have $P \in \calE_2$ if the following ranges are satisfied. First, $a \in [0, \frac{n}{3}]$ has its full range. Second, $b \leq \frac{n}{6}$, so that there are at least as many agents preferring $R_6 = (2 \succ 1 \succ 3)$ than $R_2 = (2 \succ 3 \succ 1)$, which entails $3 \notin \EW{\topRank{P}}$. Third, $c > \frac{n}{6}$, so that there are more agents preferring $R_3 = (3 \succ 2 \succ 1)$ than $R_4 = (3 \succ 1 \succ 2)$, which entails $2 \in \EW{\topRank{P}}$. Like in Lemma \ref{lem:sub_12}, the value per agent and probability of each ranking is summarized by Table \ref{tab:character_profiles_E1}.
Put together, we get the equation
\begin{align}
    & \Pr\nolimits_{P \sim \pi^n}(P \in \mathcal{E}_2) \times \mathbb{E}_{P \sim \pi^n}[\ADS(P)~|~P \in \mathcal{E}_2] \notag \\
    & = \sum_{a=0}^{\frac{n}{3}} \sum_{b=0}^{\bfloor{\frac{n}{6}}} \sum_{c=\bfloor{\frac{n}{6}}+1}^{\frac{n}{3}} \mathcal{P}_{\vec{\pi},n}(a,b,c) \cdot \mathcal{V}^2_{\vec{u},n}(a,b,c) \label{eq:three_tie_1}
\end{align}
where we define
\begin{align*}
    \mathcal{P}_{\vec{\pi},n}(a,b,c) & = \binom{n}{a, b, c, \frac{n}{3}-c, \frac{n}{3}-a, \frac{n}{3}-b} \pi_1^a \pi_2^b \pi_3^{c} \pi_4^{\frac{n}{3}-c} \pi_5^{\frac{n}{3}-a} \pi_6^{\frac{n}{3}-b} \\
    & = \binom{n}{\frac{n}{3}, \frac{n}{3}, \frac{n}{3}} \frac{1}{3^n}   \times  \binom{\frac{n}{3}}{a} \tilde{\pi}_1^a \tilde{\pi}_5^{\frac{n}{3}-a} \times \binom{\frac{n}{3}}{b} \tilde{\pi}_2^b \tilde{\pi}_6^{\frac{n}{3}-b} \times \binom{\frac{n}{3}}{c} \tilde{\pi}_3^c \tilde{\pi}_4^{\frac{n}{3}-c}
\end{align*}
and 
\begin{align*}
    \mathcal{V}^2_{\vec{u},n}(a, b, c) = & \begin{pmatrix}
        a, & b, & c, & \frac{n}{3}-c, & \frac{n}{3}-a, & \frac{n}{3}-b \end{pmatrix} \\
    & \cdot \begin{pmatrix}
        u_1 - u_2, & -u_1+u_3, & -u_2+u_3, & u_2-u_3, & u_1-u_3, & -u_1+u_2 \end{pmatrix} \\
    & = (u_2 - u_3) \left( \frac{2n}{3} -a -b-2c \right).
\end{align*}
This equation uses the following definitions:
\begin{itemize}
    \item $\tilde{\pi}_1 = \frac{\pi_1}{\pi_1+\pi_5}$ and $\tilde{\pi}_5 = \frac{\pi_5}{\pi_1+\pi_5}$,
    \item $\tilde{\pi}_2 = \frac{\pi_2}{\pi_2+\pi_6}$ and $\tilde{\pi}_6 = \frac{\pi_6}{\pi_2+\pi_6}$,
    \item $\tilde{\pi}_3 = \frac{\pi_3}{\pi_3+\pi_4}$ and $\tilde{\pi}_4 = \frac{\pi_4}{\pi_3+\pi_4}$
\end{itemize}
which we recall from Definition \ref{dfn:tilde_definitions}, where the denominators are each $\frac{1}{3}$.
We forego writing $(u_2 - u_3) > 0$ throughout this proof, by assumption on $\vec{u}$, for ease of notation.
We simplify Equation (\ref{eq:three_tie_1}) by first employing the following lemma, which is analogous to Stirling's approximation presented in Proposition \ref{prop:stirling} but with three components.
\setcounter{lem}{14}
\begin{lem}
\[
    \binom{3n}{n, n, n} \frac{1}{3^{3n}} = \Theta \left( \frac{1}{n} \right).
\]
\end{lem}

The proof may be found in Appendix \ref{apx:helper_lemmas}. We will substitute $\Theta \left( \frac{1}{n} \right)$ for now, and return to its un-simplified form in Step 4, below. Equation (\ref{eq:three_tie_1}) may then be written as
\begin{align}
    & \Theta \left( \frac{1}{n} \right) \sum_{c=\bfloor{\frac{n}{6}}+1}^{\frac{n}{3}}  \binom{\frac{n}{3}}{c} \tilde{\pi}_3^c \tilde{\pi}_4^{\frac{n}{3}-c} \sum_{b=0}^{\bfloor{\frac{n}{6}}} \binom{\frac{n}{3}}{b} \tilde{\pi}_2^b \tilde{\pi}_6^{\frac{n}{3}-b}
    \sum_{a=0}^{\frac{n}{3}} \binom{\frac{n}{3}}{a} \tilde{\pi}_1^a \tilde{\pi}_5^{\frac{n}{3}-a} \left( \frac{2n}{3} -a -b-2c \right) \notag \\
    & = \Theta \left( \frac{1}{n} \right) \sum_{c=\bfloor{\frac{n}{6}}+1}^{\frac{n}{3}}  \binom{\frac{n}{3}}{c} \tilde{\pi}_3^c \tilde{\pi}_4^{\frac{n}{3}-c} \sum_{b=0}^{\bfloor{\frac{n}{6}}} \binom{\frac{n}{3}}{b} \tilde{\pi}_2^b \tilde{\pi}_6^{\frac{n}{3}-b}
    \left( \frac{(2 - \tilde{\pi}_1)n}{3}-b-2c \right) \label{eq:three_tie_2}
\end{align}
by definition of binomial probability and expectation. 
This may be simplified to
\begin{align}
    & \Theta \left( \frac{1}{n} \right) \bigg(  \frac{(2 - \tilde{\pi}_1) n}{3} \sum_{c=\bfloor{\frac{n}{6}}+1}^{\frac{n}{3}}  \binom{\frac{n}{3}}{c} \tilde{\pi}_3^c \tilde{\pi}_4^{\frac{n}{3}-c} \sum_{b=0}^{\bfloor{\frac{n}{6}}} \binom{\frac{n}{3}}{b} \tilde{\pi}_2^b \tilde{\pi}_6^{\frac{n}{3}-b} \notag \\
    & - 2 \sum_{c=\bfloor{\frac{n}{6}}+1}^{\frac{n}{3}}  \binom{\frac{n}{3}}{c} \tilde{\pi}_3^c \tilde{\pi}_4^{\frac{n}{3}-c} c  \sum_{b=0}^{\bfloor{\frac{n}{6}}} \binom{\frac{n}{3}}{b} \tilde{\pi}_2^b \tilde{\pi}_6^{\frac{n}{3}-b} \notag \\
    &  - \sum_{c=\bfloor{\frac{n}{6}}+1}^{\frac{n}{3}}  \binom{\frac{n}{3}}{c} \tilde{\pi}_3^c \tilde{\pi}_4^{\frac{n}{3}-c} \sum_{b=0}^{\bfloor{\frac{n}{6}}} \binom{\frac{n}{3}}{b} \tilde{\pi}_2^b \tilde{\pi}_6^{\frac{n}{3}-b} b \bigg). \label{eq:three_tie_3}
\end{align}
We can see Equation (\ref{eq:three_tie_3}) consists of several separable summations each of the template $\sum_{t \in T} \binom{m}{t} p^t (1-p)^{m-t}$ or $\sum_{t \in T} \binom{m}{t} p^t (1-p)^{m-t} t$, corresponding to some contiguous domain $T \subseteq [m]$ for a binomial random variable $Bin(m,p)$. 
By Lemma \ref{lem:bin_theorems_approx} in Appendix \ref{apx:concentration_inequalities}, it follows that each summation of Equation (\ref{eq:three_tie_3}) is either $\Theta(1) \pm \calO \left( \frac{1}{\sqrt{m}} \right)$ or $\Theta(m) \pm \calO \left( \sqrt{m} \right)$, if $mp \in T$, and $\calO \left( e^{-\Theta(m) } \right)$ otherwise. For instance, we have
\begin{align*}
    \sum_{c=\bfloor{\frac{n}{6}}+1}^{\frac{n}{3}}  \binom{\frac{n}{3}}{c} \tilde{\pi}_3^c \tilde{\pi}_4^{\frac{n}{3}-c} = \begin{cases} 
        \frac{1}{2} \pm \calO \left( \frac{1}{\sqrt{n}} \right), & \tilde{\pi}_3 = \tilde{\pi}_4 \\
        1 - \calO \left( e^{-\Theta(n)} \right), & \tilde{\pi}_3 > \tilde{\pi}_4 \\
        \calO \left( e^{-\Theta(n)} \right), & \tilde{\pi}_3 < \tilde{\pi}_4
    \end{cases}
\end{align*}
and
\begin{align*}
    \sum_{c=\bfloor{\frac{n}{6}}+1}^{\frac{n}{3}}  \binom{\frac{n}{3}}{c} \tilde{\pi}_3^c \tilde{\pi}_4^{\frac{n}{3}-c} c = \begin{cases} 
        \frac{n}{12} \pm \calO \left( \frac{1}{\sqrt{n}} \right), & \tilde{\pi}_3 = \tilde{\pi}_4 \\
        \frac{\tilde{\pi}_3 n}{3} \left(  1- \calO \left( e^{-\Theta(n)} \right) \right), & \tilde{\pi}_3 > \tilde{\pi}_4 \\
        \calO \left( e^{-\Theta(n)} \right), & \tilde{\pi}_3 < \tilde{\pi}_4.
    \end{cases}
\end{align*}
The $b$-summations, in terms of $\tilde{\pi}_3$ and $\tilde{\pi}_4$ are similar, except with the inequality signs reversed.
It is therefore clear that if either $\tilde{\pi}_3 < \tilde{\pi}_4$ or $\tilde{\pi}_2 > \tilde{\pi}_6$, then Equation (\ref{eq:three_tie_3}) is $\calO \left( e^{-\Theta(n)} \right)$. This leaves four cases. First, if $\tilde{\pi}_3 = \tilde{\pi}_4$ and $\tilde{\pi}_2 = \tilde{\pi}_6$, then Equation (\ref{eq:three_tie_3}) is
\begin{align}
    & \Theta \left( \frac{1}{n} \right) \bigg( \frac{(2 - \tilde{\pi}_1)n}{3} \left( \frac{1}{2} \pm \calO \left( \frac{1}{\sqrt{n}} \right)  \right) \left( \frac{1}{2} \pm \calO \left( \frac{1}{\sqrt{n}} \right)  \right) \notag \\
    & \quad \quad - 2 \left( \frac{n}{12} \pm \calO (\sqrt{n})  \right) \left( \frac{1}{2} \pm \calO \left( \frac{1}{\sqrt{n}} \right)  \right) \notag \\
    & \quad \quad - \left( \frac{1}{2} \pm \calO \left( \frac{1}{\sqrt{n}} \right)  \right) \left( \frac{n}{12} \pm \calO (\sqrt{n})  \right)
    \bigg) \notag \\
    & =  \Theta \left( \frac{1}{n} \right) \left( n \left( \frac{1 - 2 \tilde{\pi}_1}{24} \right) \pm \calO(\sqrt{n}) \right) \notag \\
    & = \frac{1}{8} (\pi_5 - \pi_1) \Theta(1) \pm \calO \left( \frac{1}{\sqrt{n}} \right) \label{eq:three_tie_4A}
\end{align}
making use of the fact that $\pi_1 + \pi_5 = \frac{1}{3}$.
Second, if $\tilde{\pi}_3 = \tilde{\pi}_4$ and $\tilde{\pi}_2 < \tilde{\pi}_6$, then Equation (\ref{eq:three_tie_3}) is
\begin{align}
    & \Theta \left( \frac{1}{n} \right) \bigg( \frac{(2 - \tilde{\pi}_1)n}{3} \left( \frac{1}{2} \pm \calO \left( \frac{1}{\sqrt{n}} \right)  \right) \left( 1 - \calO \left( e^{-\Theta(n)} \right)  \right) \notag \\
    & \quad \quad - 2 \left( \frac{n}{12} \pm \calO (\sqrt{n})  \right) \left( 1 - \calO \left( e^{-\Theta(n)} \right)  \right) \notag \\
    & \quad \quad - \left( \frac{1}{2} \pm \calO \left( \frac{1}{\sqrt{n}} \right)  \right) \frac{\tilde{\pi}_2 n}{3} \left(  1 - \calO \left( e^{-\Theta(n)} \right) \right)
    \bigg) \notag \\
    & = \Theta \left( \frac{1}{n} \right) \left( n \left( \frac{2 - \tilde{\pi}_1}{6} - \frac{1}{6} - \frac{ \tilde{\pi}_2}{6} \right) \pm \calO(\sqrt{n}) \right) \notag \\
    & = \frac{1}{2} (\pi_6 - \pi_1) \Theta(1) \pm \calO \left( \frac{1}{\sqrt{n}} \right). \label{eq:three_tie_4B}
\end{align}
Third, if $\tilde{\pi}_3 > \tilde{\pi}_4$ and $\tilde{\pi}_2 = \tilde{\pi}_6$, then Equation (\ref{eq:three_tie_3}) is
\begin{align}
    & \Theta \left( \frac{1}{n} \right) \bigg( \frac{(2 - \tilde{\pi}_1)n}{3} \left( 1 - \calO \left( e^{-\Theta(n)} \right) \right) \left( \frac{1}{2} \pm \calO \left( \frac{1}{\sqrt{n}} \right)  \right) \notag \\
    & \quad \quad - \frac{2 \tilde{\pi}_3 n}{3} \left( 1 - \calO \left( e^{-\Theta(n)} \right)  \right) \left( \frac{1}{2} \pm \calO \left( \frac{1}{\sqrt{n}} \right)  \right) \notag \\
    & \quad \quad - \left( 1 - \calO \left( e^{-\Theta(n)} \right)  \right) \left( \frac{n}{12} \pm \calO (\sqrt{n})  \right)
    \bigg) \notag \\
    & = \Theta \left( \frac{1}{n} \right) \left( n \left( \frac{2 - \tilde{\pi}_1}{6} - \frac{\tilde{\pi}_3}{3}- \frac{1}{12} \right) \pm \calO(\sqrt{n}) \right) \notag \\
    & = \frac{1}{4} (2\pi_5 + \pi_4 - 3\pi_3) \Theta(1) \pm \calO \left( \frac{1}{\sqrt{n}} \right). \label{eq:three_tie_4C}
\end{align}
Finally, if $\tilde{\pi}_3 > \tilde{\pi}_4$ and $\tilde{\pi}_2 < \tilde{\pi}_6$, then Equation (\ref{eq:three_tie_3}) is
\begin{align}
    & \Theta \left( \frac{1}{n} \right) \bigg( \frac{(2 - \tilde{\pi}_1)n}{3} \left( 1 - \calO \left( e^{-\Theta(n)} \right) \right) \left(1 - \calO \left( e^{-\Theta(n)} \right)  \right) \notag \\
    & \quad \quad - \frac{2 \tilde{\pi}_3 n}{3} \left( 1 - \calO \left( e^{-\Theta(n)} \right)  \right) \left( 1 - \calO \left( e^{-\Theta(n)} \right)  \right) \notag \\
    & \quad \quad - \left( 1 - \calO \left( e^{-\Theta(n)} \right)  \right) \frac{\tilde{\pi}_2 n}{3} \left( 1 - \calO \left( e^{-\Theta(n)} \right) \right)
    \bigg) \notag \\
    & = \Theta \left( \frac{1}{n} \right) \left( n \left( \frac{2 -\tilde{\pi}_1 - \tilde{\pi}_2 - 2\tilde{\pi}_3 }{3}  \right) \pm \calO(\sqrt{n}) \right) \notag \\
    & = (\pi_4 + \pi_5 - \pi_1 - \pi_3) \Theta(1) \pm \calO \left( \frac{1}{\sqrt{n}} \right). \label{eq:three_tie_4D}
\end{align}
Recall that each $\Theta(1)$ in Equations (\ref{eq:three_tie_4A}), (\ref{eq:three_tie_4B}), (\ref{eq:three_tie_4C}), and (\ref{eq:three_tie_4D})
is actually an instance of $\binom{n}{\frac{n}{3}, \frac{n}{3}, \frac{n}{3}} \frac{n}{3^n}$, following Lemma \ref{lem:stirling_trinom}.
We make use of this fact and continue with these equations in Step 4, below.
This concludes the $\calE_2$ case of Lemma \ref{lem:three_tie_priority1}.
\end{paragraph}

\begin{paragraph}{Step 2: Characterize the $\calE_3$ case.}
We prove this case in the same way as Step 1 ($\calE_2$) above, keeping the same variable nomenclature but adjusting the ranges as needed. That is, we have $P \in \calE_3$ if the following ranges are satisfied. First, $a \in [0, \frac{n}{3}]$ has its full range. Second, $b > \frac{n}{6}$, so that there are more agents preferring $R_2 = (2 \succ 3 \succ 1)$ than $R_6 = (2 \succ 1 \succ 3)$, which entails $3 \in \EW{\topRank{P}}$. Third, $c \leq \frac{n}{6}$, so that there are at least as many agents preferring $R_4 = (3 \succ 1 \succ 2)$ than $R_3 = (3 \succ 2 \succ 1)$, which entails $2 \notin \EW{\topRank{P}}$. Like in Lemma \ref{lem:sub_13}, the value per agent and probability of each ranking is summarized by Table \ref{tab:character_profiles_E3}.
Put together, we get the equation
\begin{align}
    & \Pr\nolimits_{P \sim \pi^n}(P \in \mathcal{E}_3) \times \mathbb{E}_{P \sim \pi^n}[\ADS(P)~|~P \in \mathcal{E}_3] \notag \\
    & = \sum_{a=0}^{\frac{n}{3}} \sum_{b=\bfloor{\frac{n}{6}}+1}^{\frac{n}{3}} \sum_{c=0}^{\bfloor{\frac{n}{6}}} \mathcal{P}_{\vec{\pi},n}(a,b,c) \cdot \mathcal{V}^3_{\vec{u},n}(a,b,c) \label{eq:three_tie_5}
\end{align}
where $\mathcal{P}_{\vec{\pi},n}$ is the same as in Step 1, and 
\begin{align*}
    \mathcal{V}^3_{\vec{u},n}(a, b, c) = & \begin{pmatrix}
        a, & b, & c, & \frac{n}{3}-c, & \frac{n}{3}-a, & \frac{n}{3}-b \end{pmatrix} \\
    & \cdot \begin{pmatrix}
        u_1 - u_3, & -u_2+u_3, & -u_1+u_3, & -u_1+u_2, & u_1-u_2, & u_2-u_3 \end{pmatrix} \\
    & = (u_2 - u_3) \left( \frac{n}{3} +a -2b -c \right).
\end{align*}
By Lemma \ref{lem:stirling_trinom}, this leads to 
\begin{align}
    & \Theta \left( \frac{1}{n} \right) \sum_{b=\bfloor{\frac{n}{6}}+1}^{\frac{n}{3}} \binom{\frac{n}{3}}{b} \tilde{\pi}_2^b \tilde{\pi}_6^{\frac{n}{3}-b} \sum_{c=0}^{\bfloor{\frac{n}{6}}}  \binom{\frac{n}{3}}{c} \tilde{\pi}_3^c \tilde{\pi}_4^{\frac{n}{3}-c}
    \sum_{a=0}^{\frac{n}{3}} \binom{\frac{n}{3}}{a} \tilde{\pi}_1^a \tilde{\pi}_5^{\frac{n}{3}-a} \left( \frac{n}{3} +a -2b-c \right) \notag \\
    & = \Theta \left( \frac{1}{n} \right) \sum_{b=\bfloor{\frac{n}{6}}+1}^{\frac{n}{3}} \binom{\frac{n}{3}}{b} \tilde{\pi}_2^b \tilde{\pi}_6^{\frac{n}{3}-b} \sum_{c=0}^{\bfloor{\frac{n}{6}}}  \binom{\frac{n}{3}}{c} \tilde{\pi}_3^c \tilde{\pi}_4^{\frac{n}{3}-c}
    \left( \frac{(1+\tilde{\pi}_1)n}{3}-2b-c \right) \label{eq:three_tie_6}
\end{align}
by definition of binomial probability and expectation. This may be simplified to
\begin{align}
    & \Theta \left( \frac{1}{n} \right) \bigg(  \frac{(1 + \tilde{\pi}_1) n}{3} \sum_{b=\bfloor{\frac{n}{6}}+1}^{\frac{n}{3}} \binom{\frac{n}{3}}{b} \tilde{\pi}_2^b \tilde{\pi}_6^{\frac{n}{3}-b} \sum_{c=0}^{\bfloor{\frac{n}{6}}}  \binom{\frac{n}{3}}{c} \tilde{\pi}_3^c \tilde{\pi}_4^{\frac{n}{3}-c} \notag \\
    & - 2 \sum_{b=\bfloor{\frac{n}{6}}+1}^{\frac{n}{3}} \binom{\frac{n}{3}}{b} \tilde{\pi}_2^b \tilde{\pi}_6^{\frac{n}{3}-b} b \sum_{c=0}^{\bfloor{\frac{n}{6}}}  \binom{\frac{n}{3}}{c} \tilde{\pi}_3^c \tilde{\pi}_4^{\frac{n}{3}-c}  \notag \\
    & - \sum_{b=\bfloor{\frac{n}{6}}+1}^{\frac{n}{3}} \binom{\frac{n}{3}}{b} \tilde{\pi}_2^b \tilde{\pi}_6^{\frac{n}{3}-b} \sum_{c=0}^{\bfloor{\frac{n}{6}}}  \binom{\frac{n}{3}}{c} \tilde{\pi}_3^c \tilde{\pi}_4^{\frac{n}{3}-c} c   
    \bigg). \label{eq:three_tie_7}
\end{align}
It is clear that if either $\tilde{\pi}_2 < \tilde{\pi}_6$ or $\tilde{\pi}_3 > \tilde{\pi}_4$, then Equation (\ref{eq:three_tie_7}) is $\calO \left( e^{-\Theta(n)} \right)$. This leaves four cases. First, if $\tilde{\pi}_2 = \tilde{\pi}_6$ and $\tilde{\pi}_3 = \tilde{\pi}_4$, then Equation (\ref{eq:three_tie_7}) is
\begin{align}
    & \Theta \left( \frac{1}{n} \right) \bigg( \frac{(1 + \tilde{\pi}_1)n}{3} \left( \frac{1}{2} \pm \calO \left( \frac{1}{\sqrt{n}} \right)  \right) \left( \frac{1}{2} \pm \calO \left( \frac{1}{\sqrt{n}} \right)  \right) \notag \\
    & \quad \quad - 2 \left( \frac{n}{12} \pm \calO (\sqrt{n})  \right) \left( \frac{1}{2} \pm \calO \left( \frac{1}{\sqrt{n}} \right)  \right) \notag \\
    & \quad \quad - \left( \frac{1}{2} \pm \calO \left( \frac{1}{\sqrt{n}} \right)  \right) \left( \frac{n}{12} \pm \calO (\sqrt{n})  \right)
    \bigg) \notag \\
    & =  \Theta \left( \frac{1}{n} \right) \left( n \left( \frac{-1 + 2 \tilde{\pi}_1}{24} \right) \pm \calO(\sqrt{n}) \right) \notag \\
    & = \frac{1}{8} (\pi_1 - \pi_5) \Theta(1) \pm \calO \left( \frac{1}{\sqrt{n}} \right). \label{eq:three_tie_8A}
\end{align}
Second, if $\tilde{\pi}_2 = \tilde{\pi}_6$ and $\tilde{\pi}_3 < \tilde{\pi}_4$, then Equation (\ref{eq:three_tie_7}) is
\begin{align}
    & \Theta \left( \frac{1}{n} \right) \bigg( \frac{(1 + \tilde{\pi}_1)n}{3} \left( \frac{1}{2} \pm \calO \left( \frac{1}{\sqrt{n}} \right)  \right) \left( 1 - \calO \left( e^{-\Theta(n)} \right)  \right) \notag \\
    & \quad \quad - 2 \left( \frac{n}{12} \pm \calO (\sqrt{n})  \right) \left( 1 - \calO \left( e^{-\Theta(n)} \right)  \right) \notag \\
    & \quad \quad - \left( \frac{1}{2} \pm \calO \left( \frac{1}{\sqrt{n}} \right)  \right) \frac{\tilde{\pi}_3 n}{3} \left(  1 - \calO \left( e^{-\Theta(n)} \right) \right)
    \bigg) \notag \\
    & = \Theta \left( \frac{1}{n} \right) \left( n \left( \frac{1 + \tilde{\pi}_1}{6} - \frac{1}{6} - \frac{ \tilde{\pi}_3}{6} \right) \pm \calO(\sqrt{n}) \right) \notag \\
    & = \frac{1}{2}(\pi_1 - \pi_3) \Theta(1) \pm \calO \left( \frac{1}{\sqrt{n}} \right). \label{eq:three_tie_8B}
\end{align}
Third, if $\tilde{\pi}_2 > \tilde{\pi}_6$ and $\tilde{\pi}_3 = \tilde{\pi}_4$, then Equation (\ref{eq:three_tie_7}) is
\begin{align}
    & \Theta \left( \frac{1}{n} \right) \bigg( \frac{(1 + \tilde{\pi}_1)n}{3} \left( 1 - \calO \left( e^{-\Theta(n)} \right) \right) \left( \frac{1}{2} \pm \calO \left( \frac{1}{\sqrt{n}} \right)  \right) \notag \\
    & \quad \quad - \frac{2 \tilde{\pi}_2 n}{3} \left( 1 - \calO \left( e^{-\Theta(n)} \right)  \right) \left( \frac{1}{2} \pm \calO \left( \frac{1}{\sqrt{n}} \right)  \right) \notag \\
    & \quad \quad - \left( 1 - \calO \left( e^{-\Theta(n)} \right)  \right) \left( \frac{n}{12} \pm \calO (\sqrt{n})  \right)
    \bigg) \notag \\
    & = \Theta \left( \frac{1}{n} \right) \left( n \left( \frac{1 + \tilde{\pi}_1}{6} - \frac{\tilde{\pi}_2}{3}- \frac{1}{12} \right) \pm \calO(\sqrt{n}) \right) \notag \\
    & = \frac{1}{4}(2\pi_1 + \pi_6 - 3 \pi_2) \Theta(1) \pm \calO \left( \frac{1}{\sqrt{n}} \right). \label{eq:three_tie_8C}
\end{align}
Finally, if $\tilde{\pi}_2 > \tilde{\pi}_6$ and $\tilde{\pi}_3 < \tilde{\pi}_4$, then Equation (\ref{eq:three_tie_7}) is
\begin{align}
    & \Theta \left( \frac{1}{n} \right) \bigg( \frac{(1 + \tilde{\pi}_1)n}{3} \left( 1 - \calO \left( e^{-\Theta(n)} \right) \right) \left(1 - \calO \left( e^{-\Theta(n)} \right)  \right) \notag \\
    & \quad \quad - \frac{2 \tilde{\pi}_2 n}{3} \left( 1 - \calO \left( e^{-\Theta(n)} \right)  \right) \left( 1 - \calO \left( e^{-\Theta(n)} \right)  \right) \notag \\
    & \quad \quad - \left( 1 - \calO \left( e^{-\Theta(n)} \right)  \right) \frac{\tilde{\pi}_3 n}{3} \left( 1 - \calO \left( e^{-\Theta(n)} \right) \right)
    \bigg) \notag \\
    & = \Theta \left( \frac{1}{n} \right) \left( n \left( \frac{1 +\tilde{\pi}_1 - 2\tilde{\pi}_2 - \tilde{\pi}_3 }{3}  \right) \pm \calO(\sqrt{n}) \right) \notag \\
    & = (\pi_1 + \pi_6 - \pi_2 - \pi_3) \Theta(1) \pm \calO \left( \frac{1}{\sqrt{n}} \right). \label{eq:three_tie_8D}
\end{align}
As with the $\calE_2$ case above, each $\Theta(1)$ in  Equations (\ref{eq:three_tie_8A}), (\ref{eq:three_tie_8B}), (\ref{eq:three_tie_8C}), and (\ref{eq:three_tie_8D})
is actually an instance of $\binom{n}{\frac{n}{3}, \frac{n}{3}, \frac{n}{3}} \frac{n}{3^n}$.
We continue with these equations in Step 4, below. This concludes the $\calE_3$ case of Lemma \ref{lem:three_tie_priority1}.
\end{paragraph}

\begin{paragraph}{Step 3: Characterize the $\calE_{2,3}$ case.}
We keep the same variable nomenclature as the above steps, but adjust the ranges as needed. That is, we have $P \in \calE_{2,3}$ if the following ranges are satisfied. 
First, $a \in [0, \frac{n}{3}]$ has its full range. Second, $b > \frac{n}{6}$, so that there are more agents preferring $R_2 = (2 \succ 3 \succ 1)$ than $R_6 = (2 \succ 1 \succ 3)$, which entails $3 \in \EW{\topRank{P}}$. Third, $c > \frac{n}{6}$, so that there are more agents preferring $R_3 = (3 \succ 2 \succ 1)$ than $R_4 = (3 \succ 1 \succ 2)$, which entails $2 \in \EW{\topRank{P}}$.

Recall that the definition of adversarial loss for a preference profile $P$, against truthful vote profile $\topRank{P}$, is $\ADS_{\vec{u}}(P) = \SW{\vec{u}}{P}{f(\topRank{P})} - \min\nolimits_{c \in \EW{\topRank{P}}} \SW{\vec{u}}{P}{c}$. 
Since $|\EW{\topRank{P}}| = 2$ for this case, we must apply nuance in determining $\ADS(P)$, depending on number of agents with each ranking in $P$ (i.e., the values of $a$, $b$, and $c$). That is, the loss is the maximum of $\mathcal{V}^2_{\vec u, n}(a,b,c)$ and $\mathcal{V}^3_{\vec u, n}(a,b,c)$:
\begin{align*}
    & (u_2 - u_3) \max \left\{ \frac{2n}{3}-a-b-2c, \frac{n}{3}+a-2b-c \right\} \\
    & = (u_2 - u_3) \left( \frac{n}{3} +a -2b -c + \max\left\{\frac{n}{3}-2a+b-c, 0\right\}\right).
\end{align*}
It is easy to verify that $\frac{n}{3} - 2a +b -c \geq 0$, within the already-specified ranges, as long as $a \in [0, \frac{n}{4}]$, $b \in [\frac{n}{6}{, \frac{n}{3}}]$, and $c \in [\frac{n}{6}, \min\{\frac{n}{3}, \frac{n}{3}-2a+b\}]$. Therefore $\Pr\nolimits_{P \sim \pi^n}(P \in \mathcal{E}_{2,3}) \times \mathbb{E}_{P \sim \pi^n}[\ADS(P)~|~P \in \mathcal{E}_{2,3}]$ is the sum of Equations
\begin{align}
    (u_2 - u_3) \sum_{a=0}^{\frac{n}{3}} \sum_{b=\bfloor{\frac{n}{6}}+1}^{\frac{n}{3}} \sum_{c=\bfloor{\frac{n}{6}}+1}^{\frac{n}{3}} \mathcal{P}_{\vec{\pi},n}(a,b,c) \left( \frac{n}{3} +a -2b -c \right)
    \label{eq:three_tie_10a}
\end{align}
and
\begin{align}
    (u_2 - u_3) \sum_{a=0}^{\bfloor{\frac{n}{4}}} \sum_{b=\bfloor{\frac{n}{6}}+1}^{\frac{n}{3}} \sum_{c=\bfloor{\frac{n}{6}}+1}^{\min\{\frac{n}{3}, \frac{n}{3}-2a+b\}} \mathcal{P}_{\vec{\pi},n}(a,b,c) \left( \frac{n}{3} -2a +b -c \right). \label{eq:three_tie_10b}
\end{align}
We first solve Equation (\ref{eq:three_tie_10a}) using the same techniques as above. By Lemma \ref{lem:stirling_trinom}, this is
\begin{align}
    & \Theta \left( \frac{1}{n} \right) \sum_{b=\bfloor{\frac{n}{6}}+1}^{\frac{n}{3}} \binom{\frac{n}{3}}{b} \tilde{\pi}_2^b \tilde{\pi}_6^{\frac{n}{3}-b} \sum_{c=\bfloor{\frac{n}{6}}+1}^{\frac{n}{3}}  \binom{\frac{n}{3}}{c} \tilde{\pi}_3^c \tilde{\pi}_4^{\frac{n}{3}-c}
    \sum_{a=0}^{\frac{n}{3}} \binom{\frac{n}{3}}{a} \tilde{\pi}_1^a \tilde{\pi}_5^{\frac{n}{3}-a} \left( \frac{n}{3} +a -2b-c \right) \notag \\
    & = \Theta \left( \frac{1}{n} \right) \sum_{b=\bfloor{\frac{n}{6}}+1}^{\frac{n}{3}} \binom{\frac{n}{3}}{b} \tilde{\pi}_2^b \tilde{\pi}_6^{\frac{n}{3}-b} \sum_{c=\bfloor{\frac{n}{6}}+1}^{\frac{n}{3}}  \binom{\frac{n}{3}}{c} \tilde{\pi}_3^c \tilde{\pi}_4^{\frac{n}{3}-c}
    \left( \frac{(1+\tilde{\pi}_1)n}{3}-2b-c \right) \notag \\
    & = \Theta \left( \frac{1}{n} \right) \bigg(  \frac{(1 + \tilde{\pi}_1) n}{3} \sum_{b=\bfloor{\frac{n}{6}}+1}^{\frac{n}{3}} \binom{\frac{n}{3}}{b} \tilde{\pi}_2^b \tilde{\pi}_6^{\frac{n}{3}-b} \sum_{c=\bfloor{\frac{n}{6}}+1}^{\frac{n}{3}}  \binom{\frac{n}{3}}{c} \tilde{\pi}_3^c \tilde{\pi}_4^{\frac{n}{3}-c} \notag \\
    & \quad \quad - 2 \sum_{b=\bfloor{\frac{n}{6}}+1}^{\frac{n}{3}} \binom{\frac{n}{3}}{b} \tilde{\pi}_2^b \tilde{\pi}_6^{\frac{n}{3}-b} b \sum_{c=\bfloor{\frac{n}{6}}+1}^{\frac{n}{3}}  \binom{\frac{n}{3}}{c} \tilde{\pi}_3^c \tilde{\pi}_4^{\frac{n}{3}-c}  \notag \\
    & \quad \quad - \sum_{b=\bfloor{\frac{n}{6}}+1}^{\frac{n}{3}} \binom{\frac{n}{3}}{b} \tilde{\pi}_2^b \tilde{\pi}_6^{\frac{n}{3}-b} \sum_{c=\bfloor{\frac{n}{6}}+1}^{\frac{n}{3}}  \binom{\frac{n}{3}}{c} \tilde{\pi}_3^c \tilde{\pi}_4^{\frac{n}{3}-c} c   
    \bigg). \label{eq:three_tie_11}
\end{align}
by definition of binomial probability and expectation.
It is clear that if either $\tilde{\pi}_2 < \tilde{\pi}_6$ or $\tilde{\pi}_3 < \tilde{\pi}_4$, then Equation (\ref{eq:three_tie_11}) is $\calO \left( e^{-\Theta(n)} \right)$. This leaves four cases. First, if $\tilde{\pi}_2 = \tilde{\pi}_6$ and $\tilde{\pi}_3 = \tilde{\pi}_4$, then Equation (\ref{eq:three_tie_11}) is
\begin{align}
    \frac{1}{8} (\pi_1 - \pi_5) \Theta(1) \pm \calO \left( \frac{1}{\sqrt{n}} \right) \label{eq:three_tie_12A}
\end{align}
by similar reasoning as we attained Equation (\ref{eq:three_tie_8A}). Second, if $\tilde{\pi}_2 = \tilde{\pi}_6$ and $\tilde{\pi}_3 > \tilde{\pi}_4$, then Equation (\ref{eq:three_tie_11}) is
\begin{align}
    \frac{1}{2} (\pi_1 - \pi_3) \Theta(1) \pm \calO \left( \frac{1}{\sqrt{n}} \right) \label{eq:three_tie_12B}
\end{align}
by similar reasoning as we attained Equation (\ref{eq:three_tie_8B}). 
Third, if $\tilde{\pi}_2 > \tilde{\pi}_6$ and $\tilde{\pi}_3 = \tilde{\pi}_4$, then Equation (\ref{eq:three_tie_11}) is
\begin{align}
    \frac{1}{4}(2\pi_1 + \pi_6 - 3 \pi_2) \Theta(1) \pm \calO \left( \frac{1}{\sqrt{n}} \right) \label{eq:three_tie_12C}
\end{align}
by similar reasoning as we attained Equation (\ref{eq:three_tie_8C}). 
Finally, if $\tilde{\pi}_2 > \tilde{\pi}_6$ and $\tilde{\pi}_3 > \tilde{\pi}_4$, then Equation (\ref{eq:three_tie_11}) is
\begin{align}
    (\pi_1 + \pi_6 - \pi_2 - \pi_3) \Theta(1) \pm \calO \left( \frac{1}{\sqrt{n}} \right) \label{eq:three_tie_12D}
\end{align}
by similar reasoning as we attained Equation (\ref{eq:three_tie_8D}).

Now consider Equation (\ref{eq:three_tie_10b}). As described above, this may be written as a separable combination of summations of the template $\sum_{t \in T} \binom{m}{t} p^t (1-p)^{m-t}$ or $\sum_{t \in T} \binom{m}{t} p^t (1-p)^{m-t} t$, corresponding to some contiguous domain $T \subseteq [m]$ for a binomial random variable $Bin(m,p)$. 
By Lemma \ref{lem:bin_theorems_approx}, it follows that each summation is either $\Theta(1) \pm \calO \left( \frac{1}{\sqrt{m}} \right)$ or $\Theta(m) \pm \calO \left( \sqrt{m} \right)$, if $mp \in T$, and $\calO \left( e^{-\Theta(m) } \right)$ otherwise. This observation will enable us to deduce the conditions on the probability distribution $(\pi_1, \ldots, \pi_6)$ required for Equation (\ref{eq:three_tie_10b}) to be either $\Theta(1)$, $\calO \left( \frac{1}{\sqrt{n}} \right)$, or $\calO \left( e^{-\Theta(n)} \right)$. Clearly, the equation is lower bounded by zero. 
Specifically, let $\tau = \tilde{\pi}_5 + \tilde{\pi}_2 - \tilde{\pi}_3 - \tilde{\pi}_1$. We will prove that, as long as $\tilde{\pi}_1 \in (0, \frac{3}{4}]$ and $\tilde{\pi}_2, \tilde{\pi}_3 \in [\frac{1}{2}, 1)$, Equation (\ref{eq:three_tie_10b}) is
\begin{align}
    \begin{cases}
    \Theta(1), & \tau > 0 \\
    \calO \left( \frac{1}{\sqrt{n}} \right), & \tau = 0 \\
    \calO \left( e^{-\Theta(n)} \right), & \tau < 0.
    \end{cases} \label{eq:three_tie_13}
\end{align}
Otherwise (i.e., if $\tilde{\pi}_1 > \frac{3}{4}$, $\tilde{\pi}_2 < \frac{1}{2}$, or $\tilde{\pi}_3 < \frac{1}{2}$), then Equation (\ref{eq:three_tie_10b}) is $\calO \left( e^{-\Theta(n)} \right)$ by Lemma \ref{lem:bin_theorems_approx}. 
This is proved as follows.

We begin by proving the $\tau < 0$ case.
Without loss of generality, let us ignore the $\left( \frac{n}{3} -2a +b -c \right)$ factor of Equation (\ref{eq:three_tie_10b}) and instead focus on the equation
\begin{align}
    & \sum_{a=\bfloor{\frac{(\tilde{\pi}_1 - \epsilon_a)n}{3}}}^{\bfloor{\frac{(\tilde{\pi}_1 + \epsilon_a)n}{3}}} \binom{\frac{n}{3}}{a} \tilde{\pi}_1^a \tilde{\pi}_5^{\frac{n}{3}-a} \sum_{b=\bfloor{\frac{n}{6}}+1}^{\frac{n}{3}} \binom{\frac{n}{3}}{b} \tilde{\pi}_2^b \tilde{\pi}_6^{\frac{n}{3}-b}
    \sum_{c=\bfloor{\frac{n}{6}}+1}^{\min\{\frac{n}{3}, \frac{n}{3}-2a+b\}}  \binom{\frac{n}{3}}{c} \tilde{\pi}_3^c \tilde{\pi}_4^{\frac{n}{3}-c}
    \label{eq:three_tie_12_temp2}
\end{align}
for some $\epsilon_a \in (0, \min\{\tilde{\pi}_1, \frac{3}{4}-\tilde{\pi}_1\})$. \footnote{If $\tilde{\pi}_1 = 0.75$ then the proof of the $\tau < 0$ case continues as stated with only the lower-bound on the $a$-summation. That is, we sum over $a \in \left[ \bfloor{\frac{(\tilde{\pi}_1 - \epsilon_a)n}{3}}, \bfloor{\frac{\tilde{\pi}_1 n}{3}} \right]$.} 
We note the following observations. Clearly, if any of the $a$-, $b$-, or $c$-summations are exponentially small, then Equation (\ref{eq:three_tie_12_temp2}) is exponentially small. It follows from Lemma \ref{lem:bin_theorems_approx} that the $a$-summation of Equation (\ref{eq:three_tie_12_temp2}) is proportional to $\Theta(1)$. By similar reasoning, the $a$-summation with range $a \in [0, \frac{n}{4}] \backslash [\frac{(\tilde{\pi}_1 - \epsilon_a)n}{3}, \frac{(\tilde{\pi}_1 + \epsilon_a)n}{3}]$ that is present in Equation (\ref{eq:three_tie_10b}), but not Equation (\ref{eq:three_tie_12_temp2}), is exponentially small. Likewise, for any pair $(a,b)$ such that $\frac{\tilde{\pi}_3 n}{3} - (\frac{n}{3}-2a+b) = \Omega(n)$, it follows that the $c$-summation of Equation (\ref{eq:three_tie_12_temp2}) is exponentially small. We must identify the ranges of $a$ and $b$ for which this is not the case.

Let $\epsilon_a = - \frac{\tilde{\pi}_1 \tau}{4} > 0$. Given that $\tilde{\pi}_1 \in (0, \frac{3}{4}]$ and $\tilde{\pi}_2, \tilde{\pi}_3 \in [\frac{1}{2}, 1)$, we recognize that $\tau \in [-1, 1.5]$ which ensures $|\epsilon_a| < \tilde{\pi}_1$. Then we have $\frac{n}{3}-2a+b \geq \frac{\tilde{\pi}_3 n}{3}$ over \emph{any} $a \in [\frac{(\tilde{\pi}_1 - \epsilon_a)n}{3}, \frac{(\tilde{\pi}_1 + \epsilon_a)n}{3}]$ as long as
\begin{align*}
    b & \geq \frac{n}{3} (\tilde{\pi}_3 -1) +2a \\
    & \geq \frac{n}{3} \left( \tilde{\pi}_3 -1 + 2 \tilde{\pi}_1 - 2 \epsilon_a \right) \\
    & = -\frac{2 \epsilon_a n}{3} + \frac{n}{3} \left( \tilde{\pi}_3 + 2 \tilde{\pi}_1 -1 \right) \\
    & = -\frac{\tilde{\pi}_1 \tau n}{6} + \frac{(\tilde{\pi}_2 - \tau) n}{3} \\
    & = \frac{\tilde{\pi}_2 n}{3} + \Omega(n).
\end{align*}
Therefore, the $b$-summation of Equation (\ref{eq:three_tie_12_temp2}) is exponentially small; it cannot be $\Theta(1)$ as long as both the $a$- and $c$- summations are. This proves that Equation (\ref{eq:three_tie_10b}) is exponentially small when $\tau < 0$.

Now let $\tau = 0$ and consider the equation
\begin{align}
    & \Theta \left( \frac{1}{n} \right) \sum_{a=0}^{\bfloor{\frac{\tilde{\pi}_1 n}{3}}} \binom{\frac{n}{3}}{a} \tilde{\pi}_1^a \tilde{\pi}_5^{\frac{n}{3}-a} \sum_{b=\bfloor{ \frac{\tilde{\pi}_2 n}{3}}+1}^{\frac{n}{3}} \binom{\frac{n}{3}}{b} \tilde{\pi}_2^b \tilde{\pi}_6^{\frac{n}{3}-b}
    \sum_{c=\bfloor{\frac{n}{6}}+1}^{\min\{\frac{n}{3}, \frac{n}{3}-2a+b\}}  \binom{\frac{n}{3}}{c} \tilde{\pi}_3^c \tilde{\pi}_4^{\frac{n}{3}-c}
    \left( \frac{n}{3} -2a +b -c \right).
    \label{eq:three_tie_12b}
\end{align}
Clearly we have 
\begin{align*}
    \frac{n}{3} -2a +b  & \geq \frac{n}{3} \left( 1 -2\tilde{\pi}_1 + \tilde{\pi}_2 \right) \\
    & = \frac{\tilde{\pi}_3 n}{3}
\end{align*}
for all $a$ and $b$ within their respective ranges. Therefore each of the $a$-, $b$-, and $c$-summations of Equation (\ref{eq:three_tie_12b}) are $\Theta(1)$. It follows that Equation (\ref{eq:three_tie_10b}), when $a  > \bfloor{\frac{\tilde{\pi}_1 n}{3}}$ or $b < \bfloor{\frac{\tilde{\pi}_2 n}{3}}$, is exponentially small. It remains to determine precise bounds for the asymptotic rate of Equation (\ref{eq:three_tie_12b}). 
Specifically, Equation (\ref{eq:three_tie_12b}) can be written as
\begin{align}
    & \Theta \left( \frac{1}{n} \right) \sum_{a=0}^{\bfloor{\frac{\tilde{\pi}_1 n}{3}}} \binom{\frac{n}{3}}{a} \tilde{\pi}_1^a \tilde{\pi}_5^{\frac{n}{3}-a} A_a \label{eq:three_tie_14}
\end{align}
where we define
\begin{align*}
    A_a = \sum_{b=\bfloor{ \frac{\tilde{\pi}_2 n}{3}}+1}^{\frac{n}{3}} \binom{\frac{n}{3}}{b} \tilde{\pi}_2^b \tilde{\pi}_6^{\frac{n}{3}-b} (B_{a,b} + B'_{a,b})
\end{align*}
with
\begin{align*}
    B_{a,b} & = \sum_{c=\bfloor{\frac{n}{6}}+1}^{\bfloor{\frac{\tilde{\pi}_3 n}{3}}}  \binom{\frac{n}{3}}{c} \tilde{\pi}_3^c \tilde{\pi}_4^{\frac{n}{3}-c} \left( \frac{n}{3} -2a +b -c \right) 
\end{align*}
and
\begin{align*}
    B'_{a,b} & = \sum_{c=\bfloor{\frac{\tilde{\pi}_3 n}{3}}+1}^{\min\{\frac{n}{3}, \frac{n}{3}-2a+b\}} \binom{\frac{n}{3}}{c} \tilde{\pi}_3^c \tilde{\pi}_4^{\frac{n}{3}-c} \left( \frac{n}{3} -2a +b -c \right).
\end{align*}
This may be simplified as
\begin{align*}
    B_{a,b} & = \left( \frac{n}{3} -2a +b \right) \left( \frac{1}{2} \pm \calO \left( \frac{1}{\sqrt{n}} \right) \right) - \left( \frac{\tilde{\pi}_3 n}{6}  \pm \calO (\sqrt{n}) \right) \\
    & = \frac{(1-\tilde{\pi}_3)n}{6} -a + \frac{b}{2} \pm \calO (\sqrt{n})
\end{align*}
Meanwhile, 
\begin{align*}
    0 & \leq B'_{a,b} \\
    & \leq \left( \frac{n}{3} -2a+b - \frac{ \tilde{\pi}_3 n}{3}  \right) \sum_{c=\bfloor{\frac{\tilde{\pi}_3 n}{3}}+1}^{\min\{\frac{n}{3}, \frac{n}{3}-2a+b\}} \binom{\frac{n}{3}}{c} \tilde{\pi}_3^c \tilde{\pi}_4^{\frac{n}{3}-c} \\
    & \leq \left( \frac{n}{3} -2a+b - \frac{ \tilde{\pi}_3 n}{3}  \right) \sum_{c=\bfloor{\frac{\tilde{\pi}_3 n}{3}}+1}^{\frac{n}{3}} \binom{\frac{n}{3}}{c} \tilde{\pi}_3^c \tilde{\pi}_4^{\frac{n}{3}-c} \\
    & = \frac{(1-\tilde{\pi}_3)n}{6} -a + \frac{b}{2} \pm \calO (\sqrt{n}).
\end{align*}
by Lemma \ref{lem:bin_theorems_approx}. Let us set aside $B'_{a,b}$ for the moment and continue with $A_a$ only in terms of $B_{a,b}$. This entails
\begin{align*}
    A_a & = \left( \frac{(1-\tilde{\pi}_3)n}{6} -a \pm \calO (\sqrt{n}) \right) \left( \frac{1}{2} \pm \calO \left( \frac{1}{\sqrt{n}} \right) \right)  + \left( \frac{\tilde{\pi}_2 n}{12} \pm \calO(\sqrt{n}) \right) \\
    & = \frac{(1 -  \tilde{\pi}_3 + \tilde{\pi}_2) n}{12} -\frac{a}{2} \pm \calO (\sqrt{n})
\end{align*}
by Lemma \ref{lem:bin_theorems_approx}. Therefore, Equation (\ref{eq:three_tie_14}) is
\begin{align}
    & \Theta \left( \frac{1}{n} \right) \left( \left( \frac{(1-\tilde{\pi}_3 + \tilde{\pi}_2)n}{12} \pm \calO (\sqrt{n}) \right) \left( \frac{1}{2} \pm \calO \left( \frac{1}{\sqrt{n}} \right) \right) 
    - \left( \frac{\tilde{\pi}_1 n}{12} \pm \calO(\sqrt{n}) \right) \right) \notag \\
    & = (1 - \tilde{\pi}_3 + \tilde{\pi}_2 - 2\tilde{\pi}_1) \Theta(1) \pm \calO \left( \frac{1}{\sqrt{n}} \right) \notag \\
    &  = \tau \Theta(1) \pm \calO \left( \frac{1}{\sqrt{n}} \right) \notag \\
    & = \calO \left( \frac{1}{\sqrt{n}} \right) \label{eq:three_tie_15}
\end{align}
since $\tau=0$ by assumption and the objective is non-negative. Since $B'_{a,b}$ has the same form as $B_{a,b}$, as determined above, it does not affect this conclusion. 
\footnote{
The stated proof holds for $\tilde{\pi}_3 > \frac{1}{2}$. If $\tilde{\pi}_3 = \frac{1}{2}$, then $B_{a,b} = 0$, but the upper bound on $B'_{a,b}$ still holds. Therefore, this does not affect our $\calO \left( \frac{1}{\sqrt{n}} \right)$ conclusion.
}
This proves the stated asymptotic rate for Equation (\ref{eq:three_tie_10b}) when $\tau = 0$.

Finally, consider $\tau > 0$ and the equation
\begin{align}
    & \Theta \left( \frac{1}{n} \right) \sum_{a=0}^{\bfloor{\min\{\frac{\tilde{\pi} n}{3} + \tau \frac{n}{6}, \frac{n}{4}\}}} \binom{\frac{n}{3}}{a} \tilde{\pi}_1^a \tilde{\pi}_5^{\frac{n}{3}-a}
    \sum_{b=\bfloor{\max\{\frac{n}{6}, \frac{\tilde{\pi}_2 n}{3} - \tau \frac{n}{3} + 2 \max\{a - \frac{\tilde{\pi}_1 n}{3}, 0\} }}^{\frac{n}{3}} \binom{\frac{n}{3}}{b} \tilde{\pi}_2^b \tilde{\pi}_6^{\frac{n}{3}-b} \notag \\
    & \quad \quad \times \sum_{c=\bfloor{\frac{n}{6}}+1}^{\min\{\frac{n}{3}, \frac{n}{3}-2a+b\}}  \binom{\frac{n}{3}}{c} \tilde{\pi}_3^c \tilde{\pi}_4^{\frac{n}{3}-c} \left( \frac{n}{3}-2a+b-c \right).
    \label{eq:three_tie_12c}
\end{align}
When $a = \frac{\tilde{\pi}_1 n}{3}$ and $b = \frac{\tilde{\pi}_2 n}{3}$, this entails
\begin{align*}
    \frac{n}{3} -2a +b & = \frac{n}{3}  (1 - 2 \tilde{\pi}_1 + \tilde{\pi}_2)
    = \frac{(\tilde{\pi}_3 + \tau) n}{3}.
\end{align*}
Hence, $\tau$ represents the amount of \emph{slack} that the $c$-summation in Equation (\ref{eq:three_tie_12c}) has, in terms of $a$ and $b$, before $\frac{n}{3} -2a +b$ goes below $ \frac{\tilde{\pi}_3 n}{3}$ and the $c$-summation becomes exponentially small. This slack can be taken up by as much as $\tau \frac{n}{6}$ above $\frac{\tilde{\pi}_1 n}{3}$ in the $a$-summation or $\tau \frac{n}{3}$ below $\frac{\tilde{\pi}_2 n}{3}$ in the $b$-summation, as represented by Equation (\ref{eq:three_tie_12c}).

It is easy to see that Equation (\ref{eq:three_tie_14}) is included in Equation (\ref{eq:three_tie_12c}).\footnote{In this case, $B_{a,b}$ defined above has value regardless of whether $\tilde{\pi}_3 = 0.5$ or not.} Therefore its asymptotic rate is at least $\tau \Theta(1) \pm \calO \left( \frac{1}{\sqrt{n}} \right) = \Theta(1)$ by Equation (\ref{eq:three_tie_15}). Moreover, it is easy to see that Equation (\ref{eq:three_tie_12c}) is upper-bounded by $\Theta(1)$, following Lemma \ref{lem:bin_theorems_approx}. This proves the stated asymptotic rate for Equation (\ref{eq:three_tie_10b}) when $\tau > 0$.

This concludes the $\calE_{2,3}$ case of Lemma \ref{lem:three_tie_priority1}.
\end{paragraph}

\begin{paragraph}{Step 4: Putting the pieces together.}
To finish the proof, we tie our results about the $\calE_2$, $\calE_3$, and $\calE_{2,3}$ cases together. This entails the sum of Equations (\ref{eq:three_tie_3}), (\ref{eq:three_tie_7}), (\ref{eq:three_tie_11}), and (\ref{eq:three_tie_13}) subject to their respective conditions on the probability distribution $(\pi_1, \ldots, \pi_6)$. Recall that $\Theta(1)$, in many of the simplified versions of equations following (\ref{eq:three_tie_3}), (\ref{eq:three_tie_7}) and (\ref{eq:three_tie_11}), was a stand-in for $\binom{n}{\frac{n}{3}, \frac{n}{3}, \frac{n}{3}} \frac{n}{3^n}$. This enables us to combine several $\Theta(1)$-like terms together. 
\begin{table}[t]
    \centering
    \begin{tabular}{|c||c|c|c|}
        \hline
         & $\tilde{\pi}_2 = \tilde{\pi}_6$ & $\tilde{\pi}_2 > \tilde{\pi}_6$ & $\tilde{\pi}_2 < \tilde{\pi}_6$ \\
          \hline \hline
            $\tilde{\pi}_3 = \tilde{\pi}_4$ & 
            \makecell{$\pi_1 - \pi_5$ \\ by Eqns. (\ref{eq:three_tie_4A}), (\ref{eq:three_tie_8A}), and (\ref{eq:three_tie_12A})} &  
            \makecell{$2 \pi_1 + \pi_6 - 3 \pi_2$ \\ by Eqns.  (\ref{eq:three_tie_8C}) and (\ref{eq:three_tie_12C})} & 
            \makecell{$\pi_6 - \pi_1$ \\ by Eqn. (\ref{eq:three_tie_4B})} \\[0.25em]
          \hline
            $\tilde{\pi}_3 > \tilde{\pi}_4$ & 
            \makecell{$2\pi_1 + 2\pi_5 + \pi_4 - 5\pi_3$ \\ by Eqns. (\ref{eq:three_tie_4C}) and (\ref{eq:three_tie_12B})} &  
            \makecell{$\pi_1 + \pi_6 - \pi_2 - \pi_3$ \\ by Eqn. (\ref{eq:three_tie_12D})} & 
            \makecell{$\pi_4 + \pi_5 - \pi_1 - \pi_3$ \\ by Eqn. (\ref{eq:three_tie_4D})} \\[0.25em]
          \hline 
            $\tilde{\pi}_3 < \tilde{\pi}_4$ & 
            \makecell{$\pi_1 - \pi_3$ \\ by Eqn. (\ref{eq:three_tie_8B})} &  
            \makecell{$\pi_1 + \pi_6 - \pi_2 - \pi_3$ \\ by Eqn. (\ref{eq:three_tie_8D})} & 
            \makecell{N/A} \\
         \hline
    \end{tabular}
    \caption{Constants in front of $\Theta(1)$ term resulting from Equations (\ref{eq:three_tie_3}), (\ref{eq:three_tie_7}), (\ref{eq:three_tie_11}), for certain conditions on $\pi$.}
    \label{tab:complexity_three_sides_A}
\end{table}

Our conclusion is therefore $\pm \calO \left( e^{-\Theta(n)} \right)$ if $\tilde{\pi}_2 < \tilde{\pi}_6$ and $\tilde{\pi}_3 < \tilde{\pi}_4$. Otherwise, it is
\[
f(\pi_1, \ldots, \pi_6) \Theta(1) \pm \calO \left( \frac{1}{\sqrt{n}} \right) + g_n(\pi_1, \ldots, \pi_6)
\]
where $f(\pi_1, \ldots, \pi_6)$ is determined by Table \ref{tab:complexity_three_sides_A} and
\begin{align*}
    g_n(\pi_1, \ldots, \pi_6) = \begin{cases}
    \Theta(1), & \pi_2 + \pi_5 > \pi_1 + \pi_3 \\
    \calO \left( \frac{1}{\sqrt{n}} \right), & \pi_2 + \pi_5 = \pi_1 + \pi_3 \\
    \calO \left( e^{-\Theta(n)} \right), & \pi_2 + \pi_5 < \pi_1 + \pi_3.
    \end{cases}
\end{align*}
The $\tilde{\pi}_1 > 0.75$ case is covered by the exponential case here. This concludes the proof of Lemma \ref{lem:three_tie_priority1}.
\end{paragraph}
\end{proof}

%% file: EC_appendix/apx_page_3c.tex
\subsection{Case when alternative $2$ wins}
\label{apx:three_ties_P3}

\begin{table}[t]
    \centering
    \begin{tabular}{|c||c|c|c|}
        \hline
         & $\pi_1 = \pi_5$ & $\pi_1 > \pi_5$ & $\pi_1 < \pi_5$ \\
          \hline \hline
            $\pi_3 = \pi_4$ & 
            $\pi_6 - \pi_2$ &  
            $\pi_2 - \pi_4$ & 
            $3 \pi_1 - 2 \pi_2 - \pi_5$  \\[0.25em]
          \hline
            $\pi_3 > \pi_4$ & 
            $\pi_1 + 2 \pi_3 - 2\pi_2 - \pi_5$ &  
            N/A & 
            $2 \pi_1 - \pi_2 - \pi_4$ \\[0.25em]
          \hline 
            $\pi_3 < \pi_4$ & 
            $2 \pi_2 + \pi_3 - 3\pi_4$ &  
            $\pi_1 + \pi_2 - 2 \pi_4$ & 
            $2 \pi_1 - \pi_2 - \pi_4$ \\
         \hline
    \end{tabular}
    \caption{Values of $f^2(\pi_1, \ldots \pi_6)$ given conditions on $\pi$ for Lemma \ref{lem:three_tie_priority3}.}
    \label{tab:complexity_three_sides_C_final}
\end{table}

\setcounter{lem}{8}
\begin{lem}
Suppose that $\pi_1 + \pi_5 = \pi_2 + \pi_6 = \pi_3 + \pi_4 = \frac{1}{3}$  and $\pi_i > 0, ~\forall i \in [6].$ Furthermore, let $u_1 \geq u_2 > u_3$ in $\vec{u}$. 
Then $\exists N > 0$ such that $\forall n > N$ where $n-2$ is divisible by $3$, 
\[
\EDS(\{1,2, 3\}) = \begin{cases}
    \pm \calO \left( e^{-\Theta(n)} \right), & \pi_1 > \pi_5, \pi_3 > \pi_4 \\
    f^2(\pi_1, \ldots, \pi_6) \Theta(1) + g^2_n(\pi_1, \ldots, \pi_6), & \text{otherwise}
\end{cases}
\]
otherwise, 
where $f^2(\pi_1, \ldots, \pi_6)$ is determined by Table \ref{tab:complexity_three_sides_C_final} and
\begin{align*}
    g^2_n(\pi_1, \ldots, \pi_6) = \begin{cases}
        \Theta(1), & \pi_1 + \pi_6 < \pi_2 + \pi_3, \\
        \pm \calO \left( \frac{1}{\sqrt{n}} \right), & \pi_1 + \pi_6 \geq \pi_2 + \pi_3.
    \end{cases}
\end{align*}
If $u_1 > u_2 = u_3$, then $\EDS(\{1,2, 3\})$ is $\pm \calO \left( \frac{1}{n} \right)$ if $\pi_3 \leq \pi_4$; it is $\pm \calO \left( e^{-\Theta(n)} \right)$ otherwise.
\label{lem:three_tie_priority3}
\end{lem}

\begin{proof}
This lemma is analogous to Lemma \ref{lem:three_tie_priority1} except we have that two fewer than the number of agents is divisible by $3$. To help with notation, we will consider $3n+2$ agents. This covers the case where there are $(n, n+1, n+1)$ agents truthfully voting for alternatives $1$, $2$, and $3$ respectively. For any preference profile $P$ and truthful vote profile $\topRank{P}$, it follows that $1 \in \EW{\topRank{P}}$ if $R_4 \in P$ and $P[1 \succ 2] \geq P[2 \succ 1]$, whereas $3 \in \EW{\topRank{P}}$ if $R_5 \in P$ and $P[3 \succ 2] > P[2 \succ 3]$. This yields three cases for whether either or both of these are the case. We define $\calE_1$, $\calE_3$, and $\calE_{1,3}$ as follows:
\begin{itemize}
    \item $\calE_1 = \{P \in \calL(\calA)^n~:~R_4 \in P \text{ and } P[1 \succ 2] \geq P[2 \succ 1] \text{, and either } R_5 \notin P \text{ or } P[2 \succ 3] \geq P[3 \succ 2]\}$,
    \item $\calE_3 = \{P \in \calL(\calA)^n~:~R_5 \in P \text{ and } P[3 \succ 2] > P[2 \succ 3] \text{, and either } R_3 \notin P \text{ or } P[2 \succ 1] > P[1 \succ 2]\}$,
    \item $\calE_{1,3} = \{P \in \calL(\calA)^n~:~R_4, R_5 \in P \text{ and } P[1 \succ 2] \geq P[2 \succ 1] \text{ and } P[3 \succ 2] > P[2 \succ 3]\}$.
\end{itemize}
Implicitly, we note $s_1(\topRank{P}) = s_2(\topRank{P})+1 = s_3(\topRank{P})+1$ for each of these cases.
The subscript denotes which alternatives (among $1$ and $3$, excluding $2$) appears in the equilibrium winning set $\EW{\topRank{P}}$, for ease of readability.
Let $c \in [0, n]$ and $a, b \in [0,n+1]$. We take throughout this proof:
\begin{itemize}
    \item $n$ agents with rankings either $R_1$ or $R_5$: with $c$ for $R_1$ and $n-c$ for $R_5$,
    \item $n+1$ agents with rankings either $R_2$ or $R_6$: with $a$ for $R_2$ and $n+1-a$ for $R_6$,
    \item $n+1$ agents with rankings either $R_3$ or $R_4$: with $b$ for $R_3$ and $n+1-b$ for $R_4$.
\end{itemize}

\begin{paragraph}{Step 1: Characterize the $\calE_1$ case.}
We have $P \in \calE_1$ if the following ranges are satisfied. First, $a \in [0, n+1]$ has its full range. Second, $b < \frac{n+1}{2}$, so that there are at least as many agents preferring $R_4 = (3 \succ 1 \succ 2)$ than $R_3 = (3 \succ 2 \succ 1)$, which entails $1 \in \EW{\topRank{P}}$. Third, $c \geq \frac{n}{2}$, so that there are at least as many agents preferring $R_1 = (1 \succ 2 \succ 3)$ than $R_5 = (1 \succ 3 \succ 2)$, which entails $3 \notin \EW{\topRank{P}}$. Like in Lemma \ref{lem:sub_12}, the (negated) value per agent and probability of each ranking is summarized by Table \ref{tab:character_profiles_E1}
Put together, we get the equation
\begin{align}
    & \Pr\nolimits_{P \sim \pi^n}(P \in \mathcal{E}_1) \times \mathbb{E}_{P \sim \pi^n}[\ADS(P)~|~P \in \mathcal{E}_1] \notag \\
    & = \sum_{a=0}^{n+1} \sum_{b=0}^{\bceil{\frac{n+1}{2}}-1} \sum_{c=\bceil{\frac{n}{2}}}^{n} \mathcal{P}_{\vec{\pi},n}(a,b,c) \cdot \mathcal{V}^1_{\vec{u},n}(a,b,c) \label{eq:three_tieP3_1}
\end{align}
where we define
\begin{align*}
    \mathcal{P}_{\vec{\pi},n}(a,b,c) & = \binom{3n+2}{a, b, c, n-c, n+1-a, n+1-b} \pi_2^a \pi_3^b \pi_1^{c} \pi_5^{n-c} \pi_6^{n+1-a} \pi_4^{n+1-b} \\
    & = \binom{3n+2}{n+1, n+1, n} \frac{1}{3^{3n+2}}   \times  \binom{n+1}{a} \tilde{\pi}_2^a \tilde{\pi}_6^{n+1-a} \times \binom{n+1}{b} \tilde{\pi}_3^b \tilde{\pi}_4^{n+1-b} \times \binom{n}{c} \tilde{\pi}_1^c \tilde{\pi}_5^{n-c}
\end{align*}
and 
\begin{align*}
    \mathcal{V}^1_{\vec{u},n}(a, b, c) = & \begin{pmatrix}
        a, & b, & c, & n-c, & n+1-a, & n+1-b \end{pmatrix} \\
    & \cdot \begin{pmatrix}
        u_1 - u_3,& u_2 - u_3,& -u_1 + u_2,& -u_1 + u_3,& u_1 - u_2,& -u_2 + u_3 \end{pmatrix} \\
    & = (u_2 - u_3) \left( -2n +a + 2b + c \right) + (u_1 - 2u_2 + 2u_3).
\end{align*}
This equation uses the following definitions:
\begin{itemize}
    \item $\tilde{\pi}_1 = \frac{\pi_1}{\pi_1+\pi_5}$ and $\tilde{\pi}_5 = \frac{\pi_5}{\pi_1+\pi_5}$,
    \item $\tilde{\pi}_2 = \frac{\pi_2}{\pi_2+\pi_6}$ and $\tilde{\pi}_6 = \frac{\pi_6}{\pi_2+\pi_6}$,
    \item $\tilde{\pi}_3 = \frac{\pi_3}{\pi_3+\pi_4}$ and $\tilde{\pi}_4 = \frac{\pi_4}{\pi_3+\pi_4}$
\end{itemize}
which we recall from Definition \ref{dfn:tilde_definitions}, where the denominators are each $\frac{1}{3}$. 
Lemma \ref{lem:stirling_trinom} suggests that $\binom{3n+2}{n+1, n+1, n} \frac{n}{3^{3n+2}} = \Theta \left( \frac{1}{n} \right)$. Equation (\ref{eq:three_tieP3_1}) may therefore be written as
\begin{align}
    & \Theta \left( \frac{1}{n} \right) \sum_{c=\bceil{\frac{n}{2}}}^{n}  \binom{n}{c} \tilde{\pi}_1^c \tilde{\pi}_5^{n-c} \sum_{b=0}^{\bceil{\frac{n+1}{2}}-1} \binom{n+1}{b} \tilde{\pi}_3^b \tilde{\pi}_4^{n+1-b}  \notag \\
    & \quad \quad \times \sum_{a=0}^{n+1} \binom{n+1}{a} \tilde{\pi}_2^a \tilde{\pi}_6^{n+1-a} 
    \left( (u_2 - u_3)(-2n+a+2b+c) + (u_1 - 2u_2 + 2 u_3) \right) \notag \\
    & = (u_2 - u_3) \Theta \left( \frac{1}{n} \right) \sum_{c=\bceil{\frac{n}{2}}}^{n}  \binom{n}{c} \tilde{\pi}_1^c \tilde{\pi}_5^{n-c} \sum_{b=0}^{\bceil{\frac{n+1}{2}}-1} \binom{n+1}{b} \tilde{\pi}_3^b \tilde{\pi}_4^{n+1-b} 
    \left(  (-2+\tilde{\pi}_2)n+2b+c \right) \notag \\
    & + (u_1 + (\tilde{\pi}_2-2) (u_2 - u_3)) \Theta \left( \frac{1}{n} \right) 
    \sum_{c=\bceil{\frac{n}{2}}}^{n}  \binom{n}{c} \tilde{\pi}_1^c \tilde{\pi}_5^{n-c} \sum_{b=0}^{\bceil{\frac{n+1}{2}}-1} \binom{n+1}{b} \tilde{\pi}_3^b \tilde{\pi}_4^{n+1-b} \label{eq:three_tieP3_3}
\end{align}
by definition of binomial probability and expectation. 
By Lemma \ref{lem:bin_theorems_approx}, the second term of Equation (\ref{eq:three_tieP3_3}) is proportional to $\Theta \left( \frac{1}{n} \right)$ if $\pi_1 \geq \pi_5$ and $\pi_3 \leq \pi_4$, and $\calO \left( e^{-\Theta(n)} \right)$ otherwise. We carry this finding forward to Step 4, below, and continue with the first term of Equation (\ref{eq:three_tieP3_3}) assuming that $u_2 - u_3 > 0$.

We solve this part using the same techniques as with case $\calE_2$ of Lemma \ref{lem:three_tie_priority1}. The first term of Equation (\ref{eq:three_tieP3_3}) may be simplified as
\begin{align}
    & \Theta \left( \frac{1}{n} \right) \bigg(  (-2+\tilde{\pi}_2)n \sum_{c=\bceil{\frac{n}{2}}}^{n}  \binom{n}{c} \tilde{\pi}_1^c \tilde{\pi}_5^{n-c} \sum_{b=0}^{\bceil{\frac{n+1}{2}}-1} \binom{n+1}{b} \tilde{\pi}_3^b \tilde{\pi}_4^{n+1-b} \notag \\
    & + \sum_{c=\bceil{\frac{n}{2}}}^{n}  \binom{n}{c} \tilde{\pi}_1^c \tilde{\pi}_5^{n-c} c  \sum_{b=0}^{\bceil{\frac{n+1}{2}}-1} \binom{n+1}{b} \tilde{\pi}_3^b \tilde{\pi}_4^{n+1-b} \notag \\
    &  +2 \sum_{c=\bceil{\frac{n}{2}}}^{n}  \binom{n}{c} \tilde{\pi}_1^c \tilde{\pi}_5^{n-c} \sum_{b=0}^{\bceil{\frac{n+1}{2}}-1} \binom{n+1}{b} \tilde{\pi}_3^b \tilde{\pi}_4^{n+1-b} b \bigg). \label{eq:three_tieP3_4}
\end{align}
It is clear from Lemma \ref{lem:bin_theorems_approx} that if either $\tilde{\pi}_1 < \tilde{\pi}_5$ or $\tilde{\pi}_3 > \tilde{\pi}_4$, then Equation (\ref{eq:three_tieP3_4}) is $\calO \left( e^{-\Theta(n)} \right)$. This leaves four cases. First, if $\tilde{\pi}_1 = \tilde{\pi}_5$ and $\tilde{\pi}_3 = \tilde{\pi}_4$, then Equation (\ref{eq:three_tieP3_4}) is 
\begin{align}
    & \Theta \left( \frac{1}{n} \right) \bigg( (-2+\tilde{\pi}_2)n \left( \frac{1}{2} \pm \calO \left( \frac{1}{\sqrt{n}} \right)  \right) \left( \frac{1}{2} \pm \calO \left( \frac{1}{\sqrt{n}} \right)  \right) \notag \\
    & \quad \quad + \left( \frac{n}{4} \pm \calO (\sqrt{n})  \right) \left( \frac{1}{2} \pm \calO \left( \frac{1}{\sqrt{n}} \right)  \right) \notag \\
    & \quad \quad +2 \left( \frac{1}{2} \pm \calO \left( \frac{1}{\sqrt{n}} \right)  \right) \left( \frac{n}{4} \pm \calO (\sqrt{n})  \right)
    \bigg) \notag \\
    & =  \Theta \left( \frac{1}{n} \right) \left( n \left( \frac{-2+\tilde{\pi}_2}{4} + \frac{1}{8} + \frac{1}{4} \right) \pm \calO(\sqrt{n}) \right) \notag \\
    & = \frac{3}{8} (\pi_2 - \pi_6) \Theta(1) \pm \calO \left( \frac{1}{\sqrt{n}} \right) \label{eq:three_tieP3_5A}
\end{align}
making use of the fact that $\pi_2 + \pi_6 = \frac{1}{3}$.
Second, if $\tilde{\pi}_1 = \tilde{\pi}_5$ and $\tilde{\pi}_3 < \tilde{\pi}_4$, then Equation (\ref{eq:three_tieP3_4}) is
\begin{align}
    & \Theta \left( \frac{1}{n} \right) \bigg( (-2+\tilde{\pi}_2)n \left( \frac{1}{2} \pm \calO \left( \frac{1}{\sqrt{n}} \right)  \right) \left( 1 - \calO \left( e^{-\Theta(n)} \right)  \right) \notag \\
    & \quad \quad + \left( \frac{n}{4} \pm \calO (\sqrt{n})  \right) \left( 1 - \calO \left( e^{-\Theta(n)} \right)  \right) \notag \\
    & \quad \quad +2 \left( \frac{1}{2} \pm \calO \left( \frac{1}{\sqrt{n}} \right)  \right) \tilde{\pi}_3 (n+1) \left(  1 - \calO \left( e^{-\Theta(n)} \right) \right)
    \bigg) \notag \\
    & = \Theta \left( \frac{1}{n} \right) \left( n \left( \frac{-2 + \tilde{\pi}_2}{2} + \frac{1}{4} + \tilde{\pi}_3 \right) \pm \calO(\sqrt{n}) \right) \notag \\
    & = \frac{3}{4} (2\pi_2 +  \pi_3 - 3\pi_4) \Theta(1) \pm \calO \left( \frac{1}{\sqrt{n}} \right). \label{eq:three_tieP3_5B}
\end{align}
Third, if $\tilde{\pi}_1 > \tilde{\pi}_5$ and $\tilde{\pi}_3 = \tilde{\pi}_4$, then Equation (\ref{eq:three_tieP3_4}) is
\begin{align}
    & \Theta \left( \frac{1}{n} \right) \bigg( (-2+\tilde{\pi}_2)n \left( 1 - \calO \left( e^{-\Theta(n)} \right) \right) \left( \frac{1}{2} \pm \calO \left( \frac{1}{\sqrt{n}} \right)  \right) \notag \\
    & \quad \quad + \tilde{\pi}_1 n \left( 1 - \calO \left( e^{-\Theta(n)} \right)  \right) \left( \frac{1}{2} \pm \calO \left( \frac{1}{\sqrt{n}} \right)  \right) \notag \\
    & \quad \quad +2 \left( 1 - \calO \left( e^{-\Theta(n)} \right)  \right) \left( \frac{n}{4} \pm \calO (\sqrt{n})  \right)
    \bigg) \notag \\
    & = \Theta \left( \frac{1}{n} \right) \left( n \left( \frac{-2 + \tilde{\pi}_2}{2} + \frac{\tilde{\pi}_1}{2} + \frac{1}{2} \right) \pm \calO(\sqrt{n}) \right) \notag \\
    & = \frac{3}{2}(\pi_2 - \pi_5) \Theta(1) \pm \calO \left( \frac{1}{\sqrt{n}} \right). \label{eq:three_tieP3_5C}
\end{align}
Finally, if $\tilde{\pi}_1 > \tilde{\pi}_5$ and $\tilde{\pi}_3 < \tilde{\pi}_4$, then Equation (\ref{eq:three_tieP3_4}) is
\begin{align}
    & \Theta \left( \frac{1}{n} \right) \bigg( (-2+\tilde{\pi}_2) \left( 1 - \calO \left( e^{-\Theta(n)} \right) \right) \left(1 - \calO \left( e^{-\Theta(n)} \right)  \right) \notag \\
    & \quad \quad + \tilde{\pi}_1 n \left( 1 - \calO \left( e^{-\Theta(n)} \right)  \right) \left( 1 - \calO \left( e^{-\Theta(n)} \right)  \right) \notag \\
    & \quad \quad +2 \left( 1 - \calO \left( e^{-\Theta(n)} \right)  \right) \tilde{\pi}_3 (n+1) \left( 1 - \calO \left( e^{-\Theta(n)} \right) \right)
    \bigg) \notag \\
    & = \Theta \left( \frac{1}{n} \right) \left( n \left( -2 + \tilde{\pi}_2 + \tilde{\pi}_1 + 2 \tilde{\pi}_3 \right) \pm \calO(\sqrt{n}) \right) \notag \\
    & = 3(\pi_1 + \pi_2 - 2\pi_4) \Theta(1) \pm \calO \left( \frac{1}{\sqrt{n}} \right). \label{eq:three_tieP3_5D}
\end{align}
%
%
Recall that each $\Theta(1)$ in Equations (\ref{eq:three_tieP3_5A}), (\ref{eq:three_tieP3_5B}), (\ref{eq:three_tieP3_5C}), and (\ref{eq:three_tieP3_5D})
is actually an instance of $\binom{3n+2}{n+1, n+1, n} \frac{n}{3^{3n+2}}$, following Lemma \ref{lem:stirling_trinom}.
We make use of this fact and continue with these equations in Step 4, below.
This concludes the $\calE_1$ case of Lemma \ref{lem:three_tie_priority3}.
\end{paragraph}

\begin{paragraph}{Step 2: Characterize the $\calE_3$ case.}
We prove this case in the same way as Step 1 ($\calE_1$) above, keeping the same variable nomenclature but adjusting the ranges as needed. That is, we have $P \in \calE_3$ if the following ranges are satisfied. First, $a \in [0, n+1]$ has its full range. Second, $b > \frac{n+1}{2}$, so that there are more agents preferring $R_3 = (3 \succ 2 \succ 1)$ than $R_4 = (3 \succ 1 \succ 2)$, which entails $1 \notin \EW{\topRank{P}}$. Third, $c < \frac{n}{2}$, so that there are more agents preferring $R_5 = (1 \succ 3 \succ 2)$ than $R_1 = (1 \succ 2 \succ 3)$, which entails $3 \in \EW{\topRank{P}}$. Like in Lemma \ref{lem:sub_23}, the value per agent and probability of each ranking is summarized by Table \ref{tab:character_profiles_E5}.
Put together, we get the equation 
\begin{align}
    & \Pr\nolimits_{P \sim \pi^n}(P \in \mathcal{E}_3) \times \mathbb{E}_{P \sim \pi^n}[\ADS(P)~|~P \in \mathcal{E}_3] \notag \\
    & = \sum_{a=0}^{n+1} \sum_{b=\bfloor{\frac{n+1}{2}}+1}^{n+1} \sum_{c=0}^{\bceil{\frac{n}{2}}-1} \mathcal{P}_{\vec{\pi},n}(a,b,c) \cdot \mathcal{V}^3_{\vec{u},n}(a,b,c) \label{eq:three_tieP3_6}
\end{align}
where $\mathcal{P}_{\vec{\pi},n}$ is the same as in Step 1, and 
\begin{align*}
    \mathcal{V}^3_{\vec{u},n}(a, b, c) = & \begin{pmatrix}
        a, & b, & c, & n-c, & n+1-a, & n+1-b \end{pmatrix} \\
    & \cdot \begin{pmatrix}
        u_1 - u_2,& -u_1 + u_2,& u_2 - u_3,& -u_2 + u_3,& u_1 - u_3,& -u_1 + u_3 \end{pmatrix} \\
    & = (u_2 - u_3) \left( -n-a+b+2c \right).
\end{align*}
By Lemma \ref{lem:stirling_trinom}, this leads to
\begin{align}
    & \Theta \left( \frac{1}{n} \right) \sum_{b=\bfloor{\frac{n+1}{2}}+1}^{n} \binom{n+1}{b} \tilde{\pi}_3^b \tilde{\pi}_4^{n+1-b} \sum_{c=0}^{\bceil{\frac{n}{2}}-1}  \binom{n}{c} \tilde{\pi}_1^c \tilde{\pi}_5^{n-c}  \notag \\
    & \quad \quad \times \sum_{a=0}^{n+1} \binom{n+1}{a} \tilde{\pi}_2^a \tilde{\pi}_6^{n+1-a}
    (u_2 - u_3)(-n-a+b+2c) \notag \\
    & = (u_2 - u_3) \Theta \left( \frac{1}{n} \right) \sum_{b=\bfloor{\frac{n+1}{2}}+1}^{n+1} \binom{n+1}{b} \tilde{\pi}_3^b \tilde{\pi}_4^{n+1-b} \sum_{c=0}^{\bceil{\frac{n}{2}}-1}  \binom{n}{c} \tilde{\pi}_1^c \tilde{\pi}_5^{n-c}
    \left(  (-1-\tilde{\pi}_2)n+b+2c \right) \notag \\
    & - \tilde{\pi}_2 (u_2 - u_3) \Theta \left( \frac{1}{n} \right)
    \sum_{b=\bfloor{\frac{n+1}{2}}+1}^{n+1} \binom{n+1}{b} \tilde{\pi}_3^b \tilde{\pi}_4^{n+1-b} \sum_{c=0}^{\bceil{\frac{n}{2}}-1}  \binom{n}{c} \tilde{\pi}_1^c \tilde{\pi}_5^{n-c} \label{eq:three_tieP3_7}
\end{align}
by definition of binomial probability and expectation. 
By Lemma \ref{lem:bin_theorems_approx}, the second term of Equation (\ref{eq:three_tieP3_3}) is proportional to $\Theta \left( \frac{1}{n} \right)$ if $\pi_1 \leq \pi_5$ and $\pi_3 \geq \pi_4$, and $\calO \left( e^{-\Theta(n)} \right)$ otherwise. We carry this finding forward to Step 4, below, and continue with the first term of Equation (\ref{eq:three_tieP3_6}) assuming that $u_2 - u_3 > 0$.

The first term of Equation (\ref{eq:three_tieP3_7}) may be simplified as
\begin{align}
    & \Theta \left( \frac{1}{n} \right) \bigg(  (-1-\tilde{\pi}_2)n \sum_{b=\bfloor{\frac{n+1}{2}}+1}^{n+1} \binom{n+1}{b} \tilde{\pi}_3^b \tilde{\pi}_4^{n+1-b} \sum_{c=0}^{\bceil{\frac{n}{2}}-1}  \binom{n}{c} \tilde{\pi}_1^c \tilde{\pi}_5^{n-c} \notag \\
    & + \sum_{b=\bfloor{\frac{n+1}{2}}+1}^{n+1} \binom{n+1}{b} \tilde{\pi}_3^b \tilde{\pi}_4^{n+1-b} b \sum_{c=0}^{\bceil{\frac{n}{2}}-1} \binom{n}{c} \tilde{\pi}_1^c \tilde{\pi}_5^{n-c} \notag \\
    &  + 2 \sum_{b=\bfloor{\frac{n+1}{2}}+1}^{n+1} \binom{n+1}{b} \tilde{\pi}_3^b \tilde{\pi}_4^{n+1-b} \sum_{c=0}^{\bceil{\frac{n}{2}}-1}  \binom{n}{c} \tilde{\pi}_1^c \tilde{\pi}_5^{n-c} c \bigg). \label{eq:three_tieP3_8}
\end{align}
It is clear from Lemma \ref{lem:bin_theorems_approx} that if either $\tilde{\pi}_3 < \tilde{\pi}_4$ or $\tilde{\pi}_1 > \tilde{\pi}_5$, then Equation (\ref{eq:three_tieP3_8}) is $\calO \left( e^{-\Theta(n)} \right)$. This leaves four cases. First, if $\tilde{\pi}_3 = \tilde{\pi}_4$ and $\tilde{\pi}_1 = \tilde{\pi}_5$, then Equation (\ref{eq:three_tieP3_8}) is
\begin{align}
    & \Theta \left( \frac{1}{n} \right) \bigg( (-1-\tilde{\pi}_2)n \left( \frac{1}{2} \pm \calO \left( \frac{1}{\sqrt{n}} \right)  \right) \left( \frac{1}{2} \pm \calO \left( \frac{1}{\sqrt{n}} \right)  \right) \notag \\
    & \quad \quad + \left( \frac{n}{4} \pm \calO (\sqrt{n})  \right) \left( \frac{1}{2} \pm \calO \left( \frac{1}{\sqrt{n}} \right)  \right) \notag \\
    & \quad \quad +2 \left( \frac{1}{2} \pm \calO \left( \frac{1}{\sqrt{n}} \right)  \right) \left( \frac{n}{4} \pm \calO (\sqrt{n})  \right)
    \bigg) \notag \\
    & =  \Theta \left( \frac{1}{n} \right) \left( n \left( \frac{-1-\tilde{\pi}_2}{4} + \frac{1}{8} + \frac{1}{4} \right) \pm \calO(\sqrt{n}) \right) \notag \\
    & = \frac{3}{8}(\pi_6 - \pi_2) \Theta(1) \pm \calO \left( \frac{1}{\sqrt{n}} \right) \label{eq:three_tieP3_9A}
\end{align}
Second, if $\tilde{\pi}_3 = \tilde{\pi}_4$ and $\tilde{\pi}_1 < \tilde{\pi}_5$, then Equation (\ref{eq:three_tieP3_8}) is
\begin{align}
    & \Theta \left( \frac{1}{n} \right) \bigg( (-1-\tilde{\pi}_2)n \left( \frac{1}{2} \pm \calO \left( \frac{1}{\sqrt{n}} \right)  \right) \left( 1 - \calO \left( e^{-\Theta(n)} \right)  \right) \notag \\
    & \quad \quad + \left( \frac{n}{4} \pm \calO (\sqrt{n})  \right) \left( 1 - \calO \left( e^{-\Theta(n)} \right)  \right) \notag \\
    & \quad \quad +2 \left( \frac{1}{2} \pm \calO \left( \frac{1}{\sqrt{n}} \right)  \right) \tilde{\pi}_1 n \left(  1 - \calO \left( e^{-\Theta(n)} \right) \right)
    \bigg) \notag \\
    & = \Theta \left( \frac{1}{n} \right) \left( n \left( \frac{-1-\tilde{\pi}_2}{2} + \frac{1}{4} + \tilde{\pi}_1 \right) \pm \calO(\sqrt{n}) \right) \notag \\
    & = \frac{3}{4}(3 \pi_1 -2\pi_2 - \pi_5) \Theta(1) \pm \calO \left( \frac{1}{\sqrt{n}} \right). \label{eq:three_tieP3_9B}
\end{align}
Third, if $\tilde{\pi}_3 > \tilde{\pi}_4$ and $\tilde{\pi}_1 = \tilde{\pi}_5$, then Equation (\ref{eq:three_tieP3_8}) is
\begin{align}
    & \Theta \left( \frac{1}{n} \right) \bigg( (-1-\tilde{\pi}_2)n \left( 1 - \calO \left( e^{-\Theta(n)} \right) \right) \left( \frac{1}{2} \pm \calO \left( \frac{1}{\sqrt{n}} \right)  \right) \notag \\
    & \quad \quad + \tilde{\pi}_3 (n+1) \left( 1 - \calO \left( e^{-\Theta(n)} \right)  \right) \left( \frac{1}{2} \pm \calO \left( \frac{1}{\sqrt{n}} \right)  \right) \notag \\
    & \quad \quad +2 \left( 1 - \calO \left( e^{-\Theta(n)} \right)  \right) \left( \frac{n}{4} \pm \calO (\sqrt{n})  \right)
    \bigg) \notag \\
    & = \Theta \left( \frac{1}{n} \right) \left( n \left( \frac{-1-\tilde{\pi}_1}{2} + \frac{\tilde{\pi}_3}{2} + \frac{1}{2} \right) \pm \calO(\sqrt{n}) \right) \notag \\
    & = \frac{3}{2}(\pi_3 - \pi_1) \Theta(1) \pm \calO \left( \frac{1}{\sqrt{n}} \right). \label{eq:three_tieP3_9C}
\end{align}
Finally, if $\tilde{\pi}_3 > \tilde{\pi}_4$ and $\tilde{\pi}_1 < \tilde{\pi}_5$, then Equation (\ref{eq:three_tieP3_8}) is
\begin{align}
    & \Theta \left( \frac{1}{n} \right) \bigg( (-1-\tilde{\pi}_2) \left( 1 - \calO \left( e^{-\Theta(n)} \right) \right) \left(1 - \calO \left( e^{-\Theta(n)} \right)  \right) \notag \\
    & \quad \quad + \tilde{\pi}_3 (n+1) \left( 1 - \calO \left( e^{-\Theta(n)} \right)  \right) \left( 1 - \calO \left( e^{-\Theta(n)} \right)  \right) \notag \\
    & \quad \quad +2 \left( 1 - \calO \left( e^{-\Theta(n)} \right)  \right) \tilde{\pi}_1 n \left( 1 - \calO \left( e^{-\Theta(n)} \right) \right)
    \bigg) \notag \\
    & = \Theta \left( \frac{1}{n} \right) \left( n \left( -1-\tilde{\pi}_2 + \tilde{\pi}_3 + 2 \tilde{\pi}_1  \right) \pm \calO(\sqrt{n}) \right) \notag \\
    & = 3(2 \pi_1 - \pi_2 - \pi_4) \Theta(1) \pm \calO \left( \frac{1}{\sqrt{n}} \right). \label{eq:three_tieP3_9D}
\end{align}
As with the $\calE_1$ case above, each $\Theta(1)$ in  Equations (\ref{eq:three_tieP3_9A}), (\ref{eq:three_tieP3_9B}), (\ref{eq:three_tieP3_9C}), and (\ref{eq:three_tieP3_9D})
is actually an instance of $\binom{3n+2}{n+1, n+1, n} \frac{n}{3^{3n+2}}$.
We continue with these equations in Step 4, below. This concludes the $\calE_2$ case of Lemma \ref{lem:three_tie_priority3}.
\end{paragraph}

\begin{paragraph}{Step 3: Characterize the $\calE_{1,2}$ case.}
We keep the same variable nomenclature as the above steps, but adjust the ranges as needed. That is, we have $P \in \calE_{1,2}$ if the following ranges are satisfied. 
First, $a \in [0, n+1]$ has its full range. Second, $b < \frac{n+1}{2}$, so that there are at least as many agents preferring $R_4 = (3 \succ 1 \succ 2)$ than $R_3 = (3 \succ 2 \succ 1)$, which entails $1 \in \EW{\topRank{P}}$. Third, $c < \frac{n}{2}$, so that there are  agents preferring $R_5 = (1 \succ 3 \succ 2)$ than $R_1 = (1 \succ 2 \succ 3)$, which entails $3 \in \EW{\topRank{P}}$.

Since $|\EW{\topRank{P}}| = 2$ for this case, the adversarial loss $\ADS(P)$, where $P$ is in terms of $a$, $b$, and $c$, is the maximum of $\mathcal{V}^1_{\vec u, n}(a,b,c)$ and $\mathcal{V}^3_{\vec u, n}(a,b,c)$:
\begin{align*}
    &  \max \{ (u_2 - u_3)(-2n+a+2b+c -2) + u_1,
    (u_2 - u_3)(-n-a+b+2c)\} \\
    & = (u_2 - u_3)(-n-a+b+2c)
    + \max\{(u_2 - u_3)(-n+2a+b-c-2) + u_1, 0\}.
\end{align*}
Clearly, if $u_2 = u_3$, then we have
\begin{align}
    & \Pr\nolimits_{P \sim \pi^n}(P \in \mathcal{E}_{1,2}) \times \mathbb{E}_{P \sim \pi^n}[\ADS(P)~|~P \in \mathcal{E}_{1,2}] \notag \\
    & = u_1 \sum_{a=0}^{n+1} \sum_{b=0}^{\bceil{\frac{n+1}{2}}-1} \sum_{c=0}^{\bceil{\frac{n}{2}}-1} \mathcal{P}_{\vec{\pi},n}(a,b,c)    \label{eq:three_tieP3_10_base_case}
\end{align}
which is proportional to $\Theta \left( \frac{1}{n} \right)$ if $\pi_3 \leq \pi_4$ and $\pi_1 \leq \pi_5$, and $\calO \left( e^{-\Theta(n)} \right)$ otherwise.
We carry this forward to Step 4, below. Now suppose $u_2 > u_3$.
It is easy to see that $-n +2a +b -c \geq 2 - \frac{u_1}{u_2 - u_3}$, within the already-specified ranges, as long as $a \geq 2 - \frac{u_1}{u_2 - u_3} + \min\{\frac{n}{4}, \frac{n+c-b}{2}\}$. Without loss of generality, since $\Theta(n) + 2 - \frac{u_1}{u_2 - u_3} = \Theta(n)$, we may treat this term as negligible in the subsequent analysis. Therefore, $\Pr\nolimits_{P \sim \pi^n}(P \in \mathcal{E}_{1,2}) \times \mathbb{E}_{P \sim \pi^n}[\ADS(P)~|~P \in \mathcal{E}_{1,2}]$ is the sum of Equations
\begin{align}
    (u_2 - u_3) \sum_{a=0}^{n+1} \sum_{b=0}^{\bceil{\frac{n+1}{2}}-1} \sum_{c=0}^{\bceil{\frac{n}{2}}-1} \mathcal{P}_{\vec{\pi},n}(a,b,c) (-n-a+b+2c) \label{eq:three_tieP3_11}
\end{align}
and
\begin{align}
    (u_2 - u_3) \sum_{b=0}^{\bceil{\frac{n+1}{2}}-1} \sum_{c=0}^{\bceil{\frac{n}{2}}-1} \sum_{a=\bceil{\min\{\frac{n}{4}, \frac{n+c-b}{2}\}}}^{n+1} \mathcal{P}_{\vec{\pi},n}(a,b,c) (-n+2a+b-c). \label{eq:three_tieP3_12}
\end{align}
We first solve Equation (\ref{eq:three_tieP3_11}) using the same techniques as above. By Lemma \ref{lem:bin_theorems_approx}, this is
\begin{align}
    & \Theta \left( \frac{1}{n} \right) \sum_{b=0}^{\bceil{\frac{n+1}{2}}-1} \binom{n+1}{b} \tilde{\pi}_3^b \tilde{\pi}_4^{n+1-b} 
    \sum_{c=0}^{\bceil{\frac{n}{2}}-1}  \binom{n}{c} \tilde{\pi}_1^c \tilde{\pi}_5^{n-c}
    \left(  (-1-\tilde{\pi}_2)n+b+2c \right) \notag \\
    & \quad \quad - \Theta \left( \frac{1}{n} \right) \tilde{\pi}_2 \sum_{a=0}^{n+1} \sum_{b=0}^{\bceil{\frac{n+1}{2}}-1} \sum_{c=0}^{\bceil{\frac{n}{2}}-1} \mathcal{P}_{\vec{\pi},n}(a,b,c)
    \label{eq:three_tieP3_13}
\end{align}
by definition of binomial probability and expectation. 
By Lemma \ref{lem:bin_theorems_approx}, the second term of Equation (\ref{eq:three_tieP3_13}) is proportional to $\Theta \left( \frac{1}{n} \right)$ if $\tilde{\pi}_3 \leq \tilde{\pi}_4$ and $\tilde{\pi}_1 \leq \tilde{\pi}_5$, and $\calO \left( e^{-\Theta(n)} \right)$ otherwise. We carry this finding forward to Step 4, below, and continue with the first term of Equation (\ref{eq:three_tieP3_13}).
This may be simplified as
\begin{align}
    & \Theta \left( \frac{1}{n} \right) \bigg(  (-1-\tilde{\pi}_2)n \sum_{b=0}^{\bceil{\frac{n+1}{2}}-1} \binom{n+1}{b} \tilde{\pi}_3^b \tilde{\pi}_4^{n+1-b} \sum_{c=0}^{\bceil{\frac{n}{2}}-1}  \binom{n}{c} \tilde{\pi}_1^c \tilde{\pi}_5^{n-c} \notag \\
    & + \sum_{b=0}^{\bceil{\frac{n+1}{2}}-1} \binom{n+1}{b} \tilde{\pi}_3^b \tilde{\pi}_4^{n+1-b} b \sum_{c=0}^{\bceil{\frac{n}{2}}-1} \binom{n}{c} \tilde{\pi}_1^c \tilde{\pi}_5^{n-c} \notag \\
    &  + 2 \sum_{b=0}^{\bceil{\frac{n+1}{2}}-1} \binom{n+1}{b} \tilde{\pi}_3^b \tilde{\pi}_4^{n+1-b} \sum_{c=0}^{\bceil{\frac{n}{2}}-1}  \binom{n}{c} \tilde{\pi}_1^c \tilde{\pi}_5^{n-c} c \bigg). \label{eq:three_tieP3_14}
\end{align}
It is clear from Lemma \ref{lem:bin_theorems_approx} that if either $\tilde{\pi}_3 > \tilde{\pi}_4$ or $\tilde{\pi}_1 > \tilde{\pi}_5$, then Equation (\ref{eq:three_tieP3_14}) is $\calO \left( e^{-\Theta(n)} \right)$. This leaves four cases. 
First, if $\tilde{\pi}_3 = \tilde{\pi}_4$ and $\tilde{\pi}_1 = \tilde{\pi}_5$, then Equation (\ref{eq:three_tieP3_14}) is
\begin{align}
    \frac{3}{8} (\pi_6 - \pi_2) \Theta(1) \pm \calO \left( \frac{1}{\sqrt{n}} \right) \label{eq:three_tieP3_15A}
\end{align}
by similar reasoning as we attained Equation (\ref{eq:three_tieP3_9A}). Second, if $\tilde{\pi}_3 = \tilde{\pi}_4$ and $\tilde{\pi}_1 < \tilde{\pi}_5$, then Equation (\ref{eq:three_tieP3_14}) is
\begin{align}
    \frac{3}{4} (3\pi_1 - 2\pi_2 - \pi_5) \Theta(1) \pm \calO \left( \frac{1}{\sqrt{n}} \right) \label{eq:three_tieP3_15B}
\end{align}
by similar reasoning as we attained Equation (\ref{eq:three_tieP3_9B}). 
Third, if $\tilde{\pi}_3 < \tilde{\pi}_5$ and $\tilde{\pi}_1 = \tilde{\pi}_5$, then Equation (\ref{eq:three_tieP3_14}) is
\begin{align}
    \frac{3}{2} (\pi_3 - \pi_1) \Theta(1) \pm \calO \left( \frac{1}{\sqrt{n}} \right) \label{eq:three_tieP3_15C}
\end{align}
by similar reasoning as we attained Equation (\ref{eq:three_tieP3_9C}). 
Finally, if $\tilde{\pi}_3 < \tilde{\pi}_4$ and $\tilde{\pi}_1 < \tilde{\pi}_5$, then Equation (\ref{eq:three_tieP3_11}) is
\begin{align}
    3(2 \pi_1 - \pi_2 - \pi_4) \Theta(1) \pm \calO \left( \frac{1}{\sqrt{n}} \right) \label{eq:three_tieP3_15D}
\end{align}
by similar reasoning as we attained Equation (\ref{eq:three_tieP3_9D}).

Now consider Equation (\ref{eq:three_tieP3_12}). Clearly this equation is lower bounded by zero. Let $\tau = \tilde{\pi}_1 + \tilde{\pi}_6 - \tilde{\pi}_2 - \tilde{\pi}_3$. We prove that as long as $\tilde{\pi}_2 \geq \frac{1}{4}$ and $\tilde{\pi}_1, \tilde{\pi}_3 \leq \frac{1}{2}$, Equation (\ref{eq:three_tieP3_12}) is
\begin{align}
    \begin{cases}
        \Theta(1), & \tau < 0 \\
        \calO \left( \frac{1}{\sqrt{n}} \right), & \tau = 0 \\
        \calO \left( e^{-\Theta(n)} \right), & \tau > 0.
    \end{cases} \label{eq:three_tieP3_16}
\end{align}
Otherwise (i.e., if $\tilde{\pi}_1 > \tilde{\pi}_5$, $\tilde{\pi}_2 > \tilde{\pi}_6$, or $\tilde{\pi}_2 < \frac{1}{4}$), then Equation (\ref{eq:three_tieP3_12}) is $\calO \left( e^{-\Theta(n)} \right)$ by Lemma \ref{lem:bin_theorems_approx}. This is proved using the same method as Equation (\ref{eq:three_tie_13}) was proved in Lemma \ref{lem:three_tie_priority1}, as follows.

We begin by proving the $\tau > 0$ case.
Without loss of generality, let us ignore the $\left( -n+2a+b-c \right)$ factor of Equation (\ref{eq:three_tieP3_12}) and instead focus on the equation
\begin{align}
    & \sum_{b=\bfloor{(\tilde{\pi}_3 - \epsilon_b)n}}^{\bfloor{(\tilde{\pi}_3 + \epsilon_b)n}} \binom{n+1}{b} \tilde{\pi}_3^b \tilde{\pi}_4^{n+1-b} \sum_{c=0}^{\bceil{\frac{n}{2}}-1} \binom{n}{c} \tilde{\pi}_1^c \tilde{\pi}_5^{n-c}
    \sum_{a=\bceil{\min\{\frac{n}{4}, \frac{n+c-b}{2}\}}}^{n+1}  \binom{n+1}{a} \tilde{\pi}_2^a \tilde{\pi}_6^{n+1-a}
    \label{eq:three_tieP3_17}
\end{align}
for some $\epsilon_b \in (0, \min\{\tilde{\pi}_3, \frac{1}{2}-\tilde{\pi}_3\})$. \footnote{If $\tilde{\pi}_3 = 0.5$ then the proof of the $\tau < 0$ case continues as stated with only the lower-bound on the $b$-summation. That is, we sum over $b \in \left[\bfloor{(\tilde{\pi}_3-\epsilon_b) n}, \bceil{\frac{n+1}{2}}-1 \right]$.} 
We note the following observations. Clearly, if any of the $a$-, $b$-, or $c$-summations are exponentially small, then Equation (\ref{eq:three_tieP3_17}) is exponentially small. It follows from Lemma \ref{lem:bin_theorems_approx} that the $b$-summation of Equation (\ref{eq:three_tieP3_17}) is proportional to $\Theta(1)$. By similar reasoning, the $b$-summation with range $b \in [ 0, \bceil{\frac{n+1}{2}}-1] \backslash [(\tilde{\pi}_3 - \epsilon_b)n, (\tilde{\pi}_3 + \epsilon_b)n]$ that is present in Equation (\ref{eq:three_tieP3_12}), but not Equation (\ref{eq:three_tieP3_17}), is exponentially small. Likewise, for any pair $(b,c)$ such that $\frac{n+c-b}{2} - \tilde{\pi}_2 n = \Omega(n)$, it follows that the $a$-summation of Equation (\ref{eq:three_tieP3_17}) is exponentially small. We must identify the ranges of $b$ and $c$ for which this is not the case.

Let $\epsilon_b = \frac{\tilde{\pi}_3 \tau}{4} > 0$. Given that $\tilde{\pi}_2 \in [\frac{1}{4}, 1)$ and $\tilde{\pi}_1, \tilde{\pi}_3 \in (0, \frac{1}{2}]$, we recognize that $\tau \in (-1.5, 1.5]$ which ensures $|\epsilon_b| < \tilde{\pi}_3$. Then we have $\frac{n+c-b}{2} \leq \tilde{\pi}_2 n$ over \emph{any} $b \in [(\tilde{\pi}_3 - \epsilon_b)n, (\tilde{\pi}_3 + \epsilon_b)n]$ as long as
\begin{align*}
    c & \leq n (2\tilde{\pi}_2 -1) +b \\
    & \leq n \left( 2\tilde{\pi}_2 -1 + \tilde{\pi}_3 + \epsilon_b \right) \\
    & = \epsilon_b n + n \left(2 \tilde{\pi}_2 -1 + \tilde{\pi}_3 \right) \\
    & = \frac{\tilde{\pi}_3 \tau n}{4} + (\tilde{\pi}_1 - \tau) n \\
    & = \tilde{\pi}_1 n - \Omega(n).
\end{align*}
Therefore, the $c$-summation of Equation (\ref{eq:three_tieP3_17}) is exponentially small; it cannot be $\Theta(1)$ as long as both the $b$- and $c$- summations are. This proves that Equation (\ref{eq:three_tieP3_12}) is exponentially small when $\tau > 0$.

Now let $\tau = 0$ and consider the equation
\begin{align}
    & \Theta \left( \frac{1}{n} \right) \sum_{b=0}^{\bfloor{\tilde{\pi}_3 n}} \binom{n+1}{b} \tilde{\pi}_3^b \tilde{\pi}_4^{n+1-b} 
    \sum_{c=\bfloor{\tilde{\pi}_1 n}}^{\bceil{\frac{n}{2}}-1}  \binom{n}{c} \tilde{\pi}_1^c \tilde{\pi}_5^{n-c} \notag \\
    & \quad \quad \times \sum_{a=\bceil{\min\{\frac{n}{4}, \frac{n+c-b}{2}\}}}^{n+1} \binom{n+1}{a} \tilde{\pi}_2^a \tilde{\pi}_6^{n+1-a} (-n+2a+b-c)). \label{eq:three_tieP3_18}
\end{align}
Clearly we have 
\begin{align*}
    \frac{n+c-b}{2} & \leq \frac{n+\tilde{\pi}_1 n - \tilde{\pi}_3 n}{2} \\
    & = \tilde{\pi}_2 n
\end{align*}
for all $b$ and $c$ within their respective ranges. Therefore each of the $a$-, $b$-, and $c$-summations of Equation (\ref{eq:three_tieP3_18}) are $\Theta(1)$. It follows that Equation (\ref{eq:three_tieP3_18}) is exponentially small when $b  > \bfloor{\tilde{\pi}_3 n}$ or $c < \bfloor{\tilde{\pi}_1 n}$. It remains to determine precise bounds for the asymptotic rate of Equation (\ref{eq:three_tieP3_18}). 
Specifically, Equation (\ref{eq:three_tieP3_18}) can be written as
\begin{align}
    & \Theta \left( \frac{1}{n} \right) \sum_{b=0}^{\bfloor{\tilde{\pi}_3 n}} \binom{n+1}{b} \tilde{\pi}_3^b \tilde{\pi}_4^{n+1-b} B_b \label{eq:three_tieP3_19}
\end{align}
where we define
\footnote{
The stated proof holds for $\tilde{\pi}_1 < \frac{1}{2}$. If $\tilde{\pi}_1 = \frac{1}{2}$, then we take the $c$-summation (i.e., $B_b$) to span $c \in [0, \bceil{\frac{n}{2}}-1]$. This does not affect our conclusion.
}
\begin{align*}
    B_b = \sum_{c=\bfloor{ \tilde{\pi}_1 n}}^{\bceil{\frac{n}{2}}-1} \binom{n}{c} \tilde{\pi}_1^c \tilde{\pi}_5^{n-c} (C_{b,c} + C'_{b,c})
\end{align*}
with
\begin{align*}
    C_{b,c} & = \sum_{a=\bfloor{\tilde{\pi}_2 n}}^{n+1}  \binom{n+1}{a} \tilde{\pi}_2^a \tilde{\pi}_6^{n+1-a} \left( -n+2a+b-c \right) 
\end{align*}
and
\begin{align*}
    C'_{b,c} & = \sum_{a=\bceil{\min\{\frac{n}{4}, \frac{n+c-b}{2}\}}}^{\bfloor{\tilde{\pi}_2 n}-1} \binom{n+1}{a} \tilde{\pi}_2^a \tilde{\pi}_6^{n+1-a} \left( -n+2a+b-c \right) .
\end{align*}
This may be simplified as 
\begin{align*}
    C_{b,c} & = \left( -n +b -c \right) \left( \frac{1}{2} \pm \calO \left( \frac{1}{\sqrt{n}} \right) \right) + \left( \tilde{\pi}_2 (n+1)  \pm \calO (\sqrt{n}) \right) \\
    & = \frac{(\tilde{\pi}_2 - \tilde{\pi}_6)n +b -c}{2} \pm \calO (\sqrt{n})
\end{align*}
by Lemma \ref{lem:bin_theorems_approx}.
Meanwhile, 
\begin{align*}
    0 & \leq C'_{b,c} \\
    & \leq \left( -n+2\tilde{\pi}_2 n+b-c  \right) \sum_{a=\bceil{\min\{\frac{n}{4}, \frac{n+c-b}{2}\}}}^{\bfloor{\tilde{\pi}_2 n}-1} \binom{n+1}{a} \tilde{\pi}_2^a \tilde{\pi}_6^{n+1-a} \\
    & \leq \left( (\tilde{\pi}_2 - \tilde{\pi}_6) n + b - c \right) \sum_{a=\bfloor{\tilde{\pi}_2 n}}^{n+1}  \binom{n+1}{a} \tilde{\pi}_2^a \tilde{\pi}_6^{n+1-a} \\
    & = \frac{(\tilde{\pi}_2 - \tilde{\pi}_6)n +b -c}{2}  \pm \calO (\sqrt{n}).
\end{align*}
by Lemma \ref{lem:bin_theorems_approx}. Let us set aside $C'_{b,c}$ for the moment and continue with $B_b$ only in terms of $C_{b,c}$. This entails 
\begin{align*}
    B_b & = \left( \frac{(\tilde{\pi}_2 - \tilde{\pi}_6)n +b}{2}  \pm \calO (\sqrt{n}) \right) \left( \frac{1}{2} \pm \calO \left( \frac{1}{\sqrt{n}} \right) \right)  - \left( \frac{\tilde{\pi}_1 n}{4} \pm \calO(\sqrt{n}) \right) \\
    & =  \frac{(\tilde{\pi}_2 - \tilde{\pi}_1 - \tilde{\pi}_6) n +b}{4} \pm \calO (\sqrt{n})
\end{align*}
by Lemma \ref{lem:bin_theorems_approx}. Therefore, Equation (\ref{eq:three_tieP3_19}) is 
\begin{align}
    & \Theta \left( \frac{1}{n} \right) \left( \left( \frac{(\tilde{\pi}_2 - \tilde{\pi}_1 - \tilde{\pi}_6) n}{4} \pm \calO (\sqrt{n}) \right) \left( \frac{1}{2} \pm \calO \left( \frac{1}{\sqrt{n}} \right) \right)
    + \left( \frac{\tilde{\pi}_3 n}{8} \pm \calO(\sqrt{n}) \right) \right) \notag \\
    & = (\tilde{\pi}_2 + \tilde{\pi}_3 - \tilde{\pi}_1 - \tilde{\pi}_6) \Theta(1) \pm \calO \left( \frac{1}{\sqrt{n}} \right) \notag \\
    &  = -\tau \Theta(1) \pm \calO \left( \frac{1}{\sqrt{n}} \right) \notag \\
    & = \calO \left( \frac{1}{\sqrt{n}} \right) \label{eq:three_tieP3_20}
\end{align}
since $\tau=0$ by assumption and the objective is non-negative. Since $C'_{b,c}$ has the same form as $C_{b,c}$, as determined above, it does not affect this conclusion.  
This proves the stated asymptotic rate for Equation (\ref{eq:three_tieP3_12}) when $\tau = 0$.

Finally, consider $\tau < 0$ and the equation
\begin{align}
    & \Theta \left( \frac{1}{n} \right) \sum_{b=0}^{\bfloor{(\tilde{\pi}_3 - \tau) n}} \binom{n+1}{b} \tilde{\pi}_3^b \tilde{\pi}_4^{n+1-b} 
    \sum_{c=\bfloor{(\tilde{\pi}_1+\tau) n - \max\{b - \tilde{\pi}_3 n, 0\}}}^{\bceil{\frac{n}{2}}-1}  \binom{n}{c} \tilde{\pi}_1^c \tilde{\pi}_5^{n-c} \notag \\
    & \quad \quad \times \sum_{a=\bceil{\min\{\frac{n}{4}, \frac{n+c-b}{2}\}}}^{n+1} \binom{n+1}{a} \tilde{\pi}_2^a \tilde{\pi}_6^{n+1-a} (-n+2a+b-c)).
    \label{eq:three_tieP3_21}
\end{align}
When $b = \tilde{\pi}_3 n$ and $c = \tilde{\pi}_1 n$, this entails
\begin{align*}
    \frac{n+c-b}{2} = \frac{n}{2}(1 + \tilde{\pi}_1 - \tilde{\pi}_3) 
    = (2\tilde{\pi}_2 + \tau) n.
\end{align*}
Hence, $\tau$ represents the amount of \emph{slack} that the $c$-summation in Equation (\ref{eq:three_tieP3_21}) has, in terms of $b$ and $c$, before $\frac{n+c-b}{2}$ goes above $ \tilde{\pi}_2 n$ and the $a$-summation becomes exponentially small. This slack can be taken up by as much as $(-\tau n)$ above $\tilde{\pi}_3 n$ in the $a$-summation or $\tau n$ below $\tilde{\pi}_2 n$ in the $b$-summation, as represented by Equation (\ref{eq:three_tieP3_21}).

It is easy to see that Equation (\ref{eq:three_tieP3_19}) is included in Equation (\ref{eq:three_tieP3_21}).
Therefore its asymptotic rate is at least $\tau \Theta(1) \pm \calO \left( \frac{1}{\sqrt{n}} \right) = \Theta(1)$ by Equation (\ref{eq:three_tieP3_20}). Moreover, it is easy to see that Equation (\ref{eq:three_tieP3_21}) is upper-bounded by $\Theta(1)$, following Lemma \ref{lem:bin_theorems_approx}. This proves the stated asymptotic rate for Equation (\ref{eq:three_tieP3_12}) when $\tau < 0$.

This concludes the $\calE_{1,3}$ case of Lemma \ref{lem:three_tie_priority3}.




\end{paragraph}

\begin{paragraph}{Step 4: Putting the pieces together.}
To finish the proof, we tie our results about the $\calE_1$, $\calE_3$, and $\calE_{1,3}$ cases together. 
Consider first the case where $u_2 = u_3$. Then our conclusion is $\Theta \left( \frac{1}{n} \right)$ if $\pi_3 \leq \pi_4$ and $\calO \left( e^{-\Theta(n)} \right)$ otherwise. This follows from Equations (\ref{eq:three_tieP3_3}) and (\ref{eq:three_tieP3_10_base_case}).

When $u_2 > u_3$, this entails the sum of Equations (\ref{eq:three_tieP3_4}), (\ref{eq:three_tieP3_8}), (\ref{eq:three_tieP3_14}), and (\ref{eq:three_tieP3_16}) subject to their respective conditions on the probability distribution $(\pi_1, \ldots, \pi_6)$, as well as the second terms of Equations (\ref{eq:three_tieP3_3}), (\ref{eq:three_tieP3_7}), and (\ref{eq:three_tieP3_13}). Recall that $\Theta(1)$, in many of the simplified versions of equations following (\ref{eq:three_tieP3_4}), (\ref{eq:three_tieP3_8}) and (\ref{eq:three_tieP3_14}), was a stand-in for $\binom{3n+2}{n+1, n+1, n} \frac{n}{3^{3n+2}}$. This enables us to combine several $\Theta(1)$-like terms together.

\begin{table}[t]
    \centering
    \begin{tabular}{|c||c|c|c|}
        \hline
         & $\tilde{\pi}_1 = \tilde{\pi}_5$ & $\tilde{\pi}_1 > \tilde{\pi}_5$ & $\tilde{\pi}_1 < \tilde{\pi}_5$ \\
          \hline \hline
            $\tilde{\pi}_3 = \tilde{\pi}_4$ & 
            \makecell{$\pi_6 - \pi_2$ \\ by Eqns. (\ref{eq:three_tieP3_5A}), (\ref{eq:three_tieP3_9A}), (\ref{eq:three_tieP3_15A})} &  
            \makecell{$\pi_2 - \pi_4$ \\ by Eqns. (\ref{eq:three_tieP3_5C}) and (\ref{eq:three_tieP3_15C})} & 
            \makecell{$3 \pi_1 - 2 \pi_2 - \pi_5$ \\ by Eqn. (\ref{eq:three_tieP3_9B})} \\[0.25em]
          \hline
            $\tilde{\pi}_3 > \tilde{\pi}_4$ & 
            \makecell{$\pi_1 + 2 \pi_3 - 2\pi_2 - \pi_5$ \\ by Eqns. (\ref{eq:three_tieP3_9C}) and (\ref{eq:three_tieP3_15B})} &  
            \makecell{N/A} & 
            \makecell{$2 \pi_1 - \pi_2 - \pi_4$ \\ by Eqn. (\ref{eq:three_tieP3_9D})} \\[0.25em]
          \hline 
            $\tilde{\pi}_3 < \tilde{\pi}_4$ & 
            \makecell{$2 \pi_2 + \pi_3 - 3\pi_4$ \\ by Eqn. (\ref{eq:three_tieP3_5B})} &  
            \makecell{$\pi_1 + \pi_2 - 2 \pi_4$ \\ by Eqn. (\ref{eq:three_tieP3_5D})} & 
            \makecell{$2 \pi_1 - \pi_2 - \pi_4$ \\ by Eqn.  (\ref{eq:three_tieP3_15D})} \\
         \hline
    \end{tabular}
    \caption{Constants in front of $\Theta(1)$ term resulting from Equations (\ref{eq:three_tieP3_4}), (\ref{eq:three_tieP3_8}), (\ref{eq:three_tieP3_14}), for certain conditions on $\pi$.}
    \label{tab:complexity_three_sides_C}
\end{table}

Our conclusion is therefore $\pm \calO \left( e^{-\Theta(n)} \right)$ if $\pi_1 > \pi_5$ and $\pi_3 > \pi_4$. Otherwise, it is
\begin{align*}
    f(\pi_1, \ldots, \pi_6) \Theta(1) \pm \calO \left( \frac{1}{\sqrt{n}} \right) + g_n(\pi_1, \ldots, \pi_6)
\end{align*}
where $f(\pi_1, \ldots, \pi_6)$ is determined by Table \ref{tab:complexity_three_sides_C} and
\begin{align*}
    g_n(\pi_1, \ldots, \pi_6) = \begin{cases}
        \Theta(1), & \pi_1 + \pi_6 < \pi_2 + \pi_3, \\
        \calO \left( \frac{1}{\sqrt{n}} \right), & \pi_1 + \pi_6 = \pi_2 + \pi_3,  \\
        \calO \left( e^{-\Theta(n)} \right), & \pi_1 + \pi_6 > \pi_2 + \pi_3.
    \end{cases}
\end{align*}
%
\end{paragraph}
This concludes the proof of Lemma \ref{lem:three_tie_priority3}.
\end{proof}

%% file: EC_appendix/apx_page_3b.tex
\subsection{Case when alternative $3$ wins}
\label{apx:three_ties_P2}

\begin{table}[t]
    \centering
    \begin{tabular}{|c||c|c|c|}
        \hline
         & $\pi_2 = \pi_6$ & $\pi_2 > \pi_6$ & $\pi_2 < \pi_6$ \\
          \hline \hline
            $\pi_1 = \pi_5$ & 
            $3\pi_4 - 5 \pi_3$ &  
            $\pi_2 - 3 \pi_6 - 2 \pi_1$ & 
            $\pi_2 - \pi_3$ \\[0.25em]
          \hline
            $\pi_1 > \pi_5$ & 
            $\pi_4 - \pi_3$ &  
            $\pi_2 - \pi_6 - 2 \pi_1$ & 
            $\pi_2 + \pi_5 - 2 \pi_1$ \\[0.25em]
          \hline 
            $\pi_1 < \pi_5$ & 
            $\pi_1 - \pi_3$ &  
            $\pi_2 - \pi_6 - 2 \pi_1$ & 
            N/A \\
         \hline
    \end{tabular}
    \caption{Values of $f^3(\pi_1, \ldots \pi_6)$ given conditions on $\pi$ for Lemma \ref{lem:three_tie_priority2}.}
    \label{tab:complexity_three_sides_B_final}
\end{table}

\setcounter{lem}{9}
\begin{lem}
Suppose that $\pi_1 + \pi_5 = \pi_2 + \pi_6 = \pi_3 + \pi_4 = \frac{1}{3}$  and $\pi_i > 0, ~\forall i \in [6].$ Furthermore, let $u_1 \geq u_2 > u_3$ in $\vec{u}$. 
Then $\exists N > 0$ such that $\forall n > N$ where $n-1$ is divisible by $3$,
\[
\EDS(\{1,2, 3\}) = \begin{cases}
    \pm \calO \left( e^{-\Theta(n)} \right),& \pi_1 < \pi_5, \pi_2 < \pi_6 \\
    f^3(\pi_1, \ldots, \pi_6) \Theta(1) \pm \calO \left( \frac{1}{\sqrt{n}} \right), & \text{otherwise}
\end{cases}
\]
where $f^3(\pi_1, \ldots, \pi_6)$ is determined by Table \ref{tab:complexity_three_sides_B_final}.

If $u_1 > u_2 = u_3$, then $\EDS(\{1,2, 3\})$ is $\Theta \left( \frac{1}{n} \right)$ if either $\pi_1 \leq \pi_5$ and $\pi_2 \geq \pi_6$, or $\pi_1 \geq \pi_5$ and $\pi_2 \leq \pi_6$; it is $\pm \calO \left( e^{-\Theta(n)} \right)$ otherwise.

\label{lem:three_tie_priority2}
\end{lem}

\begin{proof}
This lemma is analogous to Lemmas and \ref{lem:three_tie_priority1} \ref{lem:three_tie_priority2} except we have that one fewer than the number of agents is divisible by $3$. To help with notation, we will consider $3n+1$ agents. This covers the case where there are $(n, n, n+1)$ agents truthfully voting for alternatives $1$, $2$, and $3$ respectively. For any preference profile $P$ and truthful vote profile $\topRank{P}$, it follows that $1 \in \EW{\topRank{P}}$ if $R_6 \in P$ and $P[1 \succ 3] \geq P[3 \succ 1]$, whereas $2 \in \EW{\topRank{P}}$ if $R_1 \in P$ and $P[2 \succ 3] \geq P[3 \succ 2]$. This yields three cases for whether either or both of these are the case. We define $\calE_1$, $\calE_2$, and $\calE_{1,2}$ as follows:
\begin{itemize}
    \item $\calE_1 = \{P \in \calL(\calA)^n~:~R_6 \in P \text{ and } P[1 \succ 3] \geq P[3 \succ 1] \text{, and either } R_1 \notin P \text{ or } P[3 \succ 2] > P[2 \succ 3]\}$,
    \item $\calE_2 = \{P \in \calL(\calA)^n~:~R_1 \in P \text{ and } P[2 \succ 3] \geq P[3 \succ 2] \text{, and either } R_6 \notin P \text{ or } P[3 \succ 1] > P[1 \succ 3]\}$,
    \item $\calE_{1,2} = \{P \in \calL(\calA)^n~:~R_6, R_1 \in P \text{ and } P[1 \succ 3] \geq P[3 \succ 1] \text{ and } P[2 \succ 3] \geq P[3 \succ 2]\}$.
\end{itemize}
Implicitly, we note $s_1(\topRank{P}) = s_2(\topRank{P}) = s_3(\topRank{P})+1$ for each of these cases.
The subscript denotes which alternatives (among $1$ and $2$, excluding $3$) appears in the equilibrium winning set $\EW{\topRank{P}}$, for ease of readability.
Let $a \in [0,n+1]$ and $b, c \in [0, n]$. We take throughout this proof:
\begin{itemize}
    \item $n$ agents with rankings either $R_1$ or $R_5$: with $b$ for $R_1$ and $n-b$ for $R_5$,
    \item $n$ agents with rankings either $R_2$ or $R_6$: with $c$ for $R_2$ and $n-c$ for $R_6$,
    \item $n+1$ agents with rankings either $R_3$ or $R_4$: with $a$ for $R_3$ and $n+1-a$ for $R_4$.
\end{itemize}

\begin{paragraph}{Step 1: Characterize the $\calE_1$ case.}
We have $P \in \calE_1$ if the following ranges are satisfied. First, $a \in [0, n+1]$ has its full range. Second, $b \leq \frac{n}{2}$, so that there are at least as many agents preferring $R_5 = (1 \succ 3 \succ 2)$ than $R_1 = (1 \succ 2 \succ 3)$, which entails $2 \notin \EW{\topRank{P}}$. Third, $c > \frac{n}{2}$, so that there are more agents preferring $R_6 = (2 \succ 1 \succ 3)$ than $R_2 = (2 \succ 3 \succ 1)$, which entails $1 \in \EW{\topRank{P}}$. Like in Lemma \ref{lem:sub_13}, the (negated) value per agent and probability of each ranking is summarized by Table \ref{tab:character_profiles_E3}.
Put together, we get the equation
\begin{align}
    & \Pr\nolimits_{P \sim \pi^n}(P \in \mathcal{E}_2) \times \mathbb{E}_{P \sim \pi^n}[\ADS(P)~|~P \in \mathcal{E}_2] \notag \\
    & = \sum_{a=0}^{n+1} \sum_{b=0}^{\bfloor{\frac{n}{2}}} \sum_{c=\bfloor{\frac{n}{2}}+1}^{n} \mathcal{P}_{\vec{\pi},n}(a,b,c) \cdot \mathcal{V}^1_{\vec{u},n}(a,b,c) \label{eq:three_tieP2_1}
\end{align}
where we define
\begin{align*}
    \mathcal{P}_{\vec{\pi},n}(a,b,c) & = \binom{3n+1}{a, b, c, n-c, n+1-a, n-b} \pi_3^a \pi_1^b \pi_2^{c} \pi_6^{n-c} \pi_4^{n+1-a} \pi_5^{n-b} \\
    & = \binom{3n+1}{n+1, n, n} \frac{1}{3^{3n+1}}   \times  \binom{n+1}{a} \tilde{\pi}_3^a \tilde{\pi}_4^{n+1-a} \times \binom{n}{b} \tilde{\pi}_1^b \tilde{\pi}_5^{n-b} \times \binom{n}{c} \tilde{\pi}_2^c \tilde{\pi}_6^{n-c}
\end{align*}
and 
\begin{align*}
    \mathcal{V}^1_{\vec{u},n}(a, b, c) = & \begin{pmatrix}
        a, & b, & c, & n-c, & n+1-a, & n-b \end{pmatrix} \\
    & \cdot \begin{pmatrix}
        u_1 - u_3,& -u_1 + u_3,& u_2 - u_3,& -u_2 + u_3,& u_1 - u_2,& -u_1 + u_2 \end{pmatrix} \\
    & = (u_2 - u_3) \left( -n-a-b+2c \right) + (u_1 - u_2).
\end{align*}
This equation uses the following definitions:
\begin{itemize}
    \item $\tilde{\pi}_1 = \frac{\pi_1}{\pi_1+\pi_5}$ and $\tilde{\pi}_5 = \frac{\pi_5}{\pi_1+\pi_5}$,
    \item $\tilde{\pi}_2 = \frac{\pi_2}{\pi_2+\pi_6}$ and $\tilde{\pi}_6 = \frac{\pi_6}{\pi_2+\pi_6}$,
    \item $\tilde{\pi}_3 = \frac{\pi_3}{\pi_3+\pi_4}$ and $\tilde{\pi}_4 = \frac{\pi_4}{\pi_3+\pi_4}$
\end{itemize}
which we recall from Definition \ref{dfn:tilde_definitions}, where the denominators are each $\frac{1}{3}$. 
Lemma \ref{lem:stirling_trinom} suggests that $\binom{3n+1}{n+1, n, n} \frac{n}{3^{3n+1}} = \Theta \left( \frac{1}{n} \right)$. Equation (\ref{eq:three_tieP2_1}) may therefore be written as
\begin{align}
    & \Theta \left( \frac{1}{n} \right) \sum_{c=\bfloor{\frac{n}{2}}+1}^{n}  \binom{n}{c} \tilde{\pi}_2^c \tilde{\pi}_6^{n-c} \sum_{b=0}^{\bfloor{\frac{n}{2}}} \binom{n}{b} \tilde{\pi}_1^b \tilde{\pi}_5^{n-b}  \notag \\
    & \quad \quad \times \sum_{a=0}^{n+1} \binom{n+1}{a} \tilde{\pi}_3^a \tilde{\pi}_4^{n+1-a}
    \left( (u_2 - u_3)(-n-a-b+2c) + (u_1 - u_2) \right) \notag \\
    & = (u_2 - u_3) \Theta \left( \frac{1}{n} \right) \sum_{c=\bfloor{\frac{n}{2}}+1}^{n}  \binom{n}{c} \tilde{\pi}_2^c \tilde{\pi}_6^{n-c} \sum_{b=0}^{\bfloor{\frac{n}{2}}} \binom{n}{b} \tilde{\pi}_1^b \tilde{\pi}_5^{n-b}
    \left(  (-1-\tilde{\pi}_3)n-b+2c \right) \notag \\
    & + (- \tilde{\pi}_3 (u_2 - u_3) + u_1 - u_2) \Theta \left( \frac{1}{n} \right)
    \sum_{c=\bfloor{\frac{n}{2}}+1}^{n}  \binom{n}{c} \tilde{\pi}_2^c \tilde{\pi}_6^{n-c} \sum_{b=0}^{\bfloor{\frac{n}{2}}} \binom{n}{b} \tilde{\pi}_1^b \tilde{\pi}_5^{n-b} \label{eq:three_tieP2_3}
\end{align}
by definition of binomial probability and expectation. 
By Lemma \ref{lem:bin_theorems_approx}, the second term of Equation (\ref{eq:three_tieP2_3}) is proportional to $\Theta \left( \frac{1}{n} \right)$ if $\pi_2 \geq \pi_6$ and $\pi_1 \leq \pi_5$, and $\calO \left( e^{-\Theta(n)} \right)$ otherwise. We carry this finding forward to Step 4, below, and continue with the first term of Equation (\ref{eq:three_tieP2_3}) assuming that $u_2 - u_3 > 0$.

We solve this part using the same techniques as with case $\calE_2$ of Lemma \ref{lem:three_tie_priority2}. The first term of Equation (\ref{eq:three_tieP2_3}) may be simplified as
\begin{align}
    & \Theta \left( \frac{1}{n} \right) \bigg(  (-1-\tilde{\pi}_3)n \sum_{c=\bfloor{\frac{n}{2}}+1}^{n}  \binom{n}{c} \tilde{\pi}_2^c \tilde{\pi}_6^{n-c} \sum_{b=0}^{\bfloor{\frac{n}{2}}} \binom{n}{b} \tilde{\pi}_1^b \tilde{\pi}_5^{n-b} \notag \\
    & + 2 \sum_{c=\bfloor{\frac{n}{2}}+1}^{n}  \binom{n}{c} \tilde{\pi}_2^c \tilde{\pi}_6^{n-c} c  \sum_{b=0}^{\bfloor{\frac{n}{2}}} \binom{n}{b} \tilde{\pi}_1^b \tilde{\pi}_5^{n-b} \notag \\
    &  - \sum_{c=\bfloor{\frac{n}{2}}+1}^{n}  \binom{n}{c} \tilde{\pi}_2^c \tilde{\pi}_6^{n-c} \sum_{b=0}^{\bfloor{\frac{n}{2}}} \binom{n}{b} \tilde{\pi}_1^b \tilde{\pi}_5^{n-b} b \bigg). \label{eq:three_tieP2_4}
\end{align}
It is clear from Lemma \ref{lem:bin_theorems_approx} that if either $\tilde{\pi}_2 < \tilde{\pi}_6$ or $\tilde{\pi}_1 > \tilde{\pi}_5$, then Equation (\ref{eq:three_tieP2_4}) is $\calO \left( e^{-\Theta(n)} \right)$. This leaves four cases. First, if $\tilde{\pi}_2 = \tilde{\pi}_6$ and $\tilde{\pi}_1 = \tilde{\pi}_5$, then Equation (\ref{eq:three_tieP2_4}) is
\begin{align}
    & \Theta \left( \frac{1}{n} \right) \bigg( (-1-\tilde{\pi}_3)n \left( \frac{1}{2} \pm \calO \left( \frac{1}{\sqrt{n}} \right)  \right) \left( \frac{1}{2} \pm \calO \left( \frac{1}{\sqrt{n}} \right)  \right) \notag \\
    & \quad \quad + 2 \left( \frac{n}{4} \pm \calO (\sqrt{n})  \right) \left( \frac{1}{2} \pm \calO \left( \frac{1}{\sqrt{n}} \right)  \right) \notag \\
    & \quad \quad - \left( \frac{1}{2} \pm \calO \left( \frac{1}{\sqrt{n}} \right)  \right) \left( \frac{n}{4} \pm \calO (\sqrt{n})  \right)
    \bigg) \notag \\
    & =  \Theta \left( \frac{1}{n} \right) \left( n \left( \frac{-1-\tilde{\pi}_3}{4} + \frac{1}{4} - \frac{1}{8} \right) \pm \calO(\sqrt{n}) \right) \notag \\
    & = \frac{3}{8} (\pi_4 - 3 \pi_3) \Theta(1) \pm \calO \left( \frac{1}{\sqrt{n}} \right) \label{eq:three_tieP2_5A}
\end{align}
making use of the fact that $\pi_3 + \pi_4 = \frac{1}{3}$.
Second, if $\tilde{\pi}_2 = \tilde{\pi}_6$ and $\tilde{\pi}_1 < \tilde{\pi}_5$, then Equation (\ref{eq:three_tieP2_4}) is
\begin{align}
    & \Theta \left( \frac{1}{n} \right) \bigg( (-1-\tilde{\pi}_3)n \left( \frac{1}{2} \pm \calO \left( \frac{1}{\sqrt{n}} \right)  \right) \left( 1 - \calO \left( e^{-\Theta(n)} \right)  \right) \notag \\
    & \quad \quad + 2 \left( \frac{n}{4} \pm \calO (\sqrt{n})  \right) \left( 1 - \calO \left( e^{-\Theta(n)} \right)  \right) \notag \\
    & \quad \quad - \left( \frac{1}{2} \pm \calO \left( \frac{1}{\sqrt{n}} \right)  \right) \tilde{\pi}_1 n \left(  1 - \calO \left( e^{-\Theta(n)} \right) \right)
    \bigg) \notag \\
    & = \Theta \left( \frac{1}{n} \right) \left( n \left( \frac{-1-\tilde{\pi}_3}{2} + \frac{1}{2} - \frac{\tilde{\pi}_1}{2} \right) \pm \calO(\sqrt{n}) \right) \notag \\
    & = \frac{3}{2} (\pi_1 - \pi_3) \Theta(1) \pm \calO \left( \frac{1}{\sqrt{n}} \right). \label{eq:three_tieP2_5B}
\end{align}
Third, if $\tilde{\pi}_2 > \tilde{\pi}_6$ and $\tilde{\pi}_1 = \tilde{\pi}_5$, then Equation (\ref{eq:three_tieP2_4}) is
\begin{align}
    & \Theta \left( \frac{1}{n} \right) \bigg( (-1-\tilde{\pi}_1)n \left( 1 - \calO \left( e^{-\Theta(n)} \right) \right) \left( \frac{1}{2} \pm \calO \left( \frac{1}{\sqrt{n}} \right)  \right) \notag \\
    & \quad \quad + 2 \tilde{\pi}_2 n \left( 1 - \calO \left( e^{-\Theta(n)} \right)  \right) \left( \frac{1}{2} \pm \calO \left( \frac{1}{\sqrt{n}} \right)  \right) \notag \\
    & \quad \quad - \left( 1 - \calO \left( e^{-\Theta(n)} \right)  \right) \left( \frac{n}{4} \pm \calO (\sqrt{n})  \right)
    \bigg) \notag \\
    & = \Theta \left( \frac{1}{n} \right) \left( n \left( \frac{-1-\tilde{\pi}_1}{2} + \tilde{\pi}_2 - \frac{1}{4} \right) \pm \calO(\sqrt{n}) \right) \notag \\
    & = \frac{3}{4} (\pi_2 -3 \pi_6 - 2 \pi_1) \Theta(1) \pm \calO \left( \frac{1}{\sqrt{n}} \right). \label{eq:three_tieP2_5C}
\end{align}
Finally, if $\tilde{\pi}_2 > \tilde{\pi}_6$ and $\tilde{\pi}_1 < \tilde{\pi}_5$, then Equation (\ref{eq:three_tieP2_4}) is
\begin{align}
    & \Theta \left( \frac{1}{n} \right) \bigg( (-1-\tilde{\pi}_1) \left( 1 - \calO \left( e^{-\Theta(n)} \right) \right) \left(1 - \calO \left( e^{-\Theta(n)} \right)  \right) \notag \\
    & \quad \quad + 2 \tilde{\pi}_2 n \left( 1 - \calO \left( e^{-\Theta(n)} \right)  \right) \left( 1 - \calO \left( e^{-\Theta(n)} \right)  \right) \notag \\
    & \quad \quad - \left( 1 - \calO \left( e^{-\Theta(n)} \right)  \right) \tilde{\pi}_1 n \left( 1 - \calO \left( e^{-\Theta(n)} \right) \right)
    \bigg) \notag \\
    & = \Theta \left( \frac{1}{n} \right) \left( n \left( -1-\tilde{\pi}_1 + 2 \tilde{\pi}_2 - \tilde{\pi}_1 \right) \pm \calO(\sqrt{n}) \right) \notag \\
    & = 3 (\pi_2 - \pi_6 - 2 \pi_1) \Theta(1) \pm \calO \left( \frac{1}{\sqrt{n}} \right). \label{eq:three_tieP2_5D}
\end{align}
%
%
Recall that each $\Theta(1)$ in Equations (\ref{eq:three_tieP2_5A}), (\ref{eq:three_tieP2_5B}), (\ref{eq:three_tieP2_5C}), and (\ref{eq:three_tieP2_5D})
is actually an instance of $\binom{3n+1}{n+1, n, n} \frac{n}{3^{3n+1}}$, following Lemma \ref{lem:stirling_trinom}.
We make use of this fact and continue with these equations in Step 4, below.
This concludes the $\calE_1$ case of Lemma \ref{lem:three_tie_priority2}.
\end{paragraph}

\begin{paragraph}{Step 2: Characterize the $\calE_2$ case.}
We prove this case in the same way as Step 1 ($\calE_1$) above, keeping the same variable nomenclature but adjusting the ranges as needed. That is, we have $P \in \calE_2$ if the following ranges are satisfied. First, $a \in [0, n+1]$ has its full range. Second, $b > \frac{n}{2}$, so that there are more agents preferring $R_1 = (1 \succ 2 \succ 3)$ than $R_5 = (1 \succ 3 \succ 2)$, which entails $2 \in \EW{\topRank{P}}$. Third, $c \leq \frac{n}{2}$, so that there are at least as many agents preferring $R_6 = (2 \succ 1 \succ 3)$ than $R_2 = (2 \succ 3 \succ 1)$, which entails $1 \notin \EW{\topRank{P}}$. Like in Lemma \ref{lem:sub_23}, the (negated) value per agent and probability of each ranking is summarized by Table \ref{tab:character_profiles_E5}.
Put together, we get the equation
\begin{align}
    & \Pr\nolimits_{P \sim \pi^n}(P \in \mathcal{E}_3) \times \mathbb{E}_{P \sim \pi^n}[\ADS(P)~|~P \in \mathcal{E}_3] \notag \\
    & = \sum_{a=0}^{n+1} \sum_{b=\bfloor{\frac{n}{2}}+1}^{n} \sum_{c=0}^{\bfloor{\frac{n}{2}}} \mathcal{P}_{\vec{\pi},n}(a,b,c) \cdot \mathcal{V}^2_{\vec{u},n}(a,b,c) \label{eq:three_tieP2_6}
\end{align}
where $\mathcal{P}_{\vec{\pi},n}$ is the same as in Step 1, and 
\begin{align*}
    \mathcal{V}^2_{\vec{u},n}(a, b, c) = & \begin{pmatrix}
        a, & b, & c, & n-c, & n+1-a, & n-b \end{pmatrix} \\
    & \cdot \begin{pmatrix}
        u_1 - u_2,& -u_2 + u_3,& -u_1 + u_2,& -u_1 + u_3,& u_1 - u_3,& u_2 - u_3 \end{pmatrix} \\
    & = (u_2 - u_3) \left( n -a -2b +c \right) + (u_1 - u_3).
\end{align*}
By Lemma \ref{lem:stirling_trinom}, this leads to
\begin{align}
    & \Theta \left( \frac{1}{n} \right) \sum_{b=\bfloor{\frac{n}{2}}+1}^{n} \binom{n}{b} \tilde{\pi}_1^b \tilde{\pi}_5^{n-b} \sum_{c=0}^{\bfloor{\frac{n}{2}}}  \binom{n}{c} \tilde{\pi}_2^c \tilde{\pi}_6^{n-c}  \notag \\
    & \quad \quad \times \sum_{a=0}^{n+1} \binom{n+1}{a} \tilde{\pi}_3^a \tilde{\pi}_4^{n+1-a}
    \left( (u_2 - u_3)(n-a-2b+c) + (u_1 - u_3) \right) \notag \\
    & = (u_2 - u_3) \Theta \left( \frac{1}{n} \right) \sum_{b=\bfloor{\frac{n}{2}}+1}^{n} \binom{n}{b} \tilde{\pi}_1^b \tilde{\pi}_5^{n-b} \sum_{c=0}^{\bfloor{\frac{n}{2}}}  \binom{n}{c} \tilde{\pi}_2^c \tilde{\pi}_6^{n-c} 
    \left(  (1-\tilde{\pi}_3)n-2b+c \right) \notag \\
    & + (- \tilde{\pi}_3 (u_2 - u_3) + u_1 - u_3) \Theta \left( \frac{1}{n} \right) 
    \sum_{b=\bfloor{\frac{n}{2}}+1}^{n} \binom{n}{b} \tilde{\pi}_1^b \tilde{\pi}_5^{n-b} \sum_{c=0}^{\bfloor{\frac{n}{2}}}  \binom{n}{c} \tilde{\pi}_2^c \tilde{\pi}_6^{n-c} \label{eq:three_tieP2_7}
\end{align}
by definition of binomial probability and expectation. 
By Lemma \ref{lem:bin_theorems_approx}, the second term of Equation (\ref{eq:three_tieP2_7}) is proportional to $\Theta \left( \frac{1}{n} \right)$ if $\pi_1 \geq \pi_5$ and $\pi_2 \leq \pi_6$, and $\calO \left( e^{-\Theta(n)} \right)$ otherwise. We carry this finding forward to Step 4, below, and continue with the first term of Equation (\ref{eq:three_tieP2_6}) assuming that $u_2 - u_3 > 0$.

The first term of Equation (\ref{eq:three_tieP2_7}) may be simplified as
\begin{align}
    & \Theta \left( \frac{1}{n} \right) \bigg(  (1-\tilde{\pi}_3)n \sum_{b=\bfloor{\frac{n}{2}}+1}^{n} \binom{n}{b} \tilde{\pi}_1^b \tilde{\pi}_5^{n-b} \sum_{c=0}^{\bfloor{\frac{n}{2}}}  \binom{n}{c} \tilde{\pi}_2^c \tilde{\pi}_6^{n-c} \notag \\
    & - 2 \sum_{b=\bfloor{\frac{n}{2}}+1}^{n} \binom{n}{b} \tilde{\pi}_1^b \tilde{\pi}_5^{n-b} b \sum_{c=0}^{\bfloor{\frac{n}{2}}} \binom{n}{c} \tilde{\pi}_2^c \tilde{\pi}_6^{n-c} \notag \\
    &  + \sum_{b=\bfloor{\frac{n}{2}}+1}^{n} \binom{n}{b} \tilde{\pi}_1^b \tilde{\pi}_5^{n-b} \sum_{c=0}^{\bfloor{\frac{n}{2}}}  \binom{n}{c} \tilde{\pi}_2^c \tilde{\pi}_6^{n-c} c \bigg). \label{eq:three_tieP2_8}
\end{align}
It is clear from Lemma \ref{lem:bin_theorems_approx} that if either $\tilde{\pi}_1 < \tilde{\pi}_5$ or $\tilde{\pi}_2 > \tilde{\pi}_6$, then Equation (\ref{eq:three_tieP2_8}) is $\calO \left( e^{-\Theta(n)} \right)$. This leaves four cases. First, if $\tilde{\pi}_1 = \tilde{\pi}_5$ and $\tilde{\pi}_2 = \tilde{\pi}_6$, then Equation (\ref{eq:three_tieP2_8}) is
\begin{align}
    & \Theta \left( \frac{1}{n} \right) \bigg( (1-\tilde{\pi}_3)n \left( \frac{1}{2} \pm \calO \left( \frac{1}{\sqrt{n}} \right)  \right) \left( \frac{1}{2} \pm \calO \left( \frac{1}{\sqrt{n}} \right)  \right) \notag \\
    & \quad \quad - 2 \left( \frac{n}{4} \pm \calO (\sqrt{n})  \right) \left( \frac{1}{2} \pm \calO \left( \frac{1}{\sqrt{n}} \right)  \right) \notag \\
    & \quad \quad + \left( \frac{1}{2} \pm \calO \left( \frac{1}{\sqrt{n}} \right)  \right) \left( \frac{n}{4} \pm \calO (\sqrt{n})  \right)
    \bigg) \notag \\
    & =  \Theta \left( \frac{1}{n} \right) \left( n \left( \frac{1 - \tilde{\pi}_3}{4} - \frac{1}{4} + \frac{1}{8} \right) \pm \calO(\sqrt{n}) \right) \notag \\
    & = \frac{3}{8} (\pi_4 - \pi_3) \Theta(1) \pm \calO \left( \frac{1}{\sqrt{n}} \right) \label{eq:three_tieP2_9A}
\end{align}
Second, if $\tilde{\pi}_1 = \tilde{\pi}_5$ and $\tilde{\pi}_2 < \tilde{\pi}_6$, then Equation (\ref{eq:three_tieP2_8}) is
\begin{align}
    & \Theta \left( \frac{1}{n} \right) \bigg( (1-\tilde{\pi}_3)n \left( \frac{1}{2} \pm \calO \left( \frac{1}{\sqrt{n}} \right)  \right) \left( 1 - \calO \left( e^{-\Theta(n)} \right)  \right) \notag \\
    & \quad \quad - 2 \left( \frac{n}{4} \pm \calO (\sqrt{n})  \right) \left( 1 - \calO \left( e^{-\Theta(n)} \right)  \right) \notag \\
    & \quad \quad + \left( \frac{1}{2} \pm \calO \left( \frac{1}{\sqrt{n}} \right)  \right) \tilde{\pi}_2 n \left(  1 - \calO \left( e^{-\Theta(n)} \right) \right)
    \bigg) \notag \\
    & = \Theta \left( \frac{1}{n} \right) \left( n \left( \frac{1-\tilde{\pi}_3}{2} - \frac{1}{2} + \frac{\tilde{\pi}_2}{2} \right) \pm \calO(\sqrt{n}) \right) \notag \\
    & = \frac{3}{2} (\pi_2 - \pi_3) \Theta(1) \pm \calO \left( \frac{1}{\sqrt{n}} \right). \label{eq:three_tieP2_9B}
\end{align}
Third, if $\tilde{\pi}_1 > \tilde{\pi}_5$ and $\tilde{\pi}_2 = \tilde{\pi}_6$, then Equation (\ref{eq:three_tieP2_8}) is
\begin{align}
    & \Theta \left( \frac{1}{n} \right) \bigg( (1-\tilde{\pi}_1)n \left( 1 - \calO \left( e^{-\Theta(n)} \right) \right) \left( \frac{1}{2} \pm \calO \left( \frac{1}{\sqrt{n}} \right)  \right) \notag \\
    & \quad \quad - 2 \tilde{\pi}_1 n \left( 1 - \calO \left( e^{-\Theta(n)} \right)  \right) \left( \frac{1}{2} \pm \calO \left( \frac{1}{\sqrt{n}} \right)  \right) \notag \\
    & \quad \quad + \left( 1 - \calO \left( e^{-\Theta(n)} \right)  \right) \left( \frac{n}{4} \pm \calO (\sqrt{n})  \right)
    \bigg) \notag \\
    & = \Theta \left( \frac{1}{n} \right) \left( n \left( \frac{1 - \tilde{\pi}_1}{2} - \tilde{\pi}_1 + \frac{1}{4} \right) \pm \calO(\sqrt{n}) \right) \notag \\
    & = \frac{3}{4} (\pi_5 - \pi_1) \Theta(1) \pm \calO \left( \frac{1}{\sqrt{n}} \right). \label{eq:three_tieP2_9C}
\end{align}
Finally, if $\tilde{\pi}_1 > \tilde{\pi}_5$ and $\tilde{\pi}_2 < \tilde{\pi}_6$, then Equation (\ref{eq:three_tieP2_8}) is
\begin{align}
    & \Theta \left( \frac{1}{n} \right) \bigg( (1-\tilde{\pi}_1) \left( 1 - \calO \left( e^{-\Theta(n)} \right) \right) \left(1 - \calO \left( e^{-\Theta(n)} \right)  \right) \notag \\
    & \quad \quad - 2 \tilde{\pi}_1 n \left( 1 - \calO \left( e^{-\Theta(n)} \right)  \right) \left( 1 - \calO \left( e^{-\Theta(n)} \right)  \right) \notag \\
    & \quad \quad + \left( 1 - \calO \left( e^{-\Theta(n)} \right)  \right) \tilde{\pi}_2 n \left( 1 - \calO \left( e^{-\Theta(n)} \right) \right)
    \bigg) \notag \\
    & = \Theta \left( \frac{1}{n} \right) \left( n \left( 1 - \tilde{\pi}_1 - 2 \tilde{\pi}_1 + \tilde{\pi}_2  \right) \pm \calO(\sqrt{n}) \right) \notag \\
    & = 3 (\pi_2 + \pi_5 - 2 \pi_1) \Theta(1) \pm \calO \left( \frac{1}{\sqrt{n}} \right). \label{eq:three_tieP2_9D}
\end{align}
As with the $\calE_1$ case above, each $\Theta(1)$ in  Equations (\ref{eq:three_tieP2_9A}), (\ref{eq:three_tieP2_9B}), (\ref{eq:three_tieP2_9C}), and (\ref{eq:three_tieP2_9D})
is actually an instance of $\binom{3n+1}{n+1, n, n} \frac{n}{3^{3n+1}}$.
We continue with these equations in Step 4, below. This concludes the $\calE_2$ case of Lemma \ref{lem:three_tie_priority2}.
\end{paragraph}

\begin{paragraph}{Step 3: Characterize the $\calE_{1,2}$ case.}
We keep the same variable nomenclature as the above steps, but adjust the ranges as needed. That is, we have $P \in \calE_{1,2}$ if the following ranges are satisfied. 
First, $a \in [0, n+1]$ has its full range. Second, $b > \frac{n}{2}$, so that there are more agents preferring $R_1 = (1 \succ 2 \succ 3)$ than $R_5 = (1 \succ 3 \succ 2)$, which entails $2 \in \EW{\topRank{P}}$. Third, $c > \frac{n}{2}$, so that there are more agents preferring $R_6 = (2 \succ 1 \succ 3)$ than $R_2 = (2 \succ 3 \succ 1)$, which entails $1 \in \EW{\topRank{P}}$.

Since $|\EW{\topRank{P}}| = 2$ for this case, the adversarial loss $\ADS(P)$, where $P$ is in terms of $a$, $b$, and $c$, is the maximum of $\mathcal{V}^1_{\vec u, n}(a,b,c)$ and $\mathcal{V}^2_{\vec u, n}(a,b,c)$:
\begin{align*}
    &  \max \{ (u_2 - u_3)(-n-a-b+2c) + (u_1 - u_2), \\
    & \quad \quad (u_2 - u_3)(n-a-2b+c) + (u_1 - u_3) \}.
\end{align*}
It is easy to see that $\mathcal{V}^1_{\vec u, n}(a,b,c) \geq \mathcal{V}^2_{\vec u, n}(a,b,c)$ along the aforementioned ranges of $a$, $b$, and $c$. Therefore
\begin{align}
    & \Pr\nolimits_{P \sim \pi^n}(P \in \mathcal{E}_{1,2}) \times \mathbb{E}_{P \sim \pi^n}[\ADS(P)~|~P \in \mathcal{E}_{1,2}] \notag \\
    & = \sum_{a=0}^{n+1} \sum_{b=\bfloor{\frac{n}{2}}+1}^{n} \sum_{c=\bfloor{\frac{n}{2}}+1}^{n} \mathcal{P}_{\vec{\pi},n}(a,b,c) \cdot \mathcal{V}^1_{\vec{u},n}(a,b,c) \notag \\
    & = (u_2 - u_3) \Theta \left( \frac{1}{n} \right) \sum_{c=\bfloor{\frac{n}{2}}+1}^{n}  \binom{n}{c} \tilde{\pi}_2^c \tilde{\pi}_6^{n-c} \sum_{b=\bfloor{\frac{n}{2}}+1}^{n} \binom{n}{b} \tilde{\pi}_1^b \tilde{\pi}_5^{n-b}
    \left(  -(1+\tilde{\pi}_3)n-b+2c \right) \notag \\
    & + (- \tilde{\pi}_3 (u_2 - u_3) + u_1 - u_2) \Theta \left( \frac{1}{n} \right)
    \sum_{c=\bfloor{\frac{n}{2}}+1}^{n}  \binom{n}{c} \tilde{\pi}_2^c \tilde{\pi}_6^{n-c} \sum_{b=\bfloor{\frac{n}{2}}+1}^{n} \binom{n}{b} \tilde{\pi}_1^b \tilde{\pi}_5^{n-b}  
    \label{eq:three_tieP2_10}
\end{align}
by definition of binomial probability and expectation, just like with Equation (\ref{eq:three_tieP2_3}) above. 
By Lemma \ref{lem:bin_theorems_approx}, the second term of Equation (\ref{eq:three_tieP2_10}) is proportional to $\Theta \left( \frac{1}{n} \right)$ if $\pi_1 \geq \pi_5$ and $\pi_2 \geq \pi_6$, and $\calO \left( e^{-\Theta(n)} \right)$ otherwise. We carry this finding forward to Step 4, below, and continue with the first term of Equation (\ref{eq:three_tieP2_10}) assuming that $u_2 - u_3 > 0$.
This may be simplified as
\begin{align}
    & \Theta \left( \frac{1}{n} \right) \bigg(  (-1-\tilde{\pi}_3)n \sum_{c=\bfloor{\frac{n}{2}}+1}^{n}  \binom{n}{c} \tilde{\pi}_2^c \tilde{\pi}_6^{n-c} \sum_{b=\bfloor{\frac{n}{2}}+1}^{n} \binom{n}{b} \tilde{\pi}_1^b \tilde{\pi}_5^{n-b} \notag \\
    & + 2 \sum_{c=\bfloor{\frac{n}{2}}+1}^{n}  \binom{n}{c} \tilde{\pi}_2^c \tilde{\pi}_6^{n-c} c  \sum_{b=\bfloor{\frac{n}{2}}+1}^{n} \binom{n}{b} \tilde{\pi}_1^b \tilde{\pi}_5^{n-b} \notag \\
    &  - \sum_{c=\bfloor{\frac{n}{2}}+1}^{n}  \binom{n}{c} \tilde{\pi}_2^c \tilde{\pi}_6^{n-c} \sum_{b=\bfloor{\frac{n}{2}}+1}^{n} \binom{n}{b} \tilde{\pi}_1^b \tilde{\pi}_5^{n-b} b \bigg). \label{eq:three_tieP2_11}
\end{align}
It is clear from Lemma \ref{lem:bin_theorems_approx} that if either $\tilde{\pi}_2 < \tilde{\pi}_6$ or $\tilde{\pi}_1 < \tilde{\pi}_5$, then Equation (\ref{eq:three_tieP2_11}) is $\calO \left( e^{-\Theta(n)} \right)$. This leaves four cases. 
First, if $\tilde{\pi}_2 = \tilde{\pi}_6$ and $\tilde{\pi}_1 = \tilde{\pi}_5$, then Equation (\ref{eq:three_tieP2_11}) is
\begin{align}
    \frac{3}{8} (\pi_4 - 3\pi_3) \Theta(1) \pm \calO \left( \frac{1}{\sqrt{n}} \right) \label{eq:three_tieP2_12A}
\end{align}
by similar reasoning as we attained Equation (\ref{eq:three_tieP2_9A}). Second, if $\tilde{\pi}_2 = \tilde{\pi}_6$ and $\tilde{\pi}_1 > \tilde{\pi}_5$, then Equation (\ref{eq:three_tieP2_11}) is
\begin{align}
    \frac{3}{2} (\pi_1 - \pi_3) \Theta(1) \pm \calO \left( \frac{1}{\sqrt{n}} \right) \label{eq:three_tieP2_12B}
\end{align}
by similar reasoning as we attained Equation (\ref{eq:three_tieP2_9B}). 
Third, if $\tilde{\pi}_2 > \tilde{\pi}_6$ and $\tilde{\pi}_1 = \tilde{\pi}_5$, then Equation (\ref{eq:three_tieP2_11}) is
\begin{align}
    \frac{3}{4} (\pi_2 -3\pi_6 - 2\pi_1) \Theta(1) \pm \calO \left( \frac{1}{\sqrt{n}} \right) \label{eq:three_tieP2_12C}
\end{align}
by similar reasoning as we attained Equation (\ref{eq:three_tieP2_9C}). 
Finally, if $\tilde{\pi}_2 > \tilde{\pi}_6$ and $\tilde{\pi}_1 > \tilde{\pi}_5$, then Equation (\ref{eq:three_tieP2_11}) is
\begin{align}
    3 (\pi_2 - \pi_6 - 2 \pi_1) \Theta(1) \pm \calO \left( \frac{1}{\sqrt{n}} \right) \label{eq:three_tieP2_12D}
\end{align}
by similar reasoning as we attained Equation (\ref{eq:three_tieP2_9D}).
This concludes the $\calE_{1,2}$ case of Lemma \ref{lem:three_tie_priority2}.
\end{paragraph}

\begin{paragraph}{Step 4: Putting the pieces together.}
To finish the proof, we tie our results about the $\calE_1$, $\calE_2$, and $\calE_{1,2}$ cases together. Consider first the case where $u_2 = u_3$. Then our conclusion is $\Theta \left( \frac{1}{n} \right)$ if either (i) $\pi_1 \leq \pi_5$ and $\pi_2 \geq \pi_6$, or (ii) $\pi_1 \geq \pi_5$ and $\pi_2 \leq \pi_6$ holds, and $\pm \calO \left( e^{-\Theta(n)} \right)$ otherwise. This follows from the second terms of Equations (\ref{eq:three_tieP2_3}) and (\ref{eq:three_tieP2_7}).

When $u_2 > u_3$, this entails the sum of Equations (\ref{eq:three_tieP2_4}), (\ref{eq:three_tieP2_8}), and (\ref{eq:three_tieP2_11}), subject to their respective conditions on the probability distribution $(\pi_1, \ldots, \pi_6)$, as well as the second terms of Equations (\ref{eq:three_tieP2_3}) and (\ref{eq:three_tieP2_7}). Recall that $\Theta(1)$, in many of the simplified versions of equations following (\ref{eq:three_tieP2_4}), (\ref{eq:three_tieP2_8}), and (\ref{eq:three_tieP2_11}), was a stand-in for $\binom{3n+1}{n+1, n, n} \frac{n}{3^{3n+1}}$. This enables us to combine several $\Theta(1)$-like terms together.

\begin{table}[t]
    \centering
    \begin{tabular}{|c||c|c|c|}
        \hline
         & $\tilde{\pi}_2 = \tilde{\pi}_6$ & $\tilde{\pi}_2 > \tilde{\pi}_6$ & $\tilde{\pi}_2 < \tilde{\pi}_6$ \\
          \hline \hline
            $\tilde{\pi}_1 = \tilde{\pi}_5$ & 
            \makecell{$3\pi_4 - 5 \pi_3$ \\ by Eqns. (\ref{eq:three_tieP2_5A}), (\ref{eq:three_tieP2_9A}), (\ref{eq:three_tieP2_12A})} &  
            \makecell{$\pi_2 - 3 \pi_6 - 2 \pi_1$ \\ by Eqns. (\ref{eq:three_tieP2_5C}) and (\ref{eq:three_tieP2_12C})} & 
            \makecell{$\pi_2 - \pi_3$ \\ by Eqn. (\ref{eq:three_tieP2_9B})} \\[0.25em]
          \hline
            $\tilde{\pi}_1 > \tilde{\pi}_5$ & 
            \makecell{$\pi_4 - \pi_3$ \\ by Eqns. (\ref{eq:three_tieP2_9C}) and (\ref{eq:three_tieP2_12B})} &  
            \makecell{$\pi_2 - \pi_6 - 2 \pi_1$ \\ by Eqn.  (\ref{eq:three_tieP2_12D})} & 
            \makecell{$\pi_2 + \pi_5 - 2 \pi_1$ \\ by Eqn. (\ref{eq:three_tieP2_9D})} \\[0.25em]
          \hline 
            $\tilde{\pi}_1 < \tilde{\pi}_5$ & 
            \makecell{$\pi_1 - \pi_3$ \\ by Eqn. (\ref{eq:three_tieP2_5B})} &  
            \makecell{$\pi_2 - \pi_6 - 2 \pi_1$ \\ by Eqn. (\ref{eq:three_tieP2_5D})} & 
            \makecell{N/A} \\
         \hline
    \end{tabular}
    \caption{Constants in front of $\Theta(1)$ term resulting from Equations (\ref{eq:three_tieP2_4}), (\ref{eq:three_tieP2_8}), (\ref{eq:three_tieP2_11}), for certain conditions on $\pi$.}
    \label{tab:complexity_three_sides_B}
\end{table}

Our conclusion is therefore $\pm \calO \left( e^{-\Theta(n)} \right)$ if $\pi_1 < \pi_5$ and $\pi_2 < \pi_6$. Otherwise, it is
\begin{align*}
    f(\pi_1, \ldots, \pi_6) \Theta(1) \pm \calO \left( \frac{1}{\sqrt{n}} \right)
\end{align*}
where $f(\pi_1, \ldots, \pi_6)$ is determined by Table \ref{tab:complexity_three_sides_B}.
%
\end{paragraph}
This concludes the proof of Lemma \ref{lem:three_tie_priority2}.
\end{proof}

%% file: EC_appendix/apx_page_4.tex
\newpage

\section{Multinomial Lemmas}
\label{apx:secondary_equations}


The appendices of this paper are organized so that the material of each appendix is used to prove the lemmas introduced in prior appendices, while they depend on the lemmas of later appendices. In particular, here, we provide the proof of two lemmas that are used in the proofs of Lemma \ref{lem:sub_12} (when $n$ is even) and Lemma \ref{lem:sub_12_odd} (when $n$ is odd) for the expected adversarial loss, conditioned on two-way ties (i.e.,  $\Pr\nolimits_{P \sim \pi^n}(\PW{P} = \{1,2\}) \times \mathbb{E}_{P \sim \pi^n}[\ADS(P)~|~\PW{P} = \{1,2\}]$). We prove Lemmas \ref{lem:soln_part_a} and \ref{lem:soln_part_b}.
As demonstrated below, these lemmas depend on several technical lemmas that appear in the next appendix, Appendix \ref{apx:collision_entropy}, including Lemmas \ref{lem:dec23_expected_demoivre}, \ref{lem:half_expected_bin_squares}, \ref{lem:real_to_const}, and \ref{lem:real_expected_demoivre}.

Note that the variable nomenclature is slightly different in this appendix than in Appendices \ref{apx:two_ties} or \ref{apx:three_ties}. Here, we demonstrate the asymptotic rate of functions resembling expected values based on a symmetric multinomial distribution $\sum_{q=1}^{\frac{n}{6}-1} \binom{n}{\frac{n}{2}-q, \frac{n}{2}-q, q, q} \pi_1^{n-2q} \pi_3^{2q} f_n(q)$ for some function $f_n(q)$. The four indices correspond to agents with rankings $R_1$ or $R_5$, $R_2$ or $R_6$, $R_3$, and $R_4$ respectively, with corresponding probabilities $(\pi_1, \pi_1, \pi_3, \pi_3)$. We continue to use $q$ as our index variable, but use the lower-case notation $\pi_1$ and $\pi_3$ instead of the upper-case $\Pi_1$ and $\Pi_3$ notation of the prior appendices.

\setcounter{lem}{12}
\begin{lem}
Let $\pi_1 \in [\frac{1}{3}, \frac{1}{2})$, $\pi_3 = \frac{1}{2} - \pi_1$, and consider $\tau \in [-1, 1]$. Then the following equality holds.
%
\begin{align*}
    & \sum_{q=1}^{\frac{n}{6}-1} \binom{n}{\frac{n}{2}-q, \frac{n}{2}-q, q, q} \pi_1^{n-2q} \pi_3^{2q} 
    \left( - \frac{1}{2} \left( \left( -\frac{n}{2} +q \right) \tau + 2q \right) + \frac{\pi_3}{\pi_1} \left( -\frac{n}{2}+q \right)  \right) 
    \\
    & 
    = \begin{cases}
        \Theta(1), & \pi_1 > \frac{2}{4+\tau} \\
        - \Theta(1), & \pi_1 < \frac{2}{4+\tau} \\
        \pm \calO \left( \frac{1}{n} \right), & \pi_1 = \frac{2}{4+\tau}.
    \end{cases}
\end{align*}

\label{lem:soln_part_a}
\end{lem}

\begin{proof}
Consider the objective
\[
\sum_{q=1}^{\frac{n}{6}-1} \binom{n}{\frac{n}{2}-q, \frac{n}{2}-q, q, q} \pi_1^{n-2q} \pi_3^{2q} f_{n, \tau}(q)
\]
where
\begin{align*}
    f_{n, \tau}(q) & = - \frac{1}{2} \left( \left( -\frac{n}{2} +q \right) \tau + 2q \right) + \frac{\pi_3}{\pi_1} \left( -\frac{n}{2}+q \right).
\end{align*}
We begin by considering the case where $\pi_1 = \frac{1}{3}$ (thus $\pi_3 = \frac{1}{6}$). Then $f_{n, \tau}(q)$ can be written as
\begin{align*}
    f_{n, \tau}(q) & = \left( \frac{n}{2} -q \right) \left( \frac{\tau + 2}{2} - \frac{\pi_3}{\pi_1} \right) - \frac{n}{2} \\
    & = \left( \frac{n}{2} -q \right) \left( \frac{\tau + 1}{2} \right) - \frac{n}{2}.
\end{align*}
It is easy to see that $f_{n, \tau}(q) \leq 0$ as long as
\begin{align*}
    \frac{n}{2} \left( \frac{\tau-1}{\tau+1 } \right) \leq q.
\end{align*}
Since $\frac{\tau-1}{\tau+1 } \leq 0$ for all $\tau \in [-1, 1]$, this holds for all $q \in (0, \frac{n}{6})$. Therefore $f_{n, \tau}(q) = -\Theta(n)$, so the objective is $-\Theta(1)$ by Lemma \ref{lem:prob_bounds}.

Now consider the case where $\pi_1 > \frac{1}{3}$ (thus $\pi_3 < \frac{1}{6}$). Then $f_{n, \tau}(q)$ can be written as
\begin{align*}
    f_{n, \tau} & = - \frac{1}{2} \left( - \pi_1 \tau n - \pi_3 (2 + \tau) n + (2 + \tau) q + 2 \pi_3 n \right) + \frac{\pi_3}{\pi_1} \left( -\pi_1 n - \pi_3 n+q \right) \\
    & = \left( \frac{\pi_3}{\pi_1} -1-\frac{\tau}{2}  \right) ( q - \pi_3 n) + \frac{n}{2} (\pi_1 \tau -4 \pi_3 ).
\end{align*}
Thus the objective is
\begin{align}
    & \left( \frac{\pi_3}{\pi_1} -1-\frac{\tau}{2}  \right) \sum_{q=1}^{\frac{n}{6}-1} \binom{n}{\frac{n}{2}-q, \frac{n}{2}-q, q, q} \pi_1^{n-2q} \pi_3^{2q} (q - \pi_3 n) \notag \\
    & \quad \quad + \frac{n}{2} (\pi_1 \tau -4 \pi_3 ) \sum_{q=1}^{\frac{n}{6}-1} \binom{n}{\frac{n}{2}-q, \frac{n}{2}-q, q, q} \pi_1^{n-2q} \pi_3^{2q} \notag \\
%
%
%
    & = \frac{\left( \frac{\pi_3}{\pi_1} -1-\frac{\tau}{2}  \right) \binom{n}{\frac{n}{2}} \sqrt{2n \pi_1 \pi_3}}{2^n} \sum_{q=1}^{\frac{n}{6}-1} \left( \binom{\frac{n}{2}}{q} (2\pi_1)^{\frac{n}{2}-q} (2\pi_3)^{q} \right)^2 \frac{(q - \pi_3 n)}{\sqrt{2 n \pi_3 \pi_1}} \notag \\
    & \quad \quad + \frac{n}{2} (\pi_1 \tau -4 \pi_3 ) \Theta \left( \frac{1}{n} \right) 
    \label{eq:multinom_eq1_line1b}
\end{align}
by 
Proposition \ref{prop:binom} for the first term and Lemma \ref{lem:prob_bounds} for the second term.

\setcounter{prop}{2}
\begin{prop}
Let $q \in \left[1, \frac{n}{6}-1 \right]$. Then
    \begin{align*}
    & \binom{n}{\frac{n}{2}-q, \frac{n}{2}-q, q, q} \pi_1^{n-2q} \pi_3^{2q} 
    = \frac{\binom{n}{\frac{n}{2}}}{2^n} \left( \binom{\frac{n}{2}}{q} (2\pi_1)^{\frac{n}{2}-q} (2\pi_3)^q \right)^2.
    \end{align*}
\end{prop}
This proposition, proved in Appendix \ref{apx:concentration_inequalities}, is useful for transforming the multinomial likelihood to a squared-binomial equivalence. The factor in front of the first term of Equation (\ref{eq:multinom_eq1_line1b}) is $\pm \Theta \left( 1 \right)$ following Stirling's approximation (Proposition \ref{prop:stirling}), where the sign depends on $\left( \frac{\pi_3}{\pi_1} -1-\frac{\tau}{2}  \right)$. Notice that the multinomial domain of $q$ is $[0, \frac{n}{6}]$ while the binomial domain of $q$ is $[0, \frac{n}{2}]$. We may therefore extend the range of the first term of Equation (\ref{eq:multinom_eq1_line1b}) by introducing a quantity that is exponentially small, by Hoeffding's inequality (Proposition \ref{prop:hoeffding}). That is, the summation is equivalent to
%
\begin{align}
    & \sum_{q=\frac{n}{6}}^{\frac{n}{2}} \left( \binom{\frac{n}{2}}{q} (2\pi_1)^{\frac{n}{2}-q} (2\pi_3)^{q} \right)^2 \frac{(q - \pi_3 n)}{\sqrt{2 n \pi_3 \pi_1}} \notag \\
    & \quad \quad + \left( \binom{\frac{n}{2}}{0} (2\pi_1)^{\frac{n}{2}-0} (2\pi_3)^{0} \right)^2 \frac{(0 - \pi_3 n)}{\sqrt{2 n \pi_3 \pi_1}} \notag \\
    & \quad \quad - \sum_{q=0}^{\frac{n}{2}} \left( \binom{\frac{n}{2}}{q} (2\pi_1)^{\frac{n}{2}-q} (2\pi_3)^{q} \right)^2 \frac{(q - \pi_3 n)}{\sqrt{2 n \pi_3 \pi_1}} \notag \\
    & = \pm \calO \left( e^{-\Theta(n)} \right) -  \sum_{q=0}^{\frac{n}{2}} \left( \binom{\frac{n}{2}}{q} (2\pi_1)^{\frac{n}{2}-q} (2\pi_3)^{q} \right)^2 \frac{(q - \pi_3 n)}{\sqrt{2 n \pi_3 \pi_1}}. \label{eq:multinom_eq1_line2}
\end{align}



\setcounter{prop}{3}
\begin{prop}[Hoeffding's Inequality]
Let $p \in (0,1)$ and $a, b \in \mathbb{R}$ such that $0 \leq a < b \leq 1$. If $p \notin [a,b]$ then
    \[
    \sum_{k=\bfloor{a n}}^{\bceil{b n}} \left( \binom{n}{k} p^{n-k} (1-p)^k \right)^2 = \calO \left( e^{-\Theta(n)} \right).
    \]
\end{prop}

Our specific use of this inequality is proved in Appendix \ref{apx:helper_lemmas}. Putting what we know from Equation (\ref{eq:multinom_eq1_line2}) back into Equation (\ref{eq:multinom_eq1_line1b}), we get
\begin{align*}
    &\pm \calO \left( e^{-\Theta(n)} \right) \pm \Theta(1) \mathcal{O}\left(\frac{1}{n}\right) + (\pi_1 \tau -4 \pi_3 ) \Theta (1) = \begin{cases}
        \Theta(1), & \pi_1 > \frac{2}{4+\tau} \\
        - \Theta(1), & \pi_1 < \frac{2}{4+\tau} \\
        \pm \calO \left( \frac{1}{n} \right), & \pi_1 = \frac{2}{4+\tau}
    \end{cases}
\end{align*}
by Lemma \ref{lem:dec23_expected_demoivre}, which is proved in Appendix \ref{apx:collision_entropy}. In that appendix, we discuss the necessary change of variables in order to apply the lemma. Simply put, we exchange $\frac{n}{2} \mapsto n$ and $2 \pi_3 \mapsto p$.

\setcounter{lem}{15}
\begin{lem}
Let $p \in (0, \frac{2}{3})$ and $S_n \sim Bin(n,p)$. Then
    \[
    \left| \sum_{k=0}^n \left( \frac{k-np}{\sqrt{np(1-p)}} \right) \Pr(S_n = k)^2 \right| = \mathcal{O} \left( \frac{1}{n} \right).
    \]
    \label{lem:dec23_expected_demoivre}
\end{lem}
Notice that the $\pi_1 = \frac{1}{3}$ case is covered in the $\pi_1 < \frac{2}{4 + \tau}$ case for any $\tau \in [-1, 1]$. Hence, we do not need to declare this as a special case in our final result.
This concludes the proof of Lemma \ref{lem:soln_part_a}.
\end{proof}






\setcounter{lem}{13}
\begin{lem}
Let $\pi_1 \in [\frac{1}{3}, \frac{1}{2})$, $\pi_3 = \frac{1}{2} - \pi_1$, and fix constants $\tau_1, \tau_2 \in \mathbb{R}$. Then $\exists N > 0$ such that for all $n > N$ that is even, 
\begin{align*}
    \Bigg| \sum_{q=1}^{\bfloor{\frac{n}{6}}-1} \binom{n}{\frac{n}{2}-q, \frac{n}{2}-q, q, q} \pi_1^{n-2q} \pi_3^{2q} \frac{ f_n(q)  2^{2q}}{\binom{2q}{q}} (\tau_1 (q - \pi_3 n) + \tau_2) \Bigg| = \mathcal{O}\left( \frac{1}{\sqrt{n}} \right).
\end{align*}
where $f_n(q) \in \left\{ \frac{-\frac{n}{2}+q}{2q+1}, \frac{n+1}{2q+1}, \frac{n+1}{\frac{n}{2}+1-q} \right\}$.
\label{lem:soln_part_b}
\end{lem}

\begin{proof}
This lemma is written in a general form to demonstrate that the objective is $\pm \calO \left( \frac{1}{\sqrt{n}} \right)$ for both terms of $\tau_1 (q - \pi_3 n)$ and $\tau_2$, regardless of whether $f_n(q)$ is $\frac{-\frac{n}{2}+q}{2q+1}$, $\frac{n+1}{2q+1}$, or $\frac{n+1}{\frac{n}{2}+1-q}$. The proof iterates through all possibilities to demonstrate that the objective under any of these six cases does, in fact, follow the stated asymptotic rate. We proceed in four steps: (i) the $\tau_2$ case, (ii) the $\tau_1 (q - \pi_3 n)$ case assuming $\pi_1 = \frac{1}{3}$, (iii) the $\tau_1 (q - \pi_3 n)$ case with $\pi_1 > \frac{1}{3}$ when $f_n(q)$ is either $\frac{-\frac{n}{2}+q}{2q+1}$ or $\frac{n+1}{2q+1}$, and (iv) the $\tau_1 (q - \pi_3 n)$ case with $\pi_1 > \frac{1}{3}$ when $f_n(q)$ is $\frac{n+1}{\frac{n}{2}+1-q}$.


\begin{paragraph}{Step 1 (the $\tau_2$ case):}
Consider, first, the $\tau_2$ term. Then the objective may be written as
\begin{align*}
    & \tau_2 \sum_{q=1}^{\bfloor{\frac{\pi_3 n}{2}}-1} \binom{n}{\frac{n}{2}-q, \frac{n}{2}-q, q, q} \pi_1^{n-2q} \pi_3^{2q} \frac{f_n(q) 2^{2q}}{\binom{2q}{q}}  \\
    & \quad \quad + \tau_2 \sum_{q=\bfloor{\frac{\pi_3 n}{2}}}^{\bfloor{\frac{n}{6}}-1} \binom{n}{\frac{n}{2}-q, \frac{n}{2}-q, 2q} \pi_1^{n-2q} (2\pi_3)^{2q} f_n(q)  \\
    & = \calO \left( e^{-\Theta(n)} \right) + \tau_2 \Theta(1) \sum_{q=\bfloor{\frac{\pi_3 n}{2}}}^{\bfloor{\frac{n}{6}}-1} \binom{n}{\frac{n}{2}-q, \frac{n}{2}-q, 2q} \pi_1^{n-2q} (2\pi_3)^{2q}  \\
    & = \pm \calO \left( \frac{1}{\sqrt{n}} \right)
\end{align*}
by Stirling's approximation and Lemma \ref{lem:prob_bounds} for the first summation, and Corollary \ref{coro:sub_others} for the second summation, since $f_n(q) = \Theta(1)$ along the domain of $q$.
\end{paragraph}

\begin{paragraph}{Step 2 (the $\tau_1$ case where $\pi_1 = \frac{1}{3}$):}
Notice that when $\pi_1 = \frac{1}{3}$, then $q - \pi_3 n$ has the same sign (above or below zero) for every $q \in (0, \frac{n}{6})$. We therefore must employ different techniques for the case where $\pi_1 = \frac{1}{3}$ than otherwise. This step addresses when this is the case. Then the objective may then be written as
\begin{align}
    & \tau_1 \sum_{q=1}^{\bfloor{\frac{\pi_3 n}{2}}-1} \binom{n}{\frac{n}{2}-q, \frac{n}{2}-q, q, q} \pi_1^{n-2q} \pi_3^{2q} \frac{f_n(q) 2^{2q}}{\binom{2q}{q}} (q - \pi_3 n) \notag \\
    & \quad \quad + \tau_2 \sum_{q=\bfloor{\frac{\pi_3 n}{2}}}^{\bfloor{\frac{n}{6}}-1} \binom{n}{\frac{n}{2}-q, \frac{n}{2}-q, 2q} \pi_1^{n-2q} (2\pi_3)^{2q} \frac{f_n(q) 2^{2q}}{\binom{2q}{q}} (q - \pi_3 n) \notag \\
    & = \calO \left( e^{-\Theta(n)} \right) \pm \Theta(1) \sum_{q=\bfloor{\frac{\pi_3 n}{2}}}^{\bfloor{\frac{n}{6}}-1} \binom{n}{\frac{n}{2}-q, \frac{n}{2}-q, q, q} \pi_1^{n-2q} \pi_3^{2q} (q - \pi_3 n) \label{eq:new_lem5_1} \\
    & = \calO \left( e^{-\Theta(n)} \right) \pm \frac{\Theta(1) \sqrt{2 \pi_1 \pi_3 n} \binom{n}{\frac{n}{2}}}{2^n} \sum_{q=\bfloor{\frac{\pi_3 n}{2}}}^{\bfloor{\frac{n}{6}}-1} \left( \binom{\frac{n}{2}}{q} (2 \pi_1)^{\frac{n}{2}-q} (2 \pi_3)^q \right)^2 \frac{(q - \pi_3 n)}{\sqrt{2 \pi_1 \pi_3 n}} \label{eq:new_lem5_2} \\
    & = \calO \left( e^{-\Theta(n)} \right) \pm \Theta(1) \calO \left( \frac{1}{\sqrt{n}} \right). \label{eq:new_lem5_3}
\end{align}
Equation (\ref{eq:new_lem5_1}) holds by Stirling's approximation and Lemma \ref{lem:prob_bounds}, since $f_n(q) = \Theta(1)$ along the domain of $q$ for the second summation. Equation (\ref{eq:new_lem5_2}) holds by Proposition \ref{prop:binom}. Equation (\ref{eq:new_lem5_3}) holds by Stirling's approximation (Proposition \ref{prop:stirling}) and the following lemma, proved in Appendix \ref{apx:collision_entropy}. 
\setcounter{lem}{16}
\begin{lem}
Let $p = \frac{2}{3}$ and $S_n \sim Bin(n,p)$. Then
\begin{align*}
    & \left| \sum_{k=\bfloor{\frac{np}{4}}}^{\bfloor{np}} \left( \frac{k-np}{\sqrt{np(1-p)}} \right) \Pr(S_n = k)^2 \right|
    = \calO \left( \frac{1}{\sqrt{n}} \right).
\end{align*}
\end{lem}
\end{paragraph}

\begin{paragraph}{Step 3 (the $\tau_1$ case when $f_n(q)$ is either $\frac{-\frac{n}{2}+q}{2q+1}$ or $\frac{n+1}{2q+1}$):}
For this step, consider the $\tau_1$ case when $\pi_1 > \frac{1}{3}$. Immediately we notice that we cannot use the same method as when $\pi_1 = \frac{1}{3}$. Now, even though $f_n(q) = \Theta(1)$, we cannot factor it out of the summation because $q - \pi_3 n$ takes both positive and negative values along the domain $q \in (0, \frac{n}{6})$. Instead, we will make use of the specific properties of $f_n(q)$. The objective may be written as
\begin{align}
    & \tau_1 \sum_{q=1}^{\bfloor{\frac{\pi_3 n}{2}}-1} \binom{n}{\frac{n}{2}-q, \frac{n}{2}-q, q, q} \pi_1^{n-2q} \pi_3^{2q} \frac{f_n(q) 2^{2q}}{\binom{2q}{q}} (q - \pi_3 n) \notag \\
    & \quad \quad + \tau_1 \frac{\binom{n}{\frac{n}{2}}}{2^n} \sum_{q=\bfloor{\frac{\pi_3 n}{2}}}^{\frac{n}{2}} \left( \binom{\frac{n}{2}}{q} (2\pi_1)^{\frac{n}{2}-q} (2\pi_3)^{q} \right)^2 \frac{ f_n(q) 2^{2q}}{\binom{2q}{q}} (q - \pi_3 n) \notag \\
    & \quad \quad - \tau_1 \frac{\binom{n}{\frac{n}{2}}}{2^n} \sum_{q=\frac{n}{6}}^{\frac{n}{2}} \left( \binom{\frac{n}{2}}{q} (2\pi_1)^{\frac{n}{2}-q} (2\pi_3)^{q} \right)^2 \frac{ f_n(q) 2^{2q}}{\binom{2q}{q}} (q - \pi_3 n) \label{eq:new_lem5_4}
\end{align}
by Proposition \ref{prop:binom}. The first summation of Equation (\ref{eq:new_lem5_4}) is clearly $\pm \calO \left( e^{-\Theta(n)} \right)$ by Lemma \ref{lem:prob_bounds}. Since
\[
\frac{\binom{n}{\frac{n}{2}}}{2^n} \frac{ f_n(q) 2^{2q}}{\binom{2q}{q}} (q - \pi_3 n) = \calO \left( \frac{1}{\sqrt{n}} \right) \Theta(1) \calO(\sqrt{n}) \calO(n) = \calO(1)
\]
by Stirling's approximation (Proposition \ref{prop:stirling}), it follows that the third summation of Equation (\ref{eq:new_lem5_4}) is
\begin{align*}
    \calO(1) \sum_{q=\frac{n}{6}}^{\frac{n}{2}} \left( \binom{\frac{n}{2}}{q} (2\pi_1)^{\frac{n}{2}-q} (2\pi_3)^{q} \right)^2
    = \calO \left( e^{-\Theta(n)} \right)
\end{align*}
by Hoeffding's inequality (Proposition \ref{prop:hoeffding}).
Now consider the second summation of Equation (\ref{eq:new_lem5_4}). We handle this in parts, depending on the state of $f_n(q)$.

First, if $f_n(q) = \frac{-\frac{n}{2}+q}{2q+1}$, then
\begin{align*}
    & f_n(q) (q - \pi_3 n) \\
    & = \frac{1}{2q+1} \left( -\pi_1 n - \pi_3 n+q \right) (q - \pi_3 n) \\
    & = \frac{1}{2q+1} \left( (q - \pi_3 n)^2 - \pi_1 n (q - \pi_3 n) \right).
\end{align*}
Following Stirling's approximation (Proposition \ref{prop:stirling}), the second summation of Equation (\ref{eq:new_lem5_4}) may therefore be written as 
\begin{align*}
    & \Theta \left( \frac{1}{\sqrt{n}} \right) \sum_{q=\bfloor{\frac{\pi_3 n}{2}}}^{\frac{n}{2}} \left( \binom{\frac{n}{2}}{q} (2\pi_1)^{\frac{n}{2}-q} (2\pi_3)^{q} \right)^2 \frac{  2^{2q}}{(2q+1) \binom{2q}{q}} (q - \pi_3 n)^2 \\
    & \quad \quad - \Theta(\sqrt{n}) \sum_{q=\bfloor{\frac{\pi_3 n}{2}}}^{\frac{n}{2}} \left( \binom{\frac{n}{2}}{q} (2\pi_1)^{\frac{n}{2}-q} (2\pi_3)^{q} \right)^2 \frac{ 2^{2q}}{(2q+1) \binom{2q}{q}} (q - \pi_3 n)  \\
    & = \Theta(1) \sum_{q=\bfloor{\frac{\pi_3 n}{2}}}^{\frac{n}{2}} \left( \binom{\frac{n}{2}}{q} (2\pi_1)^{\frac{n}{2}-q} (2\pi_3)^{q} \right)^2 \frac{2^{2q} \sqrt{2 \pi_1 \pi_3 n}}{(2q+1) \binom{2q}{q}} \left( \frac{ q - \pi_3 n}{\sqrt{2 \pi_1 \pi_3 n}} \right)^2 \\
    & \quad \quad - \Theta(\sqrt{n}) \sum_{q=\bfloor{\frac{\pi_3 n}{2}}}^{\frac{n}{2}} \left( \binom{\frac{n}{2}}{q} (2\pi_1)^{\frac{n}{2}-q} (2\pi_3)^{q} \right)^2 \frac{2^{2q} \sqrt{2 \pi_1 \pi_3 n}}{(2q+1) \binom{2q}{q}} \frac{ (q - \pi_3 n)}{\sqrt{2 \pi_1 \pi_3 n}}  \\
    & = \pm \calO \left( \frac{1}{\sqrt{n}} \right).
\end{align*}
This holds by the following two lemmas, which are described further and proved in Appendix \ref{apx:collision_entropy}. To  apply these lemma, we make the change of variables $\frac{n}{2} \mapsto n$, $q \mapsto k$, and $2 \pi_3 \mapsto p$.


\setcounter{lem}{17}
\begin{lem}
Let $p \in (0, \frac{2}{3})$ and $S_n \sim Bin(n,p)$. Then 
\begin{align*}
    & \left| \sum_{k=\bfloor{\frac{np}{2}}}^{n} \left( \frac{k-np}{\sqrt{np(1-p)}} \right)^2 \frac{2^{2k} \sqrt{np(1-p)}}{(2k+1) \binom{2k}{k}} \Pr(S_n = k)^2 \right|
    = \calO \left( \frac{1}{\sqrt{n}} \right).
\end{align*}
\end{lem}

\setcounter{lem}{18}
\begin{lem}
Let $p \in (0, \frac{2}{3})$ and $S_n \sim Bin(n,p)$. Then 
\begin{align*}
    & \left| \sum_{k=\bfloor{\frac{np}{2}}}^{n} \left( \frac{k-np}{\sqrt{np(1-p)}} \right) \frac{2^{2k} \sqrt{np(1-p)}}{(2k+1) \binom{2k}{k}} \Pr(S_n = k)^2 \right|
    = \calO \left( \frac{1}{n} \right).
\end{align*}
\end{lem}

Second, if $f_n(q) = \frac{n+1}{2q+1}$ then 
\begin{align*}
    & f_n(q) (q - \pi_3 n) \\
    & = \frac{1}{2q+1} (n+1) (q - \pi_3 n).
\end{align*}
Clearly the second summation of Equation (\ref{eq:new_lem5_4}) is $\pm \calO \left( \frac{1}{\sqrt{n}} \right)$ by the  above reasoning.
\end{paragraph}

\begin{paragraph}{Step 4 (the $\tau_1$ case when $f_n(q)$ is $\frac{n+1}{\frac{n}{2}+1-q}$):}
Finally, suppose that $f_n(q) = \frac{n+1}{\frac{n}{2}+1-q}$, assuming $\pi_1 > \frac{1}{3}$. Then the lemma's objective may be written as
\begin{align}
    & \tau_1 \sum_{q=1}^{\bfloor{\frac{n}{6}}-1} \binom{n+1}{\frac{n}{2}-q, \frac{n}{2}+1-q, q, q} \pi_1^{n-2q} \pi_3^{2q} \frac{  2^{2q}}{\binom{2q}{q}} (q - \pi_3 n) \notag \\
    & =  \frac{\tau_1}{\pi_1^2 (n+2)} \sum_{q=1}^{\bfloor{\frac{n}{6}}-1} \binom{n+2}{\frac{n}{2}+1-q, \frac{n}{2}+1-q, q, q} \pi_1^{n+2-2q} \pi_3^{2q} \frac{ (\frac{n+2}{2}-q) 2^{2q}}{\binom{2q}{q}} (q - \pi_3 (n+2) + 2 \pi_3) \notag \\
    & = \pm \calO \left( e^{-\Theta(n)} \right) \notag \\
    & \quad \quad - \frac{\tau_1 \binom{n+2}{\frac{n+2}{2}}}{\pi_1^2 (n+2) 2^{n+2}} \sum_{q=\bfloor{\frac{\pi_3 (n+2)}{2} }}^{\frac{n+2}{2}} \left( \binom{\frac{n+2}{2}}{q} (2\pi_1)^{\frac{n+2}{2}-q} (2\pi_3)^{q} \right)^2  \left( (q - \pi_3 (n+2))^2 + \pi_1 n (q - \pi_3 (n+2)) \right) \notag \\
    & \quad \quad + \frac{2 \tau_1 \pi_3 }{\pi_1} \sum_{q=1}^{\bfloor{\frac{n}{6}}-1} \binom{n+1}{\frac{n}{2}-q, \frac{n}{2}+1-q, 2q} \pi_1^{n+1-2q} (2\pi_3)^{2q}. \label{eq:new_lem5_5}
\end{align}
The exponential term in Equation (\ref{eq:new_lem5_5}) follows from Lemma \ref{lem:prob_bounds} and Hoeffding's inequality (Proposition \ref{prop:hoeffding}), by the same reasoning as Step 3 above.
Equation (\ref{eq:new_lem5_5}) may then be written as
\begin{align}
    & \pm \calO \left( e^{-\Theta(n)} \right) \pm \calO \left( \frac{1}{\sqrt{n}} \right) \notag \\
    & \quad \quad \pm \calO \left( \frac{1}{n^{1.5}} \right) \sum_{q=\bfloor{\frac{\pi_3 (n+2)}{2} }}^{\frac{n+2}{2}} \left( \binom{\frac{n+2}{2}}{q} (2\pi_1)^{\frac{n+2}{2}-q} (2\pi_3)^{q} \right)^2 (q - \pi_3 (n+2))^2 \notag \\
    & \quad \quad \pm \calO \left( \frac{1}{\sqrt{n}} \right) \sum_{q=\bfloor{\frac{\pi_3 (n+2)}{2} }}^{\frac{n+2}{2}} \left( \binom{\frac{n+2}{2}}{q} (2\pi_1)^{\frac{n+2}{2}-q} (2\pi_3)^{q} \right)^2 (q - \pi_3 (n+2)) \label{eq:new_lem5_6} \\
    & = \pm \calO \left( \frac{1}{\sqrt{n}} \right) \notag \\
    & \quad \quad \pm \calO \left( \frac{1}{\sqrt{n}} \right) \sum_{q=\bfloor{\frac{\pi_3 (n+2)}{2} }}^{\frac{n+2}{2}} \left( \binom{\frac{n+2}{2}}{q} (2\pi_1)^{\frac{n+2}{2}-q} (2\pi_3)^{q} \right)^2 \left( \frac{q - \pi_3 (n+2)}{\sqrt{2 \pi_1 \pi_3 (n+2)}} \right)^2 \notag \\
    & \quad \quad \pm \calO \left( 1 \right) \sum_{q=\bfloor{\frac{\pi_3 (n+2)}{2} }}^{\frac{n+2}{2}} \left( \binom{\frac{n+2}{2}}{q} (2\pi_1)^{\frac{n+2}{2}-q} (2\pi_3)^{q} \right)^2 \frac{(q - \pi_3 (n+2))}{\sqrt{2 \pi_1 \pi_3 (n+2)}}. \label{eq:new_lem5_7}
\end{align}
The $\calO \left( \frac{1}{\sqrt{n}} \right)$ term in Equation (\ref{eq:new_lem5_6}) follows from Corollary \ref{coro:sub_others}.
The second summation of Equation (\ref{eq:new_lem5_7}) is $\calO \left( 1 \right)$ following Lemma \ref{lem:limit_analysis_2}, after recognizing that the binomial probability mass function is point-wise smaller than one.

\setcounter{lem}{19}
\begin{lem}
Let $p \in (0, 1)$ and $S_n \sim Bin(n,p)$. Then
    \[
    \sum_{k=\bfloor{\frac{np}{2}}}^n \left( \frac{k-np}{\sqrt{np(1-p)}} \right)^2 \Pr(S_n = k) = \Theta(1).
    \]
\end{lem}

This lemma is proved in Appendix \ref{apx:normalized_expectations}.
The third summation of Equation (\ref{eq:new_lem5_7}) is $\calO \left( \frac{1}{\sqrt{n}} \right)$ by Lemma \ref{lem:dec23_expected_demoivre}.
\end{paragraph}
This concludes the proof of Lemma \ref{lem:soln_part_b}.
\end{proof}


%% file: EC_appendix/apx_page_5a.tex
\newpage

\section{Expected Collision Entropy}
\label{apx:collision_entropy}


This appendix describes the asymptotic rate that certain sequences of summations in Lemmas \ref{lem:soln_part_a} and \ref{lem:soln_part_b} converge to zero, such as this objective equation from Lemma \ref{lem:real_expected_demoivre}:
\[
\sum_{k=\bfloor{\frac{np}{2}}}^{n} \left( \frac{k-np}{\sqrt{np(1-p)}} \right) \frac{2^{2k} \sqrt{np(1-p)}}{(2k+1) \binom{2k}{k}} \Pr(S_n = k)^2
\]
where $p \in (0, \frac{2}{3})$ and $S_n \sim Bin(n,p)$.
\footnote{
We name this appendix ``Expected Collision Entropy'' for its relationship to R{\'e}yni entropy (of order $2$; see e.g., \citet{fehr2014conditional}). This is defined for binomial random variables as $- \ln \sum_{k=1}^n \left( \binom{n}{k} p^k (1-p)^{n-k} \right)^2$ which details the negative log likelihood of the two random variables being equal. This is ``expected'' because we're multiplying each collision likelihood by the standardized value $\frac{k-np}{\sqrt{np(1-p)}}$.

Interestingly, from Lemma \ref{lem:prob_bounds} and Proposition \ref{prop:binom}, we get 
\[
\frac{\binom{n}{\frac{n}{2}}}{2^n} \sum_{q=0}^{\frac{n}{6}} \left( \binom{\frac{n}{2}}{q} (2 \pi_1)^{\frac{n}{2}-q} (2 \pi_3)^{q} \right)^2 = \calO \left( \frac{1}{n} \right)
\]
while $\pi_1 \in [\frac{1}{3}, \frac{1}{2})$. By Stirling's approximation (Proposition \ref{prop:stirling}) this entails the R{\'e}yni entropy of the binomial is $\calO(\ln n)$.
}
Intuitively, this equation seems similar to the standardized expectation of a binomial random variable, which is clearly zero. However, there are two complications: the fact that we are squaring the binomial likelihood function and the presence of the value 
\[
g_{n,k} = \frac{2^{2k} \sqrt{np(1-p)}}{(2k+1) \binom{2k}{k}}
\]
in the summation. While $|g_{n,k}| = \Theta(1)$ by Lemma \ref{lem:mag_bound} (detailed below), it cannot be factored out of the summation through standard techniques because $\frac{k-np}{\sqrt{np(1-p)}}$ takes on both positive and negative values throughout the summation. One intuitively nice method, hypothetically, could partition the summation region at $k=np$, factor out $g_{n,k}$ for each part, and then add the two components back together. However, this method is specious; it would yield too imprecise of an asymptotic bound. Hence, different techniques must be used.

Our methods therefore include replacing the binomial probability with a discrete Gaussian form $\frac{1}{\sqrt{2\pi}} e^{-\frac{x^2}{2}}$, using triangle inequality, and then applying following theorem to asymptotically bound parts of the objective summations:
\setcounter{thm}{2}
\begin{thm}[\citet{Petrov+1975}, Chapter VII.1]
Let $S_n \sim Bin(n,p)$. Then
\begin{align*}    
    & \sup_{k \in [0,n]} \left| \Pr(S_n = k) - \frac{1}{\sqrt{2 \pi n p (1-p)}} e^{-\frac{1}{2} \left(\frac{k-np}{\sqrt{ n p (1-p)}}\right)^2} \right|
    = \calO \left( \frac{1}{n} \right).
    \end{align*}
\label{thm:petrov}
\end{thm}
For the rest of the objective, we make use of properties of $g_{n,k}$ and a change of variables to yield the desired claims. These concepts are described technically in the lemma proofs.

This appendix is presented in three parts.
First, we use different notation in this appendix than the prior lemmas in order to generalize these claims beyond our specific use-case. Appendix \ref{sec:expected_prelims} describes what change of variables are necessary to apply this appendix's lemmas from the notation used in Lemmas \ref{lem:soln_part_a} and \ref{lem:soln_part_b}. Appendix \ref{sec:expected_in_full} then lists and proves the three applicable lemmas: \ref{lem:dec23_expected_demoivre}, \ref{lem:real_to_const}, and \ref{lem:real_expected_demoivre}. This makes their proofs significantly more complicated. Third, Appendix \ref{apx:normalized_expectations} proves technical lemmas that are used in the aforementioned lemmas. 



\subsection{Preliminaries}
\label{sec:expected_prelims}

Let $p \in (0,\frac{2}{3})$ and $q = 1-p$. For each $n \in \mathbb{N}$ where $n p \in \mathbb{Z}_{\geq 0}$, let $S_n \sim Bin(n,p)$ and define the random variable
\[
X_n = \frac{S_n - np}{\sqrt{npq}}.
\]
The random variable $X_n$ takes on the values
\[
x_{n,k} = \frac{k-np}{\sqrt{npq}} \text{ for } 0 \leq k \leq n
\]
which are evenly spaced out by
\[
\Delta_n = \frac{1}{\sqrt{npq}}.
\]
We have
\[
\Pr(S_n = k) = \binom{n}{k} p^k q^{n-k}.
\]
Finally, we define
\[
f(x) = \frac{1}{\sqrt{2 \pi}} e^{-\frac{x^2}{2}}.
\]

In order to apply the subsequent lemmas to the claims made throughout the primary theorem, we make the following change of variables:
\[
\begin{pmatrix}
    \frac{n}{2}\\
    q\\
    2\pi_3\\
    2\pi_1
\end{pmatrix} \mapsto \begin{pmatrix}
    n\\
    k\\
    p\\
    q
\end{pmatrix}
\]
recalling that $\pi_1 + \pi_3 = \frac{1}{2}.$ Hence, we get the variable
\[
\frac{k - \pi_3 n}{\sqrt{2 \pi_1 \pi_3 n}} \mapsto x_{n,k} = \frac{k-np}{\sqrt{npq}}
\]
and a new variable definition
\[
\frac{ 2^{2k} \sqrt{2 \pi_1 \pi_3 n}}{(2k+1) \binom{2k}{k}} \mapsto g_{n,k} = \frac{2^{2k} \sqrt{npq}}{(2k+1) \binom{2k}{k}}.
\]


%% file: EC_appendix/apx_page_5b.tex
\subsection{Proof of Standardized Squared-Binomial Lemmas}
\label{sec:expected_in_full}

\setcounter{lem}{15}
\begin{lem}
    \[
    \left| \sum_{k=0}^n \left( \frac{k-np}{\sqrt{npq}} \right) \Pr(S_n = k)^2 \right| = \mathcal{O} \left( \frac{1}{n} \right).
    \]
\end{lem}

\begin{proof}

As described in the introduction to this appendix, it is clear that
\[
\sum_{k=0}^n x_{n,k} \Pr(S_n = k) = \frac{1}{\sqrt{npq}} \mathbb{E}[S_n - np] = 0.
\]
The challenge with this lemma is the presence of the squared-binomial probability in the objective. Intuitively, we would like to make a symmetry argument that, for any fixed $t > 0$, $x_{n,np-t} = \frac{-t}{\sqrt{npq}} = -\frac{t}{\sqrt{npq}} = -x_{n,np+t}$ and $\Pr(S_n = np-t) \approx \Pr(S_n = np+t)$. Hence, most terms would cancel out, except for perhaps the tails which occur with exponentially small likelihood by Hoeffding's inequality. This approach does not immediately work because $\Pr(|S_n - np| < t) \xrightarrow[]{n \rightarrow \infty} 0$ for fixed $t$. Rather, the lemma requires summing up over a range of at least size $\Theta(n)$ around the point $k=np$ (e.g., $[np-t, np+t]$ for $t = \Theta(n)$) whose likelihood of occurrence tends to $1$. However, when $t = \Theta(n)$ and $p \neq \frac{1}{2}$, we have $\frac{\Pr(S_n = np + t)}{\Pr(S_n = np - t)} \in \{ \exp(\Theta(n)), \exp(- \Theta(n))\}$, so the $x_{n,np-t}$ and $x_{n,np+t}$ terms would not cancel out.
\footnote{This approach could work by the (local) DeMoivre-Laplace Theorem for $t = \calO(\sqrt{n})$ (see e.g., \citet{feller1991introduction, carlen18demoivre}); still, it would not make this proof complete. We would not be able to bound the rate of convergence of the tails specifically enough.} 

Rather than keeping $\Pr(S_n = k)$ in our summation, which is skewed for $p \neq \frac{1}{2}$, we could replace it with the discretized Gaussian function $f(x_{n,k}) \Delta_n = \frac{1}{\sqrt{2 \pi npq}} e^{-\frac{x_{n,k}^2}{2}}$, which is symmetric about $np$. This idea comes from the central limit theorem by which we expect the $\frac{S_n - np}{\sqrt{npq}}$ to converge in distribution to the standard Gaussian. 
The Berry–Esseen theorem suggests that this convergence rate is $\calO \left( \frac{1}{\sqrt{n}} \right)$ (see e.g., \citet{durrett2019probability}), so, intuitively, the squared-probability should converge at rate $\calO \left( \frac{1}{n} \right)$. However, a direct application of Berry–Esseen-like theorems fail since they hold only for cumulative distribution functions. Proving this point-wise for $\Pr(S_n = k)$ at each $k$ and including the value-term $x_{n,k}$ in the summation for our lemma requires nuance. 

Hence, we make use of Theorem \ref{thm:petrov} \citep[Chapter VII.1]{Petrov+1975}, which bounds the point-wise difference between $\Pr(S_n = k)$ and $f(x_{n,k}) \Delta_n$ by the rate $\calO \left( \frac{1}{n} \right)$. This lemma's proof proceeds by substituting the binomial probability $\Pr(S_n = k)$ by adding and subtracting $f(x_{n,k}) \Delta_n$ to and from the objective. This allows us to bound each term using Theorem \ref{thm:petrov} and several convergence technical lemmas that are described and proved in Appendix \ref{apx:normalized_expectations}.

Notably, we replace the objective with $C_n = \sum_{k=0}^n x_{n,k}  f(x_{n,k})^2 \Delta_n^2$ where both the value and probability parts of the equation are symmetrical around the center $np$, plus some additional terms. Still, we run into integral problems by which $np$ may not be an integer.
It is easy to show that $|C_n| = \calO \left( e^{-\Theta(n)} \right)$ if $np$ is an integer by symmetry. Demonstrating the desired bound that $|C_n| = \calO \left( \frac{1}{n} \right)$ requires a handful of other steps when $np$ is not an integer. We demonstrate both cases in the proof below to build the reader's intuition. The technical details are as follows.

We start off by splitting up the objective into three parts in which we replace $\Pr(S_n = k)$ with $( \Pr(S_n = k) - $ $f(x_{n,k}) \Delta_n) + f(x_{n,k}) \Delta_n$ at each step:
\begin{align}
    & \sum_{k=0}^n x_{n,k} \Pr(S_n = k)^2 \notag \\
    & = \sum_{k=0}^n x_{n,k}  \Pr(S_n = k)    \Big( \Pr(S_n = k) - f(x_{n,k}) \Delta_n \Big)
    + \sum_{k=0}^n x_{n,k} \Pr(S_n = k) f(x_{n,k}) \Delta_n \notag \\
    & = A_n + B_n + C_n \label{eq:main6pf_1}
\end{align}
where we define
\begin{align*}
    A_n & = \sum_{k=0}^n x_{n,k}  \Pr(S_n = k)  \Big( \Pr(S_n = k) - f(x_{n,k}) \Delta_n \Big),
\end{align*}
\begin{align*}
    B_n & = \sum_{k=0}^n x_{n,k} f(x_{n,k}) \Delta_n  \Big( \Pr(S_n = k) - f(x_{n,k}) \Delta_n \Big),
\end{align*}
\begin{align*}
    C_n & = \sum_{k=0}^n x_{n,k}  f(x_{n,k})^2 \Delta_n^2. 
\end{align*}
Consider the first summation of Equation (\ref{eq:main6pf_1}). We have
\begin{align}
    |A_n| & = \left| \sum_{k=0}^n x_{n,k} \Pr(S_n = k)  \left( \Pr(S_n = k) - f(x_{n,k}) \Delta_n \right) \right| \notag \\
    & \leq \sum_{k=0}^n \left| x_{n,k} \Pr(S_n = k) \right|  
    \cdot \left| \Pr(S_n = k) - f(x_{n,k}) \Delta_n \right| \notag \\ 
    & \leq \mathcal{O} \left( \frac{1}{n} \right) \sum_{k=0}^n | x_{n,k} | \Pr(S_n = k) \label{eq:main6pf_2} \\
    & = \mathcal{O} \left( \frac{1}{n} \right) \label{eq:main6pf_3}
\end{align}
by triangle inequality. Equation (\ref{eq:main6pf_2}) follows from Theorem \ref{thm:petrov}. Equation (\ref{eq:main6pf_3}) follows from Lemma \ref{lem:limit_analysis_1}
, proved in Appendix \ref{apx:normalized_expectations}.

\setcounter{lem}{20}
\begin{lem}
    \[
    \sum_{k=0}^n \left| x_{n,k} \right| \Pr(S_n = k) = \Theta(1).    
    \]
\end{lem}

Now, for the second summation of Equation (\ref{eq:main6pf_1}), we have
\begin{align}
    |B_n| & = \left| \sum_{k=0}^n x_{n,k} f(x_{n,k}) \Delta_n  \left( \Pr(S_n = k) - f(x_{n,k}) \Delta_n \right) \right| \notag \\
    & \leq \sum_{k=0}^n \left| x_{n,k} f(x_{n,k}) \Delta_n \right| 
    \cdot  \left| \Pr(S_n = k) - f(x_{n,k}) \Delta_n \right| \notag \\ 
    & \leq \mathcal{O} \left( \frac{1}{n} \right) \sum_{k=0}^n | x_{n,k} | f(x_{n,k}) \Delta_n \label{eq:main6pf_4} \\
    & = \mathcal{O} \left( \frac{1}{n} \right) \label{eq:main6pf_5}
\end{align}
by triangle inequality. Equation (\ref{eq:main6pf_4}) follows from Theorem \ref{thm:petrov}. Equation (\ref{eq:main6pf_5}) follows 
from the following lemma.

\setcounter{lem}{21}
\begin{lem}
    The following equation is $\Theta(1)$:
\begin{enumerate}
    \item
    \[
    \sum_{k=0}^n |x_{n,k}| f(x_{n,k}) \Delta_n.
    \] 
\end{enumerate}
\end{lem}
Lemma \ref{lem:instead_of_integrals} consists of ten equations that we prove are all $\Theta(1)$ in Appendix \ref{apx:normalized_expectations}. Each equation is structured similarly and may be proved in almost an identical manner except for how the proof is initialized. Hence, for convenience and straightforwardness of this appendix, we pack all ten equations into the same lemma statement.

Finally, consider the third summation of Equation (\ref{eq:main6pf_1}). We prove that $|C_n| \leq \calO \left( \frac{1}{n} \right)$ with the following two cases, depending on whether $np$ is an integer or not. We demonstrate both cases to build the reader's intuition.

\begin{paragraph}{Case 1: $np$ is an integer.}
We have
\begin{align}
    C_n & = \sum_{k=0}^n x_{n,k} f(x_{n,k})^2 \Delta_n^2 \notag \\
    & = \sum_{k=\bfloor{\frac{np}{2}}}^{\bceil{\frac{3np}{2}}} x_{n,k} f(x_{n,k})^2 \Delta_n^2 + \sum_{k=0}^{\bfloor{\frac{np}{2}}-1} x_{n,k} f(x_{n,k})^2 \Delta_n^2 
    + \sum_{k=\bceil{\frac{3np}{2}}+1}^{n} x_{n,k} f(x_{n,k})^2 \Delta_n^2. \label{eq:lem6_cn_no_int}
\end{align}
The first summation of Equation (\ref{eq:lem6_cn_no_int}) is zero by symmetry since $np$ is assumed to be an integer. The second summation of Equation (\ref{eq:lem6_cn_no_int}) is
\[
    - \sum_{k=0}^{\bfloor{\frac{np}{2}}-1} \Theta(\sqrt{n}) \calO \left( e^{-\Theta(n)} \right) \Theta \left( \frac{1}{n} \right) = -\calO \left( e^{-\Theta(n)} \right),
\]
while the third summation similarly yields $\calO \left( e^{-\Theta(n)} \right)$.
\end{paragraph}

\begin{paragraph}{Case 2: $np$ is not an integer.}
%
%
Now suppose that $np$ is not an integer and that $np = t_n+b_n$ where $t_n \in \mathbb{N}$ and $b_n \in (0,1)$. Our approach is to split up $C_n$ into four regions: a ``positive'' region of size $npq$ above $np$, a ``negative'' region of size $npq$ below $np$, and two tails which are clearly exponentially small. We seek to point-wise align the positive and negative regions and have the terms at $k = \bfloor{np} - u$ and $k = \bceil{np} + u$, for $u \in [0, \bceil{npq}]$ approximately cancel out. We make the appropriate change of variables, which leads to Equation (\ref{eq:main6pf_6}) below. The final step is to appropriately bound the magnitude of each part of that equation by $\calO \left( \frac{1}{n} \right)$ using the Maclaurin–Cauchy integral test 
from Lemma \ref{lem:instead_of_integrals}. The aforementioned partition is as follows.
\begin{align}
    C_n & = \sum_{k=0}^n x_{n,k}  f(x_{n,k})^2 \Delta_n^2 \notag \\
    & = \sum_{k=\bfloor{np} - \bceil{npq}}^{\bfloor{np}} x_{n,k}  f(x_{n,k})^2 \Delta_n^2 
    + \sum_{k=\bceil{np}}^{\bceil{np} + \bceil{npq}} x_{n,k}  f(x_{n,k})^2 \Delta_n^2 \notag \\
    & \quad \quad + \sum_{k=0}^{\bfloor{np}-\bceil{npq}-1} x_{n,k}  f(x_{n,k})^2 \Delta_n^2 
    + \sum_{k=\bceil{np}+\bceil{npq}+1}^{n} x_{n,k}  f(x_{n,k})^2 \Delta_n^2 \notag \\
    & = \sum_{k=\bfloor{np} - \bceil{npq}}^{\bfloor{np}} \left( \frac{k-t_n-b_n}{\sqrt{npq}} \right)  
    f\left( \frac{k-t_n-b_n}{\sqrt{npq}} \right)^2 \Delta_n^2 \notag \\
    & \quad \quad + \sum_{k=\bceil{np}}^{\bceil{np} + \bceil{npq}} \left( \frac{k-t_n-b_n}{\sqrt{npq}} \right) 
    f\left( \frac{k-t_n-b_n}{\sqrt{npq}} \right)^2 \Delta_n^2 \notag \\
    & \quad \quad - \sum_{k=0}^{\bfloor{np}-\bceil{npq}-1} \Theta(n) \calO \left( e^{-\Theta(n)} \right) \Theta \left( \frac{1}{n} \right) \notag \\
    & \quad \quad + \sum_{k=\bceil{np}+\bceil{npq}+1}^{n} \Theta(n) \calO \left( e^{-\Theta(n)} \right) \Theta \left( \frac{1}{n} \right). \label{eq:main6pf_5b}
\end{align}
Notice that our partition is valid: both $np - npq = np(1-q) = np^2 \in (0, n)$ 
and $np + npq = np(2-p) \in (0,n)$.
Next, we make the change of variables $u = \bfloor{np} - k$ in the first summation of Equation (\ref{eq:main6pf_5b}) and $u = k - \bceil{np}$ in the second summation of Equation (\ref{eq:main6pf_5b}). We therefore get
\begin{align}
    & \sum_{u=0}^{\bceil{npq}} \left( \frac{-u-b_n}{\sqrt{npq}} \right)  f\left( \frac{-u-b_n}{\sqrt{npq}} \right)^2 \Delta_n^2 \notag \\
    & \quad \quad + \sum_{u=0}^{\bceil{npq}} \left( \frac{u+1-b_n}{\sqrt{npq}} \right)  f\left( \frac{u+1-b_n}{\sqrt{npq}} \right)^2 \Delta_n^2  \notag \\
    & \quad \quad \pm \calO \left( e^{-\Theta(n)} \right) \notag \\
    & = D_n + F_n +G_n \pm \calO \left( e^{-\Theta(n)} \right)
    \label{eq:main6pf_6}
\end{align}
where we define
\begin{align*}
    D_n & =  \sum_{u=0}^{\bceil{npq}}\left( \frac{u}{\sqrt{npq}} \right) 
    \left( -f\left( \frac{u+b_n}{\sqrt{npq}} \right)^2 + f\left( \frac{u+1-b_n}{\sqrt{npq}} \right)^2 \right) \Delta_n^2,
\end{align*}
\begin{align*}
    E_n = -b_n \sum_{u=0}^{\bceil{npq}}f\left( \frac{u+b_n}{\sqrt{npq}} \right)^2 \Delta_n^3,
\end{align*}
\begin{align*}
    F_n = (1-b_n) \sum_{u=0}^{\bceil{npq}}f\left( \frac{u+1-b_n}{\sqrt{npq}} \right)^2 \Delta_n^3.
\end{align*}
%
%
We made use of the fact that $f$ is an even function to get $D_n$ and $E_n$. Consider the first summation of Equation (\ref{eq:main6pf_6}). We have
\begin{align}
    |D_n| & \leq \sum_{u=0}^{\bceil{npq}}\left( \frac{u}{\sqrt{npq}} \right) 
    \left| -f\left( \frac{u}{\sqrt{npq}} \right)^2 + f\left( \frac{u+1}{\sqrt{npq}} \right)^2 \right| \Delta_n^2 \label{eq:dn_lem6_fix} \\
    & = \sum_{u=0}^{\bceil{npq}} \left( \frac{u}{\sqrt{npq}} \right)  f\left( \frac{u}{\sqrt{npq}} \right)^2 \Delta_n^2 
    - \sum_{u=0}^{\bceil{npq}} \left( \frac{u+1}{\sqrt{npq}} \right) f\left( \frac{u+1}{\sqrt{npq}} \right)^2 \Delta_n^2 \notag \\
    & \quad \quad + \sum_{u=0}^{\bceil{npq}} \left( \frac{1}{\sqrt{npq}} \right) f\left( \frac{u+1}{\sqrt{npq}} \right)^2 \Delta_n^2 \notag \\
    & = \pm \calO \left( e^{-\Theta(n)} \right) + \calO \left( \frac{1}{n} \right) \sum_{u=0}^{\bceil{npq}} f\left( \frac{u}{\sqrt{npq}} \right)^2 \Delta_n \notag \\
    & = \calO \left( \frac{1}{n} \right). \notag
\end{align}
by Lemma \ref{lem:instead_of_integrals}.2.
Equation (\ref{eq:dn_lem6_fix}) comes from the fact that $e^{-y^2}$ is decreasing for $y>0$, so $f \left(  \frac{u + c}{\sqrt{npq}} \right)^2 \leq f \left(  \frac{u}{\sqrt{npq}} \right)^2$ for $c \in (0,1)$.
%
%
%
Now consider the third summation of Equation (\ref{eq:main6pf_6}). We have
\begin{align*}
    |E_n| & \leq \calO \left( \frac{1}{n} \right) \sum_{u=0}^{\bceil{npq}}f\left( \frac{u}{\sqrt{npq}} \right)^2 \Delta_n \\
    & = \calO \left( \frac{1}{n} \right) 
\end{align*}
since $e^{-y^2}$ is decreasing for $y>0$ and by Lemma \ref{lem:instead_of_integrals}.2.
It is easy to see that $|F_n| = \calO \left(\frac{1}{n} \right)$ by similar reasoning. Collectively, this entails that $|C_n| = \calO \left(\frac{1}{n} \right)$.

\end{paragraph}
This concludes the proof of Lemma \ref{lem:dec23_expected_demoivre}.
\end{proof}



\setcounter{lem}{16}
\begin{lem}
Let $p = \frac{2}{3}$. Then
\begin{align*}
    & \left| \sum_{k=\bfloor{\frac{np}{2}}}^{\bfloor{np}} x_{n,k} \Pr(S_n = k)^2 \right|
    = \calO \left( \frac{1}{\sqrt{n}} \right).
\end{align*}
\label{lem:half_expected_bin_squares}
\end{lem}

\begin{proof}
The proof proceeds similar to Lemma \ref{lem:dec23_expected_demoivre} in that we substitute the binomial probability $\Pr(S_n = k)$ by adding and subtracting the discretized Gaussian function $f(x_{n,k}) \Delta_n$ to and from the objective. This allows us to bound each term using Theorem \ref{thm:petrov} and several convergence technical lemmas that are described and proved in Appendix \ref{apx:normalized_expectations}. The final step of this proof is significantly simpler than that in Lemma \ref{lem:dec23_expected_demoivre} since we only require an asymptotic bound of $\mathcal{O} \left( \frac{1}{\sqrt{n}} \right)$. 
We start with
\begin{align}
    & \sum_{k=\bfloor{\frac{np}{2}}}^{\bfloor{np}} x_{n,k}^2 \Pr(S_n = k)^2 \notag \\
    & = \sum_{k=\bfloor{\frac{np}{2}}}^{\bfloor{np}} x_{n,k}^2 \Pr(S_n = k)  
    \left( \Pr(S_n = k) - f(x_{n,k}) \Delta_n \right) \notag \\
    & \quad \quad + \sum_{k=\bfloor{\frac{np}{2}}}^{\bfloor{np}} x_{n,k}^2 \cdot \Pr(S_n = k) f(x_{n,k}) \Delta_n \notag \\
    & = A_n + B_n + C_n \label{eq:new_bin_sq_1}
\end{align}
where we define
\begin{align*}
    A_n & = \sum_{k=\bfloor{\frac{np}{2}}}^{\bfloor{np}} x_{n,k}^2 \Pr(S_n = k) 
    \left( \Pr(S_n = k) - f(x_{n,k}) \Delta_n \right),
\end{align*}
\begin{align*}
    B_n & = \sum_{k=\bfloor{\frac{np}{2}}}^{\bfloor{np}} x_{n,k}^2 f(x_{n,k}) \Delta_n 
    \left( \Pr(S_n = k) - f(x_{n,k}) \Delta_n \right),
\end{align*}
\begin{align*}
    C_n & = \sum_{k=\bfloor{\frac{np}{2}}}^{\bfloor{np}} x_{n,k}^2  f(x_{n,k})^2 \Delta_n^2. 
\end{align*}
Consider the first summation of Equation (\ref{eq:new_bin_sq_1}). We have
\begin{align*}
    |A_n| & = \left| \sum_{k=\bfloor{\frac{np}{2}}}^{\bfloor{np}} x_{n,k}^2 \Pr(S_n = k) 
    \left( \Pr(S_n = k) - f(x_{n,k}) \Delta_n \right) \right| \\
    & \leq \sum_{k=\bfloor{\frac{np}{2}}}^{\bfloor{np}} \left| x_{n,k}^2 \cdot g_{n,k} \Pr(S_n = k) \right| 
    \cdot  \left| \Pr(S_n = k) - f(x_{n,k}) \Delta_n \right| \\ 
    & \leq \mathcal{O} \left( \frac{1}{n} \right) \sum_{k=\bfloor{\frac{np}{2}}}^{\bfloor{np}} x_{n,k}^2 \Pr(S_n = k) \\
    & \leq \mathcal{O} \left( \frac{1}{n} \right) \sum_{k=\bfloor{\frac{np}{2}}}^{n} x_{n,k}^2 \Pr(S_n = k)  \\
    & = \mathcal{O} \left( \frac{1}{n} \right)
\end{align*}
by triangle inequality,  Theorem \ref{thm:petrov}, and Lemma \ref{lem:limit_analysis_2}.
For the second summation of Equation (\ref{eq:new_bin_sq_1}), we have
\begin{align*}
    |B_n| & = \left| \sum_{k=\bfloor{\frac{np}{2}}}^{\bfloor{np}} x_{n,k}^2 f(x_{n,k}) \Delta_n 
    \cdot  \left( \Pr(S_n = k) - f(x_{n,k}) \Delta_n \right) \right| \\
    & \leq \sum_{k=\bfloor{\frac{np}{2}}}^{\bfloor{np}} \left| x_{n,k}^2 f(x_{n,k}) \Delta_n \right| 
    \cdot  \left| \Pr(S_n = k) - f(x_{n,k}) \Delta_n \right| \\ 
    & \leq \mathcal{O} \left( \frac{1}{n} \right) \sum_{k=\bfloor{\frac{np}{2}}}^{\bfloor{np}} x_{n,k}^2 f(x_{n,k}) \Delta_n  \\
    & \leq \mathcal{O} \left( \frac{1}{n} \right) \sum_{k=\bfloor{\frac{np}{2}}}^{n} x_{n,k}^2 f(x_{n,k}) \Delta_n \\
    & = \mathcal{O} \left( \frac{1}{n} \right)
\end{align*}
by triangle inequality, Theorem \ref{thm:petrov}, and Lemma \ref{lem:instead_of_integrals}.3. 
Finally, consider the third line of Equation (\ref{eq:new_bin_sq_1}). We get
\begin{align*}
    |C_n| & = \left| \sum_{k=\bfloor{\frac{np}{2}}}^{\bfloor{np}} x_{n,k}^2 f(x_{n,k})^2 \Delta_n^2 \right| \\
    & \leq \sum_{k=\bfloor{\frac{np}{2}}}^{\bfloor{np}} |x_{n,k}^2 f(x_{n,k})^2 \Delta_n^2 | \\
    & \leq \calO \left( \frac{1}{\sqrt{n}} \right) \sum_{k=\bfloor{\frac{np}{2}}}^{\bfloor{np}} x_{n,k}^2 f(x_{n,k})^2 \Delta_n \\
    & \leq \calO \left( \frac{1}{\sqrt{n}} \right) \sum_{k=\bfloor{\frac{np}{2}}}^{n} x_{n,k}^2 f(x_{n,k})^2 \Delta_n \\
    & = \calO \left( \frac{1}{\sqrt{n}} \right)
\end{align*}
by triangle inequality and Lemma \ref{lem:instead_of_integrals}.4.
This concludes the proof of Lemma \ref{lem:half_expected_bin_squares}.
\end{proof}

\setcounter{lem}{17}
\begin{lem}
\[
    \left| \sum_{k=\bfloor{\frac{np}{2}}}^{n} x_{n,k}^2 \cdot g_{n,k} \Pr(S_n = k)^2 \right| = \mathcal{O} \left( \frac{1}{\sqrt{n}} \right).
\]
\label{lem:real_to_const}
\end{lem}

\begin{proof}

The proof is almost identical to the proof of Lemma \ref{lem:half_expected_bin_squares} since we require a loose $\calO \left( \frac{1}{\sqrt{n}} \right)$ bound, as opposed to the tight $\calO \left( \frac{1}{n} \right)$ bound of Lemma \ref{lem:dec23_expected_demoivre}. Our method is to substitute the binomial probability $\Pr(S_n = k)$ by adding and subtracting the discretized Gaussian function $f(x_{n,k}) \Delta_n$ to and from the objective. This allows us to bound each term using Theorem \ref{thm:petrov} and several convergence technical lemmas that are described and proved in Appendix \ref{apx:normalized_expectations}. The extra term $g_{n,k}$ does not affect the flow of the proof, as seen below. 
\begin{align}
    & \sum_{k=\bfloor{\frac{np}{2}}}^n x_{n,k}^2 \cdot g_{n,k} \Pr(S_n = k)^2 \notag \\
    & = \sum_{k=\bfloor{\frac{np}{2}}}^n x_{n,k}^2 \cdot g_{n,k} \Pr(S_n = k)  
    \left( \Pr(S_n = k) - f(x_{n,k}) \Delta_n \right) \notag \\
    & \quad \quad + \sum_{k=\bfloor{\frac{np}{2}}}^n x_{n,k}^2 \cdot g_{n,k} \Pr(S_n = k) f(x_{n,k}) \Delta_n \notag \\
    & = A_n + B_n + C_n \label{eq:const_proof_a}
\end{align}
where we define
\begin{align*}
    A_n & = \sum_{k=\bfloor{\frac{np}{2}}}^n x_{n,k}^2 \cdot g_{n,k} \Pr(S_n = k) 
    \left( \Pr(S_n = k) - f(x_{n,k}) \Delta_n \right),
\end{align*}
\begin{align*}
    B_n & = \sum_{k=\bfloor{\frac{np}{2}}}^n x_{n,k}^2 \cdot g_{n,k} f(x_{n,k}) \Delta_n 
    \left( \Pr(S_n = k) - f(x_{n,k}) \Delta_n \right),
\end{align*}
\begin{align*}
    C_n & = \sum_{k=\bfloor{\frac{np}{2}}}^n x_{n,k}^2 \cdot g_{n,k} f(x_{n,k})^2 \Delta_n^2. 
\end{align*}
Consider the first summation of Equation (\ref{eq:const_proof_a}). We have
\begin{align}
    |A_n| & = \left| \sum_{k=\bfloor{\frac{np}{2}}}^n x_{n,k}^2 \cdot g_{n,k} \Pr(S_n = k) 
    \left( \Pr(S_n = k) - f(x_{n,k}) \Delta_n \right) \right| \notag \\
    & \leq \sum_{k=\bfloor{\frac{np}{2}}}^n \left| x_{n,k}^2 \cdot g_{n,k} \Pr(S_n = k) \right| 
    \cdot  \left| \Pr(S_n = k) - f(x_{n,k}) \Delta_n \right| \notag \\ 
    & \leq \mathcal{O} \left( \frac{1}{n} \right) \sum_{k=\bfloor{\frac{np}{2}}}^n x_{n,k}^2 \Pr(S_n = k) \label{eq:const_proof_b} \\
    & = \mathcal{O} \left( \frac{1}{n} \right) \label{eq:const_proof_c}
\end{align}
by triangle inequality. Equation (\ref{eq:const_proof_b}) follows from Theorem \ref{thm:petrov} and since $|g_{n,k}| = \Theta(1)$ by Lemma \ref{lem:mag_bound}
, proved in Appendix \ref{apx:helper_lemmas}.

\setcounter{lem}{22}
\begin{lem}
    Let $k \in \left[\bfloor{\frac{np}{2}}, n \right]$. Then 
    $
    \left| g_{n,k} \right| = \Theta(1).
    $
\end{lem}

Equation (\ref{eq:const_proof_c}) follows from Lemma \ref{lem:limit_analysis_2}.
%
%
%
%
%
Now, for the second summation of Equation (\ref{eq:const_proof_a}), we have
\begin{align}
    |B_n| & = \left| \sum_{k=\bfloor{\frac{np}{2}}}^n x_{n,k}^2 \cdot g_{n,k} f(x_{n,k}) \Delta_n 
    \cdot  \left( \Pr(S_n = k) - f(x_{n,k}) \Delta_n \right) \right| \notag \\
    & \leq \sum_{k=\bfloor{\frac{np}{2}}}^n \left| x_{n,k}^2 \cdot g_{n,k} f(x_{n,k}) \Delta_n \right| 
    \cdot  \left| \Pr(S_n = k) - f(x_{n,k}) \Delta_n \right| \notag \\ 
    & \leq \mathcal{O} \left( \frac{1}{n} \right) \sum_{k=\bfloor{\frac{np}{2}}}^n x_{n,k}^2 f(x_{n,k}) \Delta_n \label{eq:const_proof_d} \\
    & = \mathcal{O} \left( \frac{1}{n} \right) \label{eq:const_proof_e}
\end{align}
by triangle inequality. Equation (\ref{eq:const_proof_d}) follows from Theorem \ref{thm:petrov} and since $|g_{n,k}| = \Theta(1)$ by Lemma \ref{lem:mag_bound}. Equation (\ref{eq:const_proof_e}) follows by Lemma \ref{lem:instead_of_integrals}.3. 
Finally, consider the third line of Equation (\ref{eq:const_proof_a}). We get
\begin{align*}
    |C_n| & = \left| \sum_{k = \bfloor{\frac{np}{2}}}^n x_{n,k}^2 \cdot g_{n,k} f(x_{n,k})^2 \Delta_n^2 \right| \\
    & \leq \sum_{k = \bfloor{\frac{np}{2}}}^n |x_{n,k}^2 \cdot g_{n,k} f(x_{n,k})^2 \Delta_n^2 | \\
    & \leq \calO \left( \frac{1}{\sqrt{n}} \right) \sum_{k = \bfloor{\frac{np}{2}}}^n x_{n,k}^2 f(x_{n,k})^2 \Delta_n \\
    & = \calO \left( \frac{1}{\sqrt{n}} \right)
\end{align*}
by triangle inequality, since $|g_{n,k}| = \Theta(1)$ by Lemma \ref{lem:mag_bound}, and by Lemma \ref{lem:instead_of_integrals}.4.

This concludes the proof of Lemma \ref{lem:real_to_const}.
\end{proof}



\setcounter{lem}{18}
\begin{lem}
    \[
    \left| \sum_{k=\bfloor{\frac{np}{2}}}^n x_{n,k} \cdot g_{n,k} \Pr(S_n = k)^2 \right| = \mathcal{O} \left( \frac{1}{n} \right).
    \]
    \label{lem:real_expected_demoivre}
\end{lem}

\begin{proof}

This proof proceeds in four parts. In the first part, we substitute the binomial probability $\Pr(S_n = k)$ by adding and subtracting the discretized Gaussian function $f(x_{n,k}) \Delta_n$ to and from the objective, similar to Lemmas \ref{lem:dec23_expected_demoivre} and \ref{lem:real_to_const}.
We make use of Theorem \ref{thm:petrov} for some parts, as in those lemmas, and are left with $C_n = \sum_{k = \bfloor{\frac{np}{2}}}^n x_{n,k} \cdot g_{n,k} f(x_{n,k})^2 \Delta_n^2$. 

Recall that in Lemma \ref{lem:dec23_expected_demoivre} we made a symmetry argument to bound $|C_n|$ by $\calO \left( \frac{1}{n} \right)$, while in Lemma \ref{lem:real_to_const} we factored $g_{n,k}$ and $\Delta_n$ out of the objective to yield a $\calO \left( \frac{1}{\sqrt{n}} \right)$ bound. Since $g_{n,k}$ is in this summation and we require an asymptotic bound of $\calO \left( \frac{1}{n} \right)$ for this lemma, the techniques of these lemmas used on $C_n$ are no longer valid.
In the second step to this proof, we therefore identify meaningful upper- and lower-bounds to $C_n$ in order to apply the squeeze theorem. We do this by exploiting properties of $g_{n,k}$ and identifying upper- and lower-bounds to $g_{n,k}$ that are asymptotically equivalent (see Lemma \ref{lem:wallis} below). The terms composing $C_n$ are both positive and negative on its range $k \in [\bfloor{\frac{np}{2}}, n]$. We upper-bound $C_n$ by using the upper-bound of $g_{n,k}$ on the positive portion of $C_n$ and lower-bound of $g_{n,k}$ on the negative portion of $C_n$. The opposite holds to lower-bound $C_n$. Recall by Lemma \ref{lem:mag_bound} that $|g_{n,k}| = \Theta(1)$. This bound is remarkably not precise enough to prove Lemma \ref{lem:real_expected_demoivre}. Rather, we require $g_{n,k}$'s bounds to be asymptotically equivalent to attain $\calO \left( \frac{1}{n} \right)$ bounds, making use of the stricter Lemma \ref{lem:wallis}.

After some simplification, we are left in the third step of the proof with \\ $F_n = \sum_{k = \bfloor{\frac{np}{2}}}^{\bceil{\frac{3np}{2}}} x_{n,k} \sqrt{\frac{1}{k+0.5}} f(x_{n,k})^2 \Delta_n$. This summation is now symmetrical around $np$ except for the $\calO \left( \frac{1}{\sqrt{k}} \right)$ factor in the summation and the possibility that $np$ may not be an integer. To handle the first issue, we make a symmetry argument and pair the terms at $k = np - u$ and $k = np+u$ for $u \in [0, \bfloor{\frac{np}{2}}]$. This leads to a summation similar to $\sum_{u = 0}^{\bfloor{\frac{np}{2}}} \left( \frac{u}{\sqrt{npq}} \right) f \left(\frac{u}{\sqrt{npq}} \right)^2 \left( \sqrt{\frac{1}{np+u}} - \sqrt{\frac{1}{np-u}} \right) \Delta_n$ (see Equation (\ref{eq:u_ints_7}) below). We require significant nuance to handle the case where $np$ may not be an integer, described in Step 3 below. All-in-all, this possibility does not affect the convergence rate. Finally, we show in Step 4 that $\left( \sqrt{\frac{1}{np+u}} - \sqrt{\frac{1}{np-u}} \right) = -u \calO \left( \frac{1}{n^{1.5}} \right)$, which enables us to prove Lemma \ref{lem:real_expected_demoivre}. The technical details are as follows.

\begin{paragraph}{Step 1: Substitute the binomial probability.}

We start off by splitting up the objective into three parts in which we replace $\Pr(S_n = k)$ with $\left( \Pr(S_n = k) - f(x_{n,k}) \Delta_n \right) + f(x_{n,k}) \Delta_n$ at each step:
\begin{align}
    & \sum_{k=\bfloor{\frac{np}{2}}}^n x_{n,k} \cdot g_{n,k} \Pr(S_n = k)^2 \notag \\
    & = \sum_{k=\bfloor{\frac{np}{2}}}^n x_{n,k} \cdot g_{n,k} \Pr(S_n = k) 
    \left( \Pr(S_n = k) - f(x_{n,k}) \Delta_n \right) \notag \\
    & \quad \quad + \sum_{k=\bfloor{\frac{np}{2}}}^n x_{n,k} \cdot g_{n,k} \Pr(S_n = k) f(x_{n,k}) \Delta_n \notag \\
    & = A_n + B_n + C_n \label{eq:dec23_expect_demoivre_a}
\end{align}
where we define
\begin{align*}
    A_n & = \sum_{k=\bfloor{\frac{np}{2}}}^n x_{n,k} \cdot g_{n,k} \Pr(S_n = k) 
    \left( \Pr(S_n = k) - f(x_{n,k}) \Delta_n \right);
\end{align*}
\begin{align*}
    B_n & = \sum_{k=\bfloor{\frac{np}{2}}}^n x_{n,k} \cdot g_{n,k} f(x_{n,k}) \Delta_n \left( \Pr(S_n = k) - f(x_{n,k}) \Delta_n \right);
\end{align*}
and
\begin{align*}
    C_n & = \sum_{k=\bfloor{\frac{np}{2}}}^n x_{n,k} \cdot g_{n,k} f(x_{n,k})^2 \Delta_n^2. 
\end{align*}
Consider the first summation of Equation (\ref{eq:dec23_expect_demoivre_a}). We have
\begin{align}
    |A_n| & = \left| \sum_{k=\bfloor{\frac{np}{2}}}^n x_{n,k} \cdot g_{n,k} \Pr(S_n = k) 
    \left( \Pr(S_n = k) - f(x_{n,k}) \Delta_n \right) \right| \notag \\
    & \leq \sum_{k=\bfloor{\frac{np}{2}}}^n \left| x_{n,k} \cdot g_{n,k} \Pr(S_n = k) \right| 
    \cdot  \left| \Pr(S_n = k) - f(x_{n,k}) \Delta_n \right| \notag \\ 
    & \leq \mathcal{O} \left( \frac{1}{n} \right) \sum_{k=\bfloor{\frac{np}{2}}}^n | x_{n,k} | \Pr(S_n = k) \label{eq:jan12_expect_demoivre_b} \\
    & = \mathcal{O} \left( \frac{1}{n} \right) \label{eq:jan12_expect_demoivre_c}
\end{align}
by triangle inequality. Equation (\ref{eq:jan12_expect_demoivre_b}) follows from Theorem \ref{thm:petrov} and since $|g_{n,k}| = \Theta(1)$ by Lemma \ref{lem:mag_bound}.
Equation (\ref{eq:jan12_expect_demoivre_c}) follows from Lemma \ref{lem:limit_analysis_1} and Hoeffding's inequality (Proposition \ref{prop:hoeffding}).

Now, for the second summation of Equation (\ref{eq:dec23_expect_demoivre_a}), we have
\begin{align}
    |B_n| & = \left| \sum_{k=\bfloor{\frac{np}{2}}}^n x_{n,k} \cdot g_{n,k} f(x_{n,k}) \Delta_n  \left( \Pr(S_n = k) - f(x_{n,k}) \Delta_n \right) \right| \notag \\
    & \leq \sum_{k=\bfloor{\frac{np}{2}}}^n \left| x_{n,k} \cdot g_{n,k} f(x_{n,k}) \Delta_n \right| 
    \cdot \left| \Pr(S_n = k) - f(x_{n,k}) \Delta_n \right| \notag \\ 
    & \leq \mathcal{O} \left( \frac{1}{n} \right) \sum_{k=\bfloor{\frac{np}{2}}}^n | x_{n,k} | f(x_{n,k}) \Delta_n \label{eq:jan12_expect_demoivre_d} \\
    & = \mathcal{O} \left( \frac{1}{n} \right) \label{eq:jan12_expect_demoivre_e}
\end{align}
by triangle inequality. Equation (\ref{eq:jan12_expect_demoivre_d}) follows from Theorem \ref{thm:petrov} and since $|g_{n,k}| = \Theta(1)$ by Lemma \ref{lem:mag_bound}. Equation (\ref{eq:jan12_expect_demoivre_e}) follows from Lemma \ref{lem:instead_of_integrals}.5.

\end{paragraph}

\begin{paragraph}{Step 2: Squeeze theorem using properties of $g_{n,k}$.}
Our next step is to identify meaningful bounds on the third summation of Equation (\ref{eq:dec23_expect_demoivre_a}), $C_n$, and apply the squeeze theorem. Our upper- and lower-bounds on $C_n$ follow from upper- and lower-bounds on $g_{n,k}$ in the following lemma,
described and proved in Appendix \ref{apx:stirling_and_wallis}. 

\setcounter{lem}{23}
\begin{lem}
\[
    \sqrt{\frac{2n}{2n+1}}  \sqrt{\frac{2}{\pi (2n+1)}} \leq \frac{2^{2n}}{(2n+1) \binom{2n}{n}} \leq \sqrt{\frac{2}{\pi (2n+1)}}.
\]
\end{lem}

Recall that $g_{n,k} = \frac{2^{2k} \sqrt{npq}}{(2k+1) \binom{2k}{k}}$, so by Lemma \ref{lem:wallis} we have
\begin{equation}
    \sqrt{\frac{2k}{2k+1}}  \sqrt{\frac{2 npq}{\pi (2k+1)}} \leq  g_{n,k} \leq \sqrt{\frac{2 npq}{\pi (2k+1)}}. \label{eq:wallis_applied}
\end{equation}
Notice that the terms composing $C_n$ are both positive and negative on its range $k \in [\bfloor{\frac{np}{2}}, n]$. We upper-bound $C_n$ by using the upper-bound of $g_{n,k}$ on the positive portion of $C_n$ and lower-bound of $g_{n,k}$ on the negative portion of $C_n$. The opposite holds to lower-bound $C_n$. This procedure partitions the range of $C_n$ into two parts: $k \in [\bfloor{\frac{np}{2}}, \bfloor{np}]$ and $k \in [\bceil{np}, n]$, each of which is $\pm \calO \left( \frac{1}{\sqrt{n}} \right)$. This bound is not tight enough to prove Lemma \ref{lem:real_expected_demoivre}. We therefore want to use a symmetry argument to have the terms at $k= np-u$ and $k=np+u$ for $u \in [0, \bfloor{\frac{np}{2}}]$ approximately cancel out, like in the proof of Lemma \ref{lem:dec23_expected_demoivre}, to yield a tighter bound. Lemma \ref{lem:wallis}'s bounds which are asymptotically equivalent (i.e., $\sqrt{\frac{2k}{2k+1}}  \sqrt{\frac{2 npq}{\pi (2k+1)}} \sim \sqrt{\frac{2 npq}{\pi (2k+1)}}$; see Lemma \ref{lem:analysis_helper} below) enables us to do this. This step concludes by bounding $|C_n| \leq \calO \left( \frac{1}{n} \right) + |F_n|$ where $F_n$ is a summation that covers the full range $k \in [\bfloor{\frac{np}{2}}, n]$ and includes an $\calO \left( \frac{1}{\sqrt{k}} \right)$ factor in the objective.

From the third summation of Equation (\ref{eq:dec23_expect_demoivre_a}), we get
\begin{align}
    C_n & = \sum_{k = \bfloor{\frac{np}{2}}}^n x_{n,k} \cdot g_{n,k} f(x_{n,k})^2 \Delta_n^2 \notag \\
    & = \sum_{k = \bfloor{\frac{np}{2}}}^{\bfloor{np}} x_{n,k} \cdot g_{n,k} f(x_{n,k})^2 \Delta_n^2 
    + \sum_{k = \bceil{np}}^{n} x_{n,k} \cdot g_{n,k} f(x_{n,k})^2 \Delta_n^2 \notag \\
    & \leq \sqrt{\frac{\bfloor{np}}{\bfloor{np}+1}} \sum_{k = \bfloor{\frac{np}{2}}}^{\bfloor{np}} x_{n,k} \sqrt{\frac{2npq}{\pi(2k+1)}} f(x_{n,k})^2 \Delta_n^2 
    + \sum_{k = \bceil{np}}^n x_{n,k} \sqrt{\frac{2npq}{\pi(2k+1)}} f(x_{n,k})^2 \Delta_n^2 \label{eq:jan10_demoivre_b2}
\end{align}
where the lower-bound on $g_{n,k}$ from Equation (\ref{eq:wallis_applied}) is applied to the negative portion of the summation, where $k \leq \bfloor{np}$, and the upper-bound on $g_{n,k}$ is applied to the positive portion of the summation, where $k \geq \bceil{np}$. Note that $\sqrt{\frac{2k}{2k+1}}$ is increasing in $k$, by the following lemma, so $k = \bfloor{\frac{np}{2}}$ was inputted to minimize this value over the domain $k \in [\bfloor{\frac{np}{2}},\bfloor{np}]$.

\setcounter{lem}{24}
\begin{lem}
    For any constant $t > 0$, $\sqrt{\frac{tn}{tn+1}} = 1 - \calO \left( \frac{1}{n} \right)$.
\end{lem}

Lemma \ref{lem:analysis_helper} is proved in Appendix \ref{apx:helper_lemmas}. By this lemma, Equation (\ref{eq:jan10_demoivre_b2}) is equivalent to
\begin{align}
    & \sqrt{\frac{2}{\pi}} \left( 1 - \calO \left( \frac{1}{n} \right) \right) \sum_{k = \bfloor{\frac{np}{2}}}^{\bfloor{np}} x_{n,k} \sqrt{\frac{1}{2k+1}} f(x_{n,k})^2 \Delta_n  
    + \sqrt{\frac{2}{\pi}} \sum_{k = \bceil{np}}^n x_{n,k} \sqrt{\frac{1}{2k+1}} f(x_{n,k})^2 \Delta_n \notag \\  
    & = -\calO \left( \frac{1}{n} \right) \sum_{k = \bfloor{\frac{np}{2}}}^{\bfloor{np}} x_{n,k} \sqrt{\frac{1}{2k+1}} f(x_{n,k})^2 \Delta_n 
    + \sqrt{\frac{2}{\pi}} \sum_{k = \bfloor{\frac{np}{2}}}^n x_{n,k} \sqrt{\frac{1}{2k+1}} f(x_{n,k})^2 \Delta_n \label{eq:jan10_demoivre_d}
\end{align}
We repeat this process to get a lower-bound on $C_n$: 
\begin{align}
    C_n & = \sum_{k = \bfloor{\frac{np}{2}}}^n x_{n,k} \cdot g_{n,k} f(x_{n,k})^2 \Delta_n^2 \notag \\
    & = \sum_{k = \bfloor{\frac{np}{2}}}^{\bfloor{np}} x_{n,k} \cdot g_{n,k} f(x_{n,k})^2 \Delta_n^2 
    + \sum_{k = \bceil{np}}^{n} x_{n,k} \cdot g_{n,k} f(x_{n,k})^2 \Delta_n^2 \notag \\
    & \geq \sum_{k = \bfloor{\frac{np}{2}}}^{\bfloor{np}} x_{n,k} \sqrt{\frac{2npq}{\pi(2k+1)}} f(x_{n,k})^2 \Delta_n^2  
    + \sqrt{\frac{2\bceil{np}}{2\bceil{np}+1}} 
    \sum_{k = \bceil{np}}^n x_{n,k} \sqrt{\frac{2npq}{\pi(2k+1)}} f(x_{n,k})^2 \Delta_n^2 \label{eq:jan10_demoivre_e2}
\end{align}
where the upper-bound on $g_{n,k}$ from Equation (\ref{eq:wallis_applied}) is applied to the negative portion of the summation, where $k \leq \bfloor{np}$. Likewise, the lower-bound on $g_{n,k}$ from Equation (\ref{eq:wallis_applied}) is applied to the positive portion of the summation, where $k \geq \bceil{np}$, with $k=\bceil{np}$ which is set at $\argmin_{k \in [\bceil{np}, n]} \sqrt{\frac{2k}{2k+1}}$.
By Lemma \ref{lem:analysis_helper}, Equation (\ref{eq:jan10_demoivre_e2}) is then equivalent to
\begin{align}
    & \sqrt{\frac{2}{\pi}} \sum_{k = \bfloor{\frac{np}{2}}}^{\bfloor{np}} x_{n,k} \sqrt{\frac{1}{2k+1}} f(x_{n,k})^2 \Delta_n  
    + \sqrt{\frac{2}{\pi}} \left( 1 - \calO \left( \frac{1}{n} \right) \right) \sum_{k = \bceil{np}}^n x_{n,k} \sqrt{\frac{1}{2k+1}} f(x_{n,k})^2 \Delta_n \notag \\ 
    & = \sqrt{\frac{2}{\pi}} \sum_{k = \bfloor{\frac{np}{2}}}^{n} x_{n,k} \sqrt{\frac{1}{2k+1}} f(x_{n,k})^2 \Delta_n 
    - \calO \left( \frac{1}{n} \right) \sum_{k = \bceil{np}}^n x_{n,k} \sqrt{\frac{1}{2k+1}} f(x_{n,k})^2 \Delta_n. \label{eq:jan10_demoivre_f} 
\end{align}

To assist the flow of the proof and reduce redundancy, we use a technical variant of the squeeze theorem. We have shown
\[
\text{Equation (\ref{eq:jan10_demoivre_d})} \leq C_n \leq \text{Equation (\ref{eq:jan10_demoivre_f})}.
\]
Rather than prove Equations (\ref{eq:jan10_demoivre_d}) and (\ref{eq:jan10_demoivre_f}) have the same asymptotic bounds, separately, we combine the equations as
\begin{align*}
    |C_n| & \leq \max \left\{ \Big| \text{Equation (\ref{eq:jan10_demoivre_d})} \Big|, \Big| \text{Equation (\ref{eq:jan10_demoivre_f})} \right\} \\
    & \leq \Big| \text{Equation (\ref{eq:jan10_demoivre_d})} \Big| + \Big| \text{Equation (\ref{eq:jan10_demoivre_f})} \Big|
\end{align*}
by triangle inequality.
We continue the proof with 
\begin{align}
    |C_n| &=  \left| \sum_{k = \bfloor{\frac{np}{2}}}^n x_{n,k} \cdot g_{n,k} f(x_{n,k})^2 \Delta_n^2 \right|  \notag \\
    & \leq \max \left\{ \Big| \text{Equation (\ref{eq:jan10_demoivre_d})} \Big|, \Big| \text{Equation (\ref{eq:jan10_demoivre_f})} \Big| \right\} \notag \\
    & \leq \Big| \text{Equation (\ref{eq:jan10_demoivre_d})} \Big| + \Big| \text{Equation (\ref{eq:jan10_demoivre_f})} \Big| \notag \\
    & = \left| -\calO \left( \frac{1}{n} \right) \sum_{k = \bfloor{\frac{np}{2}}}^{\bfloor{np}} x_{n,k} \sqrt{\frac{1}{2k+1}} f(x_{n,k})^2 \Delta_n  
    + \sqrt{\frac{2}{\pi}} \sum_{k = \bfloor{\frac{np}{2}}}^n x_{n,k} \sqrt{\frac{1}{2k+1}} f(x_{n,k})^2 \Delta_n \right| \notag  \\
    & \quad \quad + \left| \sqrt{\frac{2}{\pi}} \sum_{k = \bfloor{\frac{np}{2}}}^{n} x_{n,k} \sqrt{\frac{1}{2k+1}} f(x_{n,k})^2 \Delta_n 
    - \calO \left( \frac{1}{n} \right) \sum_{k = \bceil{np}}^n x_{n,k} \sqrt{\frac{1}{2k+1}} f(x_{n,k})^2 \Delta_n \right| \notag \\
    & \leq |D_n| + |E_n| + 2 \sqrt{\frac{2}{\pi}} \cdot |F_n| \label{eq:jan10_demoivre_g}
\end{align}
by triangle inequality, where we define
\begin{align*}
    D_n = \calO \left( \frac{1}{n} \right) \sum_{k = \bfloor{\frac{np}{2}}}^{\bfloor{np}} x_{n,k} \sqrt{\frac{1}{2k+1}} f(x_{n,k})^2 \Delta_n,
\end{align*}
\begin{align*}
    E_n = \calO \left( \frac{1}{n} \right) \sum_{k = \bceil{np}}^n x_{n,k} \sqrt{\frac{1}{2k+1}} f(x_{n,k})^2 \Delta_n,
\end{align*}
\begin{align*}
    F_n = \sum_{k = \bfloor{\frac{np}{2}}}^n x_{n,k} \sqrt{\frac{1}{2k+1}} f(x_{n,k})^2 \Delta_n.
\end{align*}
Consider the first summation in Equation (\ref{eq:jan10_demoivre_g}). We have
\begin{align*}
    |D_n| & = \calO \left( \frac{1}{n} \right) \left| \sum_{k = \bfloor{\frac{np}{2}}}^{\bfloor{np}} x_{n,k} \sqrt{\frac{1}{2k+1}} f(x_{n,k})^2 \Delta_n \right| \\
    & \leq \calO \left( \frac{1}{n^{1.5}} \right) \sum_{k = \bfloor{\frac{np}{2}}}^{\bfloor{np}} | x_{n,k}|  f(x_{n,k})^2 \Delta_n \\ 
    & \leq \calO \left( \frac{1}{n} \right) \sum_{k = \bfloor{\frac{np}{2}}}^{\bfloor{np}} f(x_{n,k})^2 \Delta_n \\
    & = \calO \left( \frac{1}{n} \right) 
\end{align*}
which follows by Lemma \ref{lem:instead_of_integrals}.6. 
Identical reasoning follows to upper bound the second summation in Equation (\ref{eq:jan10_demoivre_g}):
\begin{align*}
    |E_n| & = \calO \left( \frac{1}{n} \right) \left| \sum_{k = \bceil{np}}^n x_{n,k} \sqrt{\frac{1}{2k+1}} f(x_{n,k})^2 \Delta_n \right| \\
    & \leq \calO \left( \frac{1}{n} \right)
\end{align*}
by Lemma \ref{lem:instead_of_integrals}.7.

\end{paragraph}

\begin{paragraph}{Step 3: Handle $np$ may not be an integer.}

Until this point in the proof, we have demonstrated that the magnitude of the objective is bounded by $\calO \left( \frac{1}{n} \right) + |C_n|$ and that $|C_n| \leq \calO \left( \frac{1}{n} \right) + |F_n|$ where 
\[
    F_n = \sum_{k = \bfloor{\frac{np}{2}}}^n x_{n,k} \sqrt{\frac{1}{2k+1}} f(x_{n,k})^2 \Delta_n.
\]
Our aim is to bound $|F_n| \leq \calO \left( \frac{1}{n} \right)$. To accomplish this, in this step, we pair the terms at $k= np-u$ and $k=np+u$ for $u \in [0, \bfloor{\frac{np}{2}}]$, using a change of variables, to yield a summation similar to 
\[
    \sum_{u = 0}^{\bfloor{\frac{np}{2}}} \left( \frac{u}{\sqrt{npq}} \right) f \left(\frac{u}{\sqrt{npq}} \right)^2 \left( \sqrt{\frac{1}{np+u}} - \sqrt{\frac{1}{np-u}} \right) \Delta_n
\]
(see Equation (\ref{eq:u_ints_7}) below). We first show that the upper-tail is exponentially small. We then handle the nuance by which $np$ may not be an integer. This possibility does not affect the convergence rate, nor the intuition behind this change-of-variables. The reader may skip from Equation (\ref{eq:change_of_vars_1}) to Equation (\ref{eq:u_ints_7}) without losing the flow of the proof. The step concludes by bounding $|F_n| \leq \calO \left( \frac{1}{n} \right) + |L_n|$ where $L_n$ is a summation that covers $u \in [0, \bfloor{\frac{np}{2}}]$ and includes a factor similar to $\left( \sqrt{\frac{1}{np+u}} - \sqrt{\frac{1}{np-u}} \right)$ in the objective.
We proceed as follows.
\begin{align}
    |F_n| & = \left| \sum_{k = \bfloor{\frac{np}{2}}}^{n} x_{n,k} \sqrt{\frac{1}{k+0.5}} f(x_{n,k})^2 \Delta_n \right| \notag \\
    & \leq \left| \sum_{k = \bceil{\frac{3np}{2}}+1}^{n} x_{n,k} \sqrt{\frac{1}{k+0.5}} f(x_{n,k})^2 \Delta_n \right| 
    + \left| \sum_{k = \bfloor{\frac{np}{2}}}^{\bceil{\frac{3np}{2}}} x_{n,k} \sqrt{\frac{1}{k+0.5}} f(x_{n,k})^2 \Delta_n \right| \label{eq:split_apart}
\end{align}
by triangle inequality. Notice that for the first summation of Equation (\ref{eq:split_apart}),
\begin{align*}
    & \left| \sum_{k = \bceil{\frac{3np}{2}}+1}^{n} x_{n,k} \sqrt{\frac{1}{k+0.5}} f(x_{n,k})^2 \Delta_n \right| \\
    & \leq \sum_{k = \bceil{\frac{3np}{2}}+1}^{n} |x_{n,k}| \sqrt{\frac{1}{k+0.5}} f(x_{n,k})^2 \Delta_n \\
    & = \Theta(n) \Theta(\sqrt{n}) \Theta \left( \frac{1}{\sqrt{n}} \right) \calO \left( e^{-\Theta(n)} \right) \Theta \left( \frac{1}{\sqrt{n}} \right) \\
    & = \calO \left( e^{-\Theta(n)} \right) 
\end{align*}
by triangle inequality. Hence, we focus on the range $k \in \left[ \bfloor{\frac{np}{2}}, \bceil{\frac{3np}{2}} \right]$ in the second summation of Equation (\ref{eq:split_apart}):
\begin{align}
    & \sum_{k = \bfloor{\frac{np}{2}}}^{\bceil{\frac{3np}{2}}} x_{n,k} \sqrt{\frac{1}{k+0.5}} f(x_{n,k})^2 \Delta_n \notag \\
    & = \sum_{k = \bfloor{\frac{np}{2}}}^{\bfloor{np}} \left( \frac{k-np}{\sqrt{npq}} \right) \sqrt{\frac{1}{k+0.5}} f \left(\frac{k-np}{\sqrt{npq}} \right)^2 \Delta_n \notag \\
    & \quad \quad + \sum_{k = \bceil{np}}^{\bceil{\frac{3np}{2}}} \left( \frac{k-np}{\sqrt{npq}} \right) \sqrt{\frac{1}{k+0.5}} f \left(\frac{k-np}{\sqrt{npq}} \right)^2 \Delta_n. \label{eq:change_of_vars_1} 
\end{align}

For the first line of Equation (\ref{eq:change_of_vars_1}) we make the change of variables $u = \bfloor{np}-k$, which yields
\begin{align}
    & \sum_{u = 0}^{\bfloor{np}-\bfloor{\frac{np}{2}}} \left( \frac{\bfloor{np}-u-np}{\sqrt{npq}} \right) \sqrt{\frac{1}{\bfloor{np}-u+0.5}} 
    f \left( \frac{\bfloor{np}-u-np}{\sqrt{npq}} \right)^2 \Delta_n. \label{eq:u_ints_2}
\end{align}
Suppose that $np = t_n + b_n$ where $t_n \in \mathbb{N}$ and $b_n \in [0,1)$. Then Equation (\ref{eq:u_ints_2}) is
\begin{align}
    & - \sum_{u = 0}^{\bfloor{np}-\bfloor{\frac{np}{2}}} \left( \frac{u+b_n}{\sqrt{npq}} \right) \sqrt{\frac{1}{t_n-u+0.5}} ~f \left( \frac{u+b_n}{\sqrt{npq}} \right)^2 \Delta_n, \label{eq:u_ints_3}
\end{align}
making use of the fact that $f$ is an even function.
For the second line of Equation (\ref{eq:change_of_vars_1}) we make the change of variables $u = k - \bceil{np}$, which yields
\begin{align}
    & \sum_{k = 0}^{\bceil{\frac{3np}{2}} - \bceil{np}} \left( \frac{u+\bceil{np}-np}{\sqrt{npq}} \right) \sqrt{\frac{1}{\bceil{np}+u+0.5}} 
    f \left(\frac{u+\bceil{np}-np}{\sqrt{npq}} \right)^2 \Delta_n \notag \\
    & = \sum_{k = 0}^{\bceil{\frac{3np}{2}} - \bceil{np}} \left( \frac{u+1-b_n}{\sqrt{npq}} \right) \sqrt{\frac{1}{t_n+u+1.5}} 
    f \left(\frac{u+1-b_n}{\sqrt{npq}} \right)^2 \Delta_n.
    \label{eq:u_ints_4}
\end{align}
Let 
\[
    \tau_n = \min\left\{\bfloor{np}-\bfloor{\frac{np}{2}}, \bceil{\frac{3np}{2}} - \bceil{np} \right\}
\]
which is near $\frac{np}{2}$ (and is exact, if $np$ is an integer).
Putting together Equations (\ref{eq:u_ints_3}) and (\ref{eq:u_ints_4}) yields
\begin{align}
    G_n + H_n + I_n \pm \calO \left( e^{-\Theta(n)} \right) \label{eq:u_ints_5}
\end{align}
where we define
\begin{align*}
    G_n = \Delta_n^2 (1-b_n) \sum_{u=0}^{\tau_n} \frac{1}{\sqrt{t_n + u + 1.5}} f \left( \frac{u+1-b_n}{\sqrt{npq}} \right)^2,
\end{align*}
\begin{align*}
    H_n = - \Delta_n^2 b_n \sum_{u=0}^{\tau_n} \frac{1}{\sqrt{t_n - u + 0.5}} f \left( \frac{u+b_n}{\sqrt{npq}} \right)^2,
\end{align*}
\begin{align*}
    I_n = \Delta_n^2 \sum_{u=0}^{\tau_n} & \frac{u}{\sqrt{t_n + u + 1.5}} f \left( \frac{u+1-b_n}{\sqrt{npq}} \right)^2 
    - \frac{u}{\sqrt{t_n - u + 0.5}} f \left( \frac{u+b_n}{\sqrt{npq}} \right)^2.
\end{align*}
Note that the exponentially small term in Equation (\ref{eq:u_ints_5}) arises since there may be terms in-between $\tau_n$ and either $\bfloor{np}-\bfloor{\frac{np}{2}}$ or $\bceil{\frac{3np}{2}} - \bceil{np}$. Recall that these terms are near $\frac{np}{2}$. Plugging in $u = \Theta(n)$ for either Equations (\ref{eq:u_ints_3}) or (\ref{eq:u_ints_4}) yields $-\calO \left( e^{-\Theta(n)} \right)$ and $\calO \left( e^{-\Theta(n)} \right)$ respectively. 

The proof continues by bounding $|G_n|$, $|H_n|$, and $|I_n|$ by $\calO \left( \frac{1}{n} \right)$ each and respectively.
Consider the first summation of Equation (\ref{eq:u_ints_5}). Since $t_n = \Theta(n)$ by definition, we have
\begin{align*}
    |G_n| & = \calO \left( \frac{1}{n} \right) \left| \sum_{u=0}^{\tau_n}  f \left( \frac{u+1-b_n}{\sqrt{npq}} \right)^2 \Delta_n \right| \\
    & \leq \calO \left( \frac{1}{n} \right) \sum_{u=0}^{\tau_n}  f \left( \frac{u}{\sqrt{npq}} \right)^2 \Delta_n \\
    & = \calO \left( \frac{1}{n} \right)
\end{align*}
by triangle inequality and Lemma \ref{lem:instead_of_integrals}.8, making use of the fact that $e^{-y^2}$ is monotone decreasing for $y \geq 0$. A similar argument holds for $H_n$. Now consider the third summation of Equation (\ref{eq:u_ints_5}). We get
\begin{align}
    I_n = J_n + K_n + L_n \label{eq:u_ints_6}
\end{align}
where we define
\begin{align*}
    J_n & = \Delta_n^2 \sum_{u=0}^{\tau_n} \frac{u}{\sqrt{t_n + u + 1.5}} 
    \left( f \left( \frac{u+1-b_n}{\sqrt{npq}} \right)^2 - f \left( \frac{u}{\sqrt{npq}} \right)^2 \right),
\end{align*}
\begin{align*}
    K_n & = - \Delta_n^2 \sum_{u=0}^{\tau_n} \frac{u}{\sqrt{t_n - u + 0.5}} 
    \left( f \left( \frac{u+b_n}{\sqrt{npq}} \right)^2 - f \left( \frac{u}{\sqrt{npq}} \right)^2 \right),
\end{align*}
\begin{align*}
    L_n & = \Delta_n^2 \sum_{u = 0}^{\tau_n} u f \left(\frac{u}{\sqrt{npq}} \right)^2 
    \left( \frac{1}{\sqrt{t_n+u+1.5}} - \frac{1}{\sqrt{t_n-u+0.5}} \right) . 
\end{align*}
Consider the first summation of Equation (\ref{eq:u_ints_6}). We have that $|J_n|$ is equivalent to
\begin{align*}
    & \calO \left( \frac{1}{n^{1.5}} \right) \left| \sum_{u=0}^{\tau_n} u \left( f \left( \frac{u+1-b_n}{\sqrt{npq}} \right)^2 - f \left( \frac{u}{\sqrt{npq}} \right)^2 \right) \right| \\
    & \leq \calO \left( \frac{1}{n^{1.5}} \right) \sum_{u=0}^{\tau_n} u \left( f \left( \frac{u}{\sqrt{npq}} \right)^2 - f \left( \frac{u+1}{\sqrt{npq}} \right)^2 \right) \\
    & = \calO \left( \frac{1}{n^{1.5}} \right) 
    \sum_{u=0}^{\tau_n} \left( u f \left( \frac{u}{\sqrt{npq}} \right)^2 - (u+1) f \left( \frac{u+1}{\sqrt{npq}} \right)^2 \right) \notag \\
    &  \quad \quad + \calO \left( \frac{1}{n} \right) \sum_{u=0}^{\tau_n} f \left( \frac{u+1}{\sqrt{npq}} \right)^2 \Delta_n \\
    & = \calO \left( \frac{1}{n} \right) \sum_{u=0}^{\tau_n} f \left( \frac{u}{\sqrt{npq}} \right)^2 \Delta_n - \calO \left( \frac{1}{n^{1.5}} \right) \\
    & = \calO \left( \frac{1}{n} \right).
\end{align*}
where the second line is by triangle inequality and since $e^{-y^2}$ is decreasing for $y>0$; the last line is by Lemma \ref{lem:instead_of_integrals}.8. A similar argument holds for $K_n$. 
\end{paragraph}

\begin{paragraph}{Step 4: Handle $\frac{1}{\sqrt{k}}$ using paired terms.}
    
Now consider the third summation of Equation (\ref{eq:u_ints_6}):
\begin{align}
    L_n & = \sum_{u = 0}^{\tau_n} \left( \frac{u}{\sqrt{npq}} \right) f \left(\frac{u}{\sqrt{npq}} \right)^2 
    \left( \frac{1}{\sqrt{t_n+u+1.5}} - \frac{1}{\sqrt{t_n-u+0.5}} \right) \Delta_n. \label{eq:u_ints_7}
\end{align}
We next simplify the internal difference in this summation. Let $a = t_n+u+1.5$ and $b = t_n-u+0.5$. we get: 
\begin{align}
    & \frac{1}{\sqrt{a}} - \frac{1}{\sqrt{b}} \notag \\
    & = \frac{\sqrt{b} - \sqrt{a}}{\sqrt{ab}} \cdot \frac{\sqrt{b} + \sqrt{a}}{\sqrt{b} + \sqrt{a}} \notag \\
    & = \frac{b-a}{b \sqrt{a} + a \sqrt{b}} \notag \\
    & = \frac{-(2u+1)}{(t_n-u+0.5)\sqrt{(t_n+u+1.5)} + (t_n+u+1.5)\sqrt{(t_n-u+0.5)}} \notag \\
    & = -(2u+1) \cdot \calO \left( \frac{1}{n^{1.5}} \right). \label{eq:u_ints_subtract_sqrt_k}
\end{align}
Returning to Equation (\ref{eq:u_ints_7}), which is upper-bounded by zero, we get
\begin{align*}
    & - \calO \left( \frac{1}{n^{1.5}} \right) \sum_{u = 0}^{\tau_n} \left( \frac{u(2u+1)}{\sqrt{npq}} \right) f \left(\frac{u}{\sqrt{npq}} \right)^2 \Delta_n \\
    & = - \calO \left( \frac{1}{n} \right) \sum_{u = 0}^{\tau_n} \left( \frac{u}{\sqrt{npq}} \right)^2 f \left(\frac{u}{\sqrt{npq}} \right)^2 \Delta_n 
    - \calO \left( \frac{1}{n^{1.5}} \right) \sum_{u = 0}^{\tau_n} \left( \frac{u}{\sqrt{npq}} \right) f \left(\frac{u}{\sqrt{npq}} \right)^2 \Delta_n \\
    & = -\calO \left( \frac{1}{n} \right)
\end{align*}
by Lemmas \ref{lem:instead_of_integrals}.9 and \ref{lem:instead_of_integrals}.10. Hence, we get that $|L_n| \leq \calO \left( \frac{1}{n} \right)$.

This concludes the proof of Lemma \ref{lem:real_expected_demoivre}.
\end{paragraph}
\end{proof}







%% file: EC_appendix/apx_page_5c.tex


\subsection{Standardized Binomial and Gaussian Expectations}
\label{apx:normalized_expectations}



This subsection describes technical lemmas about the convergence of certain sequences of summations. These are used to support the lemmas in Appendix \ref{sec:expected_in_full}.


\setcounter{lem}{19}
\begin{lem}
Let $p \in (0, 1)$, $q=1-p$, and $S_n \sim Bin(n,p)$. Then
    \[
    \sum_{k=\bfloor{\frac{np}{2}}}^n x_{n,k}^2 \Pr(S_n = k) = \Theta(1).
    \]
    \label{lem:limit_analysis_2}
\end{lem}

\begin{proof}
The lemma is implied by the following:
\[
    \lim_{n \rightarrow \infty} \sum_{k=\bfloor{\frac{np}{2}}}^n x_{n,k}^2 \Pr(S_n = k) = 1.
\]
Let $X_n = \frac{S_n - np}{\sqrt{npq}}$. Then
we have that
\begin{align*}
    \sum_{k=\bfloor{\frac{np}{2}}}^n x_{n,k}^2 \Pr(S_n = k) = \mathbb{E}[X_n^2] - \calO \left( e^{-\Theta(n)} \right)
\end{align*}
by Hoeffding's inequality (Proposition \ref{prop:hoeffding}).
We know that
\[
\mathbb{E}[S_n^2] = n^2 p^2 + npq.
\]
This leads us to the conclusion that
\begin{align*}
    \mathbb{E}[X_n^2] & = \frac{1}{npq} \mathbb{E}[S_n^2 - 2 S_n np + n^2 p^2] \\
    & = \frac{1}{npq} \left( \mathbb{E}[S_n^2] - 2np \mathbb{E}[S_n] + n^2 p^2 \right) \\
    & = \frac{1}{npq} \left( (n^2 p^2 + npq) - 2np (np) + n^2 p^2 \right) \\
    & = 1.
\end{align*}
This concludes the proof of Lemma \ref{lem:limit_analysis_2}.
\end{proof}

\setcounter{lem}{20}
\begin{lem}
Let $p \in (0, 1)$, $q=1-p$, and $S_n \sim Bin(n,p)$. Then we have
    \[
    \sum_{k=0}^n | x_{n,k} | \Pr(S_n = k) = \Theta(1). 
    \]
    \label{lem:limit_analysis_1}
\end{lem}

\begin{proof}
The lemma is implied by the following:
\[
    \lim\nolimits_{n \rightarrow \infty} \sum_{k=0}^n | x_{n,k} | \Pr(S_n = k) = \frac{2}{\sqrt{2 \pi}}.
\]
We do not assume that $np$ is an integer. Rather, suppose $np = t_n + b_n$ where $t_n \in \mathbb{N}$ and $b_n \in [0,1)$.
The objective equation is then equal to
\begin{align}
    & \sum_{k=0}^{\bfloor{np}} | x_{n,k} | \Pr(S_n = k) + \sum_{k=\bceil{np}}^n | x_{n,k} | \Pr(S_n = k) \notag \\
    & = - \sum_{k=0}^{\bfloor{np}} \frac{k-t_n-b_n}{\sqrt{npq}} \Pr(S_n = k) 
    + \sum_{k=\bceil{np}}^n \frac{k-t_n-b_n}{\sqrt{npq}} \Pr(S_n = k) \notag \\
    & = \frac{b_n}{\sqrt{npq}} \sum_{k=0}^{\bfloor{np}} \Pr(S_n = k) 
    - \frac{b_n}{\sqrt{npq}} \sum_{k=\bceil{np}}^{n} \Pr(S_n = k) \notag \\
    & \quad \quad + \sum_{k=0}^{\bfloor{\frac{np}{2}}-1} \left| \frac{k-t_n}{\sqrt{npq}} \right| \Pr(S_n = k) 
    + \sum_{k=\bfloor{\frac{np}{2}}}^n \left| \frac{k-t_n}{\sqrt{npq}} \right| \Pr(S_n = k) \notag \\
    & = \pm \calO \left( \frac{1}{\sqrt{n}} \right) + \sum_{k=\bfloor{\frac{np}{2}}}^n \left| \frac{k-t_n}{\sqrt{npq}} \right| \Pr(S_n = k) \label{eq:lem11_start}
\end{align}
where we partitioned the lower domain of $k \in [0, \frac{\bfloor{np}}{{2}})$ and realized that it is exponentially small by Hoeffding's inequality (Proposition \ref{prop:hoeffding}).
Next, we change the remaining summation into a more convenient form.
\begin{align}
    & \sum_{k=\bfloor{\frac{np}{2}}}^n \left| \frac{k-t_n}{\sqrt{npq}} \right| \Pr(S_n = k) \notag \\
    & = \frac{1}{\sqrt{npq}} \left( - \sum_{k=\bfloor{\frac{np}{2}}}^{\bfloor{np}} (k-t_n) \Pr(S_n = k) 
    + \sum_{k=\bceil{np}}^n (k-t_n) \Pr(S_n = k) \right) \notag \\
    & = \frac{1}{\sqrt{npq}} \left( \sum_{k=\bfloor{np}}^{\bfloor{np}+\bfloor{\frac{np}{2}}} (k-t_n) \Pr(S_n = k)  
    + \sum_{k=\bceil{np}}^n (k-t_n) \Pr(S_n = k) \right) \notag \\
    & = \frac{1}{\sqrt{npq}} \left(- \sum_{k=\bfloor{np}+\bfloor{\frac{np}{2}}+1}^{n} (k-t_n) \Pr(S_n = k)
    + 2 \sum_{k=\bceil{np}}^n (k-t_n) \Pr(S_n = k) \right)  \notag \\
    & = \frac{2}{\sqrt{npq}} \sum_{k=\bfloor{np}}^n (k-t_n) \Pr(S_n = k) - \calO \left( e^{-\Theta(n)} \right) \label{eq:limit1_proof_a}
\end{align}
by 
Hoeffding's inequality (Proposition \ref{prop:hoeffding}). Let $T = \sum_{k=\bfloor{np}}^n k \binom{n}{k} p^k q^{n-k}$. Next, we have
\begin{align}
    T & = np \sum_{k=\bfloor{np}}^{n} \binom{n-1}{k-1} p^{k-1} q^{n-k} \notag \\
    & = \frac{np}{q} \sum_{k=\bfloor{np}-1}^{n-1} \binom{n-1}{k} p^{k} q^{n-k}  \notag \\
    & = \frac{np}{q} \sum_{k=\bfloor{np}-1}^{n-1} \binom{n}{k} p^{k} q^{n-k} \left( 1 - \frac{k}{n} \right) \label{eq:limit1_proof_b} \\
    & = \frac{np}{q} \sum_{k=\bfloor{np}}^{n} \binom{n}{k} p^{k} q^{n-k} \left( 1 - \frac{k}{n} \right) \notag \\
    &  + \frac{np}{q}  \left( \binom{n}{\bfloor{np}-1} p^{\bfloor{np}-1} q^{n-(\bfloor{np}-1)} 
    \left(1 - \frac{(\bfloor{np}-1)}{n} \right)   - \binom{n}{n} p^n q^{n-n} \left(1 - \frac{n}{n} \right) \right) \label{eq:limit1_proof_c}
\end{align}
where in Equation (\ref{eq:limit1_proof_b}) we used the substitution
\[
    \binom{n-1}{k} = \frac{(n-1)!}{k! (n-1-k)!} = \frac{n!}{k! (n-k)!} \cdot \frac{n-k}{n}.
\]
Notice in Equation (\ref{eq:limit1_proof_c}) that
\begin{align*}
    & \binom{n}{\bfloor{np}-1} p^{\bfloor{np}-1} q^{n-(\bfloor{np}-1)} \left(1 - \frac{(\bfloor{np}-1)}{n} \right) \\
    & = \binom{n}{\bfloor{np}} \frac{\bfloor{np}}{n-\bfloor{np}+1} p^{\bfloor{np}-1} q^{n-\bfloor{np}+1}  
    \left( \frac{n-\bfloor{np}+1}{n} \right) \\
    & = q \left( \frac{\bfloor{np}}{np} \right) \cdot \binom{n}{\bfloor{np}} p^{\bfloor{np}} q^{n-\bfloor{np}} \\
    & = \frac{q}{ \sqrt{2 \pi n p q}} \left( 1 \pm \calO \left( \frac{1}{n} \right) \right)
\end{align*}
by Lemma \ref{lem:stirling_binom}, proved in Appendix \ref{apx:helper_lemmas} 

\setcounter{lem}{25}
\begin{lem}
Let $p \in (0,1)$ and $q = 1-p$. Then we have
    \[
    \binom{n}{\bfloor{np}} p^{\bfloor{np}} q^{n-\bfloor{np}} = 
    \frac{1}{\sqrt{2 \pi n p q}} \left( 1 \pm \calO \left( \frac{1}{n} \right) \right).
    \]
\end{lem}

This gets us
\[
    T = \frac{np}{q} \left( \frac{1}{2} - \frac{T}{n} \right) + 
    \sqrt{\frac{np}{2 \pi q}} \left( 1 \pm \calO \left( \frac{1}{n} \right) \right)
\]
using the fact that $\sum_{k=\bfloor{np}}^{n} \binom{n}{k} p^{k} q^{n-k} = \frac{1}{2}$. Hence,
\[
    T = \frac{np}{2} + 
    \left( 1 \pm \calO \left( \frac{1}{n} \right) \right)
\]
so that our objective from Equation (\ref{eq:limit1_proof_a}) becomes
\begin{align*}
    & \frac{2}{\sqrt{npq}} \left( T - \frac{\bfloor{np}}{2} \right) - \calO \left( e^{-\Theta(n)} \right) \\
    & = \frac{2}{\sqrt{2 \pi}} \left( 1 \pm \calO \left( \frac{1}{n} \right) \right) \xrightarrow[]{n \rightarrow \infty} \frac{2}{\sqrt{2 \pi}}
\end{align*}
as claimed. This concludes the proof of Lemma \ref{lem:limit_analysis_1}.
\end{proof}



The following lemma consists of ten equations that we prove are all $\Theta(1)$. Each equation is structured similarly and may be proved in almost an identical manner. Hence, for convenience and straightforwardness of this appendix, we pack all ten equations into the same lemma statement.

\setcounter{lem}{21}
\begin{lem}
Let $p \in (0,1)$ and $q = 1-p$.
Let 
\[
    \tau_n = \min\left\{\bfloor{np}-\bfloor{\frac{np}{2}}, \bceil{\frac{3np}{2}} - \bceil{np} \right\}
\]
which is near $\frac{np}{2}$ (and is exact, if $np$ is an integer).
Then the following equations are each $\Theta(1)$:

\begin{enumerate}
    \item 
    \[
    \sum_{k=0}^n | x_{n,k} | f(x_{n,k}) \Delta_n;
    \]
    \item 
    \[
    \sum_{u=0}^{\bceil{npq}} f\left( \frac{u}{\sqrt{npq}} \right)^2 \Delta_n;
    \]
    \item 
    \[
    \sum_{k=\bfloor{\frac{np}{2}}}^n x_{n,k}^2 f(x_{n,k}) \Delta_n;
    \]
    \item 
    \[
    \sum_{k=\bfloor{\frac{np}{2}}}^n x_{n,k}^2 f(x_{n,k})^2 \Delta_n;
    \]
    \item 
    \[
    \sum_{k=\bfloor{\frac{np}{2}}}^n | x_{n,k} | f(x_{n,k}) \Delta_n;
    \]
    \item
    \[
    \sum_{k = \bfloor{\frac{np}{2}}}^{\bfloor{np}} f(x_{n,k})^2 \Delta_n;
    \]
    \item
    \[
    \sum_{k = \bceil{np}}^{n} f(x_{n,k})^2 \Delta_n;
    \]
    \item 
    \[
    \sum_{u=0}^{\tau_n} f\left( \frac{u}{\sqrt{npq}} \right)^2 \Delta_n;
    \]
    \item 
    \[
    \sum_{u = 0}^{\tau_n} \left( \frac{u}{\sqrt{npq}} \right) f \left(\frac{u}{\sqrt{npq}} \right)^2 \Delta_n;
    \]
    \item
    \[
    \sum_{u = 0}^{\tau_n} \left(\frac{u}{\sqrt{npq}} \right)^2 f \left(\frac{u}{\sqrt{npq}} \right)^2 \Delta_n.
    \]

\end{enumerate}
\label{lem:instead_of_integrals}
\end{lem}

\begin{proof}

Each of these equations is proved using similar methods. For conciseness, in this proof, we will demonstrate only the proofs of Equations $5$, which is in $x_{n,k}$-format, and $11$, which is in $u$-format. These equations have the largest terms in the objective summation among the $x_{n,k}$- and $u$-format equations, respectively. Therefore, proving that both Equations $5$ and $11$ are $\Theta(1)$ entails the same for the remainder of the equations. Our method is summarized as follows.

It is clear that each of these summations are non-negative and concentrated around the mean $k=np$ or $u=0$ (depending on the format). For each equation and large enough $|k-np|$ or $u$ that are $\Omega(\sqrt{n})$, the term is decreasing in $|k-np|$ or $u$. Hence, we make use of the Maclaurin–Cauchy integral test for convergence. For smaller $|k-np|$ or $u$ terms that are $\calO(\sqrt{n})$, we demonstrate convergence using the definition of the Riemann integral.
We make use of the error function $erf(x) \equiv \frac{2}{\sqrt{\pi}} \int_0^x e^{-y^2} dy$ in these proofs; $erf(x) \in (0,1)$ for $x>0$. Then $erfc(x) \equiv 1-erf(x)$ is the complementary error function.

\begin{paragraph}{Step 1: Demonstrate convergence for Equation $5$.}
We do not assume that $np$ is an integer. Rather, suppose that $np = t_n + b_n$ where $t_n \in \mathbb{N}$ and $b_n \in [0,1)$.
We partition the objective equation into four regions, as follows:
\begin{align}
    & \sum_{k=\bfloor{\frac{np}{2}}}^n x_{n,k}^2 f(x_{n,k})^2 \Delta_n \notag \\
    & = \sum_{k \in \left[\bfloor{\frac{np}{2}}, n \right] ~\backslash~ \left[\bfloor{np} - \bfloor{npq}, \bceil{np} + \bceil{npq} \right]  } x_{n,k}^2 f(x_{n,k})^2 \Delta_n  \notag \\
    & \quad \quad + \sum_{k=\bfloor{np} - \bfloor{\sqrt{npq}}}^{\bceil{np} + \bceil{\sqrt{npq}}} x_{n,k}^2 f(x_{n,k})^2 \Delta_n \notag \\
    & \quad \quad + \sum_{k=\bceil{np} + \bceil{\sqrt{npq}}}^{\bceil{np} + \bceil{npq}} x_{n,k}^2 f(x_{n,k})^2 \Delta_n \notag \\
    & \quad \quad  + \sum_{k=\bfloor{np} - \bfloor{npq}}^{\bfloor{np} - \bfloor{\sqrt{npq}}} x_{n,k}^2 f(x_{n,k})^2 \Delta_n \label{eq:instead_5_proof_1}
\end{align}
The first summation of Equation (\ref{eq:instead_5_proof_1}) is 
\[
    \Theta(n) \Theta(n) \calO \left( e^{-\Theta(n)} \right) \Theta \left( \frac{1}{\sqrt{n}} \right) = \calO \left( e^{-\Theta(n)} \right).
\]
The second summation of Equation (\ref{eq:instead_5_proof_1}) converges to
\[
    \frac{1}{2 \pi} \int_{-1}^1 y^2 e^{-y^2} dy = \frac{\sqrt{\pi} e \cdot erf(1) -2}{4 \pi e} = \Theta(1)
\]
by definition of the Riemann integral. The third summation of Equation (\ref{eq:instead_5_proof_1}) is equivalent to
\begin{align*}
    & \sum_{k= \bceil{\sqrt{npq}}}^{\bceil{npq}} \left( \frac{k + \bceil{np} - np}{\sqrt{npq}} \right)^2 f\left( \frac{k + \bceil{np} - np}{\sqrt{npq}} \right)^2 \Delta_n \\
    & = \sum_{R = 1}^{\bceil{\sqrt{npq}}-1} \sum_{r=0}^{\bceil{\sqrt{npq}}-1} \left(\frac{R \bceil{\sqrt{npq}} + r + 1 - b_n}{\sqrt{npq}} \right)^2 
    f \left(\frac{R \bceil{\sqrt{npq}} + r  + 1 - b_n}{\sqrt{npq}} \right)^2 \Delta_n
\end{align*}
which is at most
\begin{align*}
    \sum_{R = 1}^{\bceil{\sqrt{npq}}-1} R^2 f(R)^2,
\end{align*}
where we plugged in $r = -1+b_n$ since $y^2 e^{-y^2}$ is monotone decreasing along $y \geq 1$. This is taken $\bceil{\sqrt{npq}}$ times and cancels out with $\Delta_n$. Furthermore, we used the fact that $\frac{R \bceil{\sqrt{npq}}}{\sqrt{npq}} \geq R$. By the integral test for convergence, the third summation of Equation (\ref{eq:instead_5_proof_1}) converges because
\[
    \frac{1}{2\pi} \int_1^{\infty} y^2 e^{-y^2} dy = \frac{e \sqrt{\pi} \cdot erfc(1) -2}{4e} = \Theta(1)
\]
converges. The fourth summation of (\ref{eq:instead_5_proof_1}) follows by similar reasoning. Hence, Equation $5$ converges; i.e., is $\Theta(1)$ as claimed.
\end{paragraph}

\begin{paragraph}{Step 2: Demonstrate convergence for Equation $11$.}
The proof follows almost identically to that of Equation 5 of this lemma.
Recall that $\tau_n \approx \frac{np}{2}$. We partition the objective into three regions:
\begin{align}
    & \sum_{u = 0}^{\tau_n} \left(\frac{u}{\sqrt{npq}} \right)^2 f \left(\frac{u}{\sqrt{npq}} \right)^2 \Delta_n \notag \\
    & = \sum_{u = 0}^{\bceil{\sqrt{npq}}-1} \left(\frac{u}{\sqrt{npq}} \right)^2 f \left(\frac{u}{\sqrt{npq}} \right)^2 \Delta_n  
    + \sum_{u = \bceil{npq}}^{\tau_n} \left(\frac{u}{\sqrt{npq}} \right)^2 f \left(\frac{u}{\sqrt{npq}} \right)^2 \Delta_n \notag \\
    & \quad \quad + \sum_{R = 1}^{\bceil{\sqrt{npq}}-1} \sum_{r=0}^{\bceil{\sqrt{npq}}-1} \left(\frac{R \bceil{\sqrt{npq}} + r}{\sqrt{npq}} \right)^2 
    f \left(\frac{R \bceil{\sqrt{npq}} + r}{\sqrt{npq}} \right)^2 \Delta_n
    \label{eq:instead_11_proof_1}
\end{align}
The first summation of Equation (\ref{eq:instead_11_proof_1}) converges to
\[
\frac{1}{2\pi} \int_0^1 y^2 e^{-y^2} dy = \frac{e \sqrt{\pi} \cdot erf(1)-2}{4e} = \Theta(1)
\]
by definition of the Riemann integral. The second summation of Equation (\ref{eq:instead_11_proof_1}) is 
\[
    \Theta(n) \Theta(n) \calO \left( e^{-\Theta(n)} \right) \Theta \left( \frac{1}{\sqrt{n}} \right) = \calO \left( e^{-\Theta(n)} \right).
\]
The third summation of Equation (\ref{eq:instead_11_proof_1}) is at most
\begin{align}
    \sum_{R = 1}^{\bceil{\sqrt{npq}}-1} R^2 f(R^2) \label{eq:instead_11_proof_2}
\end{align}
where we plugged in $r = 0$ since $y^2 e^{-y^2}$ is monotone decreasing along $y \geq 1$. This is taken $\bceil{\sqrt{npq}}$ times and cancels out with $\Delta_n$. Furthermore, we used the fact that $\frac{R \bceil{\sqrt{npq}}}{\sqrt{npq}} \geq R$. By the integral test for convergence, the third summation of Equation (\ref{eq:instead_11_proof_2}) converges because
\[
    \frac{1}{2\pi} \int_1^{\infty} y^2 e^{-y^2} dy = \frac{e \sqrt{\pi} \cdot erfc(1) -2}{4e} = \Theta(1)
\]
converges. Hence, Equation $11$ converges (i.e., is $\Theta(1)$), as claimed.
\end{paragraph}
This concludes the proof of Lemma \ref{lem:instead_of_integrals}.
\end{proof}

%% file: EC_appendix/apx_page_6.tex
\newpage

\section{Stirling, Wallis, and Central Binomial Coefficients}
\label{apx:stirling_and_wallis}


Stirling's approximation for the factorial is as follows. 

\setcounter{prop}{1}
\begin{prop}[Stirling's approximation]
    Stirling's approximation says that $n! \sim \sqrt{2 \pi n} \left( \frac{n}{e} \right)^n$. 
    More precisely, $\forall n \geq 1$, 
    \[
    \sqrt{2 \pi n} \left( \frac{n}{e} \right)^n e^{\frac{1}{12n+1}} < n! < \sqrt{2 \pi n} \left( \frac{n}{e} \right)^n e^{\frac{1}{12n}}.
    \]
\end{prop}

Plugging in Stirling's approximation for the central binomial coefficient can demonstrate the asymptotic growth:
\begin{equation}
    \binom{2n}{n} \sim \frac{2^{2n}}{\sqrt{n \pi}}. 
\end{equation}
The error of this approximation is known to be $\mathcal{O}(\frac{1}{n})$ \citep{luke1969special}. \footnote{For an early history of the factorial, see \citet{dutka1991early}.} For completeness and usefulness in our main theorem, we demonstrate one proof for this asymptotic growth in the following lemma. This argument uses the Wallis product for $\pi$ \citep{wallis1656} \footnote{The Wallis product states that \[
\frac{\pi}{2} = \prod_{n=1}^\infty \frac{4n^2}{4n^2 -1} = \prod_{n=1}^\infty \left( \frac{2n}{2n-1} \cdot \frac{2n}{2n+1} \right).
\]} and is transposed from lecture notes by  \citet{galvin18wallis}.

\setcounter{lem}{23}
\begin{lem}
\[
    \sqrt{\frac{2n}{2n+1}}  \sqrt{\frac{2}{\pi (2n+1)}} \leq \frac{2^{2n}}{(2n+1) \binom{2n}{n}} \leq \sqrt{\frac{2}{\pi (2n+1)}}.
\]
    \label{lem:wallis}
\end{lem}

\begin{proof}
For each $n \geq 0$, define $S_n = \int_0^{\pi/2} \sin^n x dx$. We have
\[
S_0 = \frac{\pi}{2}, \quad S_1 = \int_{0}^{\pi/2} \sin x ~dx = 1,
\]
and for $n \geq 2$ we get from integration by parts (taking $u = \sin^{n-1} x$ and $dv = \sin x dx$, so that $du = (n-1) \sin^{n-2} x \cos x dx$ and $v = - \cos x$) that
\begin{align*}
    S_n & = (\sin^{n-1} x)(- cos x)|^{\pi/2}_{x=0}  \notag \\
    & \quad \quad - \int^{\pi/2}_0 -(n-1) \cos x \sin^{n-2} x \cos x dx \\
    & = (n-1) \int^{\pi/2}_0 \cos^2 x sin^{n-2} x dx \\
    & = (n-1) \int^{\pi/2}_0  (1 - \sin^2 x) \sin^{n-2} x dx \\
    & = (n-1) S_{n-2} - (n-1) S_n.
\end{align*}
This leads to the recurrence relation:
\[
    S_n = \frac{n-1}{n} S_{n-2} \quad \text{for } n \geq 2.
\]
Iterating the recurrence relation until the initial conditions are reached, we get that
\[
    S_{2n} = \left( \frac{2n-1}{2n} \right) \left( \frac{2n-3}{2n-2} \right) \ldots \left( \frac{3}{4} \right) \left( \frac{1}{2} \right) \frac{\pi}{2}
\]
and
\[
    S_{2n+1} = \left( \frac{2n}{2n+1} \right) \left( \frac{2n-2}{2n-1} \right) \ldots \left( \frac{4}{5} \right) \left( \frac{2}{3} \right) 1.
\]
Taking the ratio of these two identities and rearranging gets us that $\frac{\pi}{2}$ is equivalent to
\[
    \left( \frac{2}{1} \right) \left( \frac{2}{3} \right) \left( \frac{4}{3} \right) \left( \frac{4}{5} \right) \ldots \left( \frac{2n}{2n-1} \right) \left( \frac{2n}{2n+1} \right) \frac{S_{2n}}{S_{2n+1}}.
\]
For ease of notation, define
\[
\mathcal{W}_n = \left( \frac{2}{1} \right) \left( \frac{2}{3} \right) \left( \frac{4}{3} \right) \left( \frac{4}{5} \right) \ldots \left( \frac{2n}{2n-1} \right) \left( \frac{2n}{2n+1} \right)
\]
as the first $n$ terms of Wallis' product, so that $\frac{\pi}{2} = \mathcal{W}_n \frac{S_{2n}}{S_{2n+1}}$. Now, since $0 \leq \sin x \leq 1$ on $[0, \pi/2]$, we have also
\[
    0 \leq \sin^{2n+1} x \leq \sin^{2x} x \leq \sin^{2n-1} x,
\]
and so, integrating and using the recurrence relation, we get
\[
    0 \leq S_{2n+1} \leq S_{2n} \leq S_{2n-1} = \frac{2n+1}{2n} S_{2n+1}
\]
and so
\[
    1 \leq \frac{S_{2n}}{S_{2n+1}} \leq 1 + \frac{1}{2n}.
\]
Hence, $1 \leq \frac{\pi}{2 \mathcal{W}_n} \leq 1 + \frac{1}{2n}$; equivalently, $\frac{2}{\pi} \geq \mathcal{W}_n \geq \frac{2 (2n)}{\pi (2n+1)}$. Wallis' formula can now be used to estimate the central binomial coefficient:
\begin{align*}
    \binom{2n}{n}
    & = \frac{(2n) (2n-1) (2n-2) \ldots (3) (2) (1)}{(n) (n-1) \ldots (2) (1) \cdot (n) (n-1) \ldots (2) (1)} \\
    & = 2^n \frac{(2n) (2n-1) (2n-2) \ldots (3) (2) (1)}{(n) (n-1) \ldots (2) (1)} \\
    & = 2^{2n} \frac{(2n) (2n-1) (2n-2) \ldots (3) (2) (1)}{(2n) (2n-2) \ldots (4) (2)} \\
    & = \frac{2^{2n}}{\sqrt{2n+1}} \sqrt{ \frac{(2n+1)(2n-1)^2(2n-3)^2 \ldots (3)^2(1)}{(2n)^2(2n-2)^2 \ldots (4)^2(2)^2}} \\
    & = \frac{2^{2n}}{\sqrt{\mathcal{W}_n (2n+1)}}.
\end{align*}
Therefore:
\begin{align*}
    \sqrt{\frac{2n}{2n+1}}  \sqrt{\frac{2}{\pi (2n+1)}} & \leq \frac{2^{2n}}{(2n+1) \binom{2n}{n}} = \sqrt{\frac{\mathcal{W}_n}{2n+1}} \leq \sqrt{\frac{2}{\pi (2n+1)}}.
\end{align*}
This concludes the proof of Lemma \ref{lem:wallis}.
\end{proof}

%% file: EC_appendix/apx_page_7.tex
\newpage

\section{Concentration Inequality Lemmas}
\label{apx:concentration_inequalities}



Throughout this paper, we employ several concentration inequalities of the binomial distribution $\binom{n}{k} p^k (1-p)^{n-k}$ and symmetric-multinomial distribution $\binom{n}{\frac{n}{2}-q, \frac{n}{2}-q, q, q} \pi_1^{n-2q} \pi_3^{2q}$. In particular, we consider the probability that a binomial distribution is centered or one-sided about its mean, as well as its expectation, conditioned on these events. This is stated generally in Lemma \ref{lem:bin_theorems_approx} and more specifically in Lemma \ref{lem:new_comb} when $p=\frac{1}{2}$. 
\footnote{Note that in this paper's primary lemmas (Lemmas \ref{lem:sub_12} and \ref{lem:sub_12_odd}) we combined Lemmas \ref{lem:new_comb} and \ref{lem:bin_theorems_approx} for conciseness. In Lemma \ref{lem:new_comb}, below, we provide more detail about the $p=\frac{1}{2}$ case that is generalized as $\pm \calO \left( \frac{1}{\sqrt{n}} \right)$ and $\pm \calO(\sqrt{n})$ in the statement of Lemma \ref{lem:bin_theorems_approx}.}
Furthermore, we identify a relationship between the symmetric-multinomial and the square of the binomial distribution (Proposition \ref{prop:binom}). We go on to demonstrate concentration bounds of this distribution using Hoeffding's inequality, in Proposition \ref{prop:hoeffding}, and smoothed analysis techniques in Lemma \ref{lem:prob_bounds}.

\setcounter{lem}{10}
\begin{lem}
Let $p, b \in (0,1)$ and $S_n \sim Bin(n,p)$. Then
\begin{align*}
    \Pr(S_n \leq bn) = \begin{cases}
        1 - \calO \left( e^{-\Theta(n)} \right), & b > p \\
        \frac{1}{2} \pm \calO \left( \frac{1}{\sqrt{n}} \right),&  b = p \\
        \calO \left( e^{-\Theta(n)} \right), & b < p
    \end{cases}
\end{align*}
and
\begin{align*}
    \mathbb{E}[S_n \cdot \mathbbm{1}\{S_n \leq bn\}] = \begin{cases}
        np - \calO \left( e^{-\Theta(n)} \right), & b > p \\
        \frac{np}{2} \pm \calO \left( \sqrt{n} \right),&  b = p \\
        \calO \left( e^{-\Theta(n)} \right), & b < p.
    \end{cases}
\end{align*}
\label{lem:bin_theorems_approx}
\end{lem}

\begin{proof}
First, consider the probability version of the lemma.
The cases for $b < p$ and $b > p$ hold by a direct application of Hoeffding's inequality. The $b=p$ case holds by the Berry-Essen theorem (see e.g., \citet{durrett2019probability}). Specifically, let $\Phi(x)$ denote the cumulative distribution function of a unit Gaussian. Then
\begin{align*}
    & \left| \Pr(S_n \leq np) - \Phi(0) \right|
    = \left| \Pr \left( \frac{S_n-np}{\sqrt{np(1-p)}} \leq 0 \right) - \frac{1}{2} \right|
    \leq \calO \left( \frac{1}{\sqrt{n}} \right).
\end{align*}

We finish the proof by proving the expected-value version of the lemma. We have
\begin{align*}
    & \sum_{k=0}^{\bfloor{bn}} \binom{n}{k} p^k (1-p)^{n-k} k \\
    & = np \sum_{k=1}^{\bfloor{bn}} \binom{n-1}{k-1} p^{k-1} (1-p)^{n-k} \\
    & = np \sum_{k=0}^{\bfloor{bn}-1} \binom{n-1}{k} p^{k} (1-p)^{n-1-k}.
\end{align*}
This is $np$ multiplied by the same probability as above, yielding our claim.
This concludes the proof of Lemma \ref{lem:bin_theorems_approx}.
\end{proof}

The following identities are used to prove Lemma \ref{lem:new_comb}, below.

\setcounter{lem}{26}
\begin{lem}
The following identities hold:
\begin{enumerate}
    \item \[
        \sum_{\beta=0}^{q-1} \binom{2q-1}{\beta}\frac{1}{2^{2q-1}} = \frac{1}{2};
    \]
    \item \[
        \sum_{\beta=0}^{q-1} \beta \binom{2q-1}{\beta} \frac{1}{2^{2q-1}} = \left( \frac{2q-1}{4} \right) - \frac{q}{2^{2q}} \binom{2q-1}{q-1};
    \]
    \item \begin{equation*}
        \sum_{\beta=q+1}^{2q} \binom{2q}{\beta} \frac{1}{2^{2q}} = \frac{1}{2} - \frac{1}{2^{2q}} \binom{2q-1}{q-1}.
    \end{equation*}
\end{enumerate}
\label{lem:suppl_comb}
\end{lem}

\begin{proof}

We take these equations one at a time.

\begin{paragraph}{Equation 1.}
By symmetry, we have that 
\begin{equation*}
    \sum_{\beta=0}^{q-1} \binom{2q-1}{\beta} \frac{1}{2^{2q-1}} = \frac{2^{2q-2}}{2^{2q-1}} = \frac{1}{2}.
\end{equation*}
\end{paragraph}

\begin{paragraph}{Equation 2.}

\begin{align*}
    \sum_{\beta=0}^{q-1} \binom{2q}{\beta} \frac{1}{2^{2q}}
    & = \frac{1}{2^{2q}} \left( \sum_{\beta=1}^{q-1} \binom{2q-1}{\beta}  + \sum_{\beta=1}^{q-1} \binom{2q-1}{\beta-1} + 1 \right) \\
    & = \frac{1}{2^{2q}} \left( \sum_{\beta=0}^{q-1} \binom{2q-1}{\beta}  + \sum_{\beta=0}^{q-2} \binom{2q-1}{\beta} \right) \\
    & = \frac{1}{2^{2q}} \left( 2 \sum_{\beta=0}^{q-1} \binom{2q-1}{\beta} - \binom{2q-1}{q-1} \right) \\
    & = \frac{1}{2} - \frac{1}{2^{2q}} \binom{2q-1}{q-1}
\end{align*}
where the second row holds by Pascal's rule, the third row is by changing the second summation's base, the fourth row is by simplification, and the fifth row follows from applying Equation 1.
\end{paragraph}

\begin{paragraph}{Equation 3.}

\begin{align*}
    \sum_{\beta=0}^{q-1} \beta \binom{2q}{\beta} \frac{1}{2^{2q}} 
    & = \frac{1}{2^{2q}}\sum_{\beta=1}^{q-1} (2q) \binom{2q-1}{\beta-1}  \\
    & = \frac{q}{2^{2q-1}} \sum_{\beta=0}^{q-2} \binom{2q-1}{\beta} \\
    & = \frac{q}{2^{2q-1}} \left( \sum_{\beta=0}^{q-1} \binom{2q-1}{\beta} - \binom{2q-1}{q-1} \right) \\
    & = \frac{q}{2} - \frac{2q}{2^{2q}} \binom{2q-1}{q-1}
\end{align*}
where the second row holds since $b\binom{a}{b} = a \binom{a-1}{b-1}$ for any $a,b \in \mathbb{Z}_{\geq 0}$ and $0< b\leq a$, the third row is by changing the summation's base, the fourth row is by simplification, and the fifth row is by applying Equation 1.
\end{paragraph}

This concludes the proof of Lemma \ref{lem:suppl_comb}.
\end{proof}


\setcounter{lem}{1}
\begin{lem}
For $q \geq 1$, the following identities hold:
\begin{enumerate}
    \item
    \begin{equation*}
        \sum_{\beta=q+1}^{2q} \binom{2q}{\beta} \frac{1}{2^{2q}} = \frac{1}{2} - \frac{1}{2^{2q}} \binom{2q-1}{q-1};
    \end{equation*}
    \item
    \begin{equation*}
        \sum_{\beta=q+1}^{2q} \beta \binom{2q}{\beta} \frac{1}{2^{2q}} = \frac{q}{2};
    \end{equation*}
    \item
     \[
        \sum_{\beta=0}^{q} \binom{2q+1}{\beta}\frac{1}{2^{2q+1}} = \frac{1}{2};
    \]
    \item
    \begin{align*}
        & \sum_{\beta=0}^{q} \beta \binom{2q+1}{\beta} \frac{1}{2^{2q+1}} 
        = \left( \frac{2q+1}{4} \right) - \frac{2q+1}{2^{2q+1}} \binom{2q-1}{q-1};
    \end{align*}
    \item
     \[
        \sum_{\beta=q+1}^{2q+1} \binom{2q+1}{\beta}\frac{1}{2^{2q+1}} = \frac{1}{2};
    \]
    \item
    \begin{align*}
        & \sum_{\beta=q+1}^{2q+1} \beta \binom{2q+1}{\beta} \frac{1}{2^{2q+1}}
        = \left( \frac{2q+1}{4} \right) + \frac{2q+1}{2^{2q+1}} \binom{2q-1}{q-1};
    \end{align*}
    \item
     \[
        \sum_{\beta=0}^{q-1} \binom{2q}{\beta}\frac{1}{2^{2q}} = 
        \frac{1}{2} - \frac{1}{2^{2q+1}} \binom{2q}{q};
    \]
    \item
    \begin{align*}
        & \sum_{\beta=0}^{q-1} \beta \binom{2q}{\beta} \frac{1}{2^{2q}} = \frac{q}{2} - \frac{q}{2^{2q}} \binom{2q}{q}.
    \end{align*}
\end{enumerate}
\label{lem:new_comb}
\end{lem}

\begin{proof}

We take these equations one at a time.

\begin{paragraph}{Equation 1.}
\begin{align*}
    & \sum_{\beta=q+1}^{2q} \binom{2q}{\beta} \frac{1}{2^{2q}} \\
    & = \sum_{\beta=0}^{2q} \binom{2q}{\beta} \frac{1}{2^{2q}} - \sum_{\beta=0}^{q-1} \binom{2q}{\beta} \frac{1}{2^{2q}} - \frac{1}{2^{2q}} \binom{2q}{q} \\
    & = 1 - \left[ \frac{1}{2} - \frac{1}{2^{2q}} \binom{2q-1}{q-1} \right] - \frac{1}{2^{2q}} \binom{2q}{q} \\
    & = \frac{1}{2} + \frac{1}{2^{2q}} \left[ \binom{2q-1}{q-1} - \binom{2q}{q} \right] \\
    & = \frac{1}{2} - \frac{1}{2^{2q}} \binom{2q-1}{q-1}
\end{align*}
where the third row is by (Lemma \ref{lem:suppl_comb}, Equation 2) and the last row follows from Pascal's rule.
\end{paragraph}

\begin{paragraph}{Equation 2.}
\begin{align*}
    & \sum_{\beta=q+1}^{2q} \beta \binom{2q}{\beta} \frac{1}{2^{2q}} \\
    & = \sum_{\beta=0}^{2q} \beta \binom{2q}{\beta} \frac{1}{2^{2q}} - \sum_{\beta=0}^{q-1} \beta \binom{2q}{\beta} \frac{1}{2^{2q}} - \frac{q}{2^{2q}} \binom{2q}{q} \\
    & = q - \left[ \frac{q}{2} - \frac{2q}{2^{2q}} \binom{2q-1}{q-1} \right] - \frac{q}{2^{2q}} \binom{2q}{q} \\
    & = \frac{q}{2} + \frac{q}{2^{2q}} \left[ 2 \binom{2q-1}{q-1} - \binom{2q}{q} \right] \\
    & = \frac{q}{2}
\end{align*}
where the third row is by (Lemma \ref{lem:suppl_comb}, Equation 3) and the last row follows from Pascal's rule. 
\end{paragraph}

\begin{paragraph}{Equation 3.}
By symmetry, we have that 
\begin{equation*}
    \sum_{\beta=0}^{q} \binom{2q+1}{\beta} \frac{1}{2^{2q+1}} = \frac{2^{2q}}{2^{2q+1}} = \frac{1}{2}.
\end{equation*}
\end{paragraph}

\begin{paragraph}{Equation 4.}
We recall \citet[Claim 1]{Kavner21:strategic}:
\setcounter{claim}{0}
\begin{claim}\label{claim:comb_v2} For any $u\in\mathbb N$ and any $t \in [0,u]$, we have
\begin{equation*}
    \sum_{v=t}^{u} \binom{u}{v}(u-2v) = -t \binom{u}{t}.
\end{equation*}
\end{claim}

As a result,
\begin{align*}
    & \sum_{v=0}^{t} \binom{u}{v}(u-2v) \\
    & = \sum_{v=0}^{u} \binom{u}{v}(u-2v) - \sum_{v=0}^{t-1} \binom{u}{v}(u-2v) \\
    & = (t+1) \binom{u}{(t+1)}
\end{align*}
which implies
\begin{align*}
    & u \sum_{v=0}^{t} \binom{u}{v} - 2 \sum_{v=0}^{t} v\binom{u}{v} = (t+1) \binom{u}{(t+1)} \\
    & \Rightarrow \sum_{v=0}^t v\binom{u}{v} = \frac{u}{2} \sum_{v=0}^{t} \binom{u}{v} - \frac{(t+1)}{2} \binom{u}{(t+1)}
\end{align*}

Substituting $u \leftarrow (2q+1)$ and $t \leftarrow (q)$ into Equation 4, we get
\begin{align*}
    & \sum_{\beta=0}^{q} \beta \binom{2q+1}{\beta} \frac{1}{2^{2q+1}} \\
    & = \frac{1}{2^{2q+1}} \left( \frac{(2q+1)}{2} \sum_{\beta=0}^{q} \binom{2q+1}{\beta} - \frac{q+1}{2} \binom{2q+1}{q+1} \right) \\
    & = \frac{(2q+1)}{2^{2q+2}} \sum_{\beta=0}^{q} \binom{2q+1}{\beta} - \frac{q+1}{2^{2q+2}} \binom{2q+1}{q+1} \\
    & = \left( \frac{2q+1}{4} \right) - \frac{2q+1}{2^{2q+1}} \binom{2q-1}{q-1} 
\end{align*}
where the second row comes from applying Claim \ref{claim:comb_v2}, the third row is by simplification, and the fourth row is by applying Equation 3 and simplification of the binomial.
\end{paragraph}

\begin{paragraph}{Equation 5.}
Proof by symmetry.
\end{paragraph}

\begin{paragraph}{Equation 6.}
Recall from Equation 4 that
\begin{align*}
    & \sum_{\beta=0}^{q} \beta \binom{2q+1}{\beta} \frac{1}{2^{2q+1}}  = \left( \frac{2q+1}{4} \right) - \frac{2q+1}{2^{2q+1}} \binom{2q-1}{q-1}.
\end{align*}
The equation follows since
\[
    \sum_{\beta=0}^{2q+1} \beta \binom{2q+1}{\beta} \frac{1}{2^{2q+1}} = \left( \frac{2q+1}{2} \right)
\]
by definition of the expectation of a binomial random variable.
\end{paragraph}

\begin{paragraph}{Equation 7.}
Recall from Equation 1 that
\[
    \sum_{\beta=q+1}^{2q} \binom{2q}{\beta} \frac{1}{2^{2q}} = \frac{1}{2} - \frac{1}{2^{2q}} \binom{2q-1}{q-1}.
\]
The equation follows by recognizing that
\[
    \sum_{\beta=0}^{2q} \binom{2q}{\beta} \frac{1}{2^{2q}} = 1.
\]

\end{paragraph}

\begin{paragraph}{Equation 8.}
Recall from Equation 2 that
\[
    \sum_{\beta=q+1}^{2q} \beta \binom{2q}{\beta} \frac{1}{2^{2q}} = \frac{q}{2}.
\]
The equation follows by recognizing that
\[
    \sum_{\beta=0}^{2q} \beta \binom{2q}{\beta} \frac{1}{2^{2q}} = q.
\]

\end{paragraph}

This concludes the proof of Lemma \ref{lem:new_comb}.
\end{proof}

\setcounter{prop}{2}
\begin{prop}
Let $\pi_1 \in (0, \frac{1}{2})$, $\pi_3 = \frac{1}{2} - \pi_1$ and $q \in \left[1, \frac{n}{6}-1 \right]$. Then
    \begin{align*}
    & \binom{n}{\frac{n}{2}-q, \frac{n}{2}-q, q, q} \pi_1^{n-2q} \pi_3^{2q} 
    = \frac{\binom{n}{\frac{n}{2}}}{2^n} \left( \binom{\frac{n}{2}}{q} (2\pi_1)^{\frac{n}{2}-q} (2\pi_3)^q \right)^2.
    \end{align*}
    \label{prop:binom}
\end{prop}

\begin{proof}

First, we have:
\begin{align*}
    \binom{n}{\frac{n}{2}-q, \frac{n}{2}-q, q, q}  & = \frac{n!}{(\frac{n}{2}-q)!^2 q!^2} \\
    & = \frac{n!}{(\frac{n}{2})!^2} \frac{(\frac{n}{2})!^2}{(\frac{n}{2}-q)!^2 q!^2} \\
    & = \binom{n}{\frac{n}{2}} \binom{\frac{n}{2}}{q}^2.
\end{align*}
Second, we note:
\[
\pi_1^{n-2q} \pi_3^{2q} = \frac{1}{2^n} \left( (2\pi_1)^{\frac{n}{2}-q} (2 \pi_3)^{q} \right)^2.
\]
Proposition \ref{prop:binom} follows by combining these identities.
\end{proof}

\setcounter{prop}{3}
\begin{prop}[Hoeffding's Inequality]
Let $p \in (0,1)$ and $a, b \in \mathbb{R}$ such that $0 \leq a < b \leq 1$. If $p \notin [a,b]$ then 
    \[
    \sum_{k=\bfloor{a n}}^{\bceil{b n}} \left( \binom{n}{k} p^{n-k} (1-p)^k \right)^2 = \calO \left( e^{-\Theta(n)} \right).
    \]
    \label{prop:hoeffding}
\end{prop}

\begin{proof}
Consider first the case where $p<a$. Then
\begin{align*}
    0 \leq & \sum_{k=\bfloor{an}}^{\bceil{bn}} \left( \binom{n}{k} p^{n-k} (1-p)^k \right)^2 \\
    & \leq \sum_{k=\bfloor{an}}^{\bceil{bn}} \left( \binom{n}{k} p^{n-k} (1-p)^k \right) \\
    & = \Pr \Big( S_n - pn \geq \bfloor{an} - pn \Big) 
    - \Pr \Big( S_n - pn \geq \bceil{bn} - pn+1 \Big) \\
    & \leq \calO \left( e^{-\Theta(n)} \right)
\end{align*}
by Hoeffding's inequality, where $S_n \sim Bin(n, p)$. Proposition \ref{prop:hoeffding} follows because the case where $p>b$ is similar.
\end{proof}

The following lemma applies Theorem \ref{thm:smoothed-likelihood} (\citep[Theorem 1]{Xia2021:How-Likely}) from Appendix \ref{apx:smoothed_prelims} to prove that the likelihood an $(n,4)$-PMV fits into a set describing a two-way tie with an equal number of third-party agents. This additional constraint reduces the likelihood from $\Theta \left( \frac{1}{\sqrt{n}} \right)$, by Corollary \ref{coro:sub_others}, to $\Theta \left( \frac{1}{n} \right)$. This holds as long as $\pi_3 n$ is contained in the summation region; the likelihood is exponentially small otherwise. 
\footnote{
Note that Lemma \ref{lem:prob_bounds} was introduced in the proof of Lemma \ref{lem:sub_12_non_34_equal} with capital variables $\Pi_1$ and $\Pi_3$. For ease of readability and consistency with this appendix and smoothed analysis framework of Appendix \ref{apx:smoothed_prelims}, we will use the lowercase notation $\pi_1$ and $\pi_3$.
}


\setcounter{lem}{11}
\begin{lem}
Fix $a,b \in (0,\frac{1}{6})$, $a<b$. Let $\pi_1 \in [\frac{1}{3}, \frac{1}{2})$ and $\pi_3 = \frac{1}{2} - \pi_1$. 
    Then
\begin{align*}
    & \sum_{q=\bfloor{a n}}^{\frac{n}{6} } \binom{n}{\frac{n}{2}-q, \frac{n}{2}-q, q, q} \pi_1^{n-2q} \pi_3^{2q}
    = \begin{cases}
    \Theta\left( \frac{1}{n} \right), & \pi_3 \geq a \\
    \calO \left( e^{-\Theta(n)} \right), & \text{otherwise}
    \end{cases}
\end{align*}
and
\begin{align*}
    & \sum_{q=0}^{\bceil{b n}} \binom{n}{\frac{n}{2}-q, \frac{n}{2}-q, q, q} \pi_1^{n-2q} \pi_3^{2q}
    = \begin{cases}
    \Theta\left( \frac{1}{n} \right), & \pi_3 \leq b \\
    \calO \left( e^{-\Theta(n)} \right), & \text{otherwise}.
    \end{cases}
\end{align*}
%
%
%
\label{lem:prob_bounds}
\end{lem}

\begin{proof}

To prove the lemma, we may assume without loss of generality that $an, bn \in \mathbb{Z}_{\geq 0}$ are integers. This follows because $|x - \bfloor{x} | \leq 1 =o(n)$ for any $x \in \mathbb{R}$, so \citet{Xia2021:How-Likely}’s theorems are indifferent to the distinction between $x$ and $\bfloor{x}$. The same holds for $x$ and $\bceil{x}$.

Consider $n$ random variables $Q_1, \ldots, Q_n$, such that $Q_i \in [4]$, which are distributed identically and independently according to the distribution $\pi = (\pi_1, \pi_1, \pi_3, \pi_3)$ over the four values.
%
Let $\vec{X}_{\vec{\pi}}$ denote the corresponding $(n,4)$-PMV to $Q_1, \ldots, Q_n$ according to Definition \ref{dfn:pmv}; we have $\mu = 4$. 
Let us define the sets:
\begin{align*}
    & \calT^a = \left\{\left(\frac{n}{2}-q, \frac{n}{2}-q, q, q\right) : q \in \left[a n, \frac{n}{6}-1 \right] \right\} \\
    & \calT^b = \left\{\left(\frac{n}{2}-q, \frac{n}{2}-q, q, q\right) : q \in [1, b n] \right\}.
\end{align*}
Then we have 
\begin{align*}
    & \sum_{q=a n}^{\frac{n}{6}} \binom{n}{\frac{n}{2}-q, \frac{n}{2}-q, q, q} \pi_1^{n-2q} \pi_3^{2q} = \Pr( \vec{X}_{\vec{\pi}} \in \calT^a ) \\
    & \sum_{q=1}^{b n} \binom{n}{\frac{n}{2}-q, \frac{n}{2}-q, q, q} \pi_1^{n-2q} \pi_3^{2q} = \Pr( \vec{X}_{\vec{\pi}} \in \calT^b )
\end{align*}
which are instances of the PMV-in-polyhedron problem.
Specifically, notice that
\begin{equation}
    \calT^a = \left\{ \vec{x} \in \mathbb{R}^4 ~:~ \mathbf{A}^a	\vec{x}	\leq \vec{b}^a \right\}
    \label{eq:polytope_a}
\end{equation}
where
\begin{equation*}
    \mathbf{A}^a = \begin{pmatrix}
    1 & -1 & 0 & 0 \\
    -1 & 1 & 0 & 0 \\
    0 & 0 & 1 & -1 \\
    0 & 0 & -1 & 1 \\
    -\frac{1}{6} & -\frac{1}{6} & \frac{5}{6} & -\frac{1}{6} \\
a & a & -1+a & a 
    \end{pmatrix},
    \quad
    \vec{b}^a = \begin{pmatrix}
    0 \\ 0 \\ 0 \\ 0 \\ 0 \\ 0
    \end{pmatrix} 
\end{equation*}
and
\begin{equation}
    \calT^b = \left\{ \vec{x} \in \mathbb{R}^4 ~:~ \mathbf{A}^b	\vec{x}	\leq \vec{b}^b \right\}
    \label{eq:polytope_b}
\end{equation}
where
\begin{equation*}
    \mathbf{A}^b = \begin{pmatrix}
    1 & -1 & 0 & 0 \\
    -1 & 1 & 0 & 0 \\
    0 & 0 & 1 & -1 \\
    0 & 0 & -1 & 1 \\
    0 & 0 & -1 & 0 \\
    -b & -b & 1-b & -b
    \end{pmatrix},
    \quad
    \vec{b}^b = \begin{pmatrix}
    0 \\ 0 \\ 0 \\ 0 \\ 0 \\ 0
    \end{pmatrix}. 
\end{equation*}

For any $\pi_3 \in (0,\frac{1}{6}]$, we will demonstrate in Step 1 below that $[\calT^a]^{\mathbb{Z}}_n \neq \emptyset$ and $[\calT^a]^{\mathbb{Z}}_n \neq \emptyset$;
hence, the zero case of Theorem \ref{thm:smoothed-likelihood} does not apply. 
Next, in Step 2, we will demonstrate that $\pi \in \calT^a \iff \pi_3 \geq a$ and $\pi \in \calT^b \iff \pi_3 \leq b$; hence, the polynomial and exponential cases of the theorem apply when the respective conditions hold. In Step 3, we will finally demonstrate that $dim([\calT^a]_{\leq 0}) = dim([\calT^b]_{\leq 0}) = 2$, so that the polynomial power is $\frac{2-4}{2} = -1$ for each polyhedron.

\begin{paragraph}{Step 1: Zero case does not apply.}

It is easy to see that $\left( \frac{n}{2}-q, \frac{n}{2}-q, q, q \right) \in \calT^a$ for $q = a n$ and that $\left( \frac{n}{2}-q, \frac{n}{2}-q, q, q \right) \in \calT^b$ for $q = b n$. This holds because $a < \frac{1}{6}$ and $b > 0$ and implies that $[\calT^a]^{\mathbb{Z}}_n \neq \emptyset$ and $[\calT^b]^{\mathbb{Z}}_n \neq \emptyset$. Hence, the zero case of Theorem \ref{thm:smoothed-likelihood} does not apply.


\end{paragraph}

\begin{paragraph}{Step 2: Differentiate polynomial and exponential cases.}

The next condition of Theorem \ref{thm:smoothed-likelihood} is a comparison between $[\calT^a]_{\leq 0}$ or $[\calT^b]_{\leq 0}$ and the convex hull $CH(\Pi)$, where $\Pi = \{\pi^n\}$ is a singleton.
%
Consider the (fractional) vote profile $\pi n$ and the last row of $\mathbf{A}^a$. For $\calT^a$, we have 
\[
    \left( a, a, -1+a, a \right) \cdot \pi n = (-\pi_3 + a) n \leq 0
\]
if and only if $\pi_3 \geq a$.
It is easy to see that $\vec{v} \cdot \pi n \leq 0$ for any other row-vector $\vec{v} \in \mathbf{A}^a$. 
This holds by our assumption on $\pi$ that $\pi_1 \geq 2\pi_3> 0$, which necessitates that $\pi_3 \in (0, \frac{1}{6}]$.

Likewise, in the last row of $\mathbf{A}^b$ for the case of $\calT^b$, we have
\[
    \left( -b, -b, 1-b, -b \right) \cdot \pi n = (\pi_3 - b) n \leq 0
\]
if and only if $\pi_3 \leq b$. Similarly, $\vec{v} \cdot \pi n \leq 0$ for any other row-vector $\vec{v} \in \mathbf{A}^b$.
Therefore, the polynomial cases of Theorem \ref{thm:smoothed-likelihood} apply to $\Pr( \vec{X}_{\vec{\pi}} \in \calT^a)$ and $\Pr( \vec{X}_{\vec{\pi}} \in \calT^b )$ when the lemma's respective conditions hold; otherwise the exponential case applies.

\end{paragraph}
\begin{paragraph}{Step 3: Determine dimension of characteristic cones.}

Following the proof of Theorem 1 in \citet{Xia2021:How-Likely}, we start with the following definition.

\begin{dfn}[Equation (2) on page 99 of \citet{Schrijver1998:Theory}]
For any matrix $\mathbf{A}$ that defines a polyhedron $\calH$, let $\mathbf{A^{=}}$ denote the \emph{implicit equalities}, which is the maximal set of rows of $\mathbf{A}$ such that for all $\vec{x} \in \calH_{\leq 0}$, we have $\mathbf{A^{=}} \cdot (\vec{x})^T = (\vec{0})^T$. Let $\mathbf{A}^{+}$ denote the remaining rows of $\mathbf{A}$.
\end{dfn}

By Equation (9) on page 99 of \citet{Schrijver1998:Theory} we know that  $dim([\calT^a]_{\leq 0}) = \mu - rank([\calA^a]^=)$ and $dim([\calT^b]_{\leq 0}) = \mu - rank([\calA^b]^=)$. From Equations (\ref{eq:polytope_a}) and (\ref{eq:polytope_b}) we can deduce that
\[
\mathbf{[\mathbf{A}^a]^{=}} = \mathbf{[\mathbf{A}^b]^{=}} = \begin{pmatrix}
1 & -1 & 0 & 0 \\
-1 & 1 & 0 & 0 \\
0 & 0 & 1 & -1 \\
0 & 0 & -1 & 1
\end{pmatrix}
\]
which has rank $2$. Hence, the polynomial powers when we apply Theorem \ref{thm:smoothed-likelihood} are 
\[
\frac{(\mu - rank([\calA^a]^=))-\mu}{2} = \frac{(\mu - rank([\calA^b]^=))-\mu}{2} = -1.
\]
\end{paragraph}
This concludes the proof of Lemma \ref{lem:prob_bounds}.
\end{proof}

%% file: EC_appendix/apx_page_8.tex
\newpage

\section{Technical Lemmas}
\label{apx:helper_lemmas}

This appendix proves lemmas that are used throughout this paper. 





\setcounter{lem}{14}
\begin{lem}
\[
    \binom{3n}{n, n, n} \frac{1}{3^{3n}} = \Theta \left( \frac{1}{n} \right).
\]
\label{lem:stirling_trinom}

\end{lem}

\begin{proof}
We prove the lemma using Stirling's approximation (Proposition \ref{prop:stirling}):
\[
\sqrt{2 \pi n} \left( \frac{n}{e} \right)^n e^{\frac{1}{12n+1}} < n! < \sqrt{2 \pi n} \left( \frac{n}{e} \right)^n e^{\frac{1}{12n}}.
\]
To establish the upper bound, we have
\begin{align*}
    \binom{3n}{n, n, n} \frac{1}{3^{3n}}
    & \leq \frac{\sqrt{6 \pi n} \left( \frac{3n}{e} \right)^{3n} e^{\frac{1}{36n}} }{ \left( 3^n \sqrt{2 \pi n} \left( \frac{n}{e} \right)^{n} e^{\frac{1}{36n+1}} \right)^3} \\
    & = \frac{e^{\left( \frac{1}{36n} - \frac{1}{3(36n+1)} \right)} }{\pi n \sqrt{0.75}} \\
    & = \Theta \left( \frac{1}{n} \right) \left(1 + \calO \left( \frac{1}{n} \right) \right)
\end{align*}
by the Maclauren series of the exponential. To establish the lower bound, we have
\begin{align*}
    \binom{3n}{n, n, n} \frac{1}{3^{3n}}
    & \geq \frac{\sqrt{6 \pi n} \left( \frac{3n}{e} \right)^{3n} e^{\frac{1}{36n+1}} }{ \left( 3^n \sqrt{2 \pi n} \left( \frac{n}{e} \right)^{n} e^{\frac{1}{36n}} \right)^3} \\
    & = \frac{e^{\left( \frac{1}{36n+1} - \frac{1}{3(36n)} \right)} }{\pi n \sqrt{0.75}} \\
    & = \Theta \left( \frac{1}{n} \right) \left(1 + \calO \left( \frac{1}{n} \right) \right)
\end{align*}
by the Maclauren series of the exponential. This proves Lemma \ref{lem:stirling_trinom} by the squeeze theorem.
\end{proof}

\setcounter{lem}{22}
\begin{lem}
Let $p \in (0,1)$ and $k \in \left[\bfloor{\frac{np}{2}}, n \right]$. Then 
    \[
    \left| \frac{2^{2k} \sqrt{np(1-p)}}{(2k+1) \binom{2k}{k}} \right| = \Theta(1).
    \]
    \label{lem:mag_bound}
\end{lem}
\begin{proof}
By Lemma \ref{lem:wallis} we get that
\begin{align*}
    \sqrt{\frac{2}{3}} \sqrt{\frac{2}{\pi (2k+1)}} 
    & \leq \sqrt{\frac{2k}{2k+1}}  \sqrt{\frac{2}{\pi (2k+1)}} \\
    & \leq \frac{2^{2k}}{(2k+1) \binom{2k}{k}} \\
    & \leq \sqrt{\frac{2}{\pi (2k+1)}}.
\end{align*}
Lemma \ref{lem:mag_bound} follows since $k = \Theta(n)$ by assumption and there is an extra $\Theta(\sqrt{n})$ term in the lemma's objective.
\end{proof}

\setcounter{lem}{24}
\begin{lem}
    For any constant $t > 0$, $\sqrt{\frac{tn}{tn+1}} = 1 - \calO \left( \frac{1}{n} \right)$.
    \label{lem:analysis_helper}
\end{lem}

\begin{proof}
The lemma is implied by the following:
    \[
    \lim\nolimits_{n \rightarrow \infty} n \left( \sqrt{\frac{tn}{tn+1}} - 1 \right) = -\frac{1}{2t}.
    \]

Fix $\epsilon > 0$ and define $N = \frac{1}{2 t^2 \epsilon}$. Then $\forall n > N$,
\begin{align*}
    & \left| n \left( \sqrt{\frac{tn}{tn+1}} -1 \right) + \frac{1}{2t} \right| \\
    & = \left| n \left( \frac{\sqrt{tn} - \sqrt{tn+1}}{\sqrt{tn+1}} \right) \left( \frac{\sqrt{tn} + \sqrt{tn+1}}{\sqrt{tn} + \sqrt{tn+1}} \right) + \frac{1}{2t} \right| \\
    & = \left| \frac{-n}{\sqrt{tn+1} \left( \sqrt{tn} + \sqrt{tn+1} \right) } + \frac{1}{2t} \right| \\
    & = \frac{1}{2t} \left| \frac{-tn + 1 + \sqrt{(tn)(tn+1)}}{ tn+1 + \sqrt{(tn)(tn+1)}} \right| \\
    & \leq \frac{1}{2t} \left| \frac{-tn + 1 + tn+1}{2(tn)} \right| \\
    & = \frac{1}{2 t^2 n} \\
    & < \frac{1}{2 t^2 N} \\
    & = \epsilon.
\end{align*}
Lemma \ref{lem:analysis_helper} follows by definition of the limit.
\end{proof}

The following lemma is adapted from the proof of the local DeMoivre-Laplace theorem, demonstrated in lecture notes by \citet{carlen18demoivre}.

\setcounter{lem}{25}
\begin{lem}
Let $p \in (0,1)$ and $q = 1-p$. Then we have
    \[
    \binom{n}{\bfloor{np}} p^{\bfloor{np}} q^{n-\bfloor{np}} = 
    \frac{1}{\sqrt{2 \pi n p q}} \left( 1 \pm \calO \left( \frac{1}{n} \right) \right).
    \]
    \label{lem:stirling_binom}
\end{lem}

\begin{proof}
A more precise version of Stirling's formula for all $n \geq 1$ is
\[
\sqrt{2 \pi n} \left( \frac{n}{e} \right)^n e^{\frac{1}{12n+1}} \leq n! \leq \sqrt{2 \pi n} \left( \frac{n}{e} \right)^n e^{\frac{1}{12n}}.
\]
Taking logarithms, it follows that
\[
\left| \log n! - \frac{1}{2} \log(2 \pi n) - n \log n + n \right| \leq \frac{1}{12n}.
\]
For $n \in \mathbb{N}$ and $k \in [0,n]$ an integer, we compute
\begin{align*}
    \log \binom{n}{k}
    & = \log n! - \log k! - \log(n-k)! \\
    & \approx -\frac{1}{2} \log(2 \pi) + \left( n+\frac{1}{2} \right) \log n   - \left( k+\frac{1}{2} \right) \log k 
    - \left( n-k+\frac{1}{2} \right) \log(n-k) \\
    & = \frac{1}{2} \log \left(\frac{1}{2\pi n}\right)   - \left( k+\frac{1}{2} \right) \log \left(\frac{k}{n}\right) 
    - \left( n-k+\frac{1}{2} \right) \log \left(\frac{n-k}{n}\right)
\end{align*}
where we have used
\[
\left(n + \frac{1}{2} \right) = \left(k + \frac{1}{2} \right) + \left(n-k + \frac{1}{2} \right) - \frac{1}{2}
\]
to obtain the last line. Therefore
\begin{align}
    \log \left( \binom{n}{k} p^k q^{n-k} \right) 
    & \approx -\frac{1}{2} \log \left( 2\pi n p q \right) - \left( k+\frac{1}{2} \right) \log \left(\frac{k}{n p}\right)   - \left( n-k+\frac{1}{2} \right) \log \left(\frac{n-k}{n q}\right). \label{eq:log_approx}
\end{align}
Note that the error made in Equation (\ref{eq:log_approx}) is no greater than
\begin{equation*}
    \frac{1}{12} \left( \frac{1}{n} + \frac{1}{k} + \frac{1}{n-k} \right) = \calO \left( \frac{1}{n} \right) 
\end{equation*}
in magnitude. We do not assume that $np$ is an integer. Rather, suppose $np = t_n + b_n$ where $t_n \in \mathbb{N}$ and $b_n \in (0,1)$. Plugging in $k = \bfloor{np} = t_n$ into Equation (\ref{eq:log_approx}) yields
\begin{align}
    & -\frac{1}{2} \log \left( 2\pi n p q \right) + \left( \bfloor{np}+\frac{1}{2} \right) \log \left(\frac{np}{\bfloor{np}}\right)  
    + \left( n-\bfloor{np}+\frac{1}{2} \right) \log \left(\frac{nq}{n - \bfloor{np}}\right) \notag \\
    & = -\frac{1}{2} \log \left( 2\pi n p q \right) + \left( t_n+\frac{1}{2} \right) \log \left( 1 + \frac{b_n}{t_n} \right) 
    + \left( n-t_n+\frac{1}{2} \right) \log \left( 1 - \frac{b_n}{n - t_n}\right). \label{eq:log_too_much_1}
\end{align}
We apply the Taylor expansion for the natural logarithm, which is
\[
    \log(1+t) = t - \frac{1}{2} t^2 + \frac{1}{3} t^3 -  \calO(t^4)
\]
and converges for $|t|<1$. This is an alternating sequence, meaning that
\[
    \left| \log(1+t) - t + \frac{1}{2} t^2 \right| \leq \frac{1}{3} |t|^3.
\]
Hence, for $t = \pm \calO \left(\frac{1}{n} \right)$ from Equation (\ref{eq:log_too_much_1}), the error in approximating the logarithm is $\calO \left(\frac{1}{n^3} \right)$. Through this approximation, we get
\begin{align}
    & -\frac{1}{2} \log \left( 2\pi n p q \right) + \left( t_n+\frac{1}{2} \right) \left( \frac{b_n}{t_n} - \frac{b_n^2}{2 t_n^2} \right) 
    + \left( n-t_n+\frac{1}{2} \right) \left( - \frac{b_n}{n - t_n} + \frac{b_n^2}{2(n - t_n)^2}\right) \notag \\
    & = -\frac{1}{2} \log \left( 2\pi n p q \right) \pm \calO \left( \frac{1}{n} \right). \label{eq:log_too_much_2}
\end{align}
Lemma \ref{lem:stirling_binom}'s statement follows by noticing that $e^{\pm \calO \left( \frac{1}{n} \right) } = \left( 1 \pm \calO \left( \frac{1}{n} \right) \right) $ by the Maclaurin series of the exponential.
\end{proof}